\documentclass[preprint2]{emulateapj}

\usepackage{natbib}
\usepackage{longtable}
\usepackage[]{graphicx}
\usepackage{amsmath}
\usepackage{natbib}
\usepackage{tabularx}
\usepackage{bm}
\usepackage{hyperref}

\newcommand{\fitfigurewidth}{0.8\textwidth}

\shorttitle{Galaxy vivisection}
\shortauthors{Savorgnan \& Graham}

\begin{document}

\title{Supermassive black holes and their host spheroids \\ I. Disassembling galaxies}

\author{Giulia A.~D.~Savorgnan and Alister W.~Graham}
\affil{Centre for Astrophysics and Supercomputing, Swinburne University of Technology, Hawthorn, Victoria 3122, Australia}
\email{gsavorgn@astro.swin.edu.au}

\begin{abstract}
Several recent studies have performed galaxy decompositions 
to investigate correlations between the black hole mass and various properties of the host spheroid, 
but they have not converged on the same conclusions. 
This is because their models for the same galaxy were often significantly different 
and not consistent with each other in terms of fitted components. 
Using $3.6 \rm ~\mu m$ \emph{Spitzer} imagery, which is a superb tracer of the stellar mass (superior to the $K$-band), 
we have performed state-of-the-art multicomponent decompositions for 66 galaxies 
with directly measured black hole masses. 
Our sample is the largest to date and, 
unlike previous studies, contains a large number (17) of spiral galaxies with low black hole masses. 
We paid careful attention to the image mosaicking, sky subtraction and masking of contaminating sources.
After a scrupulous inspection of the galaxy photometry (through isophotal analysis and unsharp masking) and -- for the first time -- 2D kinematics, 
we were able to account for spheroids, large-scale, intermediate-scale and nuclear disks, bars, rings, spiral arms, halos, extended or unresolved nuclear 
sources and partially depleted cores. 
For each individual galaxy, we compared our best-fit model with previous studies, 
explained the discrepancies and identified the optimal decomposition.
Moreover, we have independently performed 1D and 2D decompositions, 
and concluded that, at least when modelling large, nearby galaxies, 1D techniques have more advantages than 2D techniques. 
Finally, we developed a prescription to estimate the uncertainties on the 1D best-fit parameters for the 66 spheroids 
that takes into account systematic errors, 
unlike popular 2D codes that only consider statistical errors.

\end{abstract}


\section{Introduction}
\label{sec:int}
Supermassive black holes and their host spheroids\footnote{By the term ``spheroid'' 
we mean either an elliptical galaxy or the bulge component of a disk galaxy, 
with no attempt at distinguishing between classical bulges and disk-like pseudo-bulges. } 
have very different sizes. 
If the event horizon of the Galactic supermassive black hole was as big as a grain of sand in the Sahara desert, 
then the black hole's gravitational sphere-of-influence would be as big as the international airport of Cairo, 
and the Galactic bulge would be as big as the Sahara desert itself.
It is thus surprising that the masses of supermassive black holes ($M_{\rm BH}$) 
scale with a number of properties of their host spheroid, 
indicating the non-gravitational origin of these correlations 
(e.g.~\citealt{dressler1989,yee1992,kormendyrichstone1995,laor1997,magorrian1998,ferraresemerritt2000,gebhardt2000,graham2001,
marconihunt2003,haringrix2004,grahamscott2015}). \\
The tightness of (i.e.~the small scatter about) the above observed black hole mass correlations has led to the idea
that black holes and host spheroids have coevolved with some sort of self-regulated growth.
Exploring the evolution of this growth with cosmic time could help identify
the driving mechanisms of the black hole -- spheroid coevolution.
Observations at $z=0$ set the local benchmark from which to measure this evolution.
Any complete theory or model describing the coevolution of spheroids and black holes must incorporate 
all of the observed scaling relations, which also have to be consistent with each other.
Modern hydrodynamical simulations, such as EAGLE \citep{schaye2015}, calibrate the feedback efficiencies 
to match the $z=0$ black hole mass -- galaxy mass relation.
The observed scaling relations can also be employed to predict the masses of black holes in other galaxies, where a direct
measure of $M_{\rm BH}$ would be extremely time consuming or simply impossible due to technological limitations.
Moreover, many accurate $M_{\rm BH}$ predictions enable one to derive 
the local black hole mass function (e.g.~\citealt{salucci1999,graham2007smbhmassfunction,shankar2009,shankar2013,fontanot2015}) 
and space density (e.g.~\citealt{grahamdriver2007smbhmassdensity,comastri2015}). 
All of these examples depend on the $z=0$ relations, 
and as such the recalibration of the black hole mass -- spheroid stellar mass ratio ($M_{\rm BH}/M_{\rm *,sph}$) in large spheroids,
from $0.1 - 0.2\%$ (e.g.~\citealt{marconihunt2003,haringrix2004}) to $0.49\%$ \citep{grahamscott2015}, 
has many substantial implications. \\
Since the early stellar dynamical detections of black holes were carried out in the '80s 
(see the references in the reviews by \citealt{kormendyrichstone1995}, \citealt{richstone1998} 
and \citealt{graham2015bulges} for pioneering papers), 
the number of (direct) black hole mass measurements has increased with time 
and it has recently become a statistically meaningful sample  
with which one can study SMBH demographics. 
It is now generally accepted that supermassive black holes reside at the center of most, if not all, 
massive spheroids, either quiescent or active. \\
Massive, early-type (E, E/S0, S0) galaxies are often composite systems. 
The knowledge that many ``elliptical'' (E) galaxies were misclassified and actually contain embedded stellar disks dates back  
at least three decades  
\citep{capaccioli1987,carter1987,rixwhite1990,rixwhite1992,bender1990,scorzabender1990,nieto1991,scorzabender1995}. 
After examining long-slit and integral field unit spectroscopic observations of morphologically classified elliptical galaxies in the Fornax cluster, 
\cite{donofrio1995}, \cite{graham1998fornax} and \cite{scott2014} concluded that only 3 bright galaxies do not harbour a disk-like component.
Larger surveys with integral field spectrographs of early-type galaxies, such as the ATLAS$^{\rm 3D}$ Project \citep{cappellari2011}, 
have further advanced this view of ``elliptical'' galaxies being all but simple and featureless pressure-supported systems.
Most of the morphologically classified elliptical galaxies in the ATLAS$^{\rm 3D}$ sample are fast rotators \citep{atlas3dIII}.
\cite{krajnovic2013} showed that ``fast rotators'' as a class are disk-galaxies or at least disk-like galaxies. 
In their magnitude-limited survey, systems 
without any signature of disk-like components (neither in the kinematics nor in the photometry) 
dominate only the most massive end (with stellar masses beyond $10^{11.5}~\rm M_\odot$) of the distribution. 
Given the prevalence of disks, it is clearly important to perform spheroid/disk decompositions, 
if one is to properly explore the black hole -- spheroid connection. 
Indeed, separating the disk light from that of the bulge has led to the discovery of the missing population of compact, massive spheroids 
in the local Universe \citep{GDS2015}.
If we are to properly understand the evolution of galaxies, we need to understand their components.\\
Measuring the photometric and structural properties of a galaxy's spheroidal component 
requires the ability to separate it from the rest of the galaxy. 
Such galaxy decomposition involves a parametric analysis that allows one to fit the surface brightness distribution
of galaxies using a combination of analytic functions (usually one function per
galaxy component, such as spheroids, disks, bars, nuclei, etc.).
The one-dimensional (1D) technique involves fitting isophotes to the galaxy image, extracting the (one-dimensional) surface 
brightness radial profile and modeling it with a combination of analytic functions.
With the two-dimensional (2D) technique, one fits analytic functions directly to the 2D images. \\
Over the past eight years, five independent studies 
\citep{grahamdriver2007,sani2011,vika2012,beifiori2012,lasker2014data,lasker2014anal} have attempted galaxy decomposition
in order to derive the spheroid parameters and explore their relation with the black hole masses. 
Interestingly, the past studies used almost the same sample of galaxies, yet they claimed some contradictory conclusions.
For example, one study \citep{grahamdriver2007} obtained a good $M_{\rm BH}-n_{\rm sph}$ correlation 
(the spheroid S\'ersic index $n_{\rm sph}$ is a measure of the central radial concentration of stars, \citealt{trujillo2001}), 
whereas the remaining four did not\footnote{\cite{savorgnan2013} showed that, 
by rejecting the most discrepant S\'ersic index measurements and averaging the remaning ones, 
a strong $M_{\rm BH}-n_{\rm sph}$ correlation was recovered. }.
In addition, \cite{lasker2014anal} 
declared that $M_{\rm BH}$ correlates equally well with the total galaxy luminosity as it does with the spheroid luminosity, 
as opposed to \cite{beifiori2012} who claimed that the spheroid mass is a better tracer of $M_{\rm BH}$ than the galaxy mass 
(see also \citealt{kormendygebhardt2001,erwingadotti2012}).
The past studies did not converge to the same conclusions 
because their best-fit models for the same galaxy were often 
significantly different and not consistent with each other in terms of fitted components. 
Moreover, none of these studies attempted an individual galaxy-by-galaxy 
comparison of their models with the previous literature. 
We have now made this comparison and performed the optimal decompositions,
using $3.6~\rm \mu m$ \emph{Spitzer} satellite imagery, 
which is an excellent proxy for the stellar mass, superior to the $K$-band \citep{junim2008,sheth2010}.
We will use these in a series of forthcoming papers to obtain improved black hole mass scaling relations 
using the largest sample (66) of galaxies to date with accurate spheroid properties. \\

This paper is structured as follows.
Section \ref{sec:data} presents the galaxy sample and imaging data-set used to conduct this study. 
Section \ref{sec:anal} describes how we performed the galaxy decompositions, 
i.e.~how we identified and modeled the sub-components that constitute our galaxies. 
In Section \ref{sec:res} we outline the results from our analysis and discuss the error analysis. 
Section \ref{sec:concl} summarizes our main conclusions. 
The individual galaxy decompositions are made available in the electronic version of this manuscript.

\section{Data}
\label{sec:data}
Our initial galaxy sample (see Table \ref{tab:sample}) consists of 75 objects for which a dynamical detection of the black hole mass 
had been reported in the literature at the time we started this project, 
and for which at least one $3.6~\rm \mu m$ \emph{Spitzer}/IRAC\footnote{IRAC is the InfraRed Array Camera onboard the \emph{Spitzer} Space Telescope.} 
observation was publicy available.
Black hole masses were drawn from the catalog of \citet{grahamscott2013} for 70 galaxies,
from \citet{rusli2013bhmassesDM} for 4 galaxies and from \citet{greenhill2003} for 1 galaxy. 
As explained in Section \ref{sec:res}, 
this initial sample was ultimately reduced to 66 galaxies for which useful spheroid parameters could be obtained.

\begin{table*}                                        
\begin{center}                                        
\caption{Galaxy sample.}                                 
\begin{tabular}{llllllllll}                           
\hline                                                
\multicolumn{1}{l}{{\bf Galaxy}} &                   
\multicolumn{1}{l}{{\bf Distance}} &                 
\multicolumn{1}{l}{{\bf $\bm{M_{\rm BH}}$}} &  
\multicolumn{1}{l}{{\bf Ref.}} &                     
\multicolumn{1}{l}{{\bf Core}} &                     
\multicolumn{1}{l}{{\bf Ref.}} &                     
\multicolumn{1}{l}{{\bf Rot.}} &                     
\multicolumn{1}{l}{{\bf Vel. map}} &                 
\multicolumn{1}{l}{{\bf 1D fit}} &                   
\multicolumn{1}{l}{{\bf 2D fit}} \\                
\multicolumn{1}{l}{} &                                
\multicolumn{1}{l}{[Mpc]} &                           
\multicolumn{1}{l}{$[10^8~\rm M_{\odot}]$} &         
\multicolumn{1}{l}{} &                                
\multicolumn{1}{l}{$([\rm arcsec])$} &                                
\multicolumn{1}{l}{} &                                
\multicolumn{1}{l}{} &                                
\multicolumn{1}{l}{} &                                
\multicolumn{1}{l}{} &                                
\multicolumn{1}{l}{} \\                             
\multicolumn{1}{l}{(1)} &                             
\multicolumn{1}{l}{(2)} &                             
\multicolumn{1}{l}{(3)} &                             
\multicolumn{1}{l}{(4)} &                             
\multicolumn{1}{l}{(5)} &                             
\multicolumn{1}{l}{(6)} &                             
\multicolumn{1}{l}{(7)} &                             
\multicolumn{1}{l}{(8)} &                             
\multicolumn{1}{l}{(9)} &                             
\multicolumn{1}{l}{(10)} \\                         
\hline                                                
Circinus   &  $4.0$  &  $0.017_{-0.003}^{+0.004}$   &  G+03  &  no?  &     &      &     &  no  &  no  \\ 
IC 1459  &  $28.4$  &  $24_{-10}^{+10}$   &  GS13  &  yes  $(0.7)$  &  R+13a  &      &     &  yes  &  yes  \\ 
IC 2560  &  $40.7$  &  $0.044_{-0.022}^{+0.044}$   &  GS13  &  no?  &     &      &     &  yes  &  no  \\ 
IC 4296  &  $40.7$  &  $11_{-2}^{+2}$   &  GS13  &  yes?  &     &      &     &  yes  &  yes  \\ 
M31  &  $0.7$  &  $1.4_{-0.3}^{+0.9}$   &  GS13  &  no   &     &      &     &  yes  &  no  \\ 
M32  &  $0.8$  &  $0.024_{-0.005}^{+0.005}$   &  GS13  &  no   &     &      &     &  no  &  no  \\ 
M49  &  $17.1$  &  $25_{-1}^{+3}$   &  R+13b  &  yes  $(1.5)$  &  DG13, R+13a  &   SLOW  &  A  &  yes  &  yes  \\ 
M59  &  $17.8$  &  $3.9_{-0.4}^{+0.4}$   &  GS13  &  no   &     &  FAST   &  A  &  yes  &  no  \\ 
M60  &  $16.4$  &  $47_{-10}^{+10}$   &  GS13  &  yes  $(2.7)$  &  DG13, R+13a  &  FAST   &  A, S  &  no  &  no  \\ 
M64  &  $7.3$  &  $0.016_{-0.004}^{+0.004}$   &  GS13  &  no?  &     &      &     &  yes  &  no  \\ 
M77  &  $15.2$  &  $0.084_{-0.003}^{+0.003}$   &  GS13  &  no   &     &      &     &  no  &  no  \\ 
M81  &  $3.8$  &  $0.74_{-0.11}^{+0.21}$   &  GS13  &  no   &     &      &     &  yes  &  no  \\ 
M84  &  $17.9$  &  $9.0_{-0.8}^{+0.9}$   &  GS13  &  yes  $(1.9)$  &  F+06  &   SLOW  &  A, S  &  yes  &  yes  \\ 
M87  &  $15.6$  &  $58.0_{-3.5}^{+3.5}$   &  GS13  &  yes  $(7.2)$  &  F+06  &   SLOW  &  A, S  &  yes  &  yes  \\ 
M89  &  $14.9$  &  $4.7_{-0.5}^{+0.5}$   &  GS13  &  yes  $(0.4)$  &  DG13, R+13a  &   SLOW  &  A  &  yes  &  no  \\ 
M94  &  $4.4$  &  $0.060_{-0.014}^{+0.014}$   &  GS13  &  no?  &     &      &     &  yes  &  no  \\ 
M96  &  $10.1$  &  $0.073_{-0.015}^{+0.015}$   &  GS13  &  no   &     &      &     &  yes  &  yes  \\ 
M104  &  $9.5$  &  $6.4_{-0.4}^{+0.4}$   &  GS13  &  yes   &  J+11  &      &     &  yes  &  no  \\ 
M105  &  $10.3$  &  $4_{-1}^{+1}$   &  GS13  &  yes  $(1.1)$  &  DG13, R+13a  &  FAST   &  A  &  yes  &  yes  \\ 
M106  &  $7.2$  &  $0.39_{-0.01}^{+0.01}$   &  GS13  &  no   &     &      &     &  yes  &  no  \\ 
NGC 0253  &  $3.5$  &  $0.10_{-0.05}^{+0.10}$   &  GS13  &  no   &     &      &     &  no  &  no  \\ 
NGC 0524  &  $23.3$  &  $8.3_{-1.3}^{+2.7}$   &  GS13  &  yes  $(0.2)$  &  R+11  &  FAST   &  A  &  yes  &  no  \\ 
NGC 0821  &  $23.4$  &  $0.39_{-0.09}^{+0.26}$   &  GS13  &  no   &     &  FAST   &  A, S  &  yes  &  yes  \\ 
NGC 1023  &  $11.1$  &  $0.42_{-0.04}^{+0.04}$   &  GS13  &  no   &     &  FAST   &  A, S  &  yes  &  yes  \\ 
NGC 1300  &  $20.7$  &  $0.73_{-0.35}^{+0.69}$   &  GS13  &  no   &     &      &     &  yes  &  no  \\ 
NGC 1316  &  $18.6$  &  $1.50_{-0.80}^{+0.75}$   &  GS13  &  no   &     &  FAST   &     &  yes  &  no  \\ 
NGC 1332  &  $22.3$  &  $14_{-2}^{+2}$   &  GS13  &  no   &     &      &     &  yes  &  no  \\ 
NGC 1374  &  $19.2$  &  $5.8_{-0.5}^{+0.5}$   &  R+13b  &  no?  &     &  FAST   &  A  &  yes  &  yes  \\ 
NGC 1399  &  $19.4$  &  $4.7_{-0.6}^{+0.6}$   &  GS13  &  yes  $(2.4)$  &  DG13, R+13a  &   SLOW  &  A  &  yes  &  no  \\ 
NGC 2273  &  $28.5$  &  $0.083_{-0.004}^{+0.004}$   &  GS13  &  no   &     &      &     &  yes  &  no  \\ 
NGC 2549  &  $12.3$  &  $0.14_{-0.13}^{+0.02}$   &  GS13  &  no   &     &  FAST   &  A  &  yes  &  yes  \\ 
NGC 2778  &  $22.3$  &  $0.15_{-0.10}^{+0.09}$   &  GS13  &  no   &     &  FAST   &  A  &  yes  &  no  \\ 
NGC 2787  &  $7.3$  &  $0.40_{-0.05}^{+0.04}$   &  GS13  &  no   &     &      &     &  yes  &  no  \\ 
NGC 2974  &  $20.9$  &  $1.7_{-0.2}^{+0.2}$   &  GS13  &  no   &     &  FAST   &  A, S  &  yes  &  yes  \\ 
NGC 3079  &  $20.7$  &  $0.024_{-0.012}^{+0.024}$   &  GS13  &  no?  &     &      &     &  yes  &  no  \\ 
NGC 3091  &  $51.2$  &  $36_{-2}^{+1}$   &  R+13b  &  yes  $(0.6)$  &  R+13a  &      &     &  yes  &  yes  \\ 
NGC 3115  &  $9.4$  &  $8.8_{-2.7}^{+10.0}$   &  GS13  &  no   &     &      & S    &  yes  &  no  \\ 
NGC 3227  &  $20.3$  &  $0.14_{-0.06}^{+0.10}$   &  GS13  &  no   &     &      &     &  yes  &  no  \\ 
NGC 3245  &  $20.3$  &  $2.0_{-0.5}^{+0.5}$   &  GS13  &  no   &     &  FAST   &  A  &  yes  &  yes  \\ 
NGC 3377  &  $10.9$  &  $0.77_{-0.06}^{+0.04}$   &  GS13  &  no   &     &  FAST   &  A, S  &  yes  &  yes  \\ 
NGC 3384  &  $11.3$  &  $0.17_{-0.02}^{+0.01}$   &  GS13  &  no   &     &  FAST   &  A  &  yes  &  no  \\ 
NGC 3393  &  $55.2$  &  $0.34_{-0.02}^{+0.02}$   &  GS13  &  no   &     &      &     &  yes  &  yes  \\ 
NGC 3414  &  $24.5$  &  $2.4_{-0.3}^{+0.3}$   &  GS13  &  no   &     &   SLOW  &  A  &  yes  &  no  \\ 
NGC 3489  &  $11.7$  &  $0.058_{-0.008}^{+0.008}$   &  GS13  &  no   &     &  FAST   &  A  &  yes  &  yes  \\ 
NGC 3585  &  $19.5$  &  $3.1_{-0.6}^{+1.4}$   &  GS13  &  no   &     &      &     &  yes  &  no  \\ 
NGC 3607  &  $22.2$  &  $1.3_{-0.5}^{+0.5}$   &  GS13  &  no   &     &  FAST   &  A  &  yes  &  yes  \\ 
NGC 3608  &  $22.3$  &  $2.0_{-0.6}^{+1.1}$   &  GS13  &  yes  $(0.2)$  &  DG13, R+13a  &   SLOW  &  A, S  &  yes  &  yes  \\ 
NGC 3842  &  $98.4$  &  $97_{-26}^{+30}$   &  GS13  &  yes  $(0.7)$  &  DG13, R+13a  &      &     &  yes  &  no  \\ 
NGC 3998  &  $13.7$  &  $8.1_{-1.9}^{+2.0}$   &  GS13  &  no   &     &  FAST   &  A  &  yes  &  no  \\ 
NGC 4026  &  $13.2$  &  $1.8_{-0.3}^{+0.6}$   &  GS13  &  no   &     &  FAST   &  A  &  yes  &  no  \\ 
NGC 4151  &  $20.0$  &  $0.65_{-0.07}^{+0.07}$   &  GS13  &  no   &     &      &     &  yes  &  no  \\ 
\hline         
\end{tabular}   
\label{tab:sample} 
\end{center}    
\end{table*}    

\begin{table*}                                        
\begin{center}                                        
\begin{tabular}{llllllllll}                           
\hline                                                
\multicolumn{1}{l}{{\bf Galaxy}} &                   
\multicolumn{1}{l}{{\bf Distance}} &                 
\multicolumn{1}{l}{{\bf $\mathbf{M_{\rm BH}}$}} &  
\multicolumn{1}{l}{{\bf Ref.}} &                     
\multicolumn{1}{l}{{\bf Core}} &                     
\multicolumn{1}{l}{{\bf Ref.}} &                     
\multicolumn{1}{l}{{\bf Rot.}} &                     
\multicolumn{1}{l}{{\bf Vel. map}} &                 
\multicolumn{1}{l}{{\bf 1D fit}} &                   
\multicolumn{1}{l}{{\bf 2D fit}} \\                
\multicolumn{1}{l}{} &                                
\multicolumn{1}{l}{[Mpc]} &                           
\multicolumn{1}{l}{$[10^8~\rm M_{\odot}]$} &         
\multicolumn{1}{l}{} &                                
\multicolumn{1}{l}{$([\rm arcsec])$} &                                
\multicolumn{1}{l}{} &                                
\multicolumn{1}{l}{} &                                
\multicolumn{1}{l}{} &                                
\multicolumn{1}{l}{} &                                
\multicolumn{1}{l}{} \\                             
\multicolumn{1}{l}{(1)} &                             
\multicolumn{1}{l}{(2)} &                             
\multicolumn{1}{l}{(3)} &                             
\multicolumn{1}{l}{(4)} &                             
\multicolumn{1}{l}{(5)} &                             
\multicolumn{1}{l}{(6)} &                             
\multicolumn{1}{l}{(7)} &                             
\multicolumn{1}{l}{(8)} &                             
\multicolumn{1}{l}{(9)} &                             
\multicolumn{1}{l}{(10)} \\                         
\hline                                                
NGC 4261  &  $30.8$  &  $5_{-1}^{+1}$   &  GS13  &  yes  $(1.6)$  &  R+11  &   SLOW  &  A  &  yes  &  yes  \\ 
NGC 4291  &  $25.5$  &  $3.3_{-2.5}^{+0.9}$   &  GS13  &  yes  $(0.3)$  &  DG13, R+13a  &      &     &  yes  &  yes  \\ 
NGC 4342  &  $23.0$  &  $4.5_{-1.5}^{+2.3}$   &  GS13  &  no   &     &  FAST   &  A  &  no  &  no  \\ 
NGC 4388  &  $17.0$  &  $0.075_{-0.002}^{+0.002}$   &  GS13  &  no?  &     &      &     &  yes  &  no  \\ 
NGC 4459  &  $15.7$  &  $0.68_{-0.13}^{+0.13}$   &  GS13  &  no   &     &  FAST   &  A  &  yes  &  no  \\ 
NGC 4473  &  $15.3$  &  $1.2_{-0.9}^{+0.4}$   &  GS13  &  no   &     &  FAST   &  A, S  &  yes  &  yes  \\ 
NGC 4486A  &  $17.0$  &  $0.13_{-0.08}^{+0.08}$   &  GS13  &  no   &     &  FAST   &  A  &  no  &  no  \\ 
NGC 4564  &  $14.6$  &  $0.60_{-0.09}^{+0.03}$   &  GS13  &  no   &     &  FAST   &  A  &  yes  &  no  \\ 
NGC 4596  &  $17.0$  &  $0.79_{-0.33}^{+0.38}$   &  GS13  &  no   &     &  FAST   &  A  &  yes  &  no  \\ 
NGC 4697  &  $11.4$  &  $1.8_{-0.1}^{+0.2}$   &  GS13  &  no   &     &  FAST   &  A, S  &  yes  &  yes  \\ 
NGC 4889  &  $103.2$  &  $210_{-160}^{+160}$   &  GS13  &  yes  $(1.7)$  &  F+97  &      &     &  yes  &  yes  \\ 
NGC 4945  &  $3.8$  &  $0.014_{-0.007}^{+0.014}$   &  GS13  &  no?  &     &      &     &  yes  &  yes  \\ 
NGC 5077  &  $41.2$  &  $7.4_{-3.0}^{+4.7}$   &  GS13  &  yes  $(0.3)$  &  T+04  &      &     &  yes  &  yes  \\ 
NGC 5128  &  $3.8$  &  $0.45_{-0.10}^{+0.17}$   &  GS13  &  no?  &     &      &     &  yes  &  no  \\ 
NGC 5576  &  $24.8$  &  $1.6_{-0.4}^{+0.3}$   &  GS13  &  no   &     &   SLOW  &  A  &  yes  &  yes  \\ 
NGC 5813  &  $31.3$  &  $6.8_{-0.7}^{+0.7}$   &  GS13  &  yes  $(0.4)$  &  DG13, R+13a  &   SLOW  &  A  &  no  &  no  \\ 
NGC 5845  &  $25.2$  &  $2.6_{-1.5}^{+0.4}$   &  GS13  &  no   &     &  FAST   &  A  &  yes  &  yes  \\ 
NGC 5846  &  $24.2$  &  $11_{-1}^{+1}$   &  GS13  &  yes   &  F+97  &   SLOW  &  A, S  &  yes  &  yes  \\ 
NGC 6251  &  $104.6$  &  $5_{-2}^{+2}$   &  GS13  &  yes?  &     &      &     &  yes  &  yes  \\ 
NGC 7052  &  $66.4$  &  $3.7_{-1.5}^{+2.6}$   &  GS13  &  yes  $(0.8)$  &  Q+00  &      &     &  yes  &  yes  \\ 
NGC 7582  &  $22.0$  &  $0.55_{-0.19}^{+0.26}$   &  GS13  &  no   &     &      &     &  no  &  no  \\ 
NGC 7619  &  $51.5$  &  $25_{-3}^{+8}$   &  R+13b  &  yes$^a$  $(0.5)$  &  DG13, R+13a  &      &     &  yes  &  no  \\ 
NGC 7768  &  $112.8$  &  $13_{-4}^{+5}$   &  GS13  &  yes   &  G+94  &      &     &  yes  &  no  \\ 
UGC 03789  &  $48.4$  &  $0.108_{-0.005}^{+0.005}$   &  GS13  &  no?  &     &      &     &  yes  &  no  \\ 
\hline         
\end{tabular}  
\tablecomments{\emph{Column (1):} Galaxy name.                       
\emph{Column (2):} Distance.                                   
\emph{Column (3):} Black hole mass.                                   
\emph{Column (4):} Reference of the black hole mass reported here (G+03 = \citealt{greenhill2003}, GS13 = \citealt{grahamscott2013}; R+13b = \citealt{rusli2013bhmassesDM}).                                   
\emph{Column (5):} Presence of a partially depleted core. 
The question mark is used when the classification has come from the velocity dispersion criteria mentioned in Section \ref{sec:corser}. 
The value of the core break radius is reported in parenthesis when available.  
\emph{Column (6):} Reference of the identification of a partially depleted core (G+94 = \citealt{grillmair1994}; F+97 = \citealt{forbes1997}; Q+00 = \citealt{quillen2000}, 
T+04 = \citealt{trujillo2004coresersicmodel}; F+06 = \citealt{ferrarese2006acsvcs}; J+11 = \citealt{jardel2011}; R+11 = \citealt{richings2011}; 
DG13 = \citealt{dullograham2013cores}; R+13a = \citealt{rusli2013}).  
\emph{Column (7):} Kinematical classification (fast/slow rotator).
\emph{Column (8):} Availability of velocity map (A = ATLAS$^{\rm 3D}$, S = SLUGGS). 
\emph{Column (9):} Completion of 1D fit. 
\emph{Column (10):} Completion of 2D fit. \\                                 
$^a$ NGC 7619 may contain an embedded disk, rather than possessing a partially depleted core (see Figure \ref{fig:n7619}).
} 
\end{center}    
\end{table*}

\subsection{\emph{Spitzer}/IRAC observations}
\subsubsection{Data acquisition}
For each of our 75 galaxies, we downloaded from the 
Spitzer Heritage Archive\footnote{\url{http://irsa.ipac.caltech.edu/applications/Spitzer/SHA/}} 
all the available $3.6~\rm \mu m$ IRAC Astronomical Observation Requests (AORs). 
Each AOR is an individual Spitzer observation sequence, and
includes a number of data frames (the individual exposures) and the calibration data.
The data frames were selected to be 
corrected Basic Calibrated Data (cBCD), produced by the IRAC Level 1 pipeline. 
This automatic pipeline takes a single ``raw'' image, removes the scattered light, 
and performs dark subtraction, flatfielding correction and flux calibration (into units of $\rm MJy~sr^{-1}$). 
The final product (the BCD) is a flux-calibrated image which has had all the well-understood instrumental signatures removed.
BCD frames are further processed through an ``artifact correction'' pipeline 
that mitigates the commonly found artifacts of stray light, saturation, ``muxbleed'' and column pulldown\footnote{Stray light 
includes scattered light from stars outside the array location as well as filter ghosts from bright stars. 
Multiplexer bleed, or ``muxbleed'', can be generated by stars, hot pixels, and particle hits.
It appears as a decaying trail of pixels, repeating every fourth column.
``Column pulldown'' is caused by a bright pixel that triggers a bias shift within its respective column, 
creating a lower background value throughout the entire column than in the surrounding columns.}.
After the artifact correction has been applied, the BCD becomes a cBCD.

\subsubsection{Mosaicking}
We performed image mosaicking using the MOPEX package \citep{makovozmarleau2005mopex}.
This enabled the production of suitably wide-field-of-view images for accurate sky background subtraction.
Individual cBCD frames with exposure time of $1~\rm sec$ were rejected.
Permanent or semi-permanent bad pixels, contained in a semi-static mask (the ``pmask''), were ignored.
Each AOR is associated to a specific pmask. 
Therefore, when multiple AORs were available for the same galaxy, 
we merged the different pmasks. 
Cosmic ray rejection was performed with the dual outlier and multi-frame techniques.
The pixel size of the mosaic was set to be the same as the input cBCD frames ($1''.22 \times 1''.22$).
For $3.6~\rm \mu m$ observations with this pixel scale, 
the photometric zeropoint magnitude is $m_{\rm zp} = 17.26 \rm~mag$.
The orientation of the mosaic was set to the average rotation angle of the input cBCD frames. 
Individual cBCD frames were combined together into a single mosaic with the default linear interpolation algorithm.

\subsubsection{Overlap correction}
\begin{figure}[h]
\begin{center}
\includegraphics[width=0.49\columnwidth]{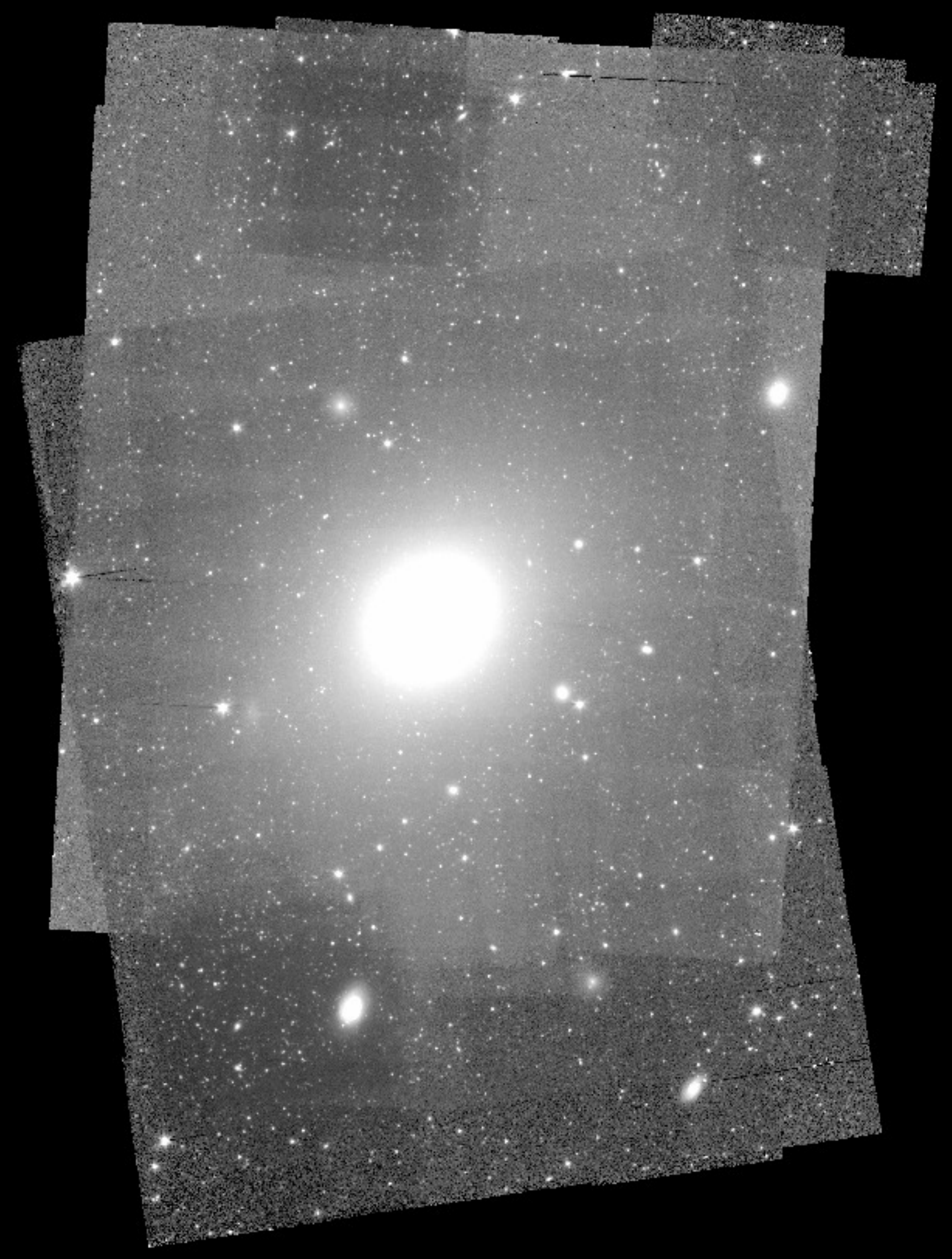}
\includegraphics[width=0.49\columnwidth]{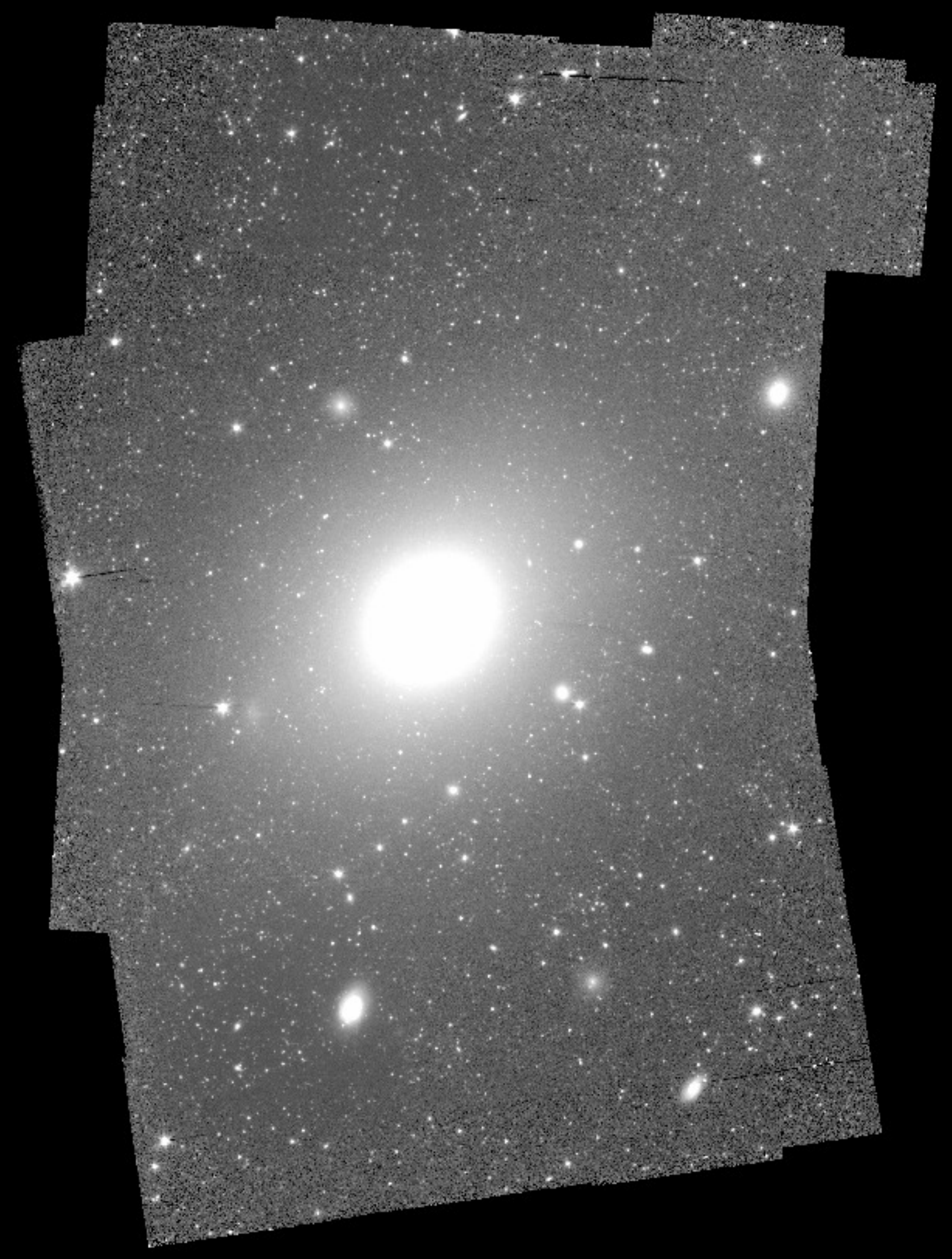}
\caption{Example of the \emph{overlap} correction. 
The image mosaic of the galaxy M49 was obtained by co-adding frames coming from 8 different AORs.
The evident patchiness (left image) was removed (right image) using the \emph{overlap} module.}
\label{fig:overlap}
\end{center}
\end{figure}

Before generating a mosaic, MOPEX can perform background matching among the individual frames
by using the \emph{overlap} module. 
This module calculates and applies an additive correction to the individual frames,
producing a consistent background across the mosaicked image.
According to the \emph{Spitzer Data Analysis 
Cookbook}\footnote{\href{http://irsa.ipac.caltech.edu/data/SPITZER/docs/dataanalysistools/cookbook/Spitzer_Data_Cookbook.pdf}
{\tt http://irsa.ipac.caltech.edu/data/SPITZER/docs/ dataanalysistools/cookbook/Spitzer\_Data\_Cookbook.pdf}}, 
the use of the \emph{overlap} module is not particularly recommended for 
$3.6~\rm \mu m$ observations. 
However, after a visual inspection of the mosaics obtained without activating the \emph{overlap} module,
we found that all the mosaics obtained from multiple AORs were affected by patchiness, 
due to bias fluctuations in the CCD array.
For this reason, we re-generated the multiple-AORs mosaics by activating the \emph{overlap} correction,
which successfully removed the ``chessboard'' pattern (see Figure \ref{fig:overlap} for an example).

\subsubsection{Sigma mosaics}
For each individual cBCD frame, the IRAC Level 1 pipeline calculates the uncertainty associated to each pixel
and produces an uncertainty frame (or sigma frame).
The initial uncertainty is estimated as the Poisson noise in electrons plus the readout noise added in quadrature
($\sigma^2 = \sigma^2_{\rm readoutnoise} + \sigma^2_{\rm Poisson}$).
This initial sigma frame is carried through the pipeline, and additional uncertainties (e.g.~dark current and flat field uncertainties) 
are added in quadrature when appropriate.
When MOPEX generates an image mosaic, it also produces the associated sigma mosaic 
by interpolating the individual uncertainty frames and co-adding them,
following the standard assumption of additive variances. 

\subsubsection{Sky subtraction}
Sky subtraction was performed manually on the image mosaics using the tasks {\tt marksky} and {\tt skyfit} 
of the IRAF\footnote{IRAF is the Image Reduction and Analysis Facility, 
distributed by the National Optical Astronomy Observatory, 
which is operated by the Association of Universities for Research in Astronomy (AURA) 
under cooperative agreement with the National Science Foundation.} 
package GALPHOT\footnote{GALPHOT was developed in the IRAF - STSDAS environment mainly by 
W. Freudling, J. Salzer, and M. P. Haynes \citep{haynes1999galphot}.}.
The task {\tt skyfit} also provided an estimate of the sky root mean square ($RMS_{\rm sky}$).

\subsubsection{Additional aestethic corrections}
\label{sec:aesth}
\begin{figure}[h]
\begin{center}
\includegraphics[width=0.49\columnwidth]{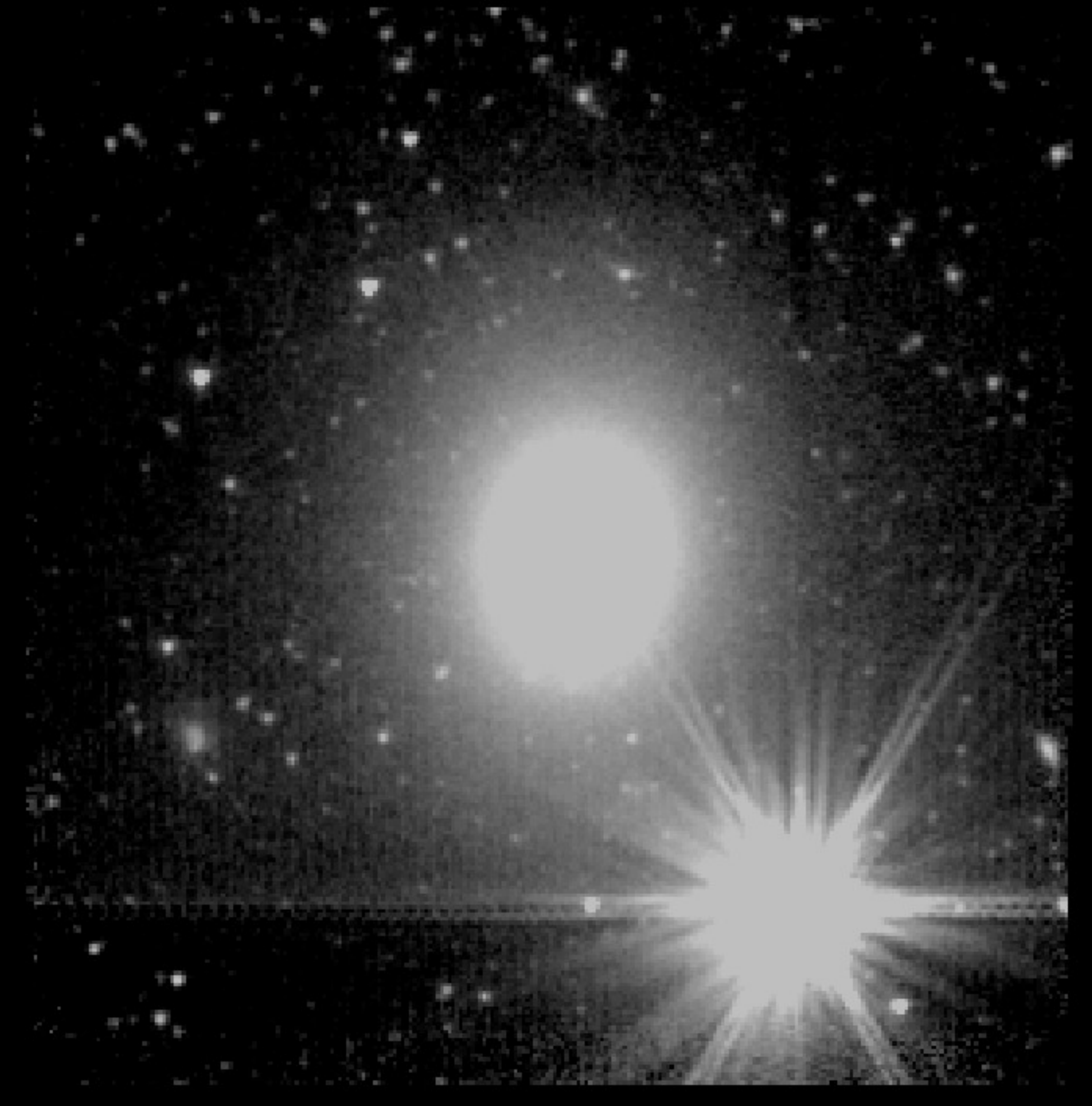}
\includegraphics[width=0.49\columnwidth]{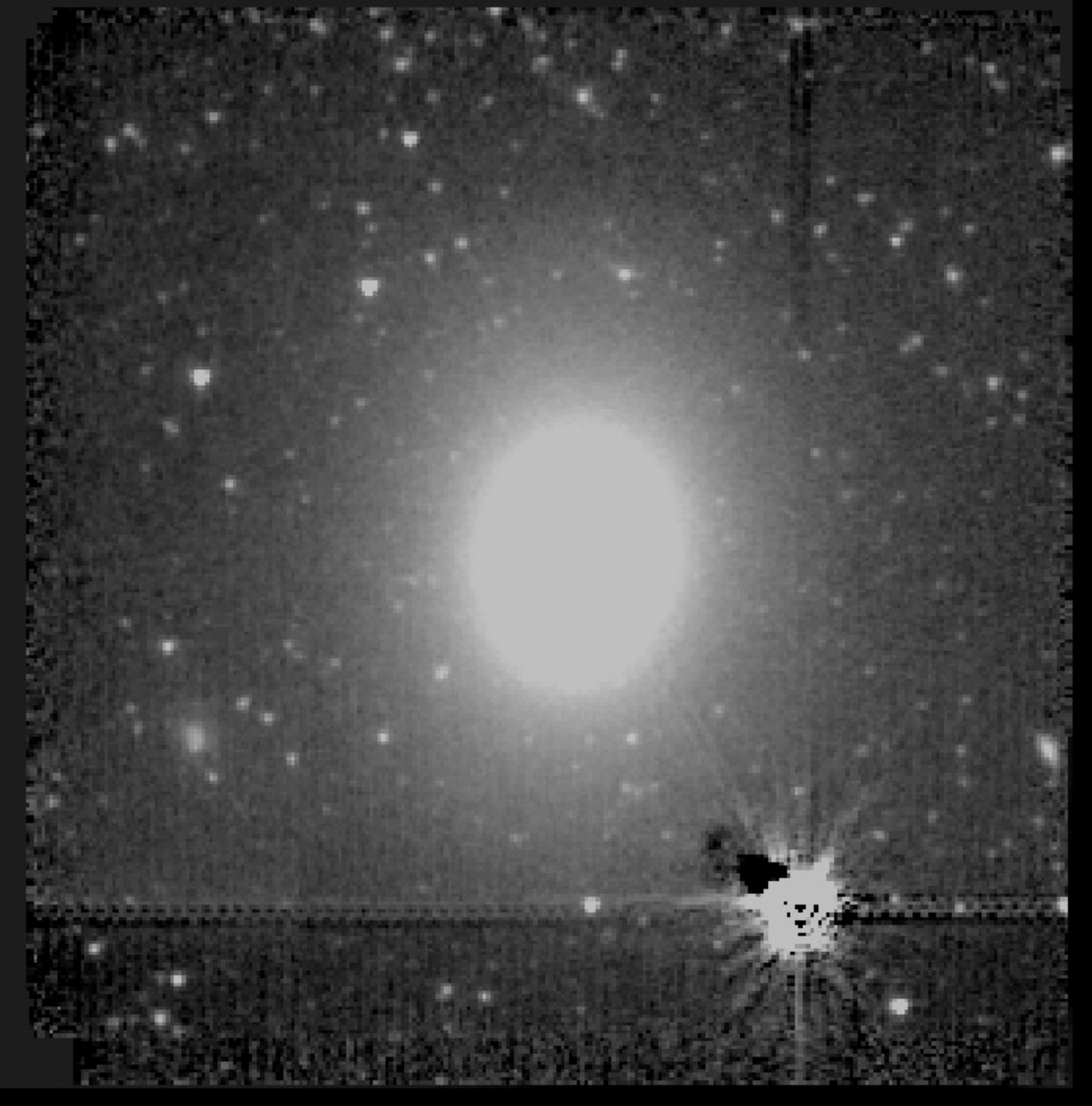}
\caption{Example of aestethic correction.
The images show the mosaic of the galaxy NGC 4459 before (left) and after (right) the partial removal 
of a bright saturated star.}
\label{fig:aesth}
\end{center}
\end{figure}

The image mosaics of 4 galaxies (NGC 0821, NGC 2974, NGC 4291, NGC 4459) were found to be affected 
by bright, highly saturated stars lying close to the target galaxies.
These stars were modeled and subtracted using the software Galfit \citep{peng2010} 
and the extended IRAC Point Response Function (PRF) image at $3.6~\rm \mu m$.
This helped remove the extended wings and spikes of the saturated stars 
(see Figure \ref{fig:aesth} for an example).

\subsubsection{Image masking}
Galactic stars and any other objects different from the target galaxy 
were masked through a two-step procedure.
First, we created an initial mask using the IRAF task 
{\tt objmasks} that identifies objects by threshold sigma detection.
Then, we refined each mask by hand, using the software 
SAOImage DS9\footnote{SAOImage DS9 development has been made possible by funding 
from the Chandra X-ray Science Center (CXC) 
and the High Energy Astrophysics Science Archive Center (HEASARC).} 
in conjunction with the IRAF task 
{\tt mskregions}. 
We identified and carefully masked not only contaminating sources located in the field of the image mosaic,
but also objects overlapping with the target galaxy, such as foreground stars, background galaxies, 
globular clusters or red giant stars.

\subsubsection{1D PSF}
A universal\footnote{Across all mosaics, the variation of the Moffat $FWHM$ is $\pm 0''.1$, 
and the variation of the Moffat $\beta$ is $\pm 2.0$.
The use of a universal PSF is justified for the following reasons: 
\emph{i}) the PSF-convolution is more sensitive to the value of the $FWHM$ than $\beta$, 
i.e. having a 50\% variation in $\beta$ is not an issue as long as the variation in $FWHM$ is small;
\emph{ii}) the use of a non-signal-to-noise-weighted fitting scheme minimizes biases from a non-accurate PSF description;
\emph{iii}) not all mosaics have enough stars suitable for the PSF characterization: 
rather than having an individual PSF for each mosaic (which would have been based on only 1-2 stars for some mosaics), 
we preferred the use of a universal PSF.}, 
average, one-dimensional Point Spread Function was characterized using the IRAF task {\tt imexamine}.
A nonlinear least squares \cite{moffat1969} profile\footnote{The \citep{moffat1969} profile 
has the following form:
\begin{equation}
\mu = \mu_{\rm 0} - 2.5 \log \Biggl[ 1 + \biggl( \frac{R}{\alpha} \biggr)^2 \Biggr]^{-\beta} ,
\end{equation}
where $R$ is the projected radius, $\mu_{\rm 0}$ is the central surface brightness, and 
$\alpha$ and $\beta$ regulate the width and the shape of the profile.} 
of fixed center and zero background was fit to the 
(background subtracted) pixels of 20 bright stars, belonging to different image mosaics.
The best-fit parameters of the 20 stars were then averaged together. 
Doing so, we obtained 
$(\alpha, \beta) = (2''.38, 4.39)$.

\subsubsection{2D PSF}
\begin{figure}[h]
\begin{center}
\includegraphics[width=1\columnwidth, trim = 240 40 50 20, clip=True]{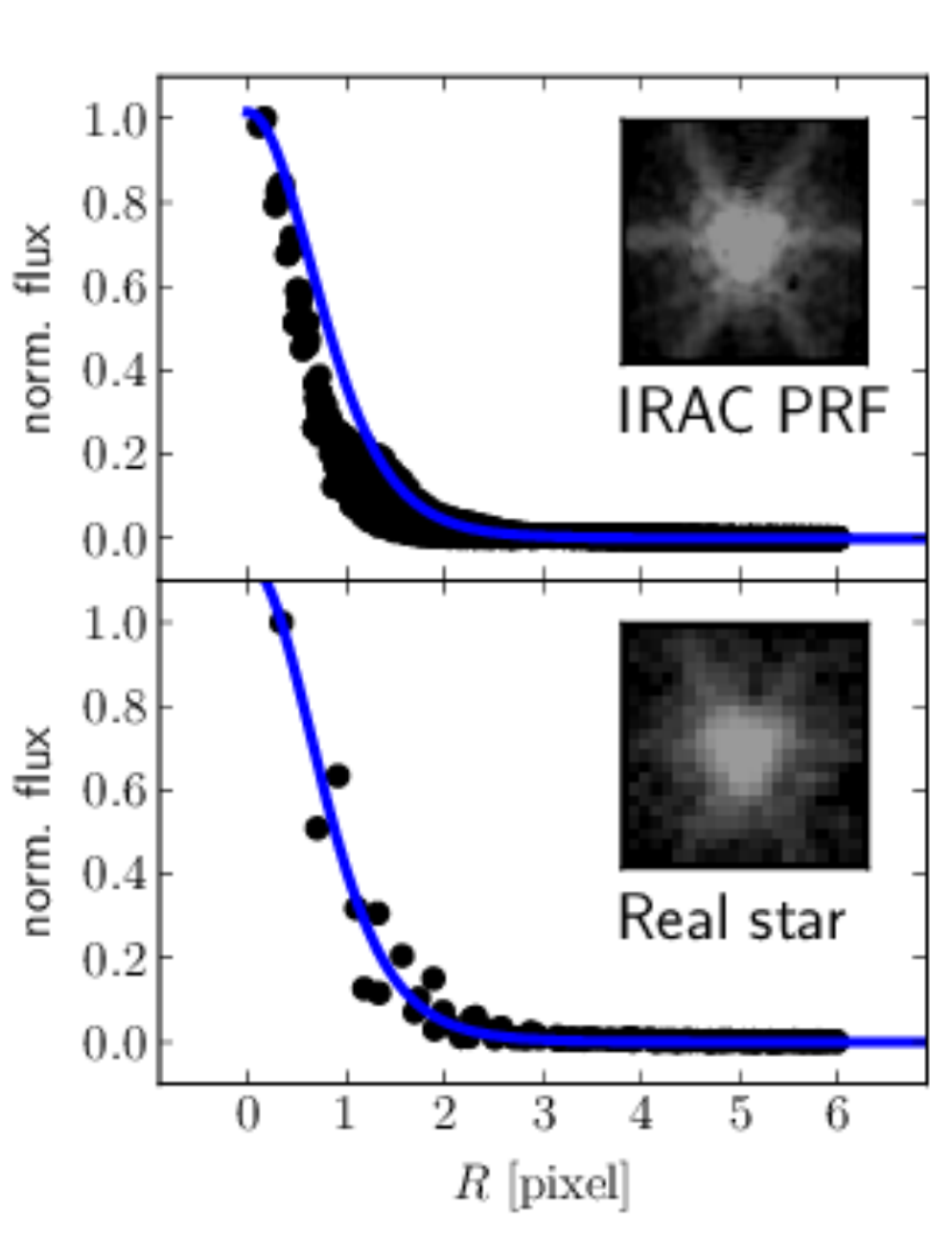}
\caption{Normalized flux versus radial distance 
from the image centroid of an observed point source (bottom panel) 
and of the IRAC instrument Point Response Function (PRF, top panel). 
The normalized flux is given in arbitrary units, 
and the radial distance is in units of pixel size of the IRAC detector (1 pixel = $1''.22$).
The blue solid line shows our 1D Moffat PSF model,
which has been normalized to intersect the data point with the largest flux. 
The $FWHM$ of the IRAC PRF is clearly smaller than that of a ``real'' star.
The inserts display the images of the observed point source and the IRAC PRF.}
\label{fig:psf}
\end{center}
\end{figure}

The IRAC support team provides users with a two-dimensional instrument Point Response Function (PRF) at $3.6~\rm \mu m$.
However, while this helped remove the extended wings of saturated stars (see Section \ref{sec:aesth}), 
we found this PRF to be inadequate for the purposes of our modelling.
In fact, the $FWHM$ of the IRAC instrument PRF ($\sim 1''.8$), as measured by the IRAF task {\tt imexamine}, 
is systematically smaller than the average $FWHM$ of ``real'' stars ($\sim 2''.0$). 
Figure \ref{fig:psf} illustrates this issue.
We also tested the IRAC PRF by providing it as the input Point Spread Function (PSF) for Galfit 
and fitting a number of stars in different image mosaics.
A visual inspection of the fit residuals confirmed that the IRAC instrument PRF is narrower than ``real'' point sources.
For this reason, we constructed our 2D PSF according to the following method 
(as directed by C. Peng, private communication).\\
We provided the IRAC instrument PRF as the input PSF for Galfit 
and simultaneously fit 7 bright stars (belonging to different mosaics), 
modeling the stars with Moffat profiles 
and constraining all the profiles to have the same $(\alpha, \beta)$, position angle and axis ratio.
The 2D PSF image was then obtained by taking the best fit Moffat model -- the same best-fit model for all 7 stars, by construction -- 
and convolving it with the IRAC instrument PRF.
The advantage of this method is to obtain a 2D PSF that is wider than the instrument PRF, 
but maintains the asymmetric features of the instrument PRF (e.g.~wings and spikes).
We then tested this 2D PSF 
on a number of stars (these stars were different from the 7 stars employed to build the 2D PSF image)
and verified that it correctly reproduces the shape of ``real'' point sources. 

\subsection{Additional data}
\label{sec:adddata}
\subsubsection{Kinematics}
\label{sec:kinem}
A kinematical classification (slow/fast rotator) is available for 34 of our 75 galaxies 
from the ATLAS$^{\rm 3D}$ survey \citep{atlas3dIII} 
and for 3 additional galaxies from \citet{scott2014}.
This classification (Table \ref{tab:sample}, column 7) 
concerns the kinematic properties of galaxies within the spectroscopic instrument's field-of-view, 
but does not contain additional information -- crucial for our analysis -- about kinematic substructures, 
such as embedded disks or kinematically decoupled components, 
which can require separate modeling.
For this reason, we also visually inspected the velocity fields of our galaxies, 
when available from the literature.
Velocity maps were taken from the ATLAS$^{\rm 3D}$ survey for 34 galaxies 
(observed with SAURON by \citealt{krajnovic2011}),
from \citet{scott2014} for 2 galaxies 
(observed with WiFeS)
and from the SLUGGS survey for 12 galaxies 
(observed with DEIMOS by \citealt{arnold2014}).
While the field-of-views of SAURON ($33'' \times 41''$) and WiFeS ($25'' \times 38''$) 
reach to about one galaxy effective radius (for our local galaxies),
observations taken with DEIMOS can probe the galaxy kinematics well beyond two effective radii.

\subsubsection{AGNs and nuclear dust}
The X-ray, UV and optical radiation 
emitted by the the accretion disks of Active Galactic Nuclei (AGNs) can stimulate infrared thermal emission 
from circumnuclear dust, if present. 
This means that if a galaxy hosts an optical AGN \emph{and} a certain amount of nuclear dust,
we may detect some non-stellar nuclear emission at $3.6~\rm \mu m$.
It is therefore important to identify which of our galaxies have both an optical AGN and circumnuclear dust.
To help with this task, we searched NED\footnote{NED is the NASA/IPAC Extragalactic Database.} 
for the individual galaxies and their associated literature.
Unsurprisingly, dusty AGNs were more frequently found in late-type spiral galaxies, 
and modelled by us with either a point source or a PSF-convolved Gaussian.

\subsubsection{S\'ersic/core-S\'ersic classification}
\label{sec:corser}
Core-S\'ersic galaxies \citep{grahamguzman2003,graham2003coresersicmodel,trujillo2004coresersicmodel} are galaxies (or spheroids) 
with partially depleted cores, 
i.e.~a central deficit of light relative to the inward extrapolation of their outer S\'ersic light profile. 
Such deficits were first noted and researched by \cite{kingminkowski1966}.
S\'ersic galaxies, instead, do not exhibit such central stellar deficits.
Partially depleted cores, as measured from high-resolution observations, have typical sizes of a few tens of parsecs
\citep{dullograham2014cores,rusli2013}.
The majority are thus unresolved in our image mosaics. 
We masked these unresolved cores (identified in high-resolution images, see Table \ref{tab:sample}) by excluding the surface brightness profile 
within 3 PSF's $FWHM$ from the galaxy center.
In case of cores with sizes exceeding the PSF's $FWHM$, we excluded the data points within the size of the core 
plus 3 PSF's $FWHM$.
The S\'ersic/core-S\'ersic classification presented in this work (Table \ref{tab:sample}, column 5)
comes from the compilation of \citet{savorgnangraham2015},
who identified partially depleted cores according to the same criteria used by \citet{grahamscott2013}.
When no high-resolution image analysis was available from the literature, 
they inferred the presence of a partially depleted core based on the stellar velocity dispersion, $\sigma$:
a galaxy is classified as core-S\'ersic if $\sigma > 270\rm~km~s^{-1}$,
or as S\'ersic if $\sigma < 166\rm~km~s^{-1}$.
This resulted in us assigning cores to just 2 galaxies using this alternative method when 
no high resolution image was available.

\section{Analysis}
\label{sec:anal}
\subsection{Isophotal analysis}
We performed an isophotal analysis of our galaxies using the IRAF task {\tt ellipse} \citep{taskellipse}, 
which fits elliptical isophotes to galaxy images.
The center of the isophotes was held fixed, while the ellipticity ($\epsilon$), the position angle ($P.A.$),  
and the amplitude of the fourth harmonic\footnote{The amplitude of the fourth harmonic deviations from perfect ellipses 
($B4$) parameterizes the diskyness ($B4>0$) or boxyness ($B4<0$) of the isophotes.} ($B4$) were allowed to vary with radius.
The step in semi-major axis length between successive ellipses was first set to increase linearly, 
and then geometrically in our second run\footnote{In the case of linear steps, 
the semi-major axis length for the next ellipse was calculated by adding $1$ pixel to the current length.
In the case of geometric steps, the semi-major axis length for the next ellipse 
was calculated as $1.1$ times the current length.}.
As a result, for each galaxy we produced respectively a ``linearly sampled'' and a ``logarithmically sampled'' 
surface brightness profile along the major-axis. 
Major-axis surface brightness profiles were additionally converted into the equivalent-axis,
i.e.~the geometric mean of the major ($a$) and minor ($b$) axis ($R_{\rm eq} = \sqrt{ab}$), 
equivalent to the circularized radius.
This resulted in 4 profiles per galaxy.
Isophotes corresponding to an intensity less than 
three times the root mean square of the sky background fluctuations ($3 \times RMS_{\rm sky}$) were ignored.
Some surface brightness profiles were truncated at our discretion before the $3 \times RMS_{\rm sky}$ limit, 
according to specific technical reasons (e.g.~contamination from light of a neighboring galaxy, 
disturbed morphology in the galaxy outskirts, etc.). 
In particular, we did not attempt to fit bends or truncations of large-scale disks 
(e.g.~\citealt{erwin2005,erwin2008,erwin2012,gutierrez2011,comeron2012,munozmateos2013,kim2014}) 
but instead truncated the surface brightness profiles before the occurrence of such features. 
Individual cases are discussed in Section \ref{sec:indgal}.

\subsection{Unsharp masking}
Unsharp masking is an image sharpening technique that is useful to reveal asymmetric structures in galaxies, 
such as bars or (inclined) embedded disks. 
First, the original galaxy image was smoothed with a Gaussian filter.
Then, the original image was divided by the smoothed one. 
The result of such an operation is the ``unsharp mask''. 
The asymmetric features revealed by this technique have sizes 
comparable to the $FWHM$ of the Gaussian kernel used for the smoothing.
Therefore, for each galaxy, we produced a set of different unsharp masks by varying the $FWHM$ of the filter,
to identify all the asymmetric features that could bias the fitting process 
and may therefore need to be considered during the galaxy modelling phase.
This information was used in combination with kinematic and AGN information discussed in Section \ref{sec:adddata}.

\subsection{1D fitting routine}
The decomposition of the surface brightness profiles was performed with software written by G. Savorgnan.
This software can fit an observed surface brightness profile with any linear combination of 
a set of analytical functions (S\'ersic, exponential, Gaussian, Moffat, Ferrer, symmetric Gaussian ring etc. 
see Appendix \ref{sec:app1} for a description of the analytical form of these profiles).
At each iteration, 
the model is numerically convolved with a Moffat filter, to account for PSF effects, 
and then matched to the data.
The minimization routine is based on the Levenberg-Marquardt algorithm.
During the fit, we deliberately did not make use of any weighting scheme on the data points that constitute the 
surface brightness profile, 
although the use of a linearly and logarithmically sampled profile effectively represents a different weighting scheme.
The all too often over-looked flaw with (signal-to-noise)-based weighting schemes 
is that they immediately become biased weighting schemes when additional components are present but not modelled.
For example, fitting only a S\'ersic model to what is actually a nucleated elliptical galaxy immediately voids 
a (signal-to-noise)-based weighting scheme 
and results in S\'ersic parameters which describe the spheroid less accurately than had no (signal-to-noise)-weighting been used.
While we have paid careful attention to the components in each galaxy, 
this is an issue that warrants the non application of (signal-to-noise)-based weighting schemes.
Over-looked partially depleted cores, or an incorrect PSF, can of course also result in (signal-to-noise)-weighted fitting schemes 
performing poorly because of the emphasis they place on matching the model to the inner data. 

\subsection{Smoothing technique}
\label{sec:smooth}
Some nearby galaxies in our sample have very large apparent sizes 
and for them we obtained surface brightness profiles more extended than $8~\rm arcmin$.
This means that their outermost (significant) isophote 
corresponds to a projected galactic radius $R$ 
of more than $240$ times the $FWHM$ of the instrumental PSF.
Such level of spatial resolution is unnecessary for the purposes of our analysis 
and, especially in the case of a clumpy star forming galaxy, 
it results in a ``noisy'' surface brightness profile.
Moreover, in the case of a linearly sampled light profile,
it significantly prolongs the computational time of the fitting routine 
(because, at each iteration, the PSF convolution is performed numerically on a large array).
To overcome this problem, we introduced a method to which we refer as the ``smoothing technique''.
This method was applied to the galaxies M31, M81, NGC 4945 and NGC 5128.
For each of these galaxies, we took the image mosaic and we convolved it with a Gaussian filter 
whose $FWHM$ was larger than the $FWHM$ of the instrumental PSF.
We then ran {\tt ellipse} on the convolved image and extracted 
``linearly sampled'' and ``logarithmically sampled'' surface brightness profiles. 
For the ``linearly sampled'' profiles, we set the radial step between contiguous isophotes to be 
comparable to the $FWHM$ of the smoothing Gaussian filter.
Doing so, we reduced the number of fitted isophotes and we also produced smoother surface brightness profiles.
Before the software fit a smoothed light profile, the model to be fit was convolved twice:
the first time to account for PSF effects, 
and the second time to account for the artificial Gaussian smoothing applied to the image mosaic.

\subsection{Identifying and modeling sub-components}
\label{sec:cpts}
\begin{figure*}[h]
\begin{center}
\includegraphics[width=\fitfigurewidth]{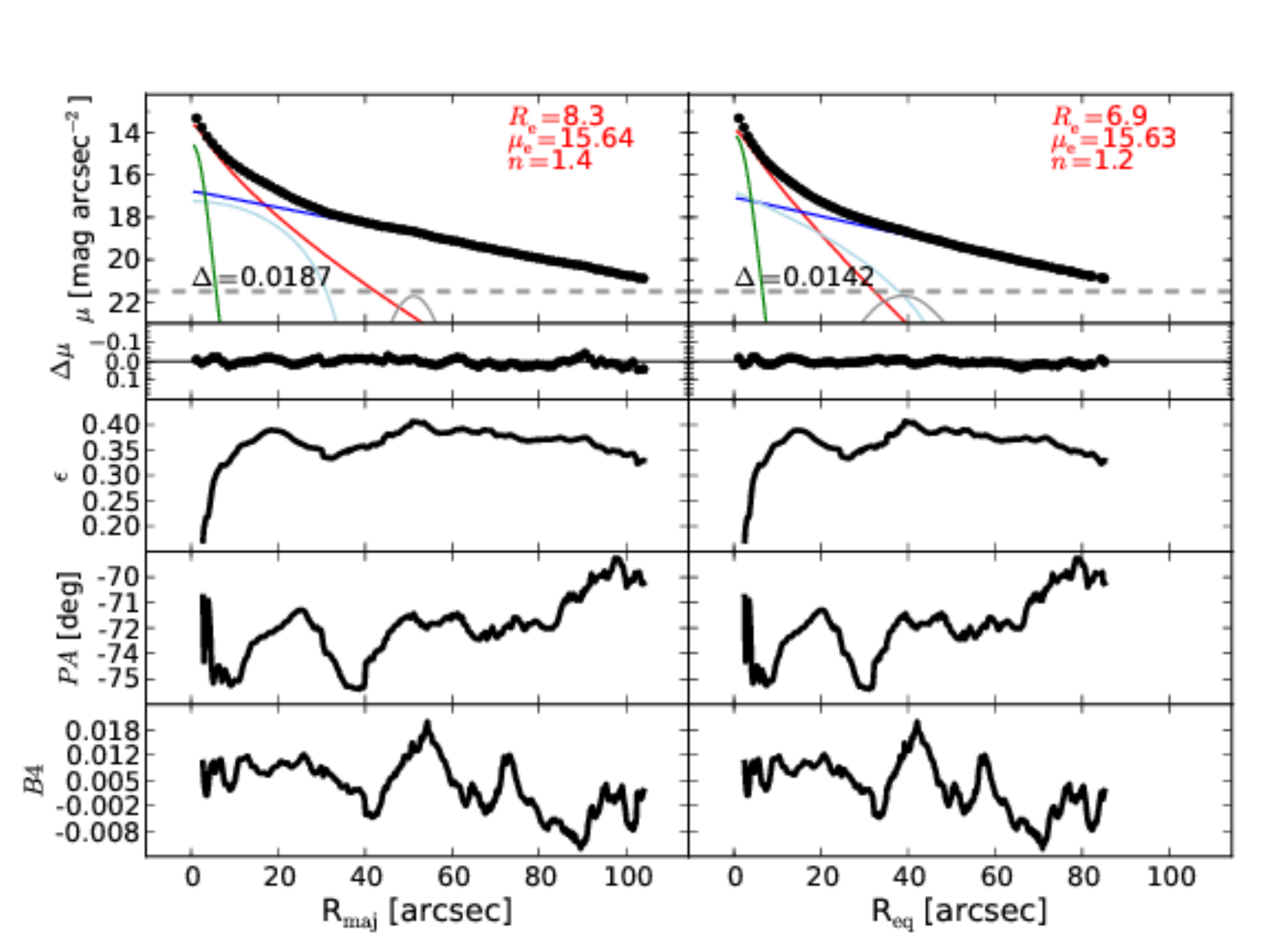}
\caption{Best-fit model, and isophotal parameters, for the galaxy NGC 2974. 
The left panels refer to the major-axis $R_{\rm maj}$, 
while the right panels refer to the equivalent-axis $R_{\rm eq}$, 
i.e.~the geometric mean of the major ($a$) and minor ($b$) axis ($R_{\rm eq} = \sqrt{ab}$).
The top panels display the galaxy surface brightness radial profiles obtained with a linear sampling. 
The black points are the observed data.  
The color lines represent the individual (PSF-convolved) model components:
red=S\'ersic (bulge), dark blue=exponential (disk), green=Gaussian (AGN), cyan=Ferrer (bar), gray=Gaussian ring (ring). 
The best-fit parameters for the S\'ersic bulge model are inset.
The total (PSF-convolved) model is shown with a black dashed line, 
but it is hard to distinguish because it almost perfectly matches the data, 
hence the residual profile is additionally shown as $\Delta \mu$ in the second row.
The horizontal grey dashed line corresponds to an intensity 
equal to three times the root mean square of the sky background fluctuations ($3 \times RMS_{\rm sky}$).
$\Delta$ denotes the rms scatter of the fit in units of $\rm mag~arcsec^{-2}$.
The lower six panels show the ellipticity ($\epsilon$), position angle ($PA$) and fourth harmonic ($B4$) radial profiles. 
Such profiles are available online for all other galaxies successfully modelled in 1D (see Table \ref{tab:fitres}).}
\label{fig:n2974ex}
\end{center}
\end{figure*}

In this Section we give a general overview of the guidelines that we followed to identify and model 
the sub-components that constitute our galaxies. 
However, given the level of accuracy and detail to which each galaxy decomposition has been performed in our analysis,
it is hard to encompass all aspects of this matter in a few paragraphs.
The modeling of each galaxy represented a particular and original problem, 
and we remand the reader to Section \ref{sec:indgal},
where we provide individual descriptions of the galaxies that we analyzed. \\
As stressed in Section \ref{sec:int}, our investigation is primarily focused on the central spheroidal components 
of galaxies. 
The objects in our sample are either early-type galaxies (elliptical+lenticular), 
or ``early-type spiral'' galaxies 
(i.e.~the morphological classification of our spiral galaxies is within \emph{Sa-Sc}, 
with the only exception of NGC 4945 which is classified as \emph{Scd}), 
therefore -- by definition -- they all have a bulge/spheroidal component, 
unlike ``late-type spiral'' galaxies that can be bulgeless (e.g.~NGC 300).   
We modeled spheroids/bulges with a S\'ersic profile, without attempting to distinguish between classical and pseudo-bulges.  \\
Disks were usually fit with the exponential model, 
although in the case of highly inclined or edge-on systems we preferred using an $n < 1$ S\'ersic function. 
\citet{pastrav2013a,pastrav2013b} showed that, due to projection effects, their simulated images of inclined galaxy disks 
are better fit by a S\'ersic function with $n < 1$ than by a pure exponential model.
The inclined, embedded disks of some ``elliptical'' galaxies were described with Ferrer functions, rather than a $n < 1$ S\'ersic function. 
This choice was partly motivated by the fact that a S\'ersic + Ferrer model is less degenerate than a S\'ersic + S\'ersic model, 
since the S\'ersic profile can assume any concave ($n > 1$) or convex ($n < 1$) curvature, 
whereas the Ferrer profile can only have a negative curvature as required for an inclined disk. \\
The presence of large-scale disks, such as those of lenticular and spiral galaxies, 
were known \emph{a priori} from the galaxy morphological classification (as listed on NED), 
although some of them were re-classified by us as having intermediate-scale embedded disks.
These were identified in a number of different ways. 
If highly inclined, they can obviously be spotted from the galaxy image or the unsharp mask.
Local maxima in the ellipticity and fourth harmonic profiles 
can provide footprints of less obvious embedded disks.
In particular, the ellipticity profile helps distinguish embedded disks from large-scale disks.
Galaxy disks typically have fixed ellipticity, reflecting their inclination to our line of sight.
On the other hand, spheroids can have their ellipticities varying with radius,
but they are usually rounder than inclined disks, thus their average ellipticities are lower than those of inclined disks.
If the ellipticity profile of a galaxy increases with radius, 
this can be ascribed to an inclined disk that becomes progressively more important over the spheroid,
whereas a radial decrease of ellipticity signifies the opposite case.
Therefore, in a situation where a disk is identified from the galaxy image, 
but its extent (large- or intermediate-scale) is ambiguous,
the shape of the ellipticity profile can be decisive. 
Another way to establish the presence of an embedded disk is to look at the velocity map of a galaxy,
following the approach of \citet{arnold2014}. 
A local angular momentum decrease with increasing radius is indicative of an intermediate-scale disk 
that fades toward larger radii. 
\cite{savorgnangraham2015ellicular} extensively discuss the topic of galaxies with intermediate-scale disks 
and show that, when these disks are misclassified and modeled as large-scale disks, 
the luminosity of the spheroidal component is underestimated, 
which makes these galaxies falsely appear as extreme outliers in the (black hole mass)-(spheroid stellar mass) diagram.  \\
Bars are usually recognizable from galaxy images and unsharp masks, 
although local maxima/minima or abrupt changes in the radial profiles of the isophotal parameters 
can provide additional evidence for less obvious bars. 
As noted, we were able to successfully fit bars with a Ferrer function 
(NGC 4151 is the only case for which we described the bar with an $n \sim 0.2$ S\'ersic model). 
Disk-like components embedded in the bulges of spiral galaxies were described with an $n \lesssim 1$ S\'ersic model or, 
in a few cases, with a Ferrer function\footnote{One advantage of choosing a Ferrer function over a S\'ersic profile 
to fit a disk-like component embedded in the bulge 
is to reduce degeneracies with the S\'ersic profile that describes the bulge.}. 
This approach is similar to that of \cite{laurikainen2010}, 
who fit bars with a Ferrer function and inner disks with a S\'ersic model. 
Given that our galaxy sample lacks ``late-type spiral'' galaxies, 
it is not surprising \citep{gadottidesouza2006} that we did not find bars with exponential profiles 
(e.g.~\citealt{elmegreenelmegreen1985,gadotti2008,kim2015}).
The bars in our sample were found to have rather flat inner profiles, 
as is commonly found for bars in ``early-type spiral'' galaxies \citep{gadottidesouza2006}.  \\
The presence of a nuclear component -- either resolved or unresolved -- was generally expected in 
(but not restricted to) galaxies that host an optical AGN and circumnuclear dust.
Nuclear stellar disks and nuclear star clusters fall into the category of nuclear components too,
but their identification can be more subtle than for AGNs.
Nuclear clusters have typical sizes of a few parsecs, 
therefore for the majority of our galaxies they are unresolved in \emph{Spitzer}/IRAC $3.6~\rm \mu m$ observations.
If an identification from high-resolution observations was available from the literature, we relied on that, 
otherwise we concluded that a galaxy was nucleated from an excess of nuclear light in the residuals of the 
fit\footnote{This conclusion was drawn after going through the following steps. 
First we identified all the sub-components of a galaxy 
(assuming that the galaxy was not nucleated), built a model accordingly and fit it to the data.
If the residuals of the fit showed a nuclear light excess, 
we repeated the fit by excluding the data points within the nuclear region.
Only after checking that the outcome of the last fit was consistent with a fit that included a small nuclear component, 
we inferred the presence of a stellar nuclear component.}.
Unresolved nuclear components were fit with our optimal Moffat PSF,
whereas resolved nuclear components were modeled with (PSF-convolved) narrow Gaussian functions 
(\citealt{wadadekar1999,ravindranath2001,peng2002} and \citealt{gadotti2008} discuss the importance of fitting nuclear components). \\
Rings were identified from galaxy images and unsharp masks, and modeled with symmetric Gaussian ring profiles 
(e.g.~\citealt{sheth2010,kim2014}). \\
As an illustration, we consider the galaxy NGC 2974, 
a spiral galaxy which has been misclassified as an elliptical galaxy in the RC3 catalog \citep{RC3}. 
This galaxy hosts a Seyfert AGN \citep{veroncettyveron2006} and filamentary dust in its center \citep{tran2001}.
NGC 2974 is classified as a fast rotator by the ATLAS$^{\rm 3D}$ survey, 
and indeed the velocity map obtained by the SLUGGS survey shows that the galaxy kinematics is rotation-dominated 
well beyond three effective radii ($R>150''$), 
as expected from a large-scale disk. 
From an inspection of the unsharp mask, we identified a ring at $R \sim 50''$, 
which might be a residual of two tightly wound spiral arms,
and an elongated bar-like component within $R \lesssim 30''$,
that is in addition to the more spherical bulge 
and produces a peak in the ellipticity and position angle profiles at $R \sim 20''$.
Our 1D galaxy decomposition for NGC 2974 (Figure \ref{fig:n2974ex}) consists of a S\'ersic bulge, an exponential large-scale disk,
a Ferrer bar, a Gaussian nuclear component (AGN) and a Gaussian ring.
Although the ring is extremely faint, it is important to account for it in the galaxy decomposition. 
A model without the ring component results in a ``steeper'' exponential profile for the disk 
(i.e.~the exponential model has a smaller scale length and a brighter central surface brightness) 
and produces bad residual structures within $R \lesssim 40''$.
Our best-fit model returns a $3.6~\rm \mu m$ bulge major-axis effective radius $R_{\rm e,sph}^{\rm maj} = 8.3\rm~arcsec$, 
equivalent-axis S\'ersic index $n_{\rm sph}^{\rm eq} = 1.2$ and apparent magnitude $m_{\rm sph} = 8.65~\rm mag$.
\cite{sani2011} modeled NGC 2974 with a S\'ersic bulge and a Gaussian nuclear component (AGN), 
but did not account for the large-scale disk.
From their best-fit 2D model, they obtained a three times larger 
$3.6~\rm \mu m$ bulge major-axis effective radius ($R_{\rm e,sph}^{\rm maj} = 27.2\rm~arcsec$), 
a 2.5 times larger S\'ersic index ($n_{\rm sph} = 3$),
and a significantly brighter apparent magnitude ($m_{\rm sph} = 7.28~\rm mag$).

\subsection{2D fits}
\label{sec:2d}
\begin{figure}[h]
\begin{center}
\includegraphics[width=\columnwidth]{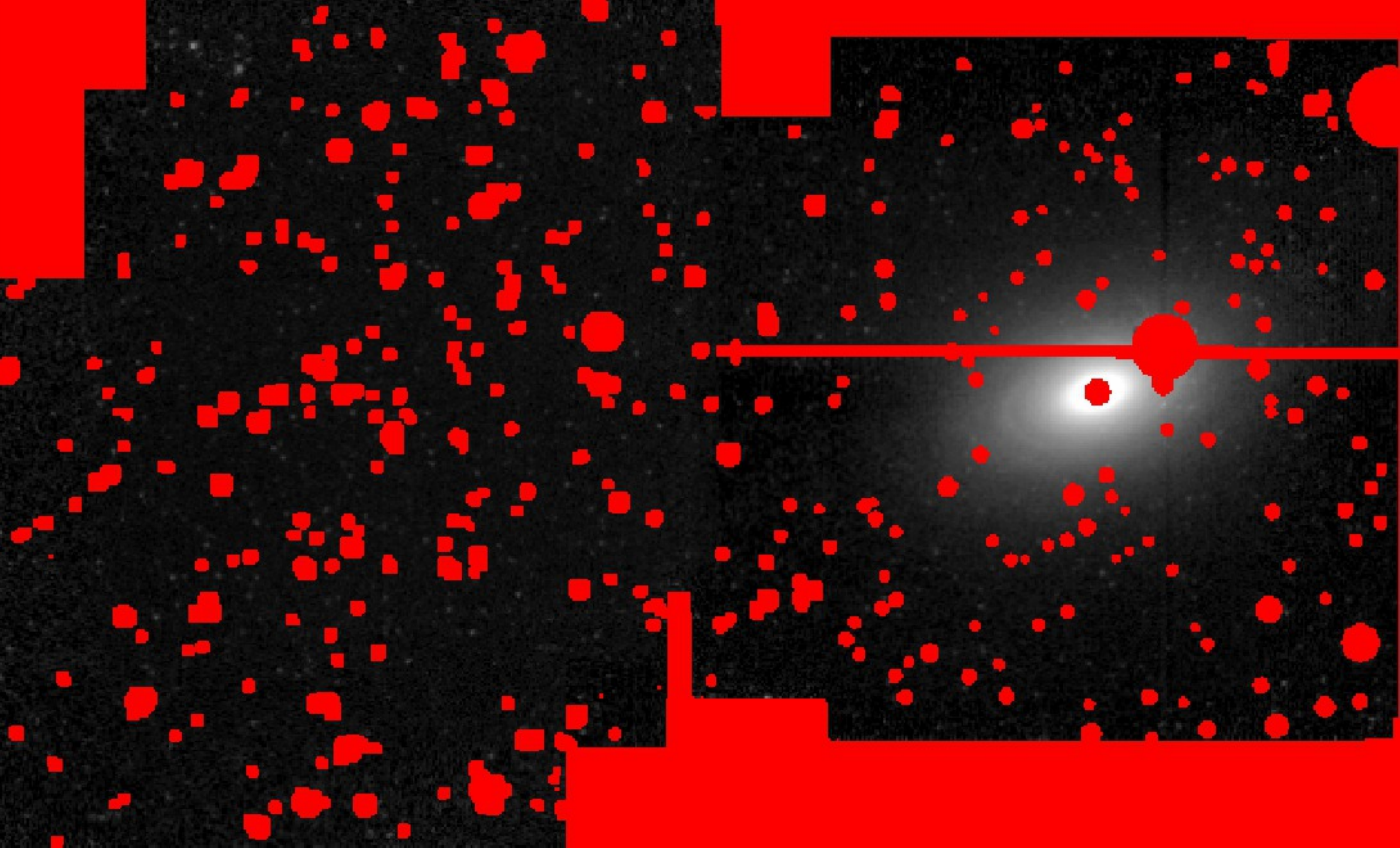}
\includegraphics[width=\columnwidth]{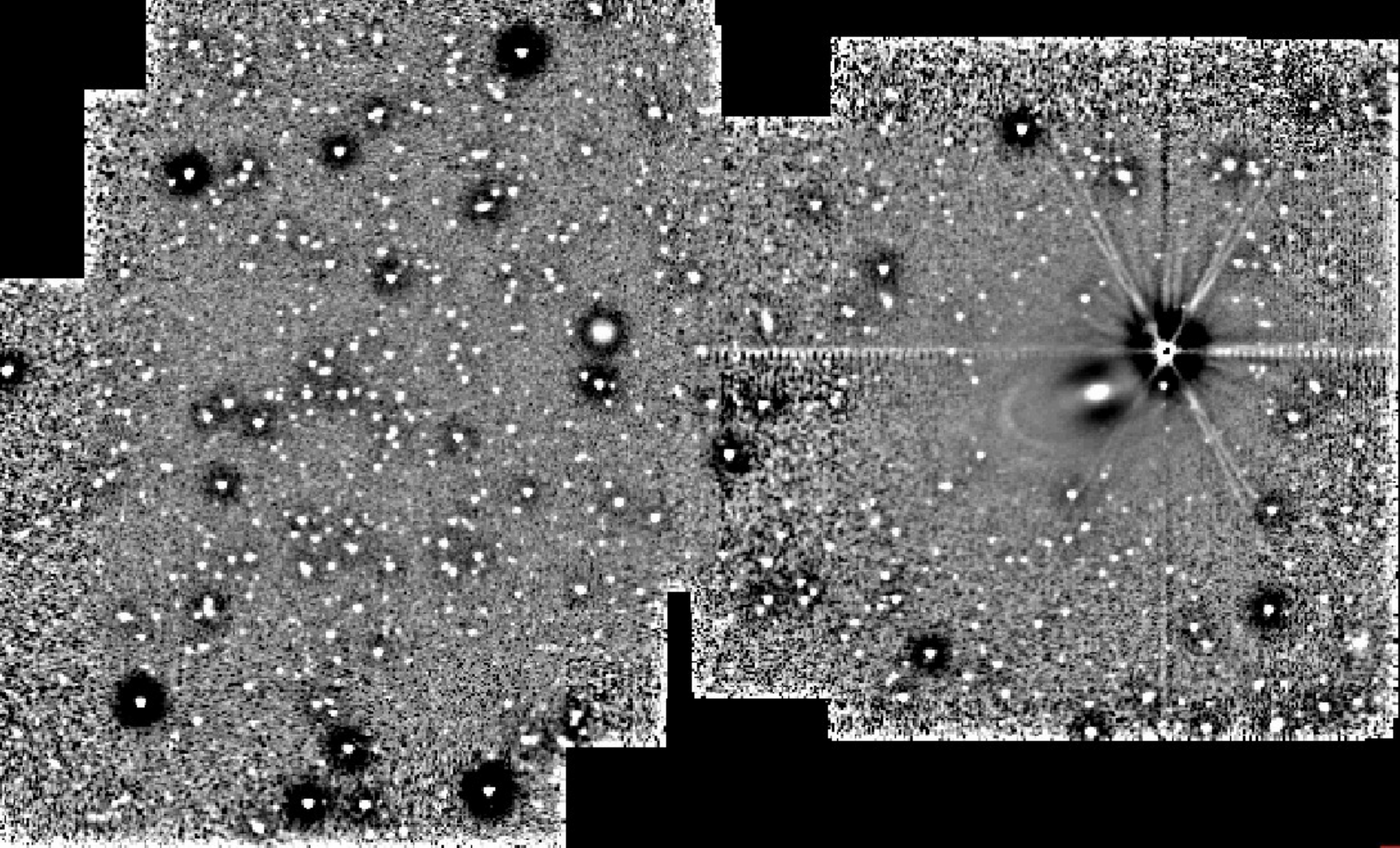}
\includegraphics[width=\columnwidth]{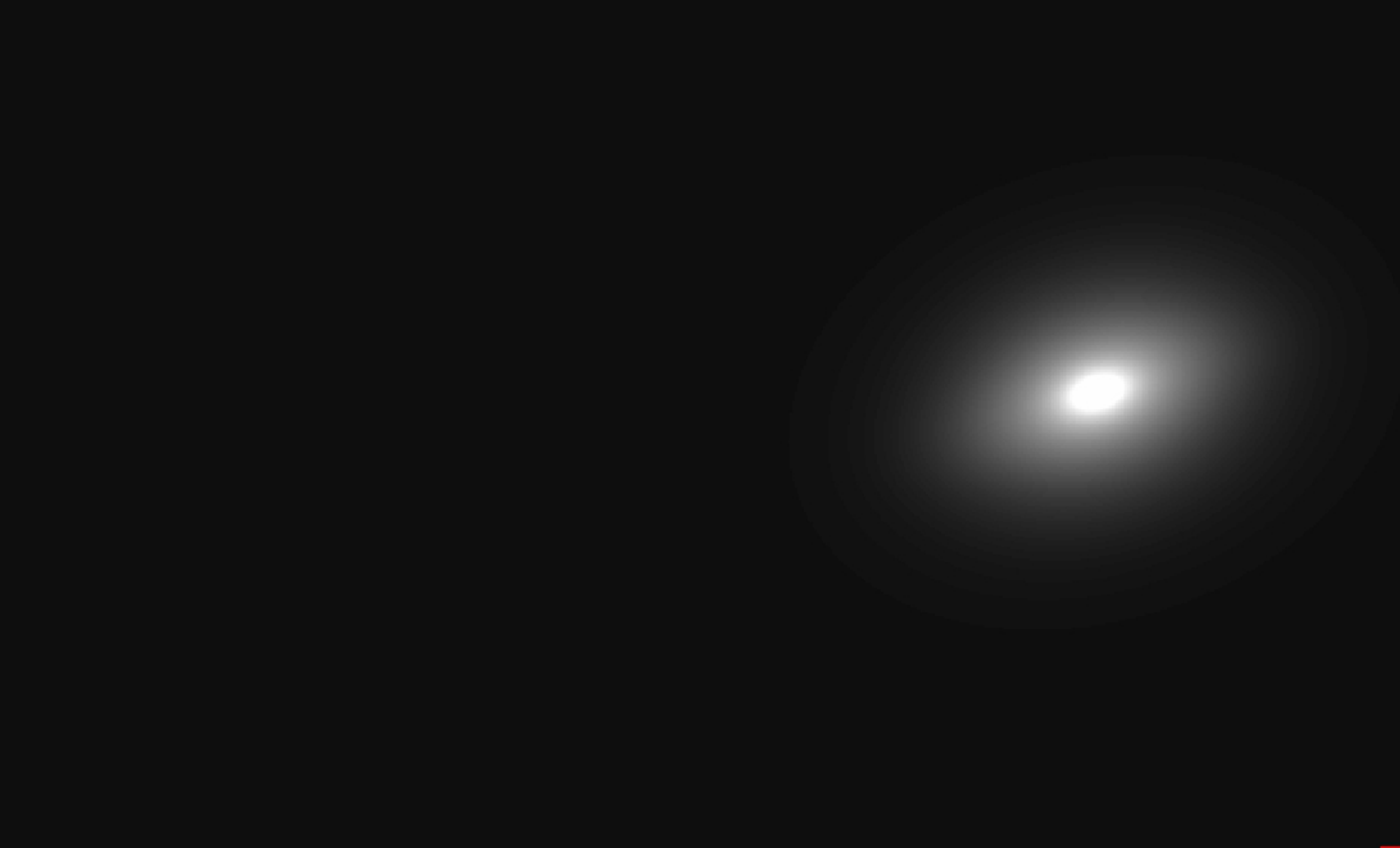}
\includegraphics[width=\columnwidth]{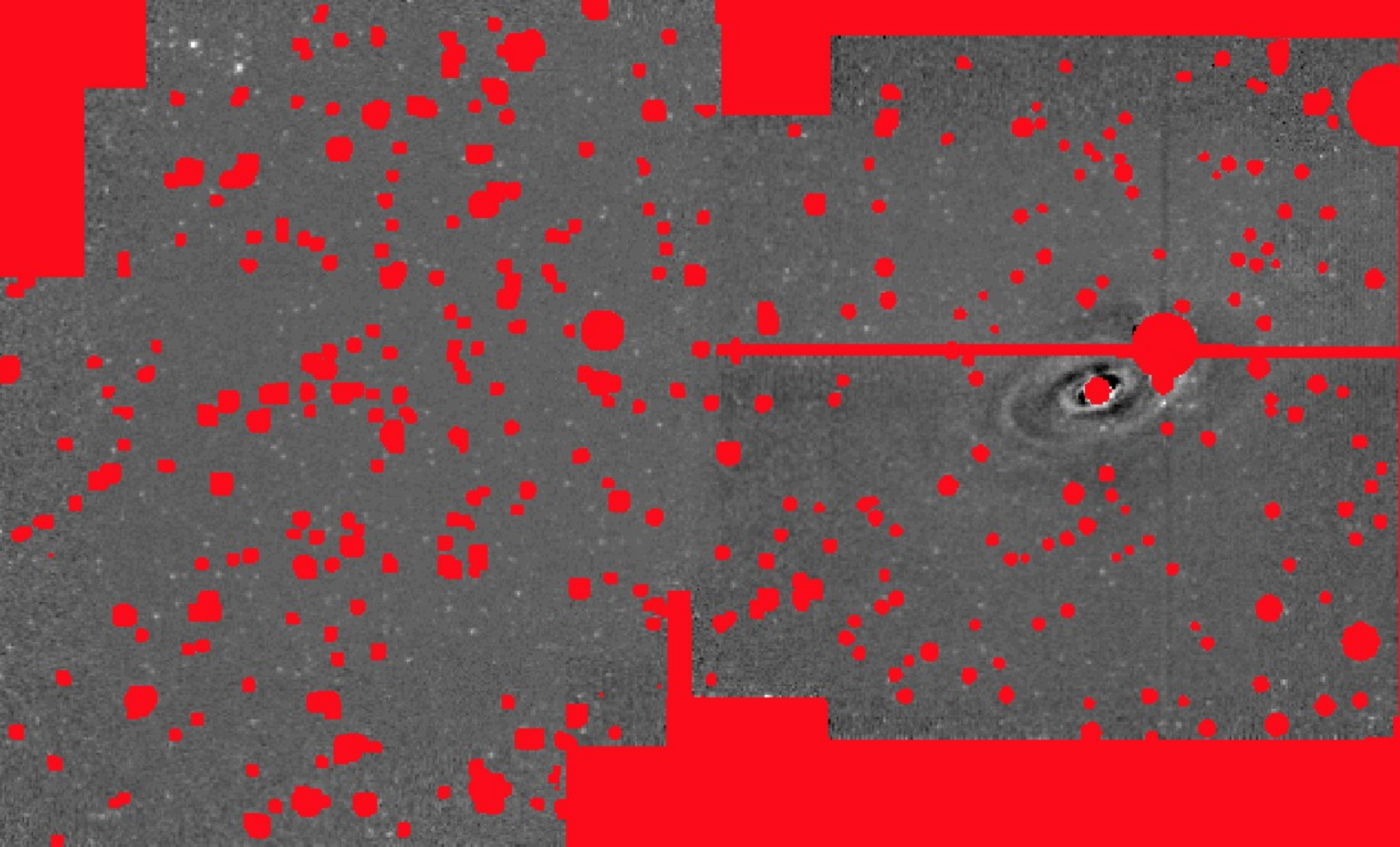}
\caption{Image of the galaxy NGC 2974 with its mask (top panel, on a logarithmic scale), 
unsharp mask (second panel, on a linear scale), 
best-fit 2D model (third panel, on a logarithmic scale), 
and residual image after the subtraction of the 2D model (bottom panel, on a linear scale; 
the residual patterns correspond to differences between the data and the model of less than 6\%).
The lefthand side of the mosaic allowed an accurate determination of the sky background level.}
\label{fig:n29742d}
\end{center}
\end{figure}

Two-dimensional decompositions were carried out using the software \textsc{Imfit} \citep{imfit}.
For each galaxy, we built a 2D model that was consistent with the corresponding 1D model in terms of number and type of components. 
The only difference between our 1D and 2D models pertains to the description of bars: 
because the Ferrer profile is not made available in \textsc{Imfit},
bars were fit with a 2D Gaussian function.  \\
The 2D decomposition of NGC 2974 is presented in Figure \ref{fig:n29742d}. 
The galaxy was modeled with a S\'ersic bulge, an exponential disk and a Gaussian bar.
The nuclear component was masked and the ring was not 
modeled\footnote{We built the 2D model first including and then omitting a gaussian ring component, 
but both models converged to the same solution, i.e.~the fit ``ignored'' the presence of the faint ring. 
This did not happen in the 1D decomposition because of the different weighting scheme used by the fitting routines.}. 
Our best-fit 2D model returns a bulge major-axis effective radius $R_{\rm e,sph}^{\rm maj} = 10.5~\rm arcsec$, 
a bulge S\'ersic index $n_{\rm sph} = 1.3$ and a $3.6~\rm \mu m$ bulge apparent magnitude $m_{\rm sph} = 8.39~\rm mag$, 
in fairly good agreement with our 1D decomposition. 
Fits and descriptions for the other galaxies are available online.

\section{Results}
\label{sec:res}
For each galaxy, after we identified its various components and built a model accordingly, 
we simultaneously performed a set of four 1D fits.
All four fits use a Moffat-convolved model.
Two fits use the major-axis surface brightness profile and the remaining two use the equivalent-axis surface brightness profile. 
For each of these pairs, we use a logarithmically sampled surface brightness profile 
and a linearly sampled surface brightness profile. 
Because our fitting routine intentionally does not employ an error weighting scheme on the data points that constitute the surface brightness profile, 
a fit to a logarithmically sampled profile puts more weight on the inner region of the galaxy 
and poorly constrains the outskirts.
On the other hand, a fit to a linearly sampled surface brightness profile equally treats inner and outer regions, 
but is more susceptible to sky-background subtraction issues. \\ 
We found that the fits are, in general, more sensitive to the choice of the initial parameters 
when using logarithmically sampled profiles than linearly sampled profiles. 
In addition, a visual examination of the residuals revealed that the quality of the fit within one galaxy effective radius 
is superior when tighter constraints are put on the galaxy outskirts.
In other words, the better quality of the residuals led us to prefer the fits that use linearly sampled surface brightness profiles,
although the results were ususally very similar,as might be expected. \\
Among the initial sample of 75 galaxies, we did not attempt to model 3 galaxies: M32, NGC 4486A and the Circinus Galaxy. 
The first two have been stripped by their massive companions, and thus have uncertain morphology. 
The Circinus Galaxy lies at only 4 degrees from the Galactic Plane, 
therefore its image mosaic is contaminated by a large number of foreground stars.
Of the remaining 72 galaxies, 
we obtained satisfactory 1D decompositions for 66, 
whereas the models of 6 galaxies were judged not reliable and were thus excluded. 
We also performed reliable 2D decompositions for 31 galaxies. \\
A galaxy-by-galaxy comparison between our best-fit models and those from the previous literature 
helped identify the optimal decompositions and past problems. 
We compared our best-fit models with those of \cite{grahamdriver2007}, \cite{sani2011}, \cite{beifiori2012}, 
\cite{vika2012} and \cite{lasker2014data}. 
We also considered the best-fit models of \cite{laurikainen2010} and \cite{rusli2013} because,   
although they did not specifically deal with black hole -- galaxy scaling relations, 
their galaxy samples significantly overlap with ours. \\
Table \ref{tab:fitres} lists the results from both the 1D and 2D fits.

\begin{table*}                                        
\begin{center}                                        
\caption{Results of galaxy decompositions.}                        
\begin{tabular}{lllllllllllll}                           
\hline                                                
 & 
\multicolumn{3}{l}{{\bf 1D Major-axis}} &                   
\multicolumn{5}{l}{{\bf 1D Equivalent-axis }} &                   
 & 
\multicolumn{3}{l}{{\bf 2D }} \\                    
\multicolumn{1}{l}{{\bf Galaxy }} &                   
\multicolumn{1}{l}{{\bf $\bm{R_{\rm e}}$ }} &                 
\multicolumn{1}{l}{{\bf $\bm{\mu_{\rm e}}$ }} &  
\multicolumn{1}{l}{{\bf $\bm{n}$ }} &			  
\multicolumn{1}{l}{{\bf $\bm{R_{\rm e}}$ }} &			  
\multicolumn{1}{l}{{\bf $\bm{\mu_{\rm e}}$ }} &			  
\multicolumn{1}{l}{{\bf $\bm{n}$ }} &			  
\multicolumn{1}{l}{{\bf $\bm{m_{\rm sph}}$ }} &			  
\multicolumn{1}{l}{{\bf $\bm{m_{\rm gal}}$ }} &			  
\multicolumn{1}{l}{{\bf Q.F. }} &			  
\multicolumn{1}{l}{{\bf $\bm{R_{\rm e}}$ }} &			  
\multicolumn{1}{l}{{\bf $\bm{n}$ }} &                   
\multicolumn{1}{l}{{\bf $\bm{m_{\rm sph}}$ }} \\                
\multicolumn{1}{l}{(1)} &                             
\multicolumn{1}{l}{(2)} &                             
\multicolumn{1}{l}{(3)} &                             
\multicolumn{1}{l}{(4)} &                             
\multicolumn{1}{l}{(5)} &                             
\multicolumn{1}{l}{(6)} &                             
\multicolumn{1}{l}{(7)} &                             
\multicolumn{1}{l}{(8)} &                             
\multicolumn{1}{l}{(9)} &                             
\multicolumn{1}{l}{(10)} &                             
\multicolumn{1}{l}{(11)} &                             
\multicolumn{1}{l}{(12)} &                             
\multicolumn{1}{l}{(13)} \\                         
\hline                                                
Circinus   \quad &   -- &   -- &   -- \quad \quad &   -- &   -- &   -- &   -- &   -- \quad \quad & 
 -- \quad \quad & 
 -- &   -- &   --    \\ 
IC 1459  \quad &  $63.1$  &  $18.49$  &  $6.6$  \quad \quad &  $57.3$  &  $18.59$  &  $7.0$  &  $6.11$  &  $6.11$  \quad \quad &  $1$  \quad \quad &  $87.5$  &  $8.3$  &  $6.04$  \\ 
IC 2560  \quad &  $6.1$  &  $16.58$  &  $0.8$  \quad \quad &  $4.5$  &  $16.48$  &  $0.6$  &  $10.77$  &  $8.29$  \quad \quad &  $2$  \quad \quad &   -- &   -- &   --    \\ 
IC 4296  \quad &  $65.9$  &  $19.32$  &  $5.8$  \quad \quad &  $68.1$  &  $19.48$  &  $6.2$  &  $6.70$  &  $6.70$  \quad \quad &  $1$  \quad \quad &  $82.3$  &  $6.6$  &  $6.66$  \\ 
M104  \quad &  $11.0$  &  $14.63$  &  $5.8$  \quad \quad &  $19.6$  &  $15.78$  &  $3.7$  &  $5.98$  &  $4.68$  \quad \quad &  $2$  \quad \quad &   -- &   -- &   --    \\ 
M105  \quad &  $57.2$  &  $17.93$  &  $5.2$  \quad \quad &  $50.9$  &  $17.84$  &  $5.3$  &  $5.77$  &  $5.77$  \quad \quad &  $2$  \quad \quad &  $73.6$  &  $7.0$  &  $5.62$  \\ 
M106  \quad &  $15.3$  &  $16.11$  &  $2.0$  \quad \quad &  $8.3$  &  $15.57$  &  $1.2$  &  $8.18$  &  $5.24$  \quad \quad &  $1$  \quad \quad &   -- &   -- &   --    \\ 
M31  \quad &  $418.6$  &  $16.80$  &  $2.2$  \quad \quad &  $173.6$  &  $15.63$  &  $1.3$  &  $1.61$  &  $-0.33$  \quad \quad &  $1$  \quad \quad &   -- &   -- &   --    \\ 
M32  \quad &   -- &   -- &   -- \quad \quad &   -- &   -- &   -- &   -- &   -- \quad \quad & 
 -- \quad \quad & 
 -- &   -- &   --    \\ 
M49  \quad &  $190.2$  &  $19.33$  &  $6.6$  \quad \quad &  $135.3$  &  $18.83$  &  $5.4$  &  $4.63$  &  $4.63$  \quad \quad &  $1$  \quad \quad &  $151.9$  &  $5.5$  &  $4.64$  \\ 
M59  \quad &  $48.0$  &  $18.02$  &  $5.5$  \quad \quad &  $90.9$  &  $19.67$  &  $8.8$  &  $6.07$  &  $5.98$  \quad \quad &  $1$  \quad \quad &   -- &   -- &   --    \\ 
M60  \quad &   -- &   -- &   -- \quad \quad &   -- &   -- &   -- &   -- &   -- \quad \quad & 
 -- \quad \quad & 
 -- &   -- &   --    \\ 
M64  \quad &  $3.8$  &  $13.38$  &  $0.8$  \quad \quad &  $4.3$  &  $13.78$  &  $1.4$  &  $7.78$  &  $5.08$  \quad \quad &  $1$  \quad \quad &   -- &   -- &   --    \\ 
M77  \quad &   -- &   -- &   -- \quad \quad &   -- &   -- &   -- &   -- &   -- \quad \quad & 
 -- \quad \quad & 
 -- &   -- &   --    \\ 
M81  \quad &  $31.0$  &  $15.22$  &  $1.7$  \quad \quad &  $33.2$  &  $15.55$  &  $2.1$  &  $4.89$  &  $3.47$  \quad \quad &  $3$  \quad \quad &   -- &   -- &   --    \\ 
M84  \quad &  $101.6$  &  $19.01$  &  $7.8$  \quad \quad &  $129.8$  &  $19.57$  &  $7.9$  &  $5.25$  &  $5.25$  \quad \quad &  $2$  \quad \quad &  $181.8$  &  $8.4$  &  $5.20$  \\ 
M87  \quad &  $203.0$  &  $19.87$  &  $10.0$  \quad \quad &  $87.1$  &  $18.26$  &  $5.9$  &  $4.97$  &  $4.97$  \quad \quad &  $2$  \quad \quad &  $88.3$  &  $4.3$  &  $5.11$  \\ 
M89  \quad &  $29.0$  &  $17.14$  &  $4.6$  \quad \quad &  $28.2$  &  $17.15$  &  $5.1$  &  $6.38$  &  $6.12$  \quad \quad &  $2$  \quad \quad &   -- &   -- &   --    \\ 
M94  \quad &  $11.4$  &  $13.73$  &  $0.9$  \quad \quad &  $8.4$  &  $13.50$  &  $1.1$  &  $6.14$  &  $4.86$  \quad \quad &  $1$  \quad \quad &   -- &   -- &   --    \\ 
M96  \quad &  $7.5$  &  $14.63$  &  $1.5$  \quad \quad &  $5.3$  &  $14.28$  &  $1.3$  &  $7.87$  &  $5.82$  \quad \quad &  $1$  \quad \quad &  $8.3$  &  $2.0$  &   $7.36$    \\ 
NGC 0253  \quad &   -- &   -- &   -- \quad \quad &   -- &   -- &   -- &   -- &   -- \quad \quad & 
 -- \quad \quad & 
 -- &   -- &   --    \\ 
NGC 0524  \quad &  $6.0$  &  $15.24$  &  $1.1$  \quad \quad &  $5.8$  &  $15.21$  &  $1.1$  &  $8.65$  &  $6.91$  \quad \quad &  $1$  \quad \quad &   -- &   -- &   --    \\ 
NGC 0821  \quad &  $36.5$  &  $18.40$  &  $5.3$  \quad \quad &  $18.9$  &  $17.83$  &  $6.1$  &  $7.85$  &  $7.59$  \quad \quad &  $3$  \quad \quad &  $33.8$  &  $2.5$  &  $7.78$  \\ 
NGC 1023  \quad &  $9.2$  &  $14.96$  &  $2.1$  \quad \quad &  $7.4$  &  $14.79$  &  $2.0$  &  $7.41$  &  $6.03$  \quad \quad &  $1$  \quad \quad &  $6.6$  &  $2.3$  &  $7.49$  \\ 
NGC 1300  \quad &  $9.9$  &  $17.62$  &  $3.8$  \quad \quad &  $8.1$  &  $17.41$  &  $3.6$  &  $9.52$  &  $7.42$  \quad \quad &  $2$  \quad \quad &   -- &   -- &   --    \\ 
NGC 1316  \quad &  $21.5$  &  $15.55$  &  $2.0$  \quad \quad &  $15.9$  &  $15.43$  &  $1.8$  &  $6.46$  &  $4.87$  \quad \quad &  $2$  \quad \quad &   -- &   -- &   --    \\ 
NGC 1332  \quad &  $34.7$  &  $17.44$  &  $5.1$  \quad \quad &  $18.0$  &  $16.47$  &  $3.7$  &  $6.85$  &  $6.79$  \quad \quad &  $3$  \quad \quad &   -- &   -- &   --    \\ 
NGC 1374  \quad &  $25.6$  &  $18.06$  &  $3.7$  \quad \quad &  $24.8$  &  $18.11$  &  $4.1$  &  $7.74$  &  $7.72$  \quad \quad &  $1$  \quad \quad &  $25.2$  &  $3.7$  &  $7.81$  \\ 
NGC 1399  \quad &  $405.2$  &  $21.80$  &  $10.0$  \quad \quad &  $338.1$  &  $21.53$  &  $10.0$  &  $5.01$  &  $4.98$  \quad \quad &  $1$  \quad \quad &   -- &   -- &   --    \\ 
NGC 2273  \quad &  $1.6$  &  $13.36$  &  $2.1$  \quad \quad &  $1.9$  &  $13.83$  &  $2.7$  &  $9.27$  &  $8.06$  \quad \quad &  $2$  \quad \quad &   -- &   -- &   --    \\ 
NGC 2549  \quad &  $6.1$  &  $15.57$  &  $2.3$  \quad \quad &  $3.1$  &  $14.54$  &  $1.5$  &  $9.20$  &  $7.85$  \quad \quad &  $1$  \quad \quad &  $5.6$  &  $2.1$  &  $8.76$  \\ 
NGC 2778  \quad &  $2.3$  &  $15.61$  &  $1.3$  \quad \quad &  $2.2$  &  $15.46$  &  $1.2$  &  $10.94$  &  $9.30$  \quad \quad &  $2$  \quad \quad &   -- &   -- &   --    \\ 
NGC 2787  \quad &  $4.8$  &  $14.86$  &  $1.1$  \quad \quad &  $3.3$  &  $14.62$  &  $1.3$  &  $9.21$  &  $7.04$  \quad \quad &  $2$  \quad \quad &   -- &   -- &   --    \\ 
NGC 2974  \quad &  $8.3$  &  $15.64$  &  $1.4$  \quad \quad &  $6.9$  &  $15.63$  &  $1.2$  &  $8.65$  &  $7.44$  \quad \quad &  $2$  \quad \quad &  $10.6$  &  $1.3$  &  $8.39$  \\ 
NGC 3079  \quad &  $6.8$  &  $14.47$  &  $1.3$  \quad \quad &  $4.3$  &  $14.48$  &  $1.1$  &  $8.57$  &  $7.13$  \quad \quad &  $2$  \quad \quad &   -- &   -- &   --    \\ 
NGC 3091  \quad &  $100.5$  &  $20.43$  &  $7.6$  \quad \quad &  $51.2$  &  $19.47$  &  $6.6$  &  $7.27$  &  $7.27$  \quad \quad &  $1$  \quad \quad &  $67.1$  &  $6.7$  &  $7.26$  \\ 
NGC 3115  \quad &  $43.6$  &  $16.67$  &  $4.4$  \quad \quad &  $34.4$  &  $16.85$  &  $5.1$  &  $5.65$  &  $5.47$  \quad \quad &  $1$  \quad \quad &   -- &   -- &   --    \\ 
NGC 3227  \quad &  $8.1$  &  $16.56$  &  $1.7$  \quad \quad &  $4.6$  &  $15.83$  &  $1.1$  &  $9.77$  &  $7.28$  \quad \quad &  $2$  \quad \quad &   -- &   -- &   --    \\ 
NGC 3245  \quad &  $4.4$  &  $14.96$  &  $2.9$  \quad \quad &  $2.4$  &  $14.00$  &  $1.7$  &  $9.11$  &  $7.66$  \quad \quad &  $1$  \quad \quad &  $1.9$  &  $1.8$  &  $9.19$  \\ 
NGC 3377  \quad &  $61.8$  &  $19.16$  &  $7.7$  \quad \quad &  $91.7$  &  $20.33$  &  $9.2$  &  $6.69$  &  $6.62$  \quad \quad &  $2$  \quad \quad &  $71.8$  &  $3.7$  &  $7.21$  \\ 
NGC 3384  \quad &  $5.5$  &  $14.21$  &  $1.6$  \quad \quad &  $5.6$  &  $14.56$  &  $1.8$  &  $7.83$  &  $6.52$  \quad \quad &  $1$  \quad \quad &   -- &   -- &   --    \\ 
NGC 3393  \quad &  $1.4$  &  $14.03$  &  $3.4$  \quad \quad &  $1.4$  &  $14.15$  &  $2.6$  &  $10.23$  &  $8.42$  \quad \quad &  $2$  \quad \quad &  $1.2$  &  $1.9$  &  $10.45$  \\ 
NGC 3414  \quad &  $28.0$  &  $18.10$  &  $4.8$  \quad \quad &  $25.5$  &  $18.08$  &  $4.5$  &  $7.60$  &  $7.53$  \quad \quad &  $1$  \quad \quad &   -- &   -- &   --    \\ 
NGC 3489  \quad &  $2.2$  &  $13.47$  &  $1.5$  \quad \quad &  $1.7$  &  $13.25$  &  $1.3$  &  $9.21$  &  $7.27$  \quad \quad &  $2$  \quad \quad &  $1.7$  &  $2.1$  &  $9.04$  \\ 
NGC 3585  \quad &  $105.0$  &  $19.13$  &  $5.2$  \quad \quad &  $86.3$  &  $19.24$  &  $6.3$  &  $5.93$  &  $5.90$  \quad \quad &  $2$  \quad \quad &   -- &   -- &   --    \\ 
NGC 3607  \quad &  $69.3$  &  $19.00$  &  $5.5$  \quad \quad &  $65.5$  &  $19.01$  &  $5.6$  &  $6.37$  &  $6.29$  \quad \quad &  $2$  \quad \quad &  $60.0$  &  $5.3$  &  $6.40$  \\ 
NGC 3608  \quad &  $47.5$  &  $18.93$  &  $5.2$  \quad \quad &  $43.4$  &  $19.00$  &  $5.7$  &  $7.25$  &  $7.25$  \quad \quad &  $2$  \quad \quad &  $62.0$  &  $7.0$  &  $7.15$  \\ 
NGC 3842  \quad &  $100.7$  &  $21.43$  &  $8.1$  \quad \quad &  $73.6$  &  $21.07$  &  $8.2$  &  $7.97$  &  $7.92$  \quad \quad &  $1$  \quad \quad &   -- &   -- &   --    \\ 
NGC 3998  \quad &  $5.8$  &  $15.15$  &  $1.2$  \quad \quad &  $4.8$  &  $14.63$  &  $1.3$  &  $8.37$  &  $7.15$  \quad \quad &  $3$  \quad \quad &   -- &   -- &   --    \\ 
NGC 4026  \quad &  $3.4$  &  $15.52$  &  $2.4$  \quad \quad &  $6.3$  &  $16.09$  &  $2.1$  &  $9.02$  &  $7.44$  \quad \quad &  $3$  \quad \quad &   -- &   -- &   --    \\ 
NGC 4151  \quad &  $7.6$  &  $15.50$  &  $1.4$  \quad \quad &  $6.8$  &  $15.26$  &  $1.9$  &  $8.10$  &  $7.06$  \quad \quad &  $2$  \quad \quad &   -- &   -- &   --    \\ 
NGC 4261  \quad &  $52.6$  &  $18.58$  &  $4.7$  \quad \quad &  $47.3$  &  $18.53$  &  $4.3$  &  $6.72$  &  $6.68$  \quad \quad &  $2$  \quad \quad &  $50.4$  &  $4.4$  &  $6.73$  \\ 
NGC 4291  \quad &  $15.0$  &  $17.14$  &  $4.2$  \quad \quad &  $15.4$  &  $17.51$  &  $5.9$  &  $7.99$  &  $7.99$  \quad \quad &  $2$  \quad \quad &  $20.8$  &  $7.7$  &  $7.91$  \\ 
NGC 4342  \quad &   -- &   -- &   -- \quad \quad &   -- &   -- &   -- &   -- &   -- \quad \quad & 
 -- \quad \quad & 
 -- &   -- &   --    \\ 
NGC 4388  \quad &  $4.6$  &  $15.89$  &  $0.6$  \quad \quad &  $4.2$  &  $15.86$  &  $1.3$  &  $9.89$  &  $7.66$  \quad \quad &  $3$  \quad \quad &   -- &   -- &   --    \\ 
NGC 4459  \quad &  $18.4$  &  $16.69$  &  $3.1$  \quad \quad &  $13.0$  &  $16.23$  &  $2.6$  &  $7.50$  &  $6.97$  \quad \quad &  $2$  \quad \quad &   -- &   -- &   --    \\ 
NGC 4473  \quad &  $45.9$  &  $17.93$  &  $2.3$  \quad \quad &  $36.9$  &  $18.10$  &  $2.9$  &  $7.04$  &  $6.82$  \quad \quad &  $2$  \quad \quad &  $49.8$  &  $3.0$  &  $7.03$  \\ 
NGC 4486A  \quad &   -- &   -- &   -- \quad \quad &   -- &   -- &   -- &   -- &   -- \quad \quad & 
 -- \quad \quad & 
 -- &   -- &   --    \\ 
NGC 4564  \quad &  $5.0$  &  $15.23$  &  $2.6$  \quad \quad &  $6.0$  &  $15.65$  &  $3.0$  &  $8.52$  &  $7.83$  \quad \quad &  $1$  \quad \quad &   -- &   -- &   --    \\ 
NGC 4596  \quad &  $6.6$  &  $15.93$  &  $2.7$  \quad \quad &  $9.0$  &  $16.44$  &  $3.0$  &  $8.43$  &  $6.98$  \quad \quad &  $1$  \quad \quad &   -- &   -- &   --    \\ 
NGC 4697  \quad &  $239.3$  &  $20.62$  &  $7.2$  \quad \quad &  $226.4$  &  $20.90$  &  $6.7$  &  $5.47$  &  $5.34$  \quad \quad &  $3$  \quad \quad &  $121.4$  &  $5.0$  &  $5.72$  \\ 
NGC 4889  \quad &  $119.7$  &  $21.01$  &  $8.1$  \quad \quad &  $60.8$  &  $20.11$  &  $6.8$  &  $7.53$  &  $7.53$  \quad \quad &  $1$  \quad \quad &  $104.3$  &  $7.8$  &  $7.43$  \\ 
NGC 4945  \quad &  $13.9$  &  $14.95$  &  $1.4$  \quad \quad &  $9.5$  &  $14.78$  &  $1.7$  &  $6.94$  &  $4.11$  \quad \quad &  $2$  \quad \quad &  $16.2$  &  $0.8$  &  $7.07$  \\ 
NGC 5077  \quad &  $23.5$  &  $17.67$  &  $4.2$  \quad \quad &  $23.0$  &  $18.01$  &  $5.7$  &  $7.62$  &  $7.62$  \quad \quad &  $1$  \quad \quad &  $30.5$  &  $6.8$  &  $7.57$  \\ 
NGC 5128  \quad &  $61.3$  &  $15.73$  &  $1.2$  \quad \quad &  $60.8$  &  $16.01$  &  $2.2$  &  $4.01$  &  $2.93$  \quad \quad &  $3$  \quad \quad &   -- &   -- &   --    \\ 
NGC 5576  \quad &  $61.5$  &  $19.41$  &  $3.3$  \quad \quad &  $49.3$  &  $19.34$  &  $3.7$  &  $7.53$  &  $7.53$  \quad \quad &  $1$  \quad \quad &  $45.9$  &  $8.3$  &  $7.19$  \\ 
NGC 5813  \quad &   -- &   -- &   -- \quad \quad &   -- &   -- &   -- &   -- &   -- \quad \quad & 
 -- \quad \quad & 
 -- &   -- &   --    \\ 
\hline         
\end{tabular}   
\label{tab:fitres} 
\end{center}    
\end{table*}    

\begin{table*}                                        
\begin{center}                                        
\begin{tabular}{lllllllllllll}                           
\hline                                                
 & 
\multicolumn{3}{l}{{\bf 1D Major-axis}} &                   
\multicolumn{5}{l}{{\bf 1D Equivalent-axis }} &                   
 & 
\multicolumn{3}{l}{{\bf 2D }} \\                    
\multicolumn{1}{l}{{\bf Galaxy }} &                   
\multicolumn{1}{l}{{\bf $\bm{R_{\rm e}}$ }} &                 
\multicolumn{1}{l}{{\bf $\bm{\mu_{\rm e}}$ }} &  
\multicolumn{1}{l}{{\bf $\bm{n}$ }} &			  
\multicolumn{1}{l}{{\bf $\bm{R_{\rm e}}$ }} &			  
\multicolumn{1}{l}{{\bf $\bm{\mu_{\rm e}}$ }} &			  
\multicolumn{1}{l}{{\bf $\bm{n}$ }} &			  
\multicolumn{1}{l}{{\bf $\bm{m_{\rm sph}}$ }} &			  
\multicolumn{1}{l}{{\bf $\bm{m_{\rm gal}}$ }} &			  
\multicolumn{1}{l}{{\bf Q.F. }} &			  
\multicolumn{1}{l}{{\bf $\bm{R_{\rm e}}$ }} &			  
\multicolumn{1}{l}{{\bf $\bm{n}$ }} &                   
\multicolumn{1}{l}{{\bf $\bm{m_{\rm sph}}$ }} \\                
\multicolumn{1}{l}{(1)} &                             
\multicolumn{1}{l}{(2)} &                             
\multicolumn{1}{l}{(3)} &                             
\multicolumn{1}{l}{(4)} &                             
\multicolumn{1}{l}{(5)} &                             
\multicolumn{1}{l}{(6)} &                             
\multicolumn{1}{l}{(7)} &                             
\multicolumn{1}{l}{(8)} &                             
\multicolumn{1}{l}{(9)} &                             
\multicolumn{1}{l}{(10)} &                             
\multicolumn{1}{l}{(11)} &                             
\multicolumn{1}{l}{(12)} &                             
\multicolumn{1}{l}{(13)} \\                         
\hline                                                
NGC 5845  \quad &  $3.6$  &  $14.79$  &  $2.5$  \quad \quad &  $3.1$  &  $14.64$  &  $2.3$  &  $9.05$  &  $8.91$  \quad \quad &  $3$  \quad \quad &  $2.8$  &  $2.4$  &  $9.09$  \\ 
NGC 5846  \quad &  $105.1$  &  $19.67$  &  $6.4$  \quad \quad &  $83.4$  &  $19.28$  &  $5.7$  &  $6.10$  &  $6.10$  \quad \quad &  $2$  \quad \quad &  $85.1$  &  $5.2$  &  $6.14$  \\ 
NGC 6251  \quad &  $41.7$  &  $19.82$  &  $6.8$  \quad \quad &  $30.1$  &  $19.31$  &  $5.6$  &  $8.35$  &  $8.35$  \quad \quad &  $1$  \quad \quad &  $39.3$  &  $7.1$  &  $8.27$  \\ 
NGC 7052  \quad &  $59.4$  &  $19.38$  &  $4.2$  \quad \quad &  $37.0$  &  $19.19$  &  $5.6$  &  $7.79$  &  $7.79$  \quad \quad &  $1$  \quad \quad &  $36.2$  &  $4.0$  &  $8.09$  \\ 
NGC 7582  \quad &   -- &   -- &   -- \quad \quad &   -- &   -- &   -- &   -- &   -- \quad \quad & 
 -- \quad \quad & 
 -- &   -- &   --    \\ 
NGC 7619  \quad &  $63.2$  &  $19.53$  &  $5.3$  \quad \quad &  $58.0$  &  $19.55$  &  $5.2$  &  $7.21$  &  $7.15$  \quad \quad &  $2$  \quad \quad &   -- &   -- &   --    \\ 
NGC 7768  \quad &  $92.9$  &  $21.37$  &  $8.4$  \quad \quad &  $42.1$  &  $20.15$  &  $6.7$  &  $8.36$  &  $8.36$  \quad \quad &  $2$  \quad \quad &   -- &   -- &   --    \\ 
UGC 03789  \quad &  $1.8$  &  $15.26$  &  $1.9$  \quad \quad &  $2.4$  &  $15.39$  &  $1.4$  &  $10.65$  &  $9.22$  \quad \quad &  $3$  \quad \quad &   -- &   -- &   --    \\ 
\hline         
\end{tabular}
\tablecomments{
\emph{Column (1):} Galaxy name.                       
\emph{Column (2-4):} Effective radius (in units of $[\rm arcsec]$),  surface brightness at the effective radius (in units of $[\rm mag~arcsec^{-2}]$)  and S\'ersic index for 1D fits along the major-axis.
\emph{Column (5-9):} Effective radius (in units of $[\rm arcsec]$),  surface brightness at the effective radius (in units of $[\rm mag~arcsec^{-2}]$),  S\'ersic index,  spheroid apparent magnitude (in units of $[\rm mag]$)  and galaxy apparent magnitude (in units of $[\rm mag]$) for 1D fits along the equivalent-axis.
\emph{Column (10):} Quality flag of the 1D fits (see Section \ref{sec:err}). 
\emph{Column (11-13):}  Effective radius (in units of $[\rm arcsec]$),  S\'ersic index,  and spheroid apparent magnitude (in units of $[\rm mag]$) for 2D fits.  
}   
\end{center}    
\end{table*}

\subsection{1D versus 2D decompositions}
\begin{figure}[h]
\begin{center}
\includegraphics[width=\columnwidth]{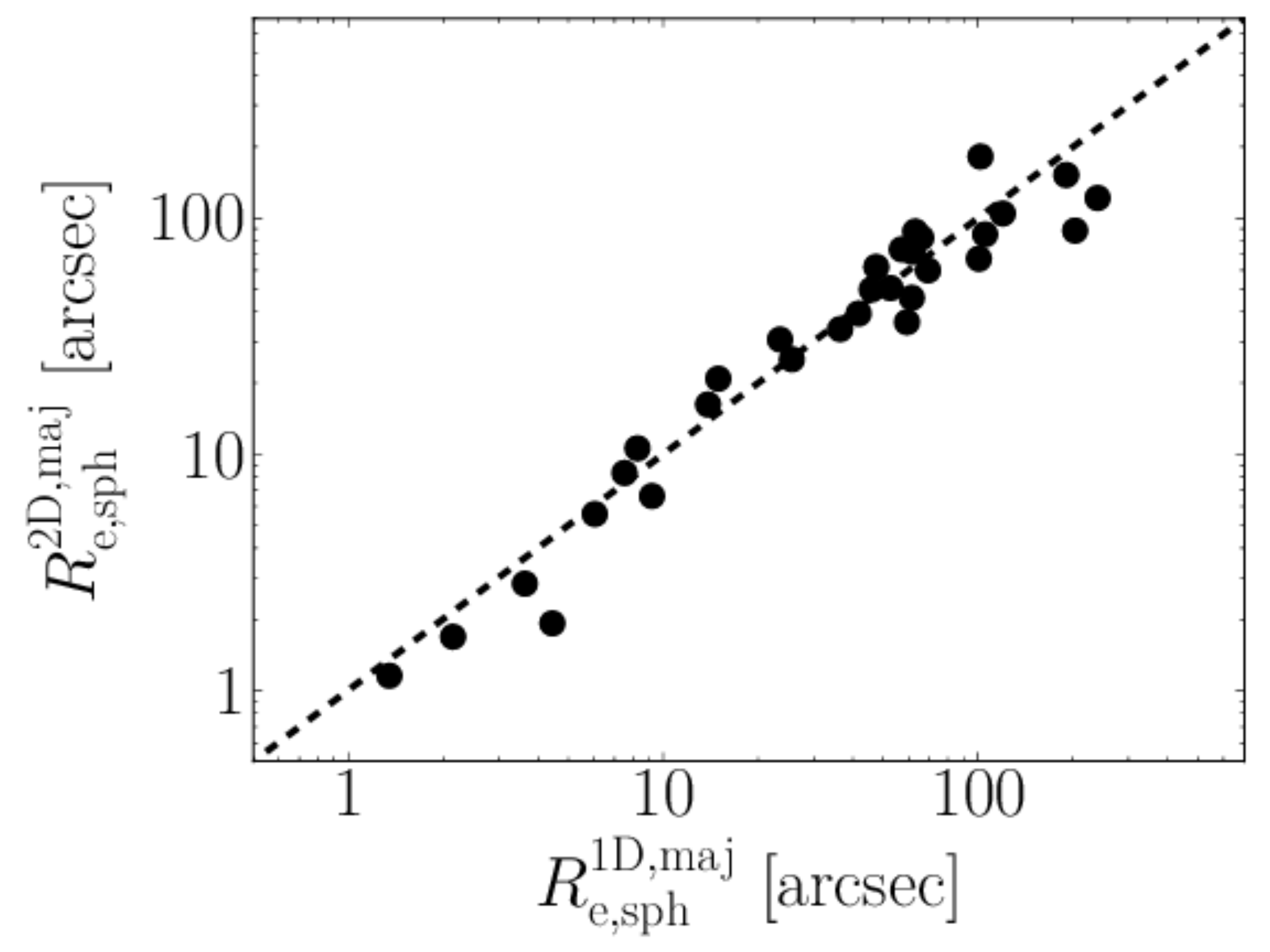} 
\caption{2D versus 1D major-axis measurements of the spheroid effective radii. 
The dashed line displays the 1:1 relation.}
\label{fig:re1d2d}
\end{center}
\end{figure}

\begin{figure}[h]
\begin{center}
\includegraphics[width=\columnwidth]{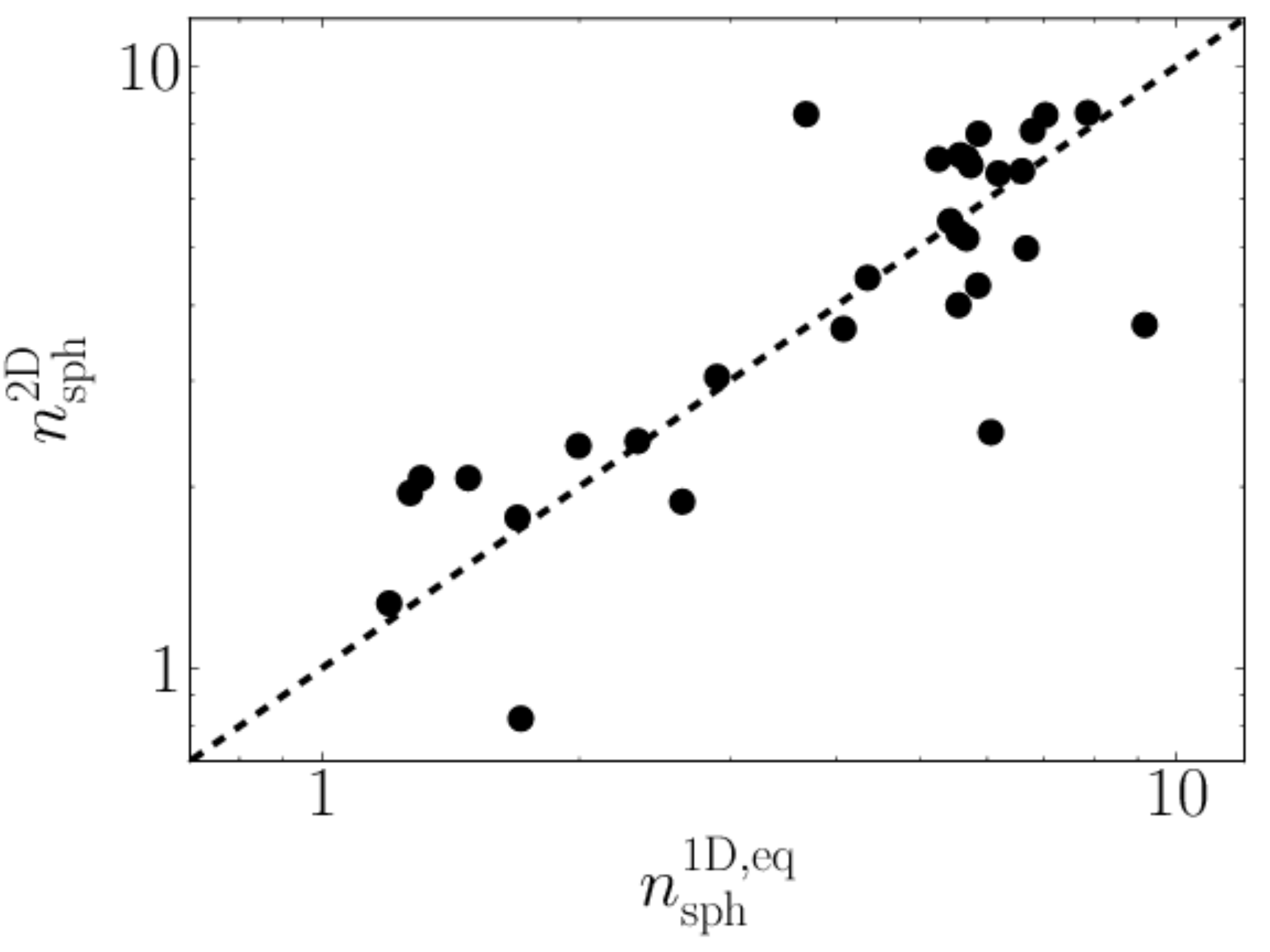} 
\caption{2D measurements of the spheroid S\'ersic indices roughly approximate the 1D equivalent-axis measurements. 
The dashed line displays the 1:1 relation.
The four most obvious outliers are NGC 821, NGC 3377, NGC 4945, and NGC 5576. 
Their individual cases are discussed in the Appendix. }
\label{fig:n1d2d}
\end{center}
\end{figure}

\begin{figure}[h]
\begin{center}
\includegraphics[width=\columnwidth]{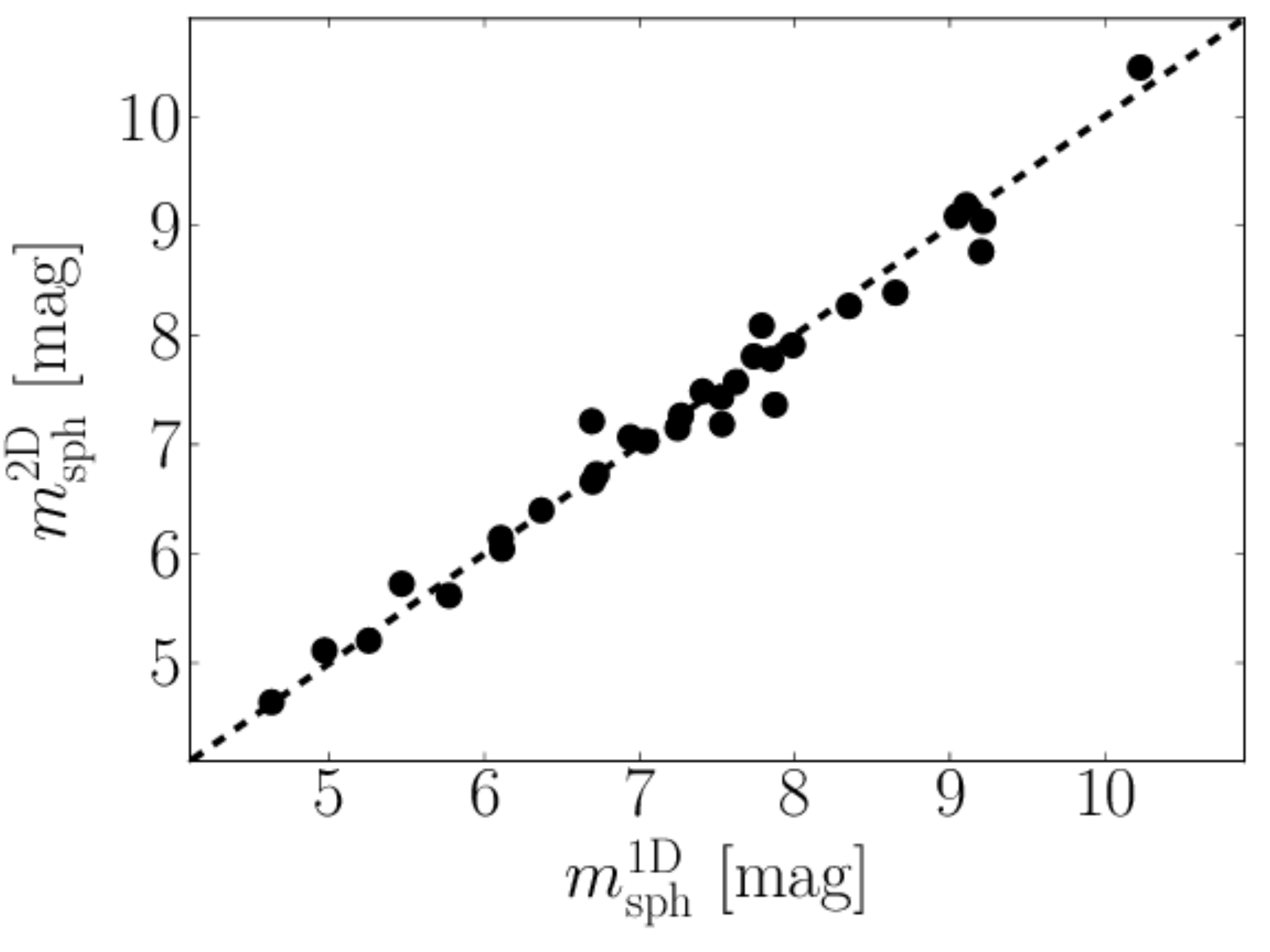} 
\caption{2D versus 1D measurements of the $3.6~\rm \mu m$ spheroid magnitudes. The dashed line displays the 1:1 relation.}
\label{fig:mag1d2d}
\end{center}
\end{figure}

Here we explore how 1D and 2D decompositions compare with each other. 
Readers not interested in our practical knowledge, 
having dealt with 1D and 2D techniques of galaxy modeling at the same time, can skip to Section \ref{sec:err}. 
We summarize our experience in the following points:
\begin{itemize}
\item A visual inspection of galaxy images and their unsharp masks is often not sufficient to accurately identify all of a galaxy's components. 
Weak bars and some embedded disks can easily be missed. 
Other (inclined) embedded disks can be confused with large-scale disks.
In this regards, the 1D isophotal analysis is extremely helpful. 
Local minima/maxima or abrupt changes in the ellipticity, position angle and fourth harmonic profiles contain precious information 
about a galaxy's constituents.
\item The ellipticity and position angle of triaxial spheroids can vary with radius.
The analytic functions used by 2D decomposition codes to fit galaxy components have fixed ellipticity and position angle, 
thus they cannot account for these radial gradients. 
This problem is overcome with 1D decomposition techniques 
because the ellipticity, and additionally the deviations from elliptical isophotes, 
are efficiently included in the equivalent-axis fit (see the Appendix of \citealt{ciambur2015}).
\item The interpretation of the residual surface brightness profile is often crucial to identify the optimal decomposition for a galaxy. 
In our experience, we found that interpreting 1D residuals was easier and more productive than 2D residuals.
\end{itemize}
Although we attempted 2D modeling for the 72 galaxies in our sample, 
more than half of the 2D decompositions were not successful or did not converge to a meaningful solution. 
This should serve as a tip to users of 2D fitting codes. 
When physically meaningful spheroid parameters are required, 
the output may not be reliable and should be inspected.
For the 31 galaxies that had successful 2D decompositions, 
we compare with their 1D parameters in Figures \ref{fig:re1d2d}, \ref{fig:n1d2d} and \ref{fig:mag1d2d}.
The agreement between 1D and 2D effective radii and magnitudes is remarkable, 
whereas a larger amount of scatter in the S\'ersic indices can be caused by the fact that 
2D measurements do not exactly correspond to 1D equivalent-axis measurements. 
No systematic effects are observed in any of these three plots, 
which indicates that 1D and 2D techniques of galaxy modeling -- when performed on the same galaxy -- 
can give consistent results. \\
In conclusion, since we found that the best-fit parameters do not depend on the decomposition method (1D or 2D) used, 
and given that we obtained more successful 1D decompositions than 2D, 
we will base our analysis on the results from the 1D fits.

\subsection{Parameter uncertainty}
\label{sec:err}
Estimating the uncertainties associated with the best-fit parameters of our 1D galaxy decompositions is not straightforward.
Monte Carlo simulations could be used for this purpose, 
but they would take into account only random errors and not unknown systematic errors.
Systematic errors include incorrect sky subtraction, 
inaccurate masking of contaminating sources, 
imprecise description of the PSF, 
erroneous choice of model components 
(for example, when failing to identify a galaxy sub-component and thus omitting it in the model, 
or when describing a galaxy sub-component with an inadequate function),
the radial extent of the surface brightness profile and its sampling.
These factors are not included in popular 2D fitting codes which report only the random errors associated with their fitted parameters.  
Moreover, when performing multi-component decomposition of high signal-to-noise images of nearby -- therefore well resolved -- galaxies, 
errors are dominated by systematics rather than Poisson noise.
For this reason, we decided to estimate the uncertainties of the spheroid best-fit parameters 
with a method that took into account systematic errors. 

\subsubsection{Goodness of the spheroid modeling}
For each of our fits, we calculated the associated $RMS$ scatter using:
\begin{equation}
\Delta = \sqrt{\frac{\sum_{\rm i=0}^{N} (\mu_{\rm i}^{\rm obs} - \mu_{\rm i}^{\rm mod})^2}{N_{\rm DOF}} } ,
\end{equation} 
where $N_{\rm DOF}$ is the number of degrees-of-freedom, 
$\mu_{\rm i}^{\rm obs}$ is the observed surface brightness 
and $\mu_{\rm i}^{\rm mod}$ is the model surface brightness at each data point $i$.
Although useful to evaluate the overall quality of a galaxy decomposition, 
the $RMS$ scatter alone cannot be used to assess the goodness of the fit for the spheroidal component only, 
unless the galaxy is a pure spheroid and has consequently been modeled with a single S\'ersic profile.
To illustrate this point with an example, 
one can imagine a situation in which a galaxy is thought to be made of a small bulge and a much more extended disk.
This galaxy is decomposed with a S\'ersic + exponential model.
The exponential function provides an excellent description of the light profile of the disk, 
whereas the S\'ersic function does not do the same for the bulge.
The residuals of the fit will then be flat and close to zero at large radii, 
where the emission of the disk dominates over that of the bulge,
while they will display significant departures at small radii, 
in correspondence with the poorly fit spheroidal component.
In the case of linear sampling, 
because the part of the surface brightness profile pertaining to the disk may contain more data points 
than the part pertaining to the bulge, 
the global $RMS$ scatter will be relatively small, 
but it obviously will not reflect the accuracy of the fit to the spheroidal component only. \\
A simple but powerful way to get a feeling of how precisely the global model and its spheroidal component have fared
is to look at the major- and equivalent-axis fits of each galaxy 
and visually inspect the structures of the residual surface brightness profile 
(i.e.~the second row in Figure \ref{fig:n2974ex}) within $\sim 1 - 2$ spheroid effective radii. 
We did this using a grade from 1 to 3, assigned according to the following criteria.
\begin{itemize}
\item [1)] A grade of 1 was given to the best fits, i.e.~fits that do not exhibit any of the problems listed below.
\item [2)] A grade of 2 was issued in the following cases. 
The residuals in correspondence with the radial extent of the spheroidal component are not randomly distributed around zero, 
being symptomatic of a S\'ersic model having a curvature (regulated by the S\'ersic index $n$) 
that does not quite match the real ``shape'' of the spheroid. 
When we identified, 
but were not able to model a galaxy sub-component with the same accuracy dedicated to the other components.
In the case of apparent inconsistencies between the model and the observed galaxy properties 
(e.g.~an embedded disk modeled with an exponential function, 
whose scale length does not quite match the size of the disk as expected from the ellipticity profile or the velocity map).
When the S\'ersic model used to describe the spheroidal component has a size -- as measured by the effective radius -- 
comparable to a few times the $FWHM$ of the PSF.
Galaxies in this category are reasonably well fit despite these issues.
\item [3)] A grade of 3 was assigned to the poorer and more anomalous fits, 
or those affected by an obvious degeneracy between the spheroid S\'ersic profile and the remaining model components 
(e.g.~when the spheroid S\'ersic index varies by as much as 50\% among the four different realizations of the fit, 
or when the output of the fit strongly depends on the choice of the initial parameters).
\end{itemize}
As a result, we classified 27 galaxies (38\% of the 72 galaxies for which we attempted a 1D decomposition) with grade 1, 
29 galaxies (40\%) with grade 2 and 10 galaxies (14\%) with grade 3.
Six galaxies could not be modelled.
We report the assigned grades in Table \ref{tab:fitres} (column 10).

\subsubsection{Uncertainties on $n_{\rm sph}$}
In Figure \ref{fig:compn}, for 58 galaxies, 
we compare the measurements of the spheroid S\'ersic index obtained by different authors 
with those obtained by us. 
For each galaxy, we computed the average value $\langle \log(n_{\rm sph}) \rangle$ of the available measurements, 
and we plot it against the scatter of the individual measurements around each spheroid's $\langle \log(n_{\rm sph}) \rangle$.
The measurements are heterogeneous, 
in the sense that they were obtained from 1D or 2D decompositions of data in different wavelengths, 
and they refer either to the major-axis, the equivalent-axis or some 2D average.
In Figure \ref{fig:compn}, the black histogram shows the distribution of the scatter around $\langle \log(n_{\rm sph}) \rangle$ 
for our measurements. 
38\% of this distribution lie within $-\sigma_1^- = -0.08~\rm dex$ and $+\sigma_1^+ = +0.06~\rm dex$, 
78\% lie within $-\sigma_2^- = -0.21~\rm dex$ and $+\sigma_2^+ = +0.17~\rm dex$,
and 92\% lie within $-\sigma_3^- = -0.25~\rm dex$ and $+\sigma_3^+ = +0.25~\rm dex$. 
We elect to use $\pm \sigma_1^\pm$, $\pm \sigma_2^\pm$ and $\pm \sigma_3^\pm$ as $1\sigma$ uncertainties 
for our measurements of $\log(n_{\rm sph})$ obtained from ``grade 1'', ``grade 2'' and ``grade 3'' fits, respectively. 
This means that, if the fit to a galaxy was classified as ``grade 1'', 
the (logarithmic) value of the spheroid S\'ersic index $\tilde{n}_{\rm sph}$ and associated uncertainties would be 
$\log(\tilde{n}_{\rm sph})^{+\sigma_1^+}_{-\sigma_1^-} = \log(\tilde{n}_{\rm sph})^{+0.06}_{-0.08}$; 
if the fit to that galaxy was classified as ``grade 2'', 
then one would have $\log(\tilde{n}_{\rm sph})^{+\sigma_2^+}_{-\sigma_2^-} = \log(\tilde{n}_{\rm sph})^{+0.17}_{-0.21}$, 
and so on.   

\subsubsection{Uncertainties on $R_{\rm e,sph}$}
The uncertainties on the spheroid effective radii were computed using the same methodology as employed for the S\'ersic indices. 
However, in Figure \ref{fig:compre} we include only major-axis measurements of $R_{\rm e,sph}$. 
The associated $1\sigma$ uncertainties for our measurements of $\log(R_{\rm e,sph})$ are 
$-\sigma_1^- = -0.11~\rm dex$ and $+\sigma_1^+ = +0.07~\rm dex$, 
$-\sigma_2^- = -0.30~\rm dex$ and $+\sigma_2^+ = +0.32~\rm dex$,
and $-\sigma_3^- = -0.39~\rm dex$ and $+\sigma_3^+ = +0.42~\rm dex$. 

\subsubsection{Uncertainties on $m_{\rm sph}$}
To estimate the uncertainties on the spheroid magnitudes, 
we compared (see Figure \ref{fig:compmag}) only those literature measurements coming from $K$-band or $3.6~\rm \mu m$ observations.
To do this, the $3.6~\rm \mu m$ magnitudes were converted into $K$-band magnitudes by applying an additive factor of $0.27\rm~mag$, 
which was estimated using the stellar population models of \cite{worthey1994},
assuming a 13 Gyr old single-burst stellar population with solar metallicity. 
The associated $1\sigma$ uncertainties for our measurements of $m_{\rm sph}$ are 
$-\sigma_1^- = -0.11~\rm mag$ and $+\sigma_1^+ = +0.18~\rm mag$, 
$-\sigma_2^- = -0.58~\rm mag$ and $+\sigma_2^+ = +0.66~\rm mag$,
and $-\sigma_3^- = -0.66~\rm mag$ and $+\sigma_3^+ = +0.88~\rm mag$. 

\subsubsection{Uncertainties on $\mu_{\rm e,sph}$}
As for the spheroid magnitudes, we estimated the uncertainties on the spheroid effective surface brightnesses $\mu_{\rm e,sph}$ 
by comparing only $K$-band or $3.6~\rm \mu m$ measurements (see Figure \ref{fig:compmue}), 
and accounting for the mean color difference of $0.27\rm~mag$. 
Not explicitly reported in the literature, effective surface brightnesses were calculated by us using:
\begin{multline}
\mu_{\rm e,sph} = m_{\rm sph} + 5\log \Bigl(R_{\rm e,sph}^{\rm maj}\sqrt{(b/a)_{\rm sph}} \Bigr) + \\
+ 2.5\log \Bigl[2\pi n_{\rm sph} e^{b_n} b_n^{-2n_{\rm sph}} \Gamma(2n_{\rm sph})\Bigr], 
\end{multline}
where $b_n$ and $\Gamma(2n_{\rm sph})$ are defined in the Appendix, 
and $(b/a)_{\rm sph}$ is the spheroid axis ratio.
While \cite{laurikainen2010} and \cite{sani2011} reported their estimates of $(b/a)_{\rm sph}$, 
\cite{vika2012} and \cite{lasker2014data} did not. 
For the last two studies, we used the values of $(b/a)_{\rm sph}$ reported by \cite{sani2011}.
The associated $1\sigma$ uncertainties for our measurements of $\mu_{\rm e,sph}$ are 
$-\sigma_1^- = -0.33~\rm mag$ and $+\sigma_1^+ = +0.57~\rm mag$, 
$-\sigma_2^- = -0.84~\rm mag$ and $+\sigma_2^+ = +1.59~\rm mag$,
and $-\sigma_3^- = -1.06~\rm mag$ and $+\sigma_3^+ = +1.74~\rm mag$. 

\subsubsection{Uncertainties on $\mu_{\rm 0,sph}$}
From the values of $\mu_{\rm e,sph}$, we derived the central surface brightnesses $\mu_{\rm 0,sph}$ from the equation
\begin{equation}
\mu_{\rm 0,sph} = \mu_{\rm e,sph} - \frac{2.5 b_n}{\ln(10)}.
\end{equation}
The associated $1\sigma$ uncertainties for our measurements of $\mu_{\rm 0,sph}$ are 
$-\sigma_1^- = -0.20~\rm mag$ and $+\sigma_1^+ = +0.64~\rm mag$, 
$-\sigma_2^- = -0.70~\rm mag$ and $+\sigma_2^+ = +1.24~\rm mag$,
and $-\sigma_3^- = -1.05~\rm mag$ and $+\sigma_3^+ = +1.57~\rm mag$.

\begin{figure}
\begin{center}
\includegraphics[width=1.1\columnwidth]{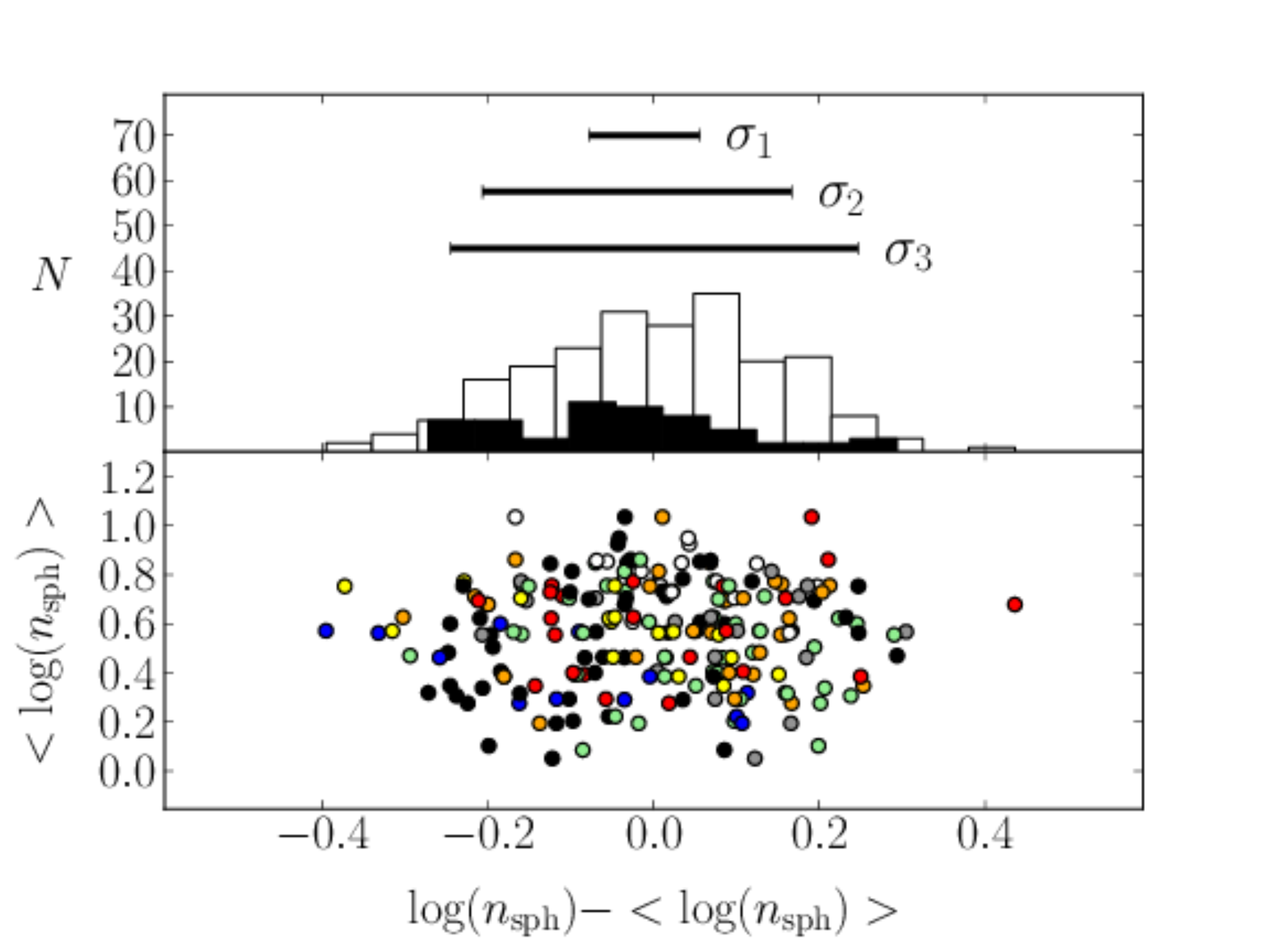} 
\caption{Bottom panel: 58 galaxies for which at least one measurement of the spheroid S\'ersic index $n_{\rm sph}$ is available 
from the literature (Table \ref{tab:lit}), in addition to that measured by us.
The average (logarithmic) value $\langle \log(n_{\rm sph}) \rangle$ is plotted against 
the difference between (the logarithm of) the individual measurements of a galaxy 
and the average (logarithmic) value for that same galaxy.
Each data point corresponds to an individual measurement from: 
\citeauthor{grahamdriver2007}~(\citeyear{grahamdriver2007}, red points);  
\citeauthor{laurikainen2010}~(\citeyear{laurikainen2010}, blue points); 
\citeauthor{sani2011}~(\citeyear{sani2011}, green points);  
\citeauthor{vika2012}~(\citeyear{vika2012}, yellow points);
\citeauthor{beifiori2012}~(\citeyear{beifiori2012}, gray points); 
\citeauthor{rusli2013}~(\citeyear{rusli2013}, white points); 
\citeauthor{lasker2014data}~(\citeyear{lasker2014data}, orange points). 
Black points are measurements obtained from the 1D fits presented in this work 
(using linearly sampled surface brightness profiles, along the major-axis).
In the top panel, the white histogram shows the distribution of $\log(n_{\rm sph}) - \langle \log(n_{\rm sph}) \rangle$ 
for all measurements, 
whereas the black histogram refers only to the measurements obtained by us.
The (asymmetric) error bars $\sigma_1$, $\sigma_2$ and $\sigma_3$ enclose 38\%, 78\% and 92\% 
of the black histogram, respectively, 
corresponding to our quality flags 1, 2 and 3 given in Table \ref{tab:fitres} (see Section \ref{sec:err} for details).
We consider these to be absolute upper limits to the uncertainty on our parameters 
given the care we have taken to minimise sources of systematic errors. }
\label{fig:compn}
\end{center}
\end{figure}

\begin{figure}
\begin{center}
\includegraphics[width=1.1\columnwidth]{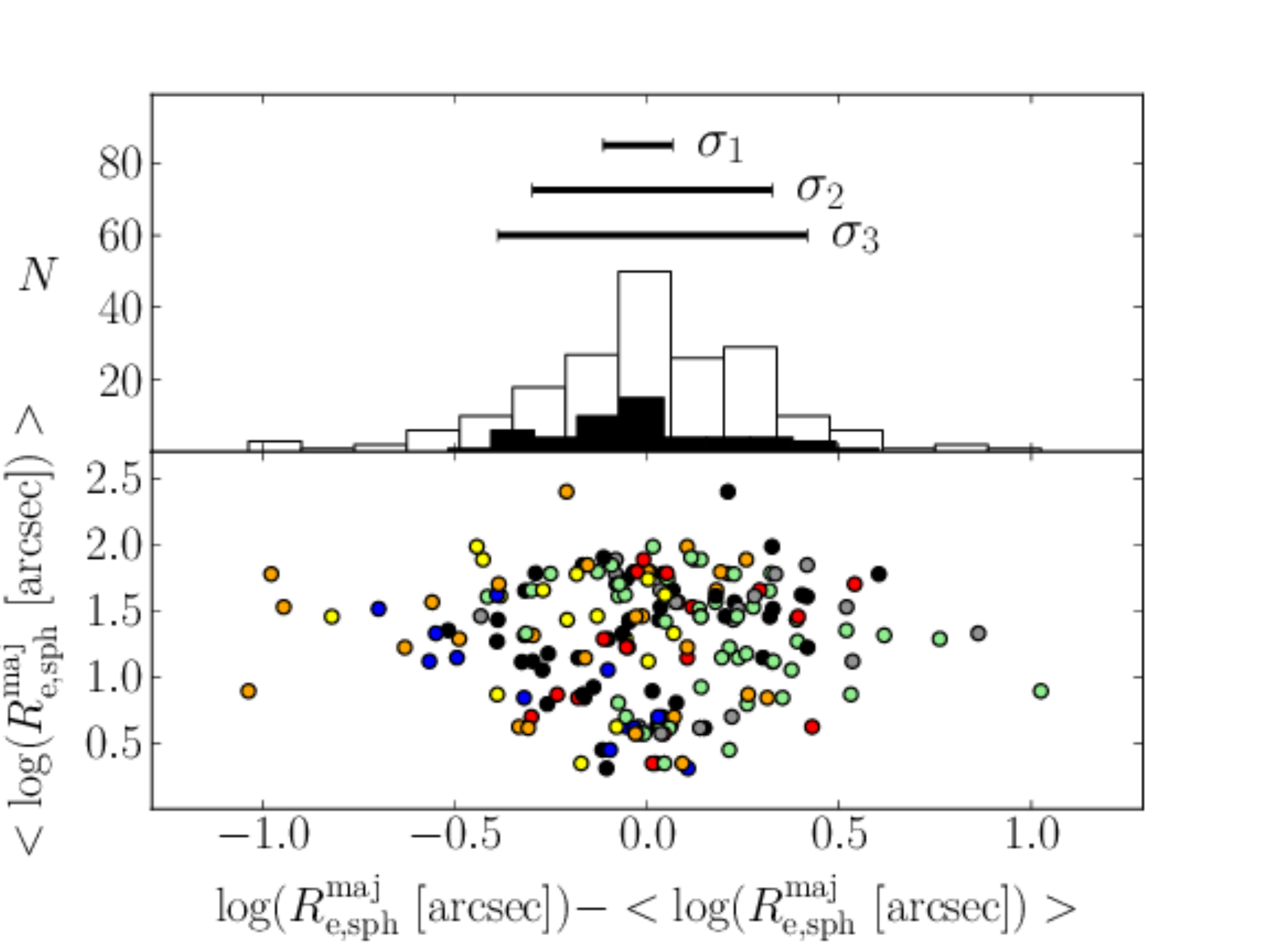} 
\caption{Bottom panel: 52 galaxies for which at least one measurement of the spheroid major-axis effective radius $R_{\rm e}^{\rm maj}$ is available 
from the literature (Table \ref{tab:lit}), in addition to that measured by us.
The average (logarithmic) value $\langle \log(R_{\rm e}^{\rm maj}) \rangle$ is plotted against 
the difference between (the logarithm of) the individual measurements of a galaxy 
and the average (logarithmic) value for that same galaxy.
See Figure \ref{fig:compn} for color description and explanation of the top panel.}
\label{fig:compre}
\end{center}
\end{figure}

\begin{figure}
\begin{center}
\includegraphics[width=1.1\columnwidth]{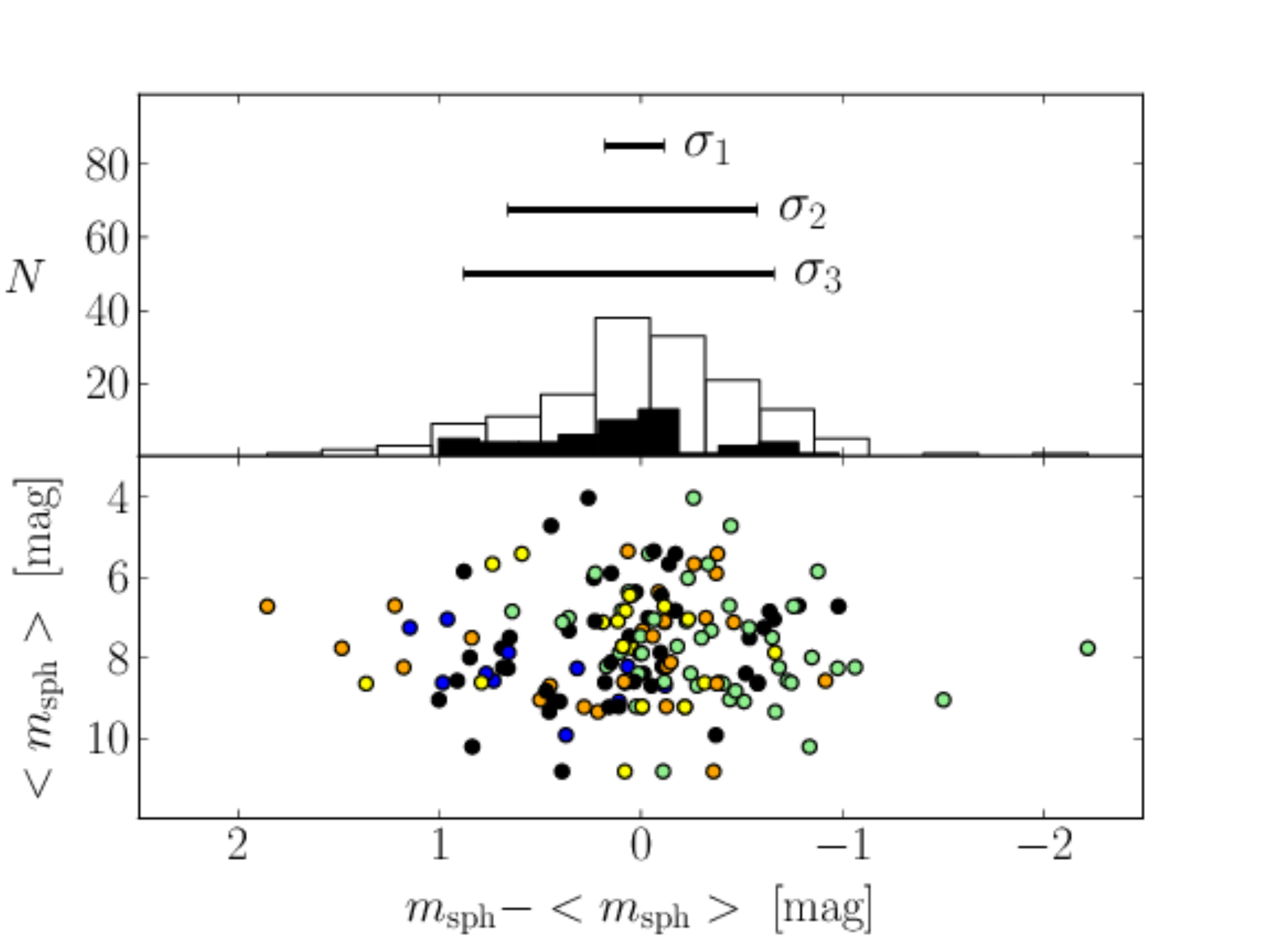} 
\caption{Bottom panel: 51 galaxies for which at least one measurement of the spheroid apparent magnitude $m_{\rm sph}$ 
-- either in the $K$-band or at $3.6~\rm \mu m$ -- is available from the literature (Table \ref{tab:lit}), in addition to that measured by us. 
The $3.6~\rm \mu m$ magnitudes were converted into $K$-band magnitudes (see Section \ref{sec:err} for details).
The average value $\langle m_{\rm sph} \rangle$ is plotted against 
the difference between the individual measurements of a galaxy 
and the average value for that same galaxy.
See Figure \ref{fig:compn} for color description and explanation of the top panel.}
\label{fig:compmag}
\end{center}
\end{figure}

\begin{figure}
\begin{center}
\includegraphics[width=1.1\columnwidth]{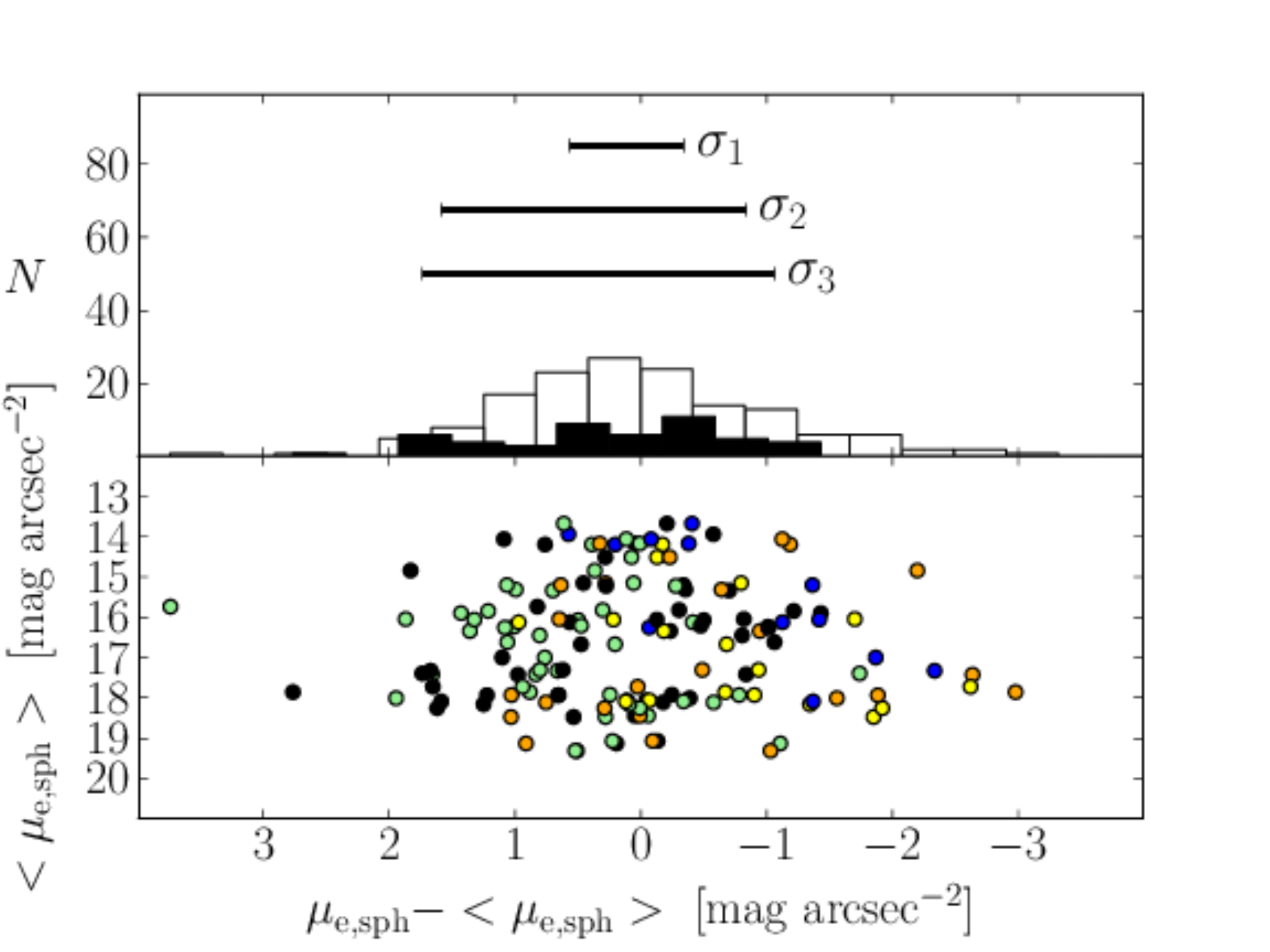} 
\caption{Bottom panel: 49 galaxies for which at least one measurement of the spheroid effective surface brightness $\mu_{\rm e,sph}$ 
-- either in the $K$-band or at $3.6~\rm \mu m$ -- is available from the literature (Table \ref{tab:lit}), in addition to that measured by us. 
The $K$-band magnitudes were converted into $3.6~\rm \mu m$ magnitudes (see Section \ref{sec:err} for details).
The average value $\langle \mu_{\rm e,sph} \rangle$ from all fits to a galaxy, 
not to be confused with the mean effective surface brightness within $R_{\rm e,sph}$, is plotted against 
the difference between the individual measurements of a galaxy 
and the average value for that same galaxy.
See Figure \ref{fig:compn} for color description and explanation of the top panel.}
\label{fig:compmue}
\end{center}
\end{figure}

\begin{figure}
\begin{center}
\includegraphics[width=1.1\columnwidth]{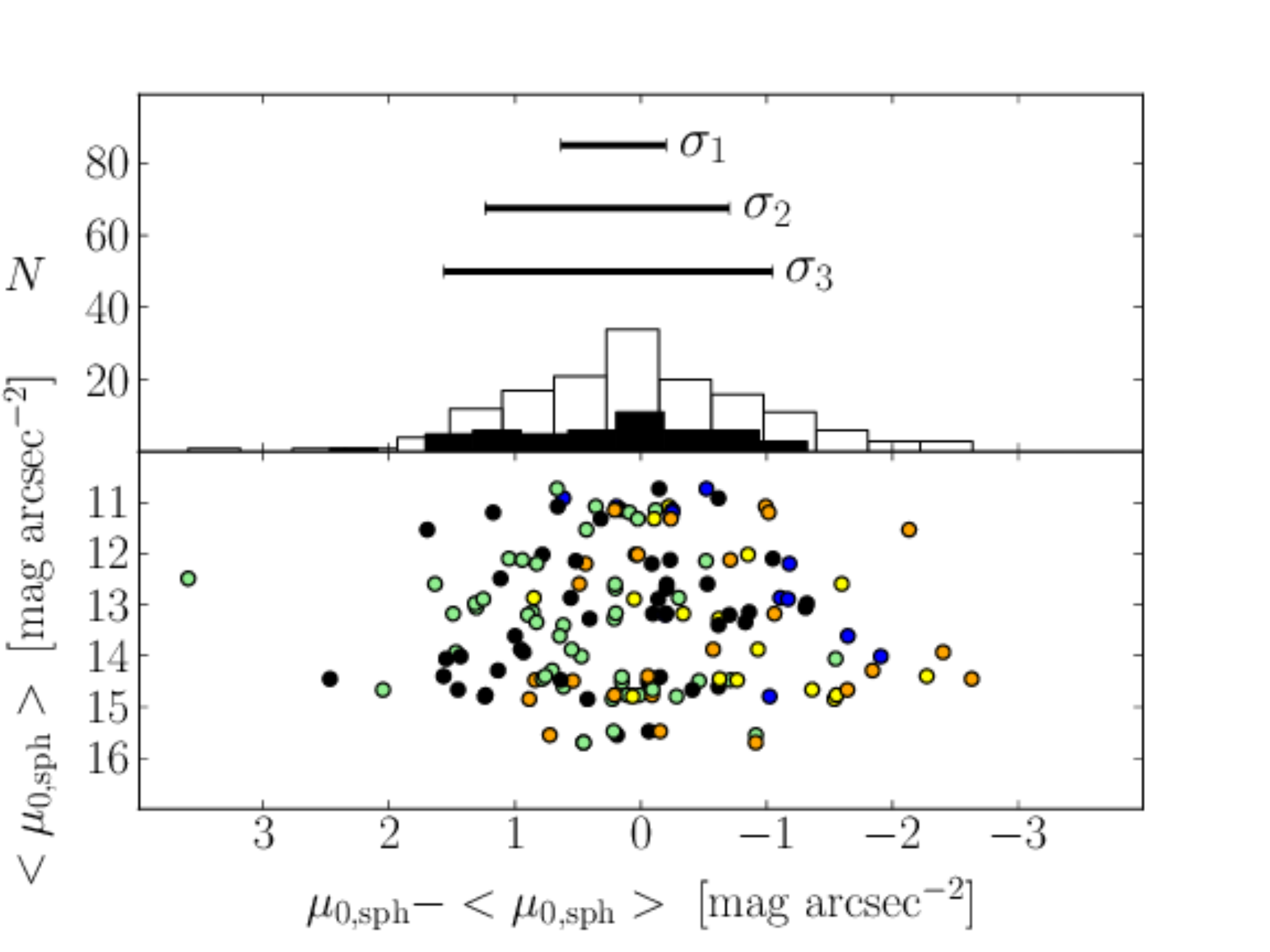} 
\caption{Bottom panel: 49 galaxies for which at least one measurement of the spheroid central surface brightness $\mu_{\rm 0,sph}$ 
-- either in the $K$-band or at $3.6~\rm \mu m$ -- is available from the literature (Table \ref{tab:lit}), in addition to that measured by us. 
The $K$-band magnitudes were converted into $3.6~\rm \mu m$ magnitudes (see Section \ref{sec:err} for details).
The average value $\langle \mu_{\rm 0,sph} \rangle$ is plotted against 
the difference between the individual measurements of a galaxy 
and the average value for that same galaxy.
See Figure \ref{fig:compn} for color description and explanation of the top panel.}
\label{fig:compmu0}
\end{center}
\end{figure}

\section{Conclusions}
\label{sec:concl}
The widespread presence of embedded components -- in particular intermediate-scale disks -- in massive early-type galaxies 
makes galaxy decomposition an essential tool to properly investigate the scaling relations between black hole masses 
and host spheroid properties. \\
Past studies often used different model components for the same galaxy, and obtained significantly discrepant results,  
which led them to draw contrasting conclusions about the black hole -- spheroid correlations.
These inconsistencies motivated our effort to refine and secure the measure of spheroid properties in a sample of 66 galaxies 
with a dynamical estimate of the black hole mass. 
Using $3.6~\rm \mu m$ \emph{Spitzer} satellite images, 
we performed state-of-the-art galaxy decompositions. 
The $3.6~\rm \mu m$ band is an excellent tracer of the stellar mass, superior to optical bands and the $K$-band. 
Considerable care has been taken in the data reduction, image mosaicking, sky subtraction and component model fitting to the galaxy light.
We have compared our best-fit models with those from the literature, 
to identify and explain discrepancies when present. 
Our analysis additionally benefited from recourse to kinematical information -- not previously used -- to aid in the identification of 
somewhat face-on disks, or the distinction between intermediate-scale disks and large-scale disks, 
missed in some past investigations and decompositions.
Table \ref{tab:lit} summarizes the main characteristics of the five past studies which, since 2007, attempted galaxy decompositions 
in order to derive black hole -- spheroid scaling relations, 
and highlights, in part, why our endeavor represents a substantial improvement over the past literature. \\

\begin{table*}
\begin{center}  				      
\caption{Summary of previous investigations of black hole mass scaling relations} 
\begin{tabular}{lllllll}
\hline
              & {\bf GD07}  & {\bf S+11}   & {\bf V+12} & {\bf B+12}   & {\bf L+14} & {\bf This work}	      \\ 
\hline
{\bf Galaxies with successful fit} & 27		     & 57	       & 25      & 19          & 35      & 66               \\
{\bf Wavelength}        	   & $R$-band	     & $3.6~\rm \mu m$ & $K$-band& $i$-band    & $K$-band& $3.6~\rm \mu m$  \\
{\bf Decomposition}     	   & 1D 	     & 2D	       & 2D	 & 2D	       & 2D	 & 1D \& 2D	    \\
{\bf Nuclear components}       	   & masked	     & modeled         & modeled & not treated & modeled & modeled/masked   \\
{\bf Partially depleted cores} 	   & masked	     & masked	       & masked  & not treated & masked  & masked	    \\ 
{\bf Bars}	        	   & excluded	     & modeled         & modeled & excluded    & modeled & modeled	    \\ 
{\bf Other components}  	   & no 	     & no 	       & no	 & no	       & yes	 & yes  	    \\ 
{\bf Kinematics}        	   & no 	     & no	       & no	 & no	       & no	 & yes  	    \\  
\hline 
\end{tabular}
\label{tab:lit} 
\tablecomments{GD07 = \cite{grahamdriver2007}, 
S+11 = \cite{sani2011}, V+12 \cite{vika2012}, B+12 = \cite{beifiori2012}, L+14 = \cite{lasker2014data}.}
\end{center}	
\end{table*}

We reveal that one-dimensional and two-dimensional techniques of galaxy decomposition return the same results when applied to the same galaxy.
However, in our practical experience, the failure rate of two-dimensional decompositions 
is a factor of two higher than the failure rate of one-dimensional decompositions,  
either because the fit does not converge or because the result is unphysical. 
A strong limitation of two-dimensional codes is their inability to accommodate the radial gradients of ellipticity and position angle 
often observed in galaxy spheroids.
The interpretation of one-dimensional residual surface brightness profiles is easier than that of two-dimensional residual images. 
A one-dimensional isophotal analysis was extremely helpful and sometimes even necessary to accurately identify galaxy components.
A correct interpretation of the residuals is fundamental to understand and determine the optimal model for a galaxy.
Given the level of detail to which each galaxy decomposition was performed, 
we believe that our analysis cannot be reproduced by current automatic routines. 
The uncertainties associated with the literatur best-fit parameters of the spheroid are dominated by systematic errors (e.g.~incorrect sky subtraction, 
inaccurate masking of contaminating sources, erroneous choice of model components), and only marginally affected by random errors. 
For this reason, we developed a method to estimate the uncertainties on the best-fit parameters that takes into account systematic errors 
(see Figures \ref{fig:compn} -- \ref{fig:compmu0}).

We will use the results from our one-dimensional galaxy decompositions to obtain improved black hole mass scaling relations. 
These results will be presented in a series of forthcoming papers.

\acknowledgments
GS acknowledges the invaluable support received from Alessandro Marconi, Eleonora Sani and Leslie Hunt 
in the early stages of this research.
GS warmly thanks Chieng Peng, Peter Erwin, Luca Cortese, Giuseppe Gavazzi, 
Bililign Dullo, Paolo Bonfini, Elisabete Lima Da Cunha 
and Gonzalo Diaz 
for useful discussion. 
We thank the anonymous referee for their thorough review and highly appreciate the comments and suggestions, 
which significantly contributed to improving the quality of the publication.  \\
This research was supported by Australian Research Council funding through grants
DP110103509 and FT110100263.
This work is based on observations made with the IRAC instrument \citep{fazio2004IRAC} 
on-board the Spitzer Space Telescope, 
which is operated by the Jet Propulsion Laboratory, 
California Institute of Technology under a contract with NASA.
This research has made use of the GOLDMine database \citep{goldmine} and the NASA/IPAC Extragalactic Database (NED) 
which is operated by the Jet Propulsion Laboratory, California Institute of Technology, 
under contract with the National Aeronautics and Space Administration.

\clearpage

  \onecolumngrid
  \subsection{Individual galaxy decompositions}
  \label{sec:indgal}

  For each galaxy, we show a figure and a table. 
  The figure illustrates our 1D model and the galaxy isophotal parameters. 
  The left panels refer to the major-axis $R_{\rm maj}$, 
  while the right panels refer to the equivalent-axis $R_{\rm eq}$, 
  i.e.~the geometric mean of the major ($a$) and minor ($b$) axis ($R_{\rm eq} = \sqrt{ab}$), 
  equivalent to a circularized profile.
  The top panels display the galaxy surface brightness ($\mu$) radial profiles obtained with a linear sampling. 
  The black points are the observed data used in the fit and the empty points are the observed data excluded from the fit.  
  The color lines represent the individual (PSF-convolved) model components:
  red=S\'ersic; dark blue=exponential; green=Gaussian; cyan=Ferrer; gray=Gaussian ring; pink=PSF. 
  The parameters for the S\'ersic spheroid model are inset.
  The total (PSF-convolved) model is shown with a black dashed line. 
  The residual profile ($data - model$) is shown as $\Delta \mu$ in the second row.
  The horizontal grey dashed line corresponds to an intensity 
  equal to three times the root mean square of the sky background fluctuations ($3 \times RMS_{\rm sky}$).
  $\Delta$ denotes the rms scatter of the fit in units of $\rm mag~arcsec^{-2}$.
  The lower six panels show the ellipticity ($\epsilon$), position angle ($PA$) and fourth harmonic ($B4$) radial profiles from {\tt ellipse}. \\
  The Tables report a comparison between our results (from both our 1D and 2D decompositions) and those obtained by the following authors: 
  GD07 = \citeauthor{grahamdriver2007} (\citeyear{grahamdriver2007}, who performed 1D fits along the major-axis), 
  L+10 = \citeauthor{laurikainen2010} (\citeyear{laurikainen2010}, who performed 2D fits), 
  S+11 = \citeauthor{sani2011} (\citeyear{sani2011}, who performed 2D fits), 
  B+12 = \citeauthor{beifiori2012} (\citeyear{beifiori2012}, who performed 2D fits), 
  V+12 = \citeauthor{vika2012} (\citeyear{vika2012}, who performed 2D fits), 
  R+13 = \citeauthor{rusli2013} (\citeyear{rusli2013}, who performed 1D fits along the equivalent-axis),
  and L+14 = \citeauthor{lasker2014data} (\citeyear{lasker2014data}, who performed 2D fits).
  Each galaxy model is the sum of its individual components, which are expressed with the following nomenclature: (analytic function)-(physical component).
  The analytic functions can be: S=S\'ersic, e=exponential, G=Gaussian, F=Ferrer, M=Moffat and PSF.
  The physical components can be: bul=bulge (or spheroid), d=disk, id=inner disk, bar, n=nucleus, l=lens or oval, r=ring, halo and spiral arms. 
  When a nuclear component or a partially depleted core have been masked, we signal them as ``m-n'' and ``m-c'', respectively.
  For example, the model ``S-bul + e-d + e-id + m-n + G-r'' reads ``S\'ersic-bulge + exponential-disk + exponential-(inner disk) + mask-nucleus + Gaussian-ring''. 
  The core-S\'ersic model used by \cite{rusli2013} is always implicitly associated with the galaxy spheroidal component. 
  \citeauthor{grahamdriver2007} excluded the innermost data points when fitting their galaxy light profiles,  
  therefore their models implicitly include ``m-n'' or ``m-c''. 
  In the table caption, we comment on the most significant discrepancies between our results and those obtained by the other studies. 

  \clearpage\newpage\noindent
  {\bf IC 1459 \\}

  \begin{figure}[h]
  \begin{center}
  \includegraphics[width=\fitfigurewidth]{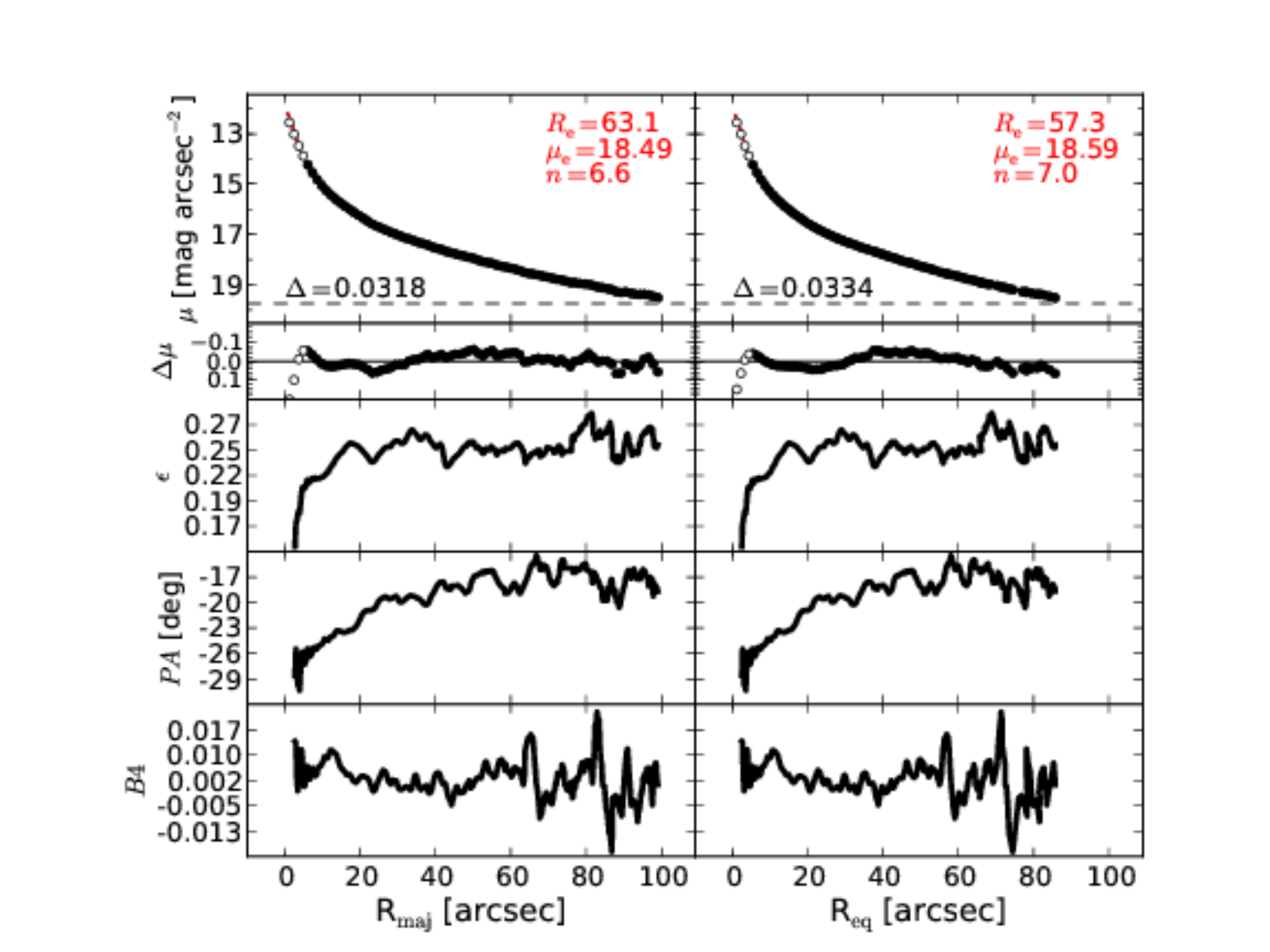}
  \caption{IC 1459: 
  An elliptical galaxy with a fast counterrotating stellar component \citep{franxillingworth1988ic1459,cappellari2002ic1459},
  nuclear dust and indications of a nuclear stellar disk \citep{forbes1994ic1459}.
  The kinematically decoupled component cannot be identified as a separate structure in photometric observations.
  This galaxy also has an unresolved partially depleted core \citep{rusli2013}.  
  Our isophotal analysis confirms a simple morphology for IC 1459, with no evident embedded components.
  After masking the innermost $6''.1$, we fit this galaxy with a S\'ersic profile.  }
  \end{center}
  \end{figure}

  \begin{table}[h]
  \small
  \caption{Best-fit parameters for the spheroidal component of IC 1459.}
  \begin{center}
  \begin{tabular}{llcc}
  \hline
  {\bf Work} & {\bf Model}   & $\bm R_{\rm e,sph}$    & $\bm n_{\rm sph}$ \\
    &  &  $[\rm arcsec]$ & \\
  \hline
  1D maj. & S-bul + m-c & $63.1$  &  $6.6$ \\
  1D eq.  & S-bul + m-c & $57.3$  &  $7.0$ \\
  2D      & S-bul + m-c & $87.5$  &  $8.3$ \\
  \hline
  S+11 2D         & S-bul + G-n   & $61.1$  &  $6.0$ \\
  R+13 1D eq.         & core-S\'ersic & $45.4$  &  $7.6$ \\
  L+14 2D         & S-bul + m-c   & $62.4$  &  $8.3$ \\
  \hline
  \end{tabular}
  \end{center}
  \label{tab:ic1459}
  \tablecomments{Our results are shown in the upper portion of the Table. 
  See Section \ref{sec:indgal} for the legend key to the other authors listed in the lower portion of the Table. 
  See Table \ref{tab:fitres} for our associated surface brightnesses and magnitudes. }
  \end{table}

  \clearpage\newpage\noindent
  {\bf IC 2560 \\}

  \begin{figure}[h]
  \begin{center}
  \includegraphics[width=\fitfigurewidth]{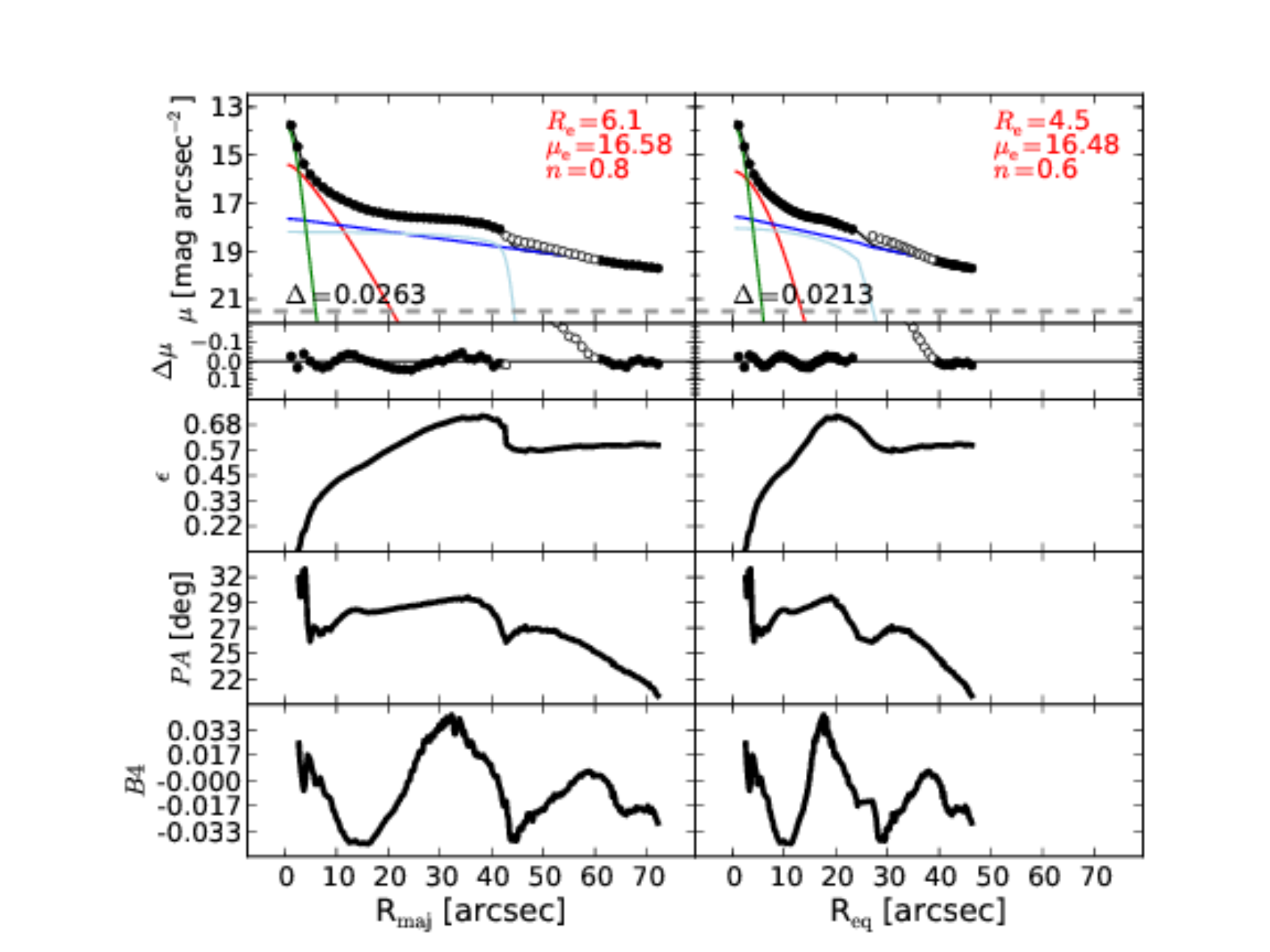}
  \caption{IC 2560: A barred spiral galaxy with a Seyfert AGN \citep{veroncettyveron2006}
  and dust within the central $2''.5$ \citep{martini2003}.
  A visual inspection of the image of IC 2560 reveals a boxy bulge and
  a large scale bar that extends out to $R_{\rm maj} \lesssim 43''$.
  The disk appears to be slightly lopsided along the direction of the bar,
  due to two non-symmetric ansae, but it becomes symmetric beyond $R_{\rm maj} \gtrsim 60''$. 
  This is why the surface brightness profile deviates from a perfect exponential
  in the radial range $42'' \lesssim R_{\rm maj} \lesssim 60''$, which is excluded from the fit.
  Motivated by the presence of a strong optical AGN and dust in the nucleus (which adds rather than obscures at $3.6\rm~\mu m$), 
  we account for an excess of non-stellar light by adding a central Gaussian component to the model.}
  \end{center}
  \end{figure}

  \begin{table}[h]
  \small
  \caption{Best-fit parameters for the spheroidal component of IC 2560.}
  \begin{center}
  \begin{tabular}{llcc}
  \hline
  {\bf Work} & {\bf Model}   & $\bm R_{\rm e,sph}$    & $\bm n_{\rm sph}$ \\
    &  &  $[\rm arcsec]$ & \\
  \hline
  1D maj. & S-bul + e-d + F-bar + G-n & $6.1$  &  $0.8$ \\
  1D eq.  & S-bul + e-d + F-bar + G-n & $4.5$  &  $0.6$ \\
  \hline
  S+11 2D         & S-bul + e-d + G-n & $27.5$  &  $2.0$ \\
  \hline
  \end{tabular}
  \end{center}
  \label{tab:ic2560}
  \tablecomments{In their model, S+11 did not account for the bar component and thus overestimated the effective radius and the S\'ersic index of the bulge.}
  \end{table}

  \clearpage\newpage\noindent
  {\bf IC 4296 \\}

  \begin{figure}[h]
  \begin{center}
  \includegraphics[width=\fitfigurewidth]{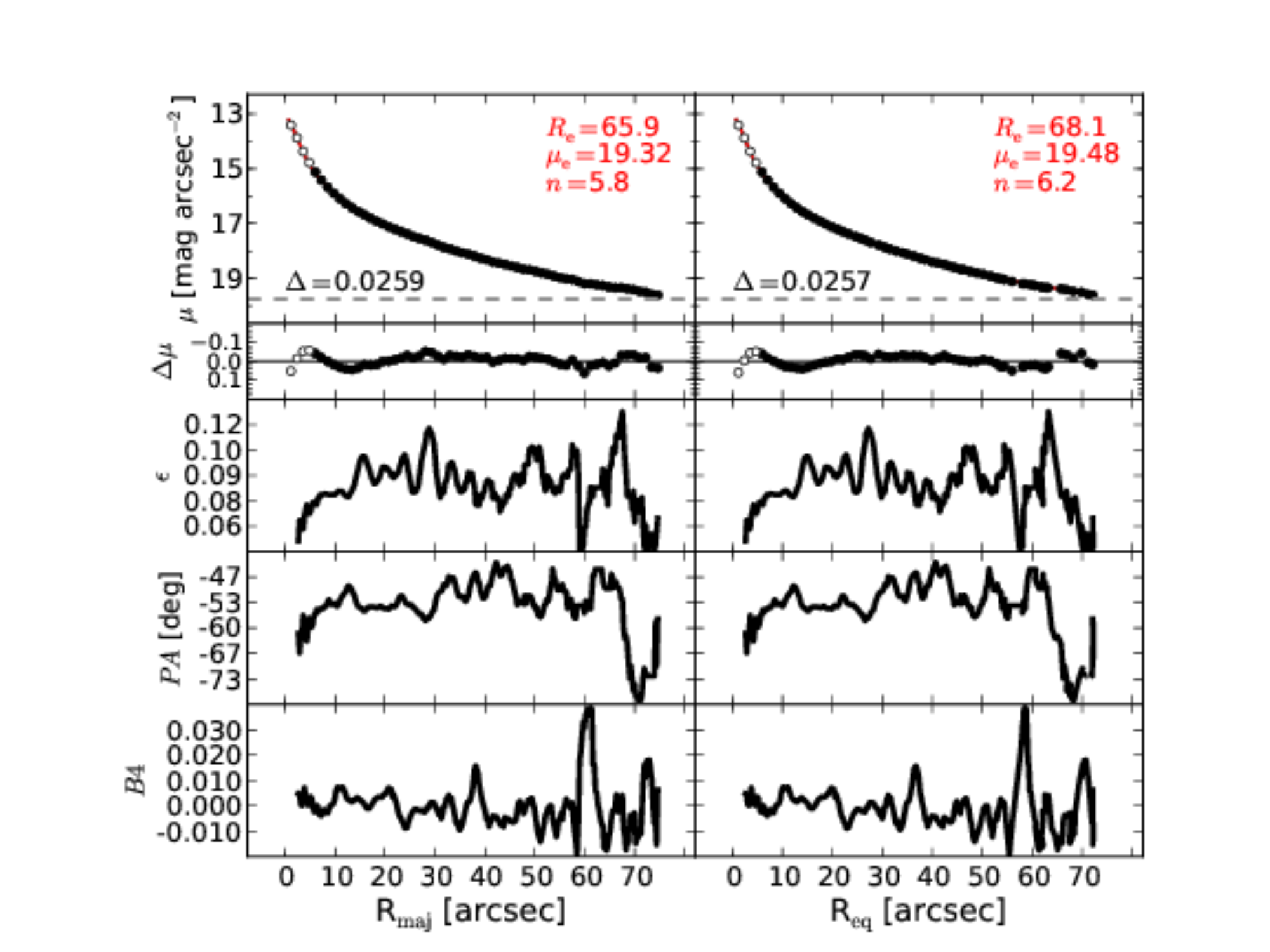}
  \caption{IC 4296:
  An elliptical galaxy. 
  Due to its high stellar velocity dispersion, this galaxy is expected to host a partially depleted core. 
  After masking the innermost $6''.1$, 
  we find that a single S\'ersic profile provides a good description of this galaxy.}
  \end{center}
  \end{figure}

  \begin{table}[h]
  \small
  \caption{Best-fit parameters for the spheroidal component of IC 4296.}
  \begin{center}
  \begin{tabular}{llcc}
  \hline
  {\bf Work} & {\bf Model}   & $\bm R_{\rm e,sph}$    & $\bm n_{\rm sph}$ \\
    &  &  $[\rm arcsec]$ & \\
  \hline
  1D maj. & S-bul + m-c & $65.9$  &  $5.8$ \\
  1D eq.  & S-bul + m-c & $68.1$  &  $6.2$ \\
  2D      & S-bul + m-c & $82.3$  &  $6.6$ \\
  \hline
  S+11 2D         & S-bul + G-n & $33.6$  &  $4.0$ \\
  L+14 2D         & S-bul + m-c & $97.8$  &  $8.2$ \\
  \hline
  \end{tabular}
  \end{center}
  \label{tab:ic4296}
  \tablecomments{
  S+11 obtained a small effective radius and S\'ersic index because they fit a nuclear component rather than masking the core. 
  }
  \end{table}

  \clearpage\newpage\noindent
  {\bf M31 -- NGC 0224 \\}

  \begin{figure}[h]
  \begin{center}
  \includegraphics[width=\fitfigurewidth]{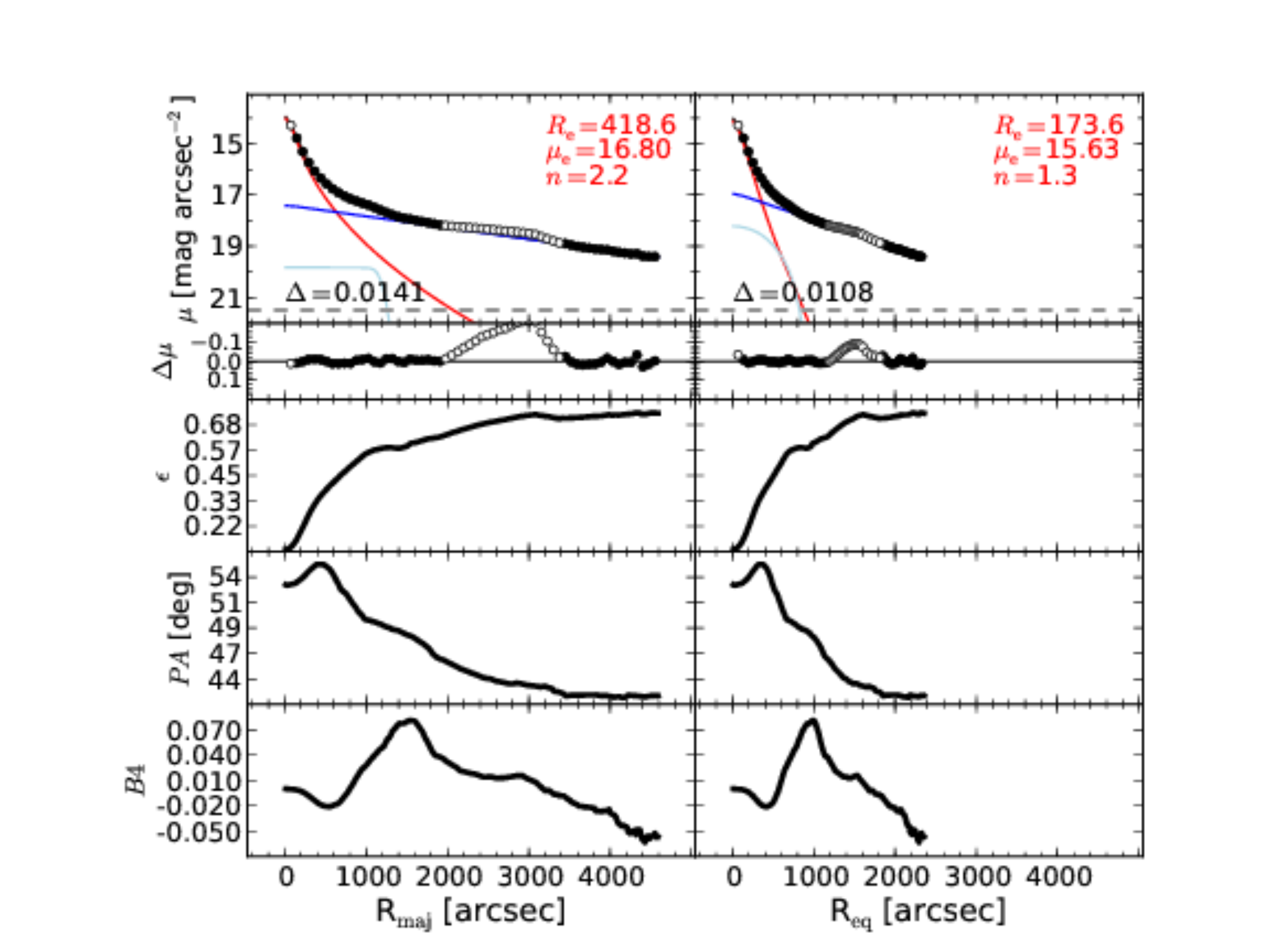}
  \caption{M31 (Andromeda galaxy):
  A spiral galaxy. 
  Although for decades this galaxy had been classified as an unbarred spiral,
  recent works have revealed the presence of a bar \citep{athanassoulabeaton2006m31,beaton2007m31,morrison2011m31}, 
  seen as the plateau at $800'' \lesssim R_{\rm maj} \lesssim 1000''$.
  M31 also features a broad ring-like structure at $R_{\rm maj} \sim 50'$ \citep{athanassoulabeaton2006m31}. 
  We applied the \emph{smoothing} technique described in Section \ref{sec:smooth} to the analysis of M31.
  The region $2000'' \lesssim R_{\rm maj} \lesssim 3400''$, where the pseudo-ring is observed, is excluded from the fit.
  A S\'ersic + exponential fit is not adequate to describe the light profile of M31,
  as the residuals of such fit display a structure in correspondence of the bar ($R_{\rm maj} \lesssim 1500''$).
  The addition of a Ferrer function to account for the bar notably improves the fit.
  }
  \end{center}
  \end{figure}

  \begin{table}[h]
  \small
  \caption{Best-fit parameters for the spheroidal component of M31.}
  \begin{center}
  \begin{tabular}{llcc}
  \hline
  {\bf Work} & {\bf Model}   & $\bm R_{\rm e,sph}$    & $\bm n_{\rm sph}$ \\
    &  &  $[\rm arcsec]$ & \\
  \hline
  1D maj. & S-bul + e-d + F-bar & $418.6$  &  $2.2$ \\
  1D eq.  & S-bul + e-d + F-bar & $173.6$  &  $1.3$ \\
  \hline
  \end{tabular}
  \end{center}
  \label{tab:m31}
  \end{table}

  \clearpage\newpage\noindent
  {\bf M49 -- NGC 4472 \\}
  
  \begin{figure}[h]
  \begin{center}
  \includegraphics[width=\fitfigurewidth]{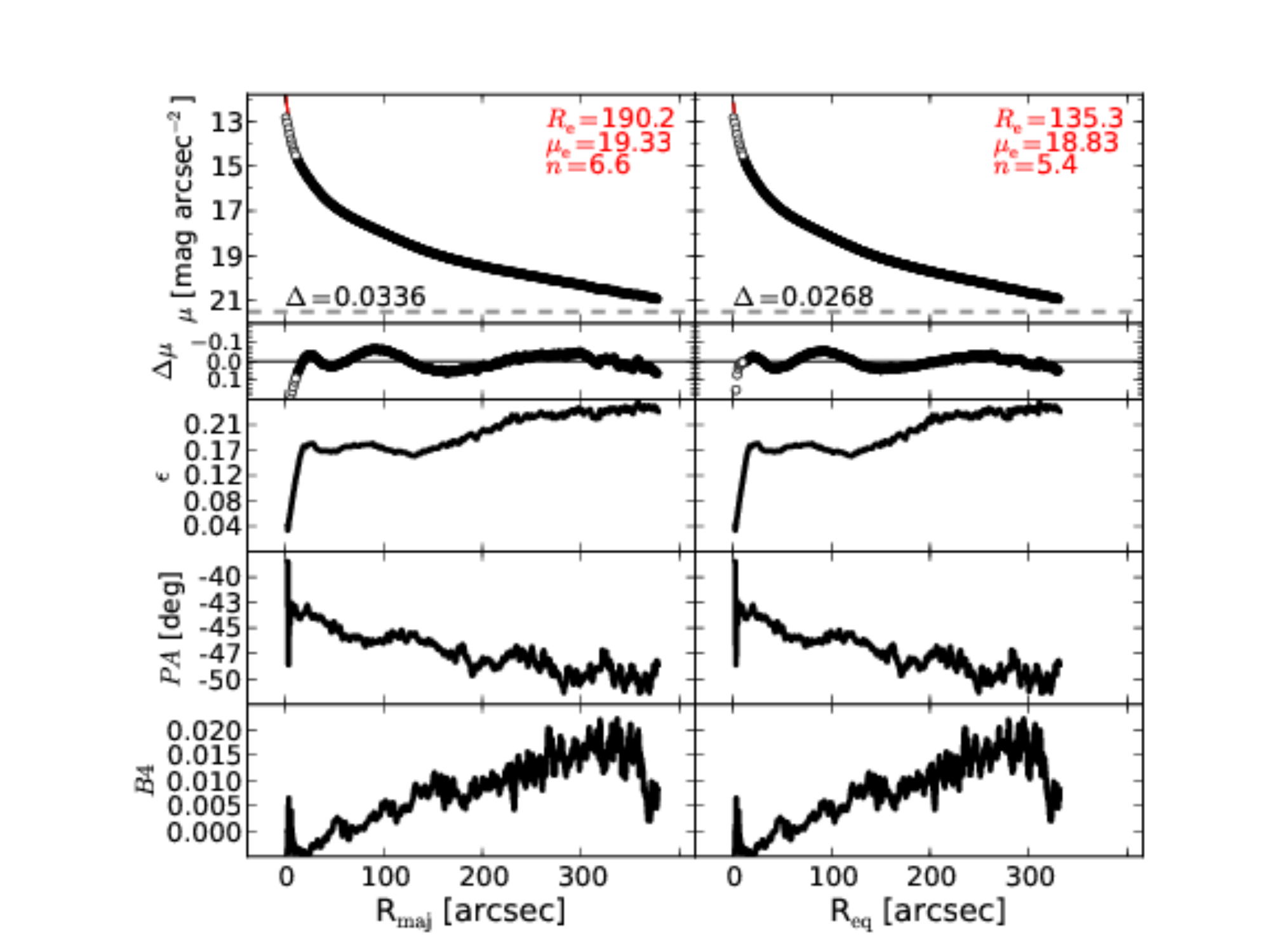}
  \caption{M49:
  The brightest member of the Virgo cluster, a giant elliptical galaxy 
  with a slightly resolved partially depleted core \citep{rusli2013}.  
  The data within the innermost $12''$ are excluded from the fit.
  We fit M49 with a single S\'ersic profile. 
  }
  \end{center}
  \end{figure}

  \begin{table}[h]
  \small
  \caption{Best-fit parameters for the spheroidal component of M49.}
  \begin{center}
  \begin{tabular}{llcc}
  \hline
  {\bf Work} & {\bf Model}   & $\bm R_{\rm e,sph}$    & $\bm n_{\rm sph}$ \\
    &  &  $[\rm arcsec]$ & \\
  \hline
  1D maj. & S-bul + m-c   & $190.2$  &  $6.6$ \\
  1D eq.  & S-bul + m-c   & $135.3$  &  $5.4$ \\
  2D      & S-bul + m-c   & $151.9$  &  $5.5$ \\
  \hline
  R+13 1D eq.         & core-S\'ersic & $199.0$  &  $5.6$ \\
  \hline
  \end{tabular}
  \end{center}
  \label{tab:m49}
  \tablecomments{
  The equivalent-axis effective radius estimated by R+13 is larger than that measured by us.
  Since their circularized light profile is almost 3 times more extended than ours, 
  it is possible that their best-fit model required a larger $R_{\rm e}$ to account for the galaxy intracluster halo light.
  }
  \end{table}

  \clearpage\newpage\noindent
  {\bf M59 -- NGC 4621 \\}

  \begin{figure}[h]
  \begin{center}
  \includegraphics[width=\fitfigurewidth]{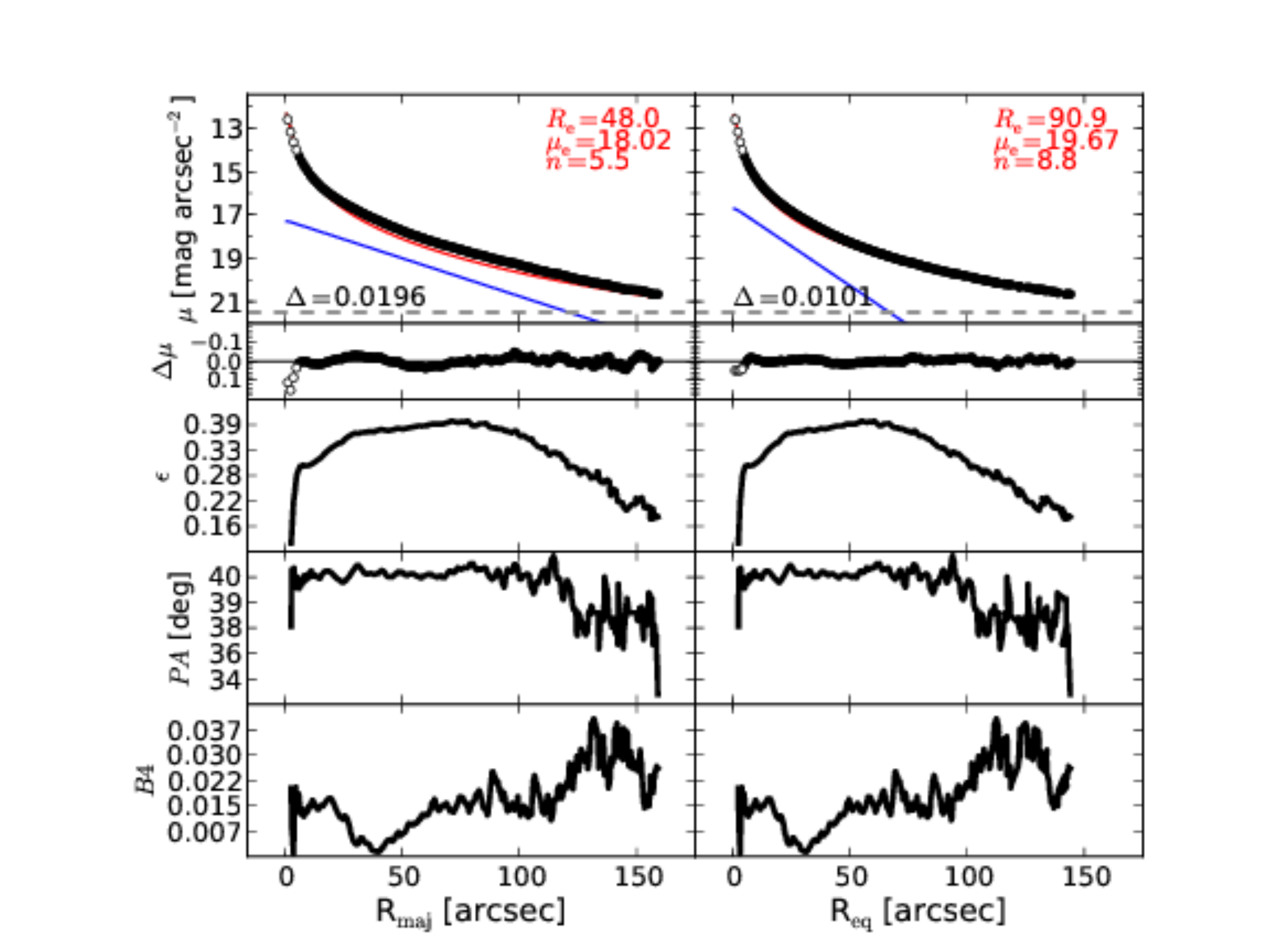}
  \caption{M59:
  An elliptical galaxy with an edge-on, intermediate-scale embedded disk \citep{scorzabender1995}
  and a thin, faint nuclear stellar disk \citep{ferrarese2006acsvcs,ledo2010}. 
  The intermediate-scale embedded disk is clearly visible in our unsharp mask image (not shown), and 
  the velocity map (ATLAS$^{\rm 3D}$) confirms the presence of this rapidly rotating component, which is modeled with an exponential function.
  We choose not to account for the nuclear stellar disk by excluding the innermost $6''.1$ from the fit.  
  }
  \end{center}
  \end{figure}

  \begin{table}[h]
  \small
  \caption{Best-fit parameters for the spheroidal component of M59.}
  \begin{center}
  \begin{tabular}{llcc}
  \hline
  {\bf Work} & {\bf Model}   & $\bm R_{\rm e,sph}$    & $\bm n_{\rm sph}$ \\
    &  &  $[\rm arcsec]$ & \\
  \hline
  1D maj. & S-bul + m-n + e-id & $48.0$  &  $5.5$ \\
  1D eq.  & S-bul + m-n + e-id & $90.9$  &  $8.8$ \\
  \hline 
  S+11 2D         & S-bul       & $70.1$  &  $5.0$ \\
  V+12 2D         & S-bul + m-c & $54.7$  &  $5.7$ \\
  \hline
  \end{tabular}
  \end{center}
  \label{tab:m59}
  \end{table}

  \clearpage\newpage\noindent
  {\bf M64 -- NGC 4826 \\}
  
  \begin{figure}[h]
  \begin{center}
  \includegraphics[width=\fitfigurewidth]{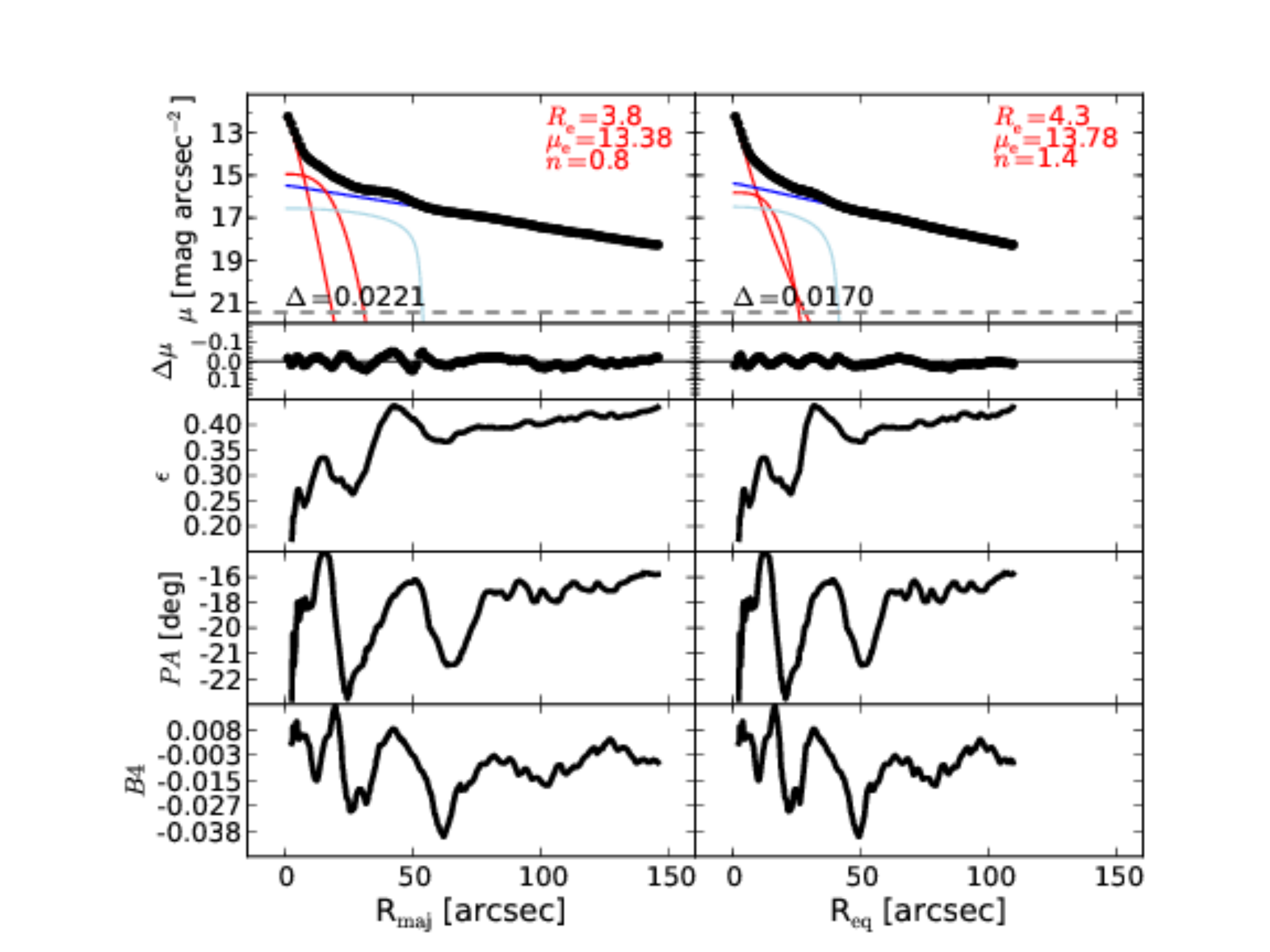}
  \caption{M64: 
  A dusty spiral galaxy with a Seyfert AGN \citep{veroncettyveron2006}. 
  A large-scale bar ($R_{\rm maj} \lesssim 50''$) can be identified in the unsharp mask 
  and produces corresponding peaks in the ellipticity, $PA$ and $B4$ profiles.
  This is modeled with a Ferrer function.
  Additional peaks in the ellipticity, $PA$ and $B4$ profiles at $R_{\rm maj} \sim 15''$ 
  signal the presence of an embedded disky component, that we describe with a low-$n$ S\'ersic function.
  Although the galaxy is dusty and hosts an AGN, we do not observe any nuclear excess of light 
  in the fit residuals and therefore we found it unnecessary to add a nuclear component to our model. 
  }
  \end{center}
  \end{figure}

  \begin{table}[h]
  \small
  \caption{Best-fit parameters for the spheroidal component of M64.}
  \begin{center}
  \begin{tabular}{llcc}
  \hline
  {\bf Work} & {\bf Model}   & $\bm R_{\rm e,sph}$    & $\bm n_{\rm sph}$ \\
    &  &  $[\rm arcsec]$ & \\
  \hline
  1D maj. & S-bul + e-d + F-bar + S-id  & $3.8$  &  $0.8$ \\
  1D eq.  & S-bul + e-d + F-bar + S-id  & $4.3$  &  $1.4$ \\
  \hline
  B+12 2D      & S-bul + e-d  & $5.0$  &  $1.5$ \\
  \hline
  \end{tabular}
  \end{center}
  \label{tab:m64}
  \end{table}

  \clearpage\newpage\noindent
  {\bf M81 -- NGC 3031 \\}

  \begin{figure}[h]
  \begin{center}
  \includegraphics[width=\fitfigurewidth]{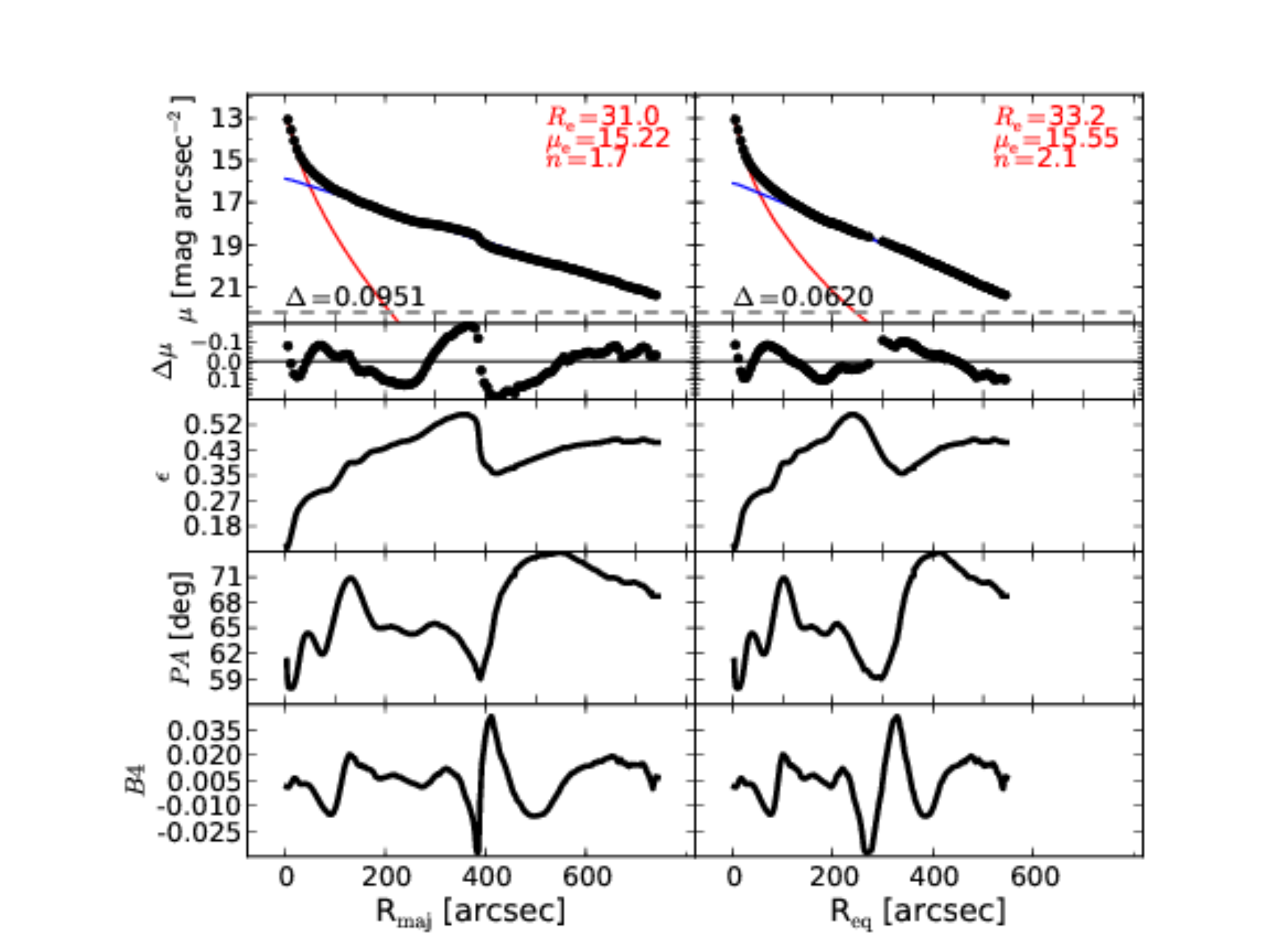}
  \caption{M81: 
  An early-type spiral galaxy.
  This galaxy features a nuclear bar at $R_{\rm maj} \lesssim 17''$ and a large-scale bar at $R_{\rm maj} \lesssim 130''$ 
  \citep{elmegreen1995m81,gutierrez2011,erwindebattista2013}. 
  We applied the \emph{smoothing} technique described in Section \ref{sec:smooth} to the analysis of M81.
  The bars are not particularly evident in the galaxy image or in the unsharp mask. 
  The bump at $R_{\rm maj} \lesssim 400''$ is due to spiral arms. 
  Attempts to account for the bars in the fit were unsuccessful. 
  For this reason, we use a simple S\'ersic + exponential model. 
  }
  \end{center}
  \end{figure}

  \begin{table}[h]
  \small
  \caption{Best-fit parameters for the spheroidal component of M81.}
  \begin{center}
  \begin{tabular}{llccc}
  \hline
  {\bf Work} & {\bf Model}   & $\bm R_{\rm e,sph}$    & $\bm n_{\rm sph}$ \\
    &  &  $[\rm arcsec]$ & \\
  \hline
  1D maj. & S-bul + e-d  & $31.0$  &  $1.7$ \\
  1D eq.  & S-bul + e-d  & $33.2$  &  $2.1$ \\
  \hline
  GD07 1D maj.      & S-bul + e-d	 & $68.1$   &  $3.3$ \\
  S+11 2D      & S-bul + e-d + G-n  & $127.3$  &  $3.0$ \\
  B+12 2D      & S-bul + e-d	 & $50.0$   &  $2.6$ \\
  \hline
  \end{tabular}
  \end{center}
  \label{tab:m81}
  \tablecomments{
  GD07 and B+12 obtained estimates of the effective radius larger than ours by a factor of two. 
  This is not particularly surprising, given the complicated surface brightness distribution of this galaxy. 
  The large measurement of the effective radius reported by S+11 is the most discrepant, 
  but the reasons for this are unclear. 
  }
  \end{table}

  \clearpage\newpage\noindent
  {\bf M84 -- NGC 4374 \\}

  \begin{figure}[h]
  \begin{center}
  \includegraphics[width=\fitfigurewidth]{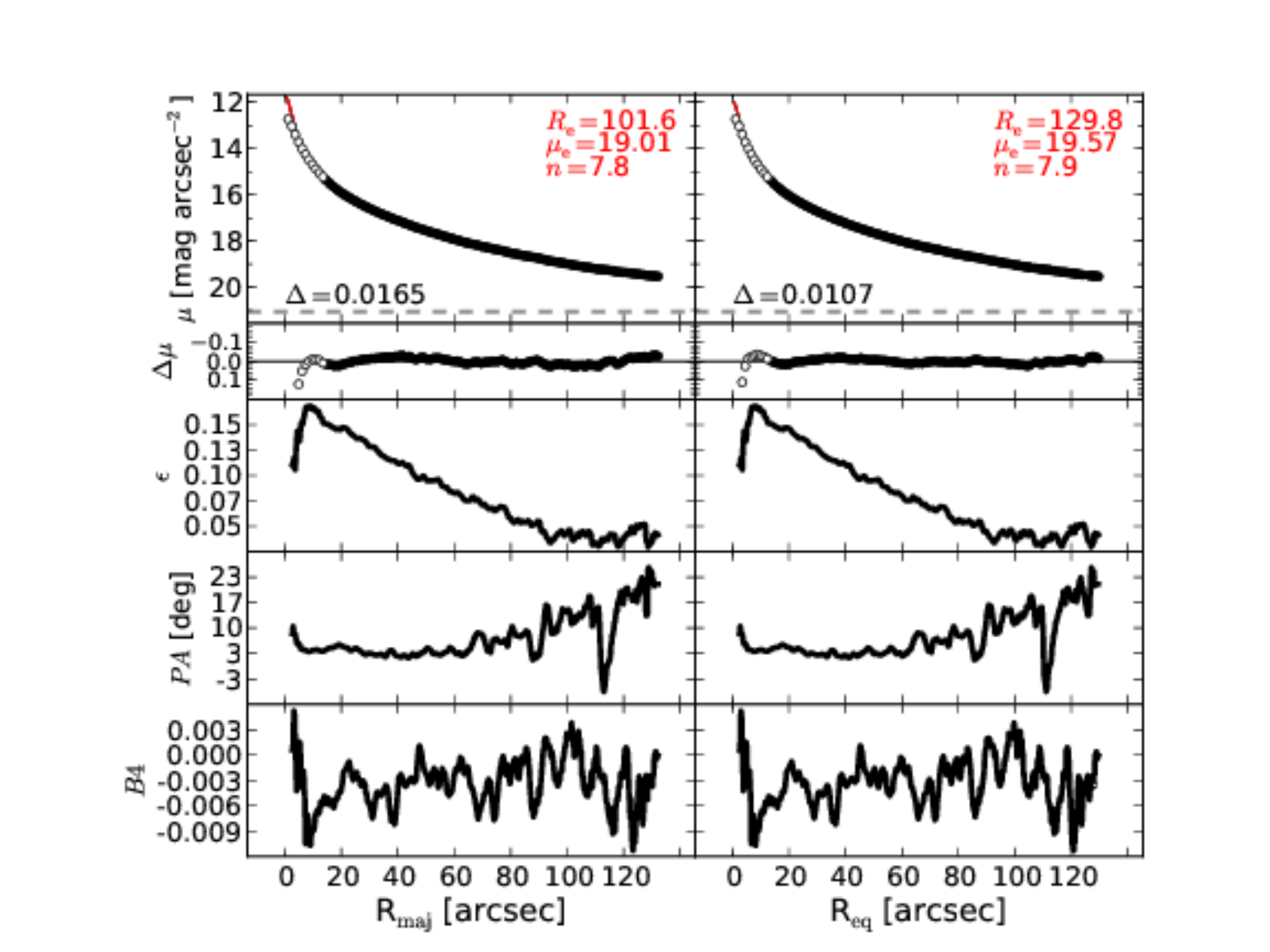}
  \caption{M84:
  An elliptical galaxy with a slightly resolved partially depleted core \citep{rusli2013}. 
  The unsharp mask reveals a faint inner component ($R_{\rm maj} \lesssim 12''$). 
  We exclude the data within $R_{\rm maj} < 13''.8$ and model the galaxy with a S\'ersic profile.
  }
  \end{center}
  \end{figure}

  \begin{table}[h]
  \small
  \caption{Best-fit parameters for the spheroidal component of M84.}
  \begin{center}
  \begin{tabular}{llcc}
  \hline
  {\bf Work} & {\bf Model}   & $\bm R_{\rm e,sph}$    & $\bm n_{\rm sph}$ \\
    &  &  $[\rm arcsec]$ & \\
  \hline
  1D maj. & S-bul + m-c   & $101.6$  &  $7.8$ \\
  1D eq.  & S-bul + m-c   & $129.8$  &  $7.9$ \\
  2D      & S-bul + m-c   & $181.8$  &  $8.4$ \\
  \hline
  GD07 1D maj.         & S-bul         & $75.1$   &  $5.6$ \\
  S+11 2D         & S-bul + G-n   & $105.9$  &  $7.0$ \\
  V+12 2D         & S-bul + m-c   & $28.7$   &  $3.5$ \\
  B+12 2D         & S-bul         & $63.6$   &  $4.1$ \\
  R+13 1D eq.         & core-S\'ersic & $126.2$  &  $7.1$ \\
  L+14 2D         & S-bul + m-c   & $139.0$  &  $8.3$ \\
  \hline
  \end{tabular}
  \end{center}
  \label{tab:m84}
  \tablecomments{
  B+12 did not mask the core and thus underestimated the effective radius and the S\'ersic index. 
  V+12 used the same model as R+13, L+14 and us, 
  but the smaller radial extent of their data led them to underestimate the effective radius and the S\'ersic index.
    }
  \end{table}

  \clearpage\newpage\noindent
  {\bf M87 -- NGC 4486 \\}

  \begin{figure}[h]
  \begin{center}
  \includegraphics[width=\fitfigurewidth]{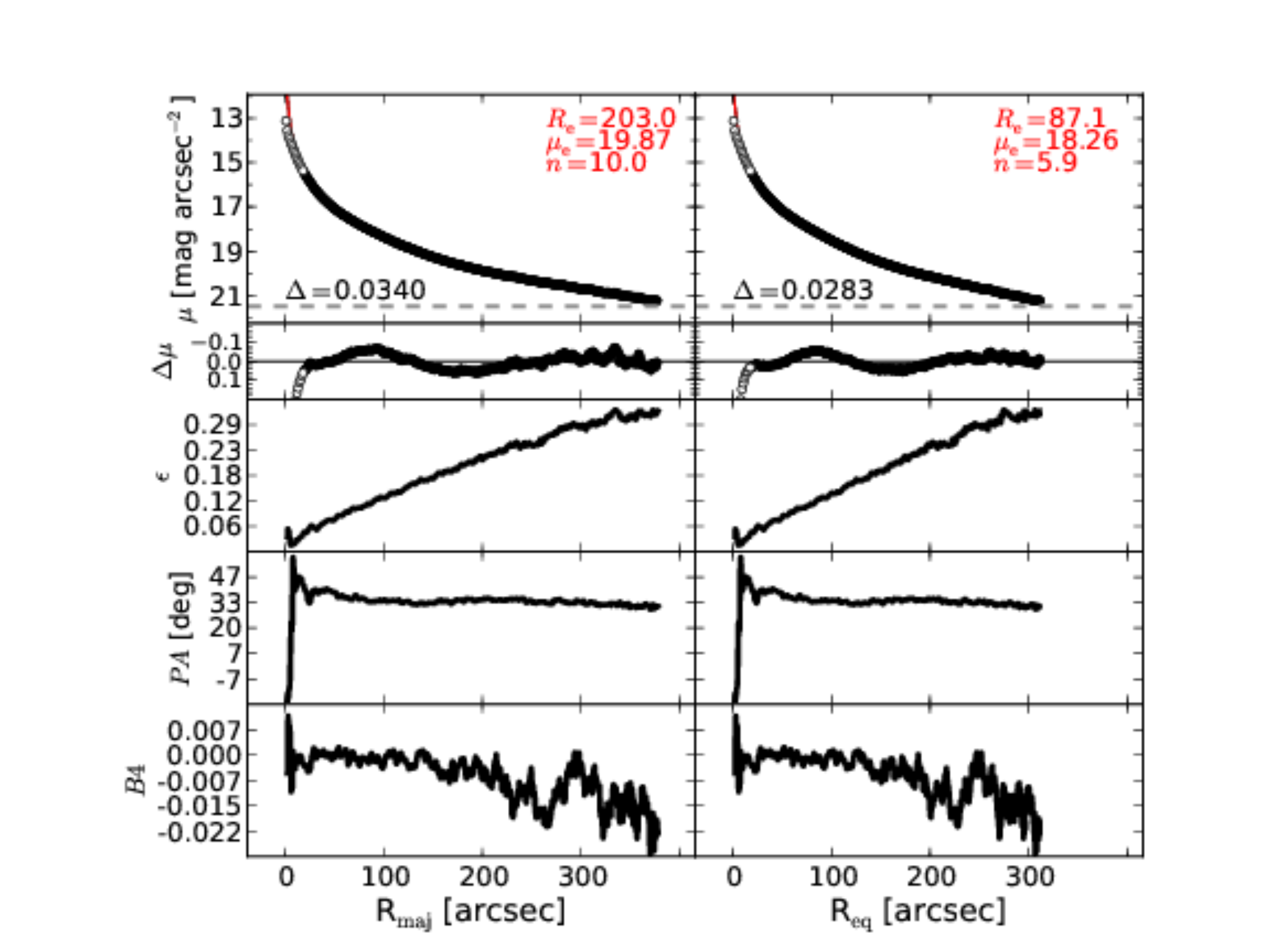}
  \caption{M87:
  A giant elliptical galaxy belonging to the Virgo cluster.
  This galaxy has a large ($> 7''$), partially depleted core \citep{ferrarese2006acsvcs}.
  The morphology of M87 is similar to that of M49.
  The innermost $18''.5$ are excluded and
  a S\'ersic model is used to fit M87. 
  We note that the effective radius obtained from our 1D major-axis fit is more than a factor of two larger than that obtained from our 2D fit.
  This is due to the fact that the 2D model has fixed ellipticity and cannot account for the strong ellipticity gradient of the galaxy, 
  which serves as a warning to some 2D fits in the literature.  
  }
  \end{center}
  \end{figure}

  \begin{table}[h]
  \small
  \caption{Best-fit parameters for the spheroidal component of M87.}
  \begin{center}
  \begin{tabular}{llcc}
  \hline
  {\bf Work} & {\bf Model}   & $\bm R_{\rm e,sph}$    & $\bm n_{\rm sph}$ \\
    &  &  $[\rm arcsec]$ & \\
  \hline
  1D maj. & S-bul + m-c  & $203.9$  &  $10.0$ \\
  1D eq.  & S-bul + m-c  & $87.1$   &  $5.9$ \\
  2D      & S-bul + m-c  & $88.3$   &  $4.3$ \\
  \hline
  GD07 1D maj.      & S-bul	    & $-$     &  $6.9$ \\
  S+11 2D      & S-bul + G-n   & $99.5$  &  $4.0$ \\
  V+12 2D      & S-bul + m-c   & $34.6$  &  $2.4$ \\
  R+13 1D eq.      & core-S\'ersic & $180.9$ &  $8.9$ \\
  L+14 2D      & S-bul + m-c   & $122.0$ &  $5.6$ \\
  \hline
  \end{tabular}
  \end{center}
  \label{tab:m87}
  \tablecomments{
  The equivalent-axis fit of R+13 returns the largest values for $R_{\rm e,sph}$ and $n_{\rm sph}$. 
  As in the case of M49, since their circularized light profile is almost three times more extended than ours, 
  it is possible that their best-fit model required a larger $R_{\rm e}$ to account for the extra intracluster halo light.
  V+12 obtained the smallest estimates of $R_{\rm e,sph}$ and $n_{\rm sph}$ because of the small radial extent of their data.
  }
  \end{table}

  \clearpage\newpage\noindent

  {\bf M89 -- NGC 4552 \\}

  \begin{figure}[h]
  \begin{center}
  \includegraphics[width=\fitfigurewidth]{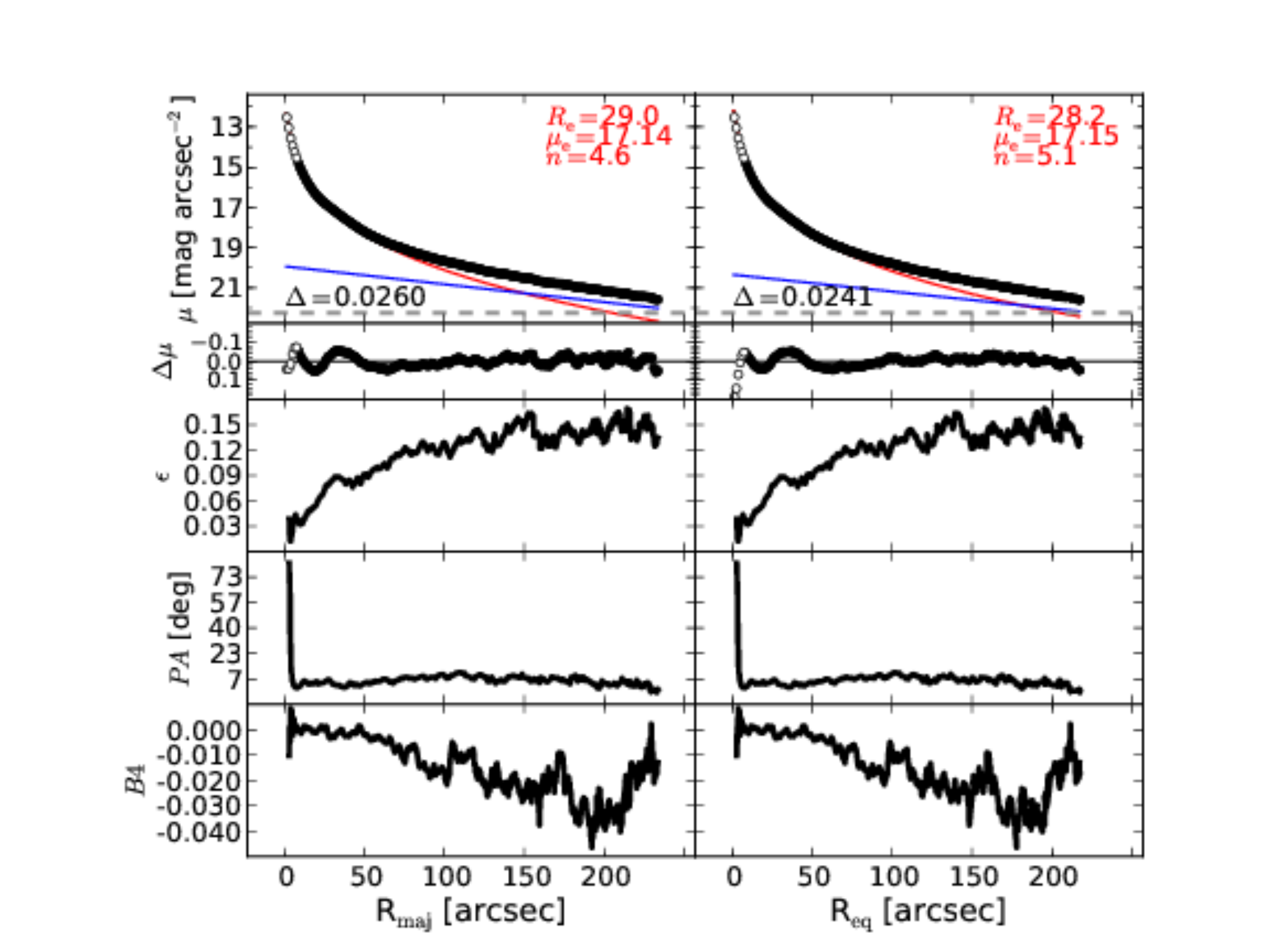}
  \caption{M89: 
  An elliptical galaxy with an unresolved partially depleted core \citep{rusli2013}. 
  The galaxy halo dominates the light at large radii, where the isophotes are significantly more elliptical than in the inner part of the galaxy. 
  The halo is modeled with an exponential function. 
  }
  \end{center}
  \end{figure}

  \begin{table}[h]
  \small
  \caption{Best-fit parameters for the spheroidal component of M89.}
  \begin{center}
  \begin{tabular}{llcc}
  \hline
  {\bf Work} & {\bf Model}   & $\bm R_{\rm e,sph}$    & $\bm n_{\rm sph}$ \\
    &  &  $[\rm arcsec]$ & \\
  \hline
  1D maj. & S-bul + e-halo  & $29.0$  &  $4.6$ \\
  1D eq.  & S-bul + e-halo  & $28.2$  &  $5.1$ \\
  \hline
  S+11 2D         & S-bul + e-d     & $24.3$  &  $4.0$ \\
  V+12 2D         & S-bul           & $16.7$  &  $3.6$ \\
  B+12 2D         & S-bul           & $45.2$  &  $4.3$ \\
  R+13 1D eq.         & S-bul + S-halo  & $19.8$  &  $3.8$ \\
  \hline
  \end{tabular}
  \end{center}
  \label{tab:m89}
  \end{table}

  \clearpage\newpage\noindent
  {\bf M94 -- NGC 4736 \\}

  \begin{figure}[h]
  \begin{center}
  \includegraphics[width=\fitfigurewidth]{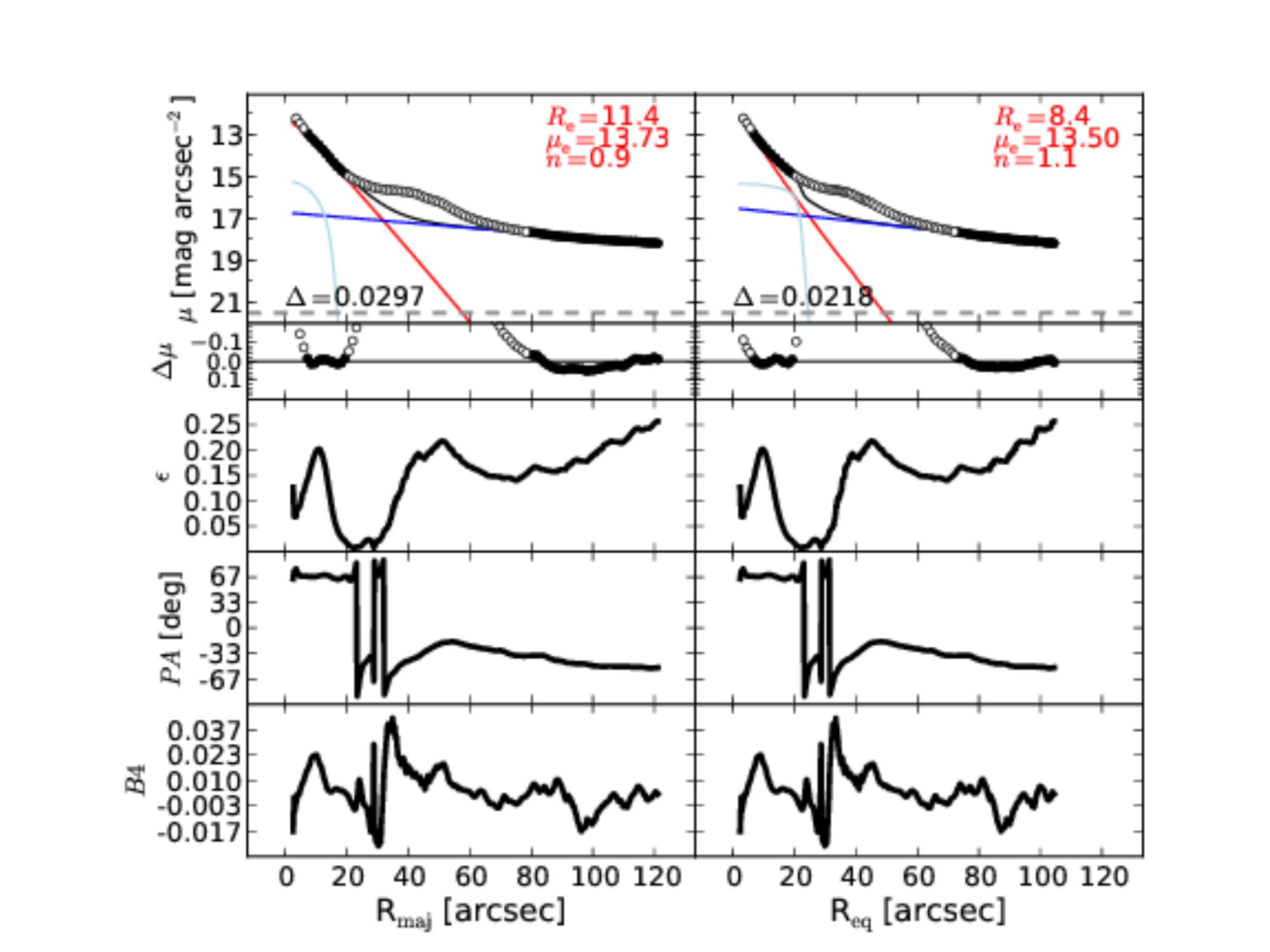}
  \caption{M94: 
  A face-on spiral galaxy, with a Seyfert AGN \citep{veroncettyveron2006} and 
  circumnuclear dust \citep{elmegreen2002m94,peeplesmartini2006m94}.
  This galaxy has a ring-like structure at $R_{\rm maj} \sim 50''$ \citep{munoztunon1989m94}.
  The ellipticity and $B4$ profiles display a local maximum at $R_{\rm maj} \sim 10''$, 
  indicating the presence of a disky component embedded in the bulge,
  as already noted by \cite{fisherdrory2010}.
  We model this component with a Ferrer function.
  We exclude the data within the innermost $6''.7$, due to the AGN contribution.
  Attempts to model the ring were unsuccessful, resulting in a degenerate model, 
  therefore we exclude the data in the range $20'' \lesssim R_{\rm maj} \lesssim 80''$.
  }
  \end{center}
  \end{figure}

  \begin{table}[h]
  \small
  \caption{Best-fit parameters for the spheroidal component of M94.}
  \begin{center}
  \begin{tabular}{llcc}
  \hline
  {\bf Work} & {\bf Model}   & $\bm R_{\rm e,sph}$    & $\bm n_{\rm sph}$ \\
    &  &  $[\rm arcsec]$ & \\
  \hline
  1D maj. & S-bul + e-d + F-bar & $11.4$  &  $0.9$ \\
  1D eq.  & S-bul + e-d + F-bar & $8.4$   &  $1.1$ \\
  \hline
  \end{tabular}
  \end{center}
  \label{tab:m94}
  \end{table}

  \clearpage\newpage\noindent
  {\bf M96 -- NGC 3368 \\}

  \begin{figure}[h]
  \begin{center}
  \includegraphics[width=\fitfigurewidth]{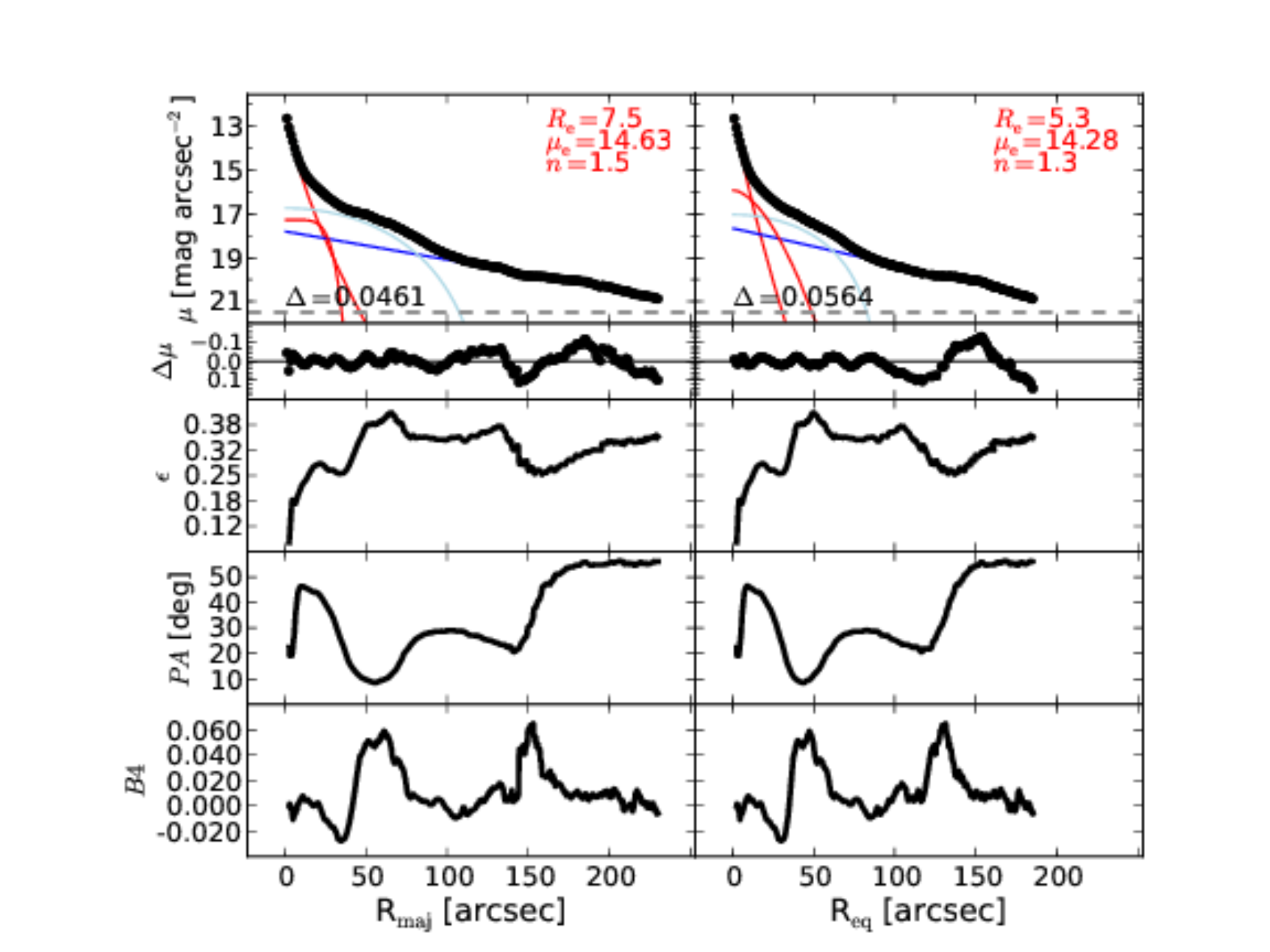}
  \caption{M96:
  A spiral galaxy with a large scale bar and no AGN activity \citep{martini2003,nowak2010n3368n3489}. 
  \cite{erwin2004} observed the presence of two bars in this galaxy. 
  \cite{nowak2010n3368n3489} reported the presence of a large pseudo-bulge and a tiny classical bulge 
  ($R_{\rm e, sph} = 1''.6$, $n_{\rm sph} = 2.3$). 
  However, they came to this conclusion by modeling only the innermost $8''.5$ of the 
  light profile of M96 with a S\'ersic-classical bulge and an exponential-pseudobulge, 
  but without subtracting the contribution of the large-scale disk and the large-scale bar.
  They also acknowledged the presence of a secondary inner bar, but did not include it in their fit. 
  M96 has a complex morphology. 
  The large-scale disk does not exhibit a perfect exponential profile, 
  mainly because of wound spiral arms resembling a ring ($R_{\rm maj} \sim 185''$),
  which however are too faint and irregular to be described with a Gaussian ring model. 
  The large-scale bar extends out to $R_{\rm maj} \lesssim 80''$ and is modeled with a Ferrer function.
  A peak at $R_{\rm maj} \sim 25''$ in the ellipticity profile reveals an inner disky component embedded in the bulge, 
  that we model with a low-$n$ S\'ersic profile.
  }
  \end{center}
  \end{figure}

  \begin{table}[h]
  \small
  \caption{Best-fit parameters for the spheroidal component of M96.}
  \begin{center}
  \begin{tabular}{llcc}
  \hline
  {\bf Work} & {\bf Model}   & $\bm R_{\rm e,sph}$    & $\bm n_{\rm sph}$ \\
    &  &  $[\rm arcsec]$ & \\
  \hline
  1D maj. & S-bul + e-d + F-bar + S-id  & $7.5$  &  $1.5$ \\
  1D eq.  & S-bul + e-d + F-bar + S-id  & $5.3$  &  $1.3$ \\
  2D      & S-bul + e-d + G-bar + G-id  & $8.3$  &  $2.0$ \\
  \hline
  S+11 2D         & S-bul + e-d + G-bar  & $45.6$  &  $1.0$ \\
  \hline
  \end{tabular}
  \end{center}
  \label{tab:m96}
  \tablecomments{
  The bulge effective radius obtained by S+11 largely exceeds our estimates 
  because their galaxy decomposition does not account for the inner component embedded in the bulge ($R_{\rm maj} \lesssim 25''$). 
  }
  \end{table}

  \clearpage\newpage\noindent

  {\bf M104 -- NGC 4594 \\} 

  \begin{figure}[h]
  \begin{center}
  \includegraphics[width=\fitfigurewidth]{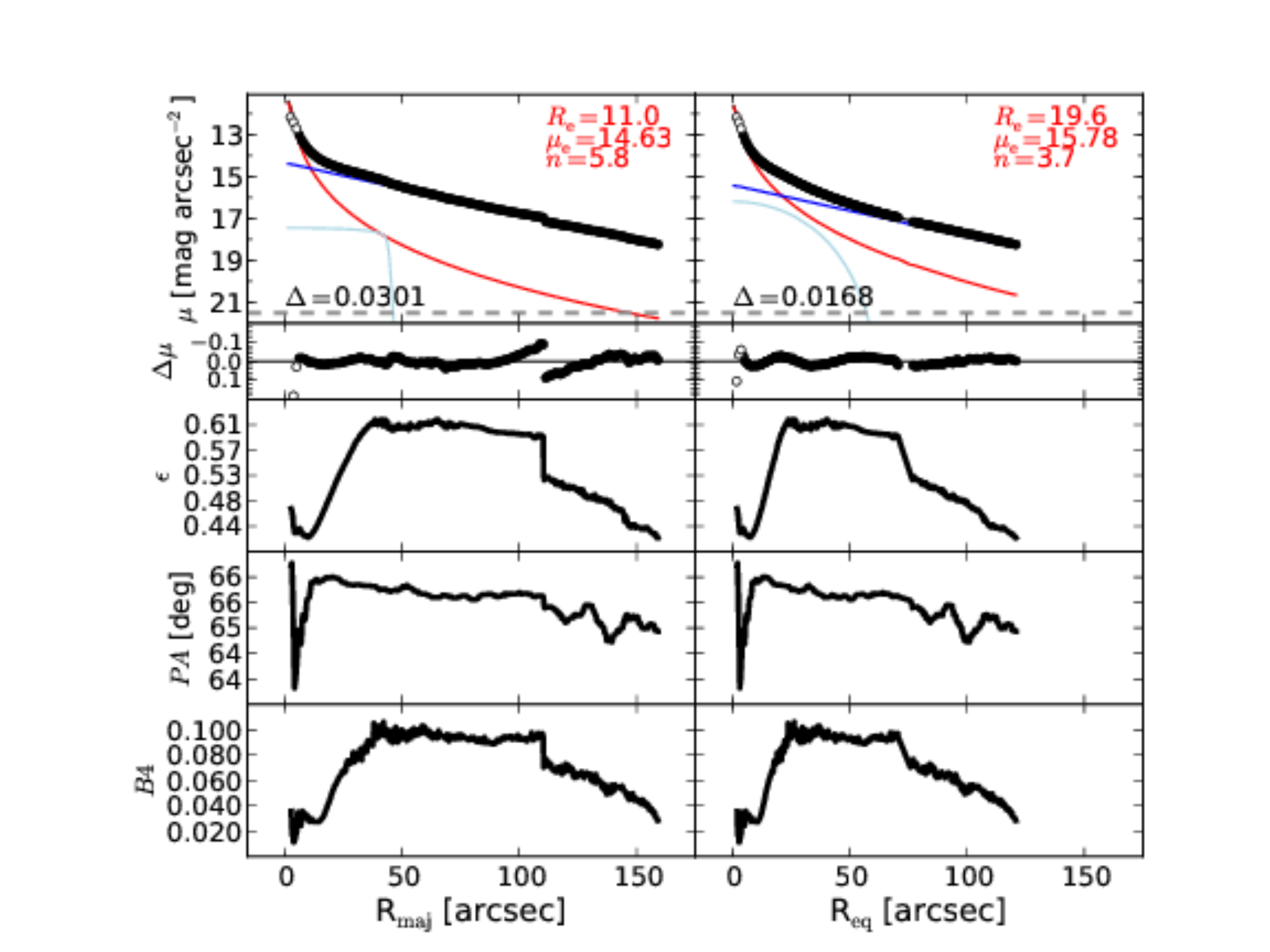}
  \caption{M104: 
  A lenticular/spiral galaxy with a partially depleted core \citep{jardel2011}. 
  The data within the innermost $6''.1$ are excluded from the fit, 
  and the most obvious elongated feature of this galaxy is modeled with a Ferrer function. 
  The ellipticity profile indicates that the disk is of intermediate-scale 
  (for a thorough analysis of this galaxy, see \citealt{gadottisanchezjanssen2012}).
  }
  \end{center}
  \end{figure}
  
  \begin{table}[h]
  \small
  \caption{Best-fit parameters for the spheroidal component of M104.}
  \begin{center}
  \begin{tabular}{llcc}
  \hline
  {\bf Work} & {\bf Model}   & $\bm R_{\rm e,sph}$    & $\bm n_{\rm sph}$ \\
    &  &  $[\rm arcsec]$ & \\
  \hline
  1D maj. & S-bul + e-d + F-bar + m-c  & $11.0$  &  $5.8$ \\
  1D eq.  & S-bul + e-d + F-bar + m-c  & $19.6$  &  $3.7$ \\
  \hline
  S+11 2D         & S-bul + e-d + G-bar + G-n  & $66.1$  &  $1.5$ \\
  \hline
  \end{tabular}
  \end{center}
  \label{tab:m104}
  \tablecomments{The bulge S\'ersic index obtained by S+11 is smaller than our estimates because 
  they fit an additional nuclear component. }
  \end{table}

  \clearpage\newpage\noindent
  {\bf M105 -- NGC 3379 \\}

  \begin{figure}[h]
  \begin{center}
  \includegraphics[width=\fitfigurewidth]{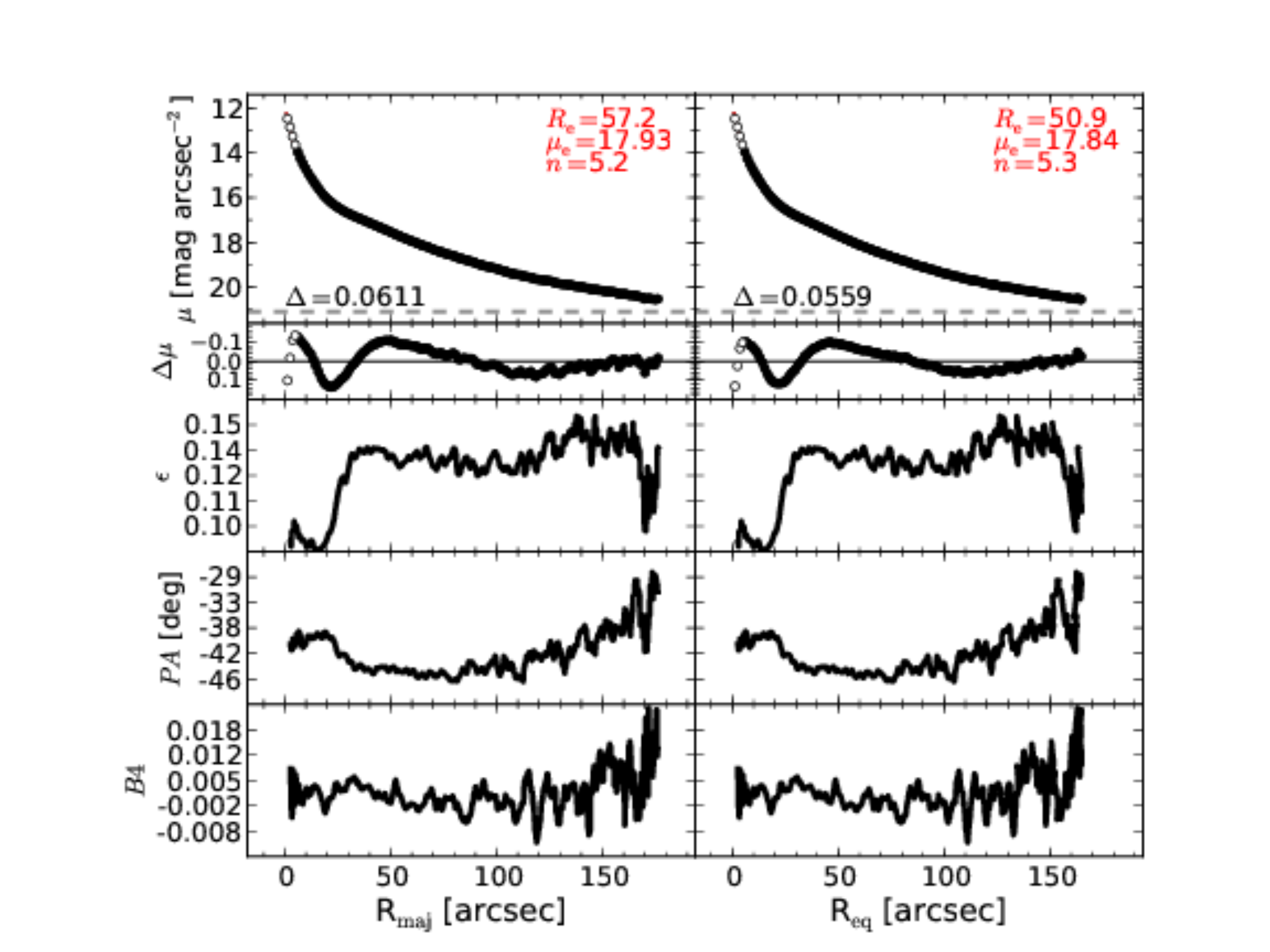}
  \caption{M105: 
  An elliptical galaxy with an unresolved partially depleted core \citep{rusli2013}. 
  M105 exhibits a mild isophotal twist and an abrupt change of ellipticity at $R_{\rm maj} \sim 30''$.
  The velocity map (ATLAS$^{\rm 3D}$) shows rotation at least within $R_{\rm maj} \lesssim 30''$, 
  but no embedded disk can be recognized in the unsharp mask 
  and the ellipticity profile is quite unusual for a spheroidal system with an inner disk.
  The data within the innermost $6''.1$ are excluded from the fit.
  A single S\'ersic model does not provide a good description of the galaxy light profile.
  However, the addition of a second function (of any analytic form) to the model does not improve the fit.
  We conclude that the galaxy may not yet be fully relaxed, and we fit it with a single S\'ersic profile. 
  }
  \end{center}
  \end{figure}

  \begin{table}[h]
  \small
  \caption{Best-fit parameters for the spheroidal component of M105.}
  \begin{center}
  \begin{tabular}{llcc}
  \hline
  {\bf Work} & {\bf Model}   & $\bm R_{\rm e,sph}$    & $\bm n_{\rm sph}$ \\
    &  &  $[\rm arcsec]$ & \\
  \hline
  1D maj. & S-bul + m-c & $57.2$  &  $5.2$ \\
  1D eq.  & S-bul + m-c & $50.9$  &  $5.3$ \\
  2D      & S-bul + m-c & $73.6$  &  $7.0$ \\
  \hline
  GD07 1D maj.         & S-bul & $58.3$  &  $4.3$ \\
  S+11 2D         & S-bul & $46.0$  &  $5.0$ \\
  R+13 1D eq.         & core-S\'ersic & $55.1$  &  $5.8$ \\
  L+14 2D         & S-bul + m-c & $96.3$  &  $9.3$ \\
  \hline
  \end{tabular}
  \end{center}
  \label{tab:m105}
  \tablecomments{
  L+14 obtained the largest values of $R_{\rm e,sph}$ and $n_{\rm sph}$, 
  possibly due to incorrect sky subtraction (see the ``upturn'' of the three outermost data points in their Figure 21).
  }
  \end{table}

  \clearpage\newpage\noindent
  {\bf M106 -- NGC 4258 \\} 
  
  \begin{figure}[h]
  \begin{center}
  \includegraphics[width=\fitfigurewidth]{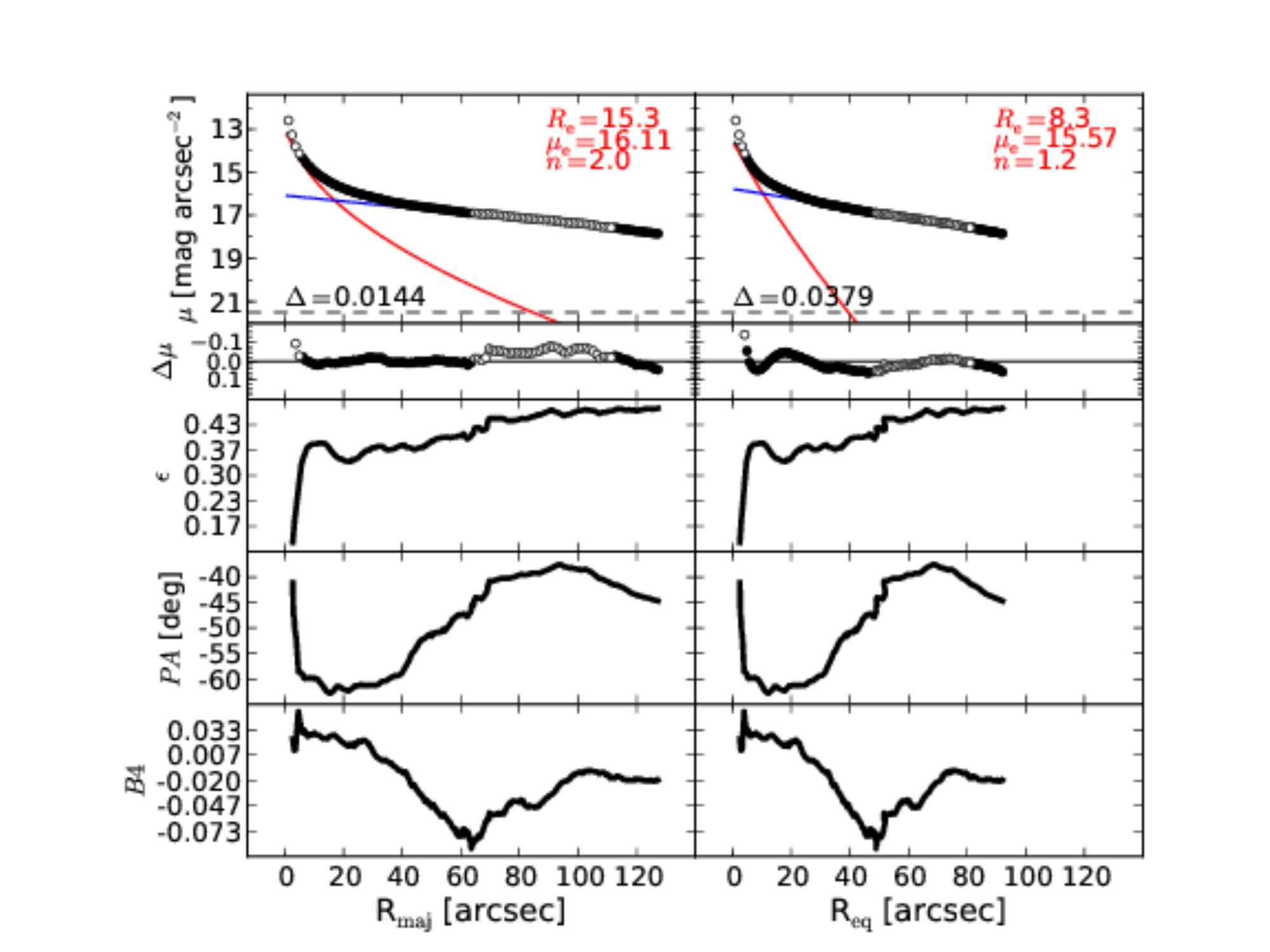}
  \caption{M106: 
  A barred spiral galaxy, harbouring a Seyfert AGN \citep{veroncettyveron2006} 
  with circumnuclear dust \citep{martini2003}. 
  We exclude from the fit the innermost $6''.1$ because it is affected by the AGN emission.
  The boxy bar, responsible for the minimum in the $B4$ profile at $R_{\rm maj} \sim 65''$, 
  is fainter than the large-scale disk and in the light profile it is only detectable as a slight swelling
  within $65'' \lesssim R_{\rm maj} \lesssim 110''$.
  This region is excluded from the fit.
  A peak in the ellipticity profile at $R_{\rm maj} \sim 10''$ suggests the presence of disky component embedded in the bulge, 
  but a model that accounts for this component is degenerate with the other fitted components.
  }
  \end{center}
  \end{figure}

  \begin{table}[h]
  \small
  \caption{Best-fit parameters for the spheroidal component of M106.}
  \begin{center}
  \begin{tabular}{llcc}
  \hline
  {\bf Work} & {\bf Model}   & $\bm R_{\rm e,sph}$    & $\bm n_{\rm sph}$ \\
    &  &  $[\rm arcsec]$ & \\
  \hline
  1D maj. & S-bul + e-d & $15.3$  &  $2.0$ \\
  1D eq.  & S-bul + e-d & $8.3$   &  $1.2$ \\
  \hline
  GD07 1D maj.     & S-bul + e-d		 & $14.9$   &  $2.0$ \\
  S+11 2D     & S-bul + e-d + G-n  	 & $111.7$  &  $2.0$ \\
  V+12 2D (1) & S-bul + e-d + PSF-n	 & $17.0$   &  $3.5$ \\
  V+12 2D (2) & S-bul + e-d + PSF-n + S-bar & $6.3$    &  $2.2$ \\
  L+14 2D     & S-bul + e-d + PSF-n + e-id + S-bar + spiral arms & $6.3$	 &  $3.3$ \\
  \hline
  \end{tabular}
  \end{center}
  \label{tab:m106}
  \tablecomments{
  The bulge effective radius obtained by S+11 largely exceedes all the other estimates, behavior noted also in NGC 3031.
  }
  \end{table}

  \clearpage\newpage\noindent
  {\bf NGC 0524 \\} 

  \begin{figure}[h]
  \begin{center}
  \includegraphics[width=\fitfigurewidth]{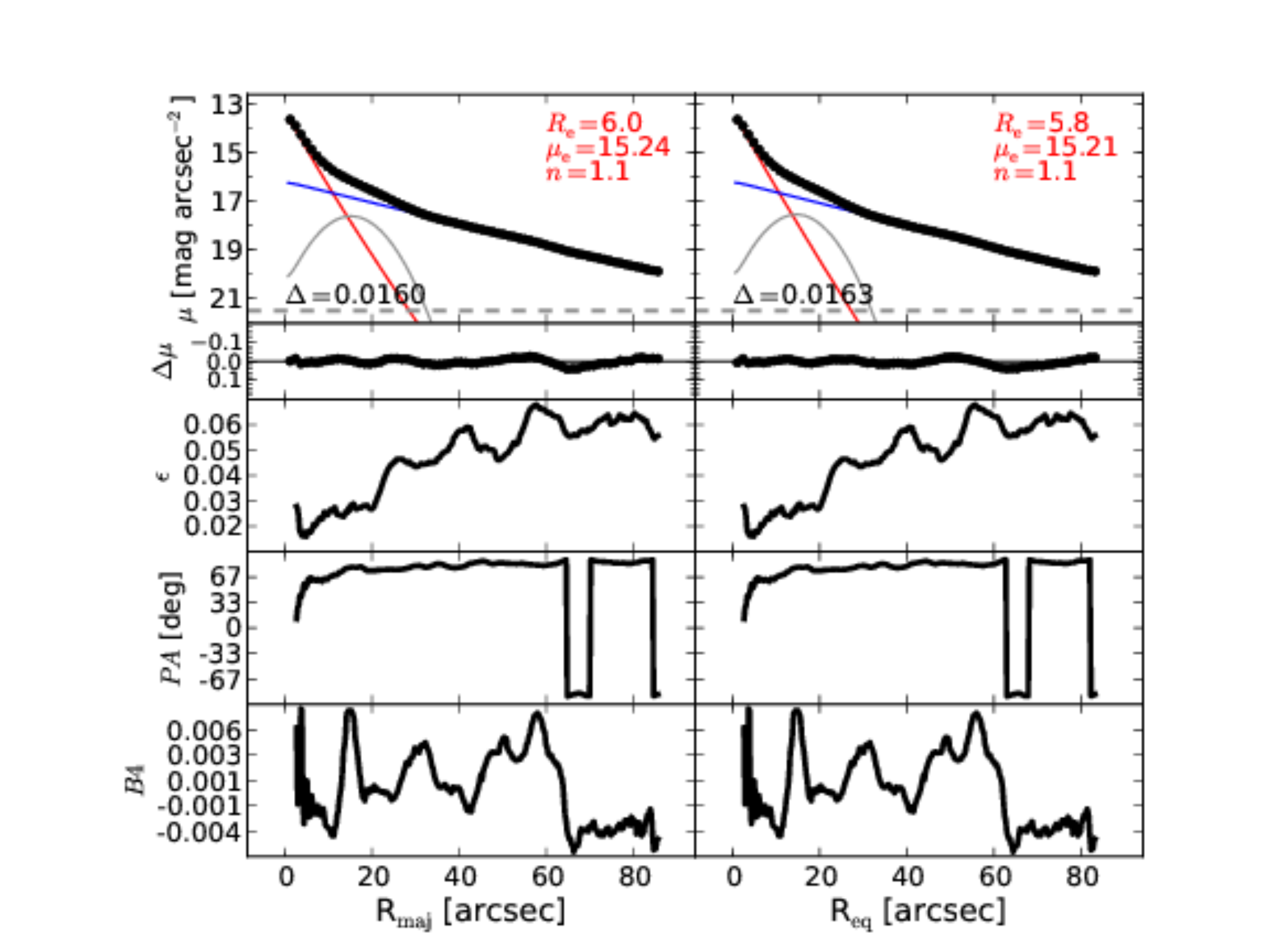}
  \caption{NGC 0524: 
  An unbarred face-on lenticular galaxy. 
  The unsharp mask of NGC 0524 reveals a faint multi-ring structure in the galaxy disk,
  with a substantial ring peaking at $R_{\rm maj} \sim 20''$.
  We account for this brightest ring using a Gaussian ring profile.
  }
  \end{center}
  \end{figure}

  \begin{table}[h]
  \small
  \caption{Best-fit parameters for the spheroidal component of NGC 0524.}
  \begin{center}
  \begin{tabular}{llcc}
  \hline
  {\bf Work} & {\bf Model}   & $\bm R_{\rm e,sph}$    & $\bm n_{\rm sph}$ \\
    &  &  $[\rm arcsec]$ & \\
  \hline
  1D maj. & S-bul + e-d + G-r & $6.0$  &  $1.1$ \\
  1D eq.  & S-bul + e-d + G-r & $5.8$  &  $1.1$ \\
  \hline
  L+10 2D      & S-bul + e-d + 2 F-l & $8.9$   &  $2.7$ \\
  S+11 2D      & S-bul + e-d	  & $26.8$  &  $3.0$ \\
  \hline
  \end{tabular}
  \end{center}
  \label{tab:n0524}
  \tablecomments{
  S+11 obtained the largest value of the effective radius because their two-component model does not account for the ring.
  Both L+10 and S+11 estimated a S\'ersic index of $\sim$3, 
  three times larger than the value obtained by us.
  }
  \end{table}

  \clearpage\newpage\noindent

  {\bf NGC 0821 \\}

  \begin{figure}[h]
  \begin{center}
  \includegraphics[width=\fitfigurewidth]{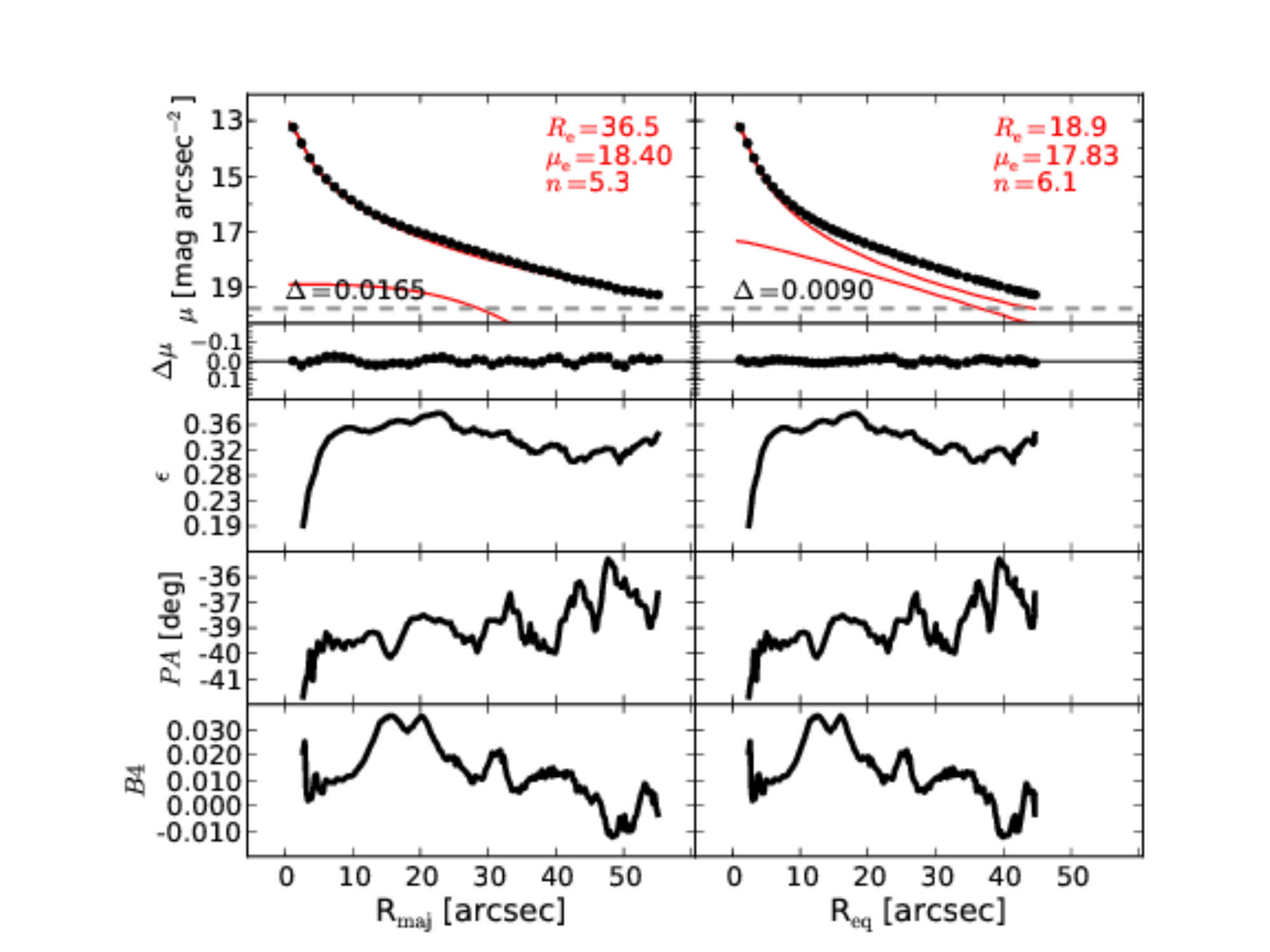}
  \caption{NGC 0821: 
  An elliptical galaxy with an edge-on embedded disk. 
  The ellipticity profile and the velocity map (ATLAS$^{\rm 3D}$, SLUGGS) show the presence of the faint intermediate-scale disk. 
  Given its edge-on inclination, the disk is modeled with a low-$n$ S\'ersic function.   
  }
  \end{center}
  \end{figure}
  
  \begin{table}[h]
  \small
  \caption{Best-fit parameters for the spheroidal component of NGC 0821.}
  \begin{center}
  \begin{tabular}{llcc}
  \hline
  {\bf Work} & {\bf Model}   & $\bm R_{\rm e,sph}$    & $\bm n_{\rm sph}$ \\
    &  &  $[\rm arcsec]$ & \\
  \hline
  1D maj. & S-bul + e-id + PSF-n & $36.8$  &  $5.4$ \\
  1D eq.  & S-bul + e-id + PSF-n & $20.7$  &  $6.0$ \\
  2D      & S-bul + e-id + m-n   & $33.8$  &  $2.5$ \\
  \hline
  GD07 1D maj.    & S-bul 		 & $44.1$   &  $4.0$ \\           
  S+11 2D    & S-bul 		 & $63.6$   &  $7.0$ \\           
  B+12 2D    & S-bul 		 & $111.3$  &  $7.7$ \\           
  L+14 2D    & S-bul + e-d + S-halo & $3.8$    &  $3.1$ \\
  \hline
  \end{tabular}
  \end{center}
  \label{tab:n0821}
  \tablecomments{
  L+14 obtained a tiny estimate of the spheroid effective radius and a small S\'ersic index because they failed to identify the extent 
  of the intermediate-scale disk. 
  Inaccurate sky subtraction could be the reason why B+12 obtained a large estimate of the effective radius.
  We could not obtain a successful 2D model that included a nuclear component, 
  therefore we opted for masking the nuclear region of the galaxy. 
  This resulted in a significantly lower S\'ersic index, that we consider underestimated. }
  \end{table}

  \clearpage\newpage\noindent
  {\bf NGC 1023 \\}
  
  \begin{figure}[h]
  \begin{center}
  \includegraphics[width=\fitfigurewidth]{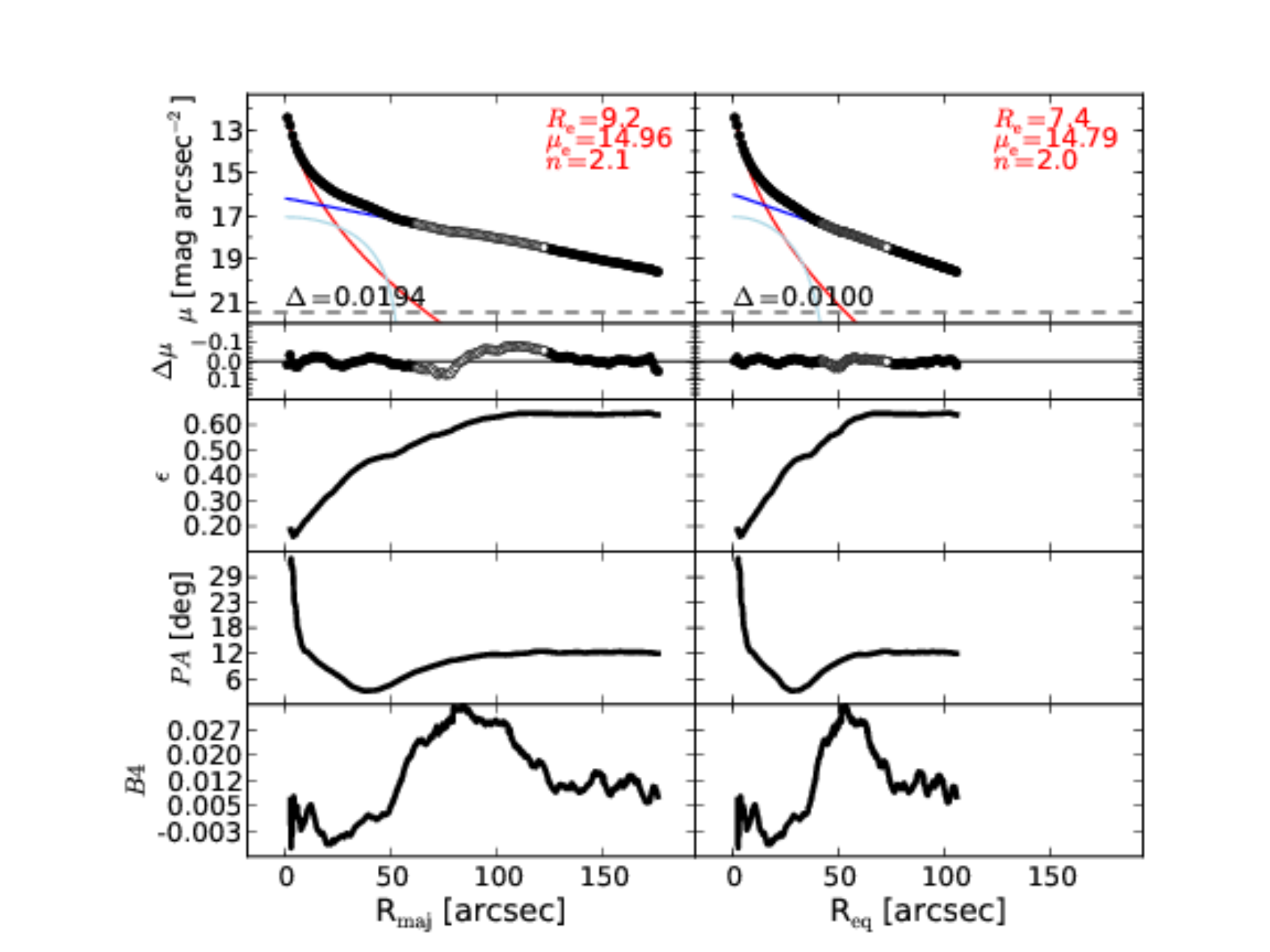}
  \caption{NGC 1023:
  A barred lenticular galaxy. 
  The bar extends out to $R_{\rm maj} \lesssim 40''$.
  Small deviations of the disk from a perfect exponential profile, 
  within $60'' \lesssim R_{\rm maj} \lesssim 125''$ (the data in this range are excluded from the fit), 
  can be ascribed to faint residual spiral arms, also noticeable in the 2D residual image. 
  The peak at $R_{\rm maj} \sim 10''$ in the $B4$ profile signals the presence of an embedded (and faint) disk.
  The 1D residuals show a structure within $R_{\rm maj} \lesssim 10''$ caused by this unsubtracted 
  component, as also noted by \cite{lasker2014data}.
  Our attempts to account for the inner disk by adding a fourth function to the model 
  did not significantly change the bulge best-fit parameters,
  thus we elect not to model this embedded disk. 
  }
  \end{center}
  \end{figure}

  \begin{table}[h]
  \small
  \caption{Best-fit parameters for the spheroidal component of NGC 1023.}
  \begin{center}
  \begin{tabular}{llcc}
  \hline
  {\bf Work} & {\bf Model}   & $\bm R_{\rm e,sph}$    & $\bm n_{\rm sph}$ \\
    &  &  $[\rm arcsec]$ & \\
  \hline
  1D maj. & S-bul + e-d + F-bar & $9.2$  &  $2.1$ \\
  1D eq.  & S-bul + e-d + F-bar & $7.4$  &  $2.0$ \\
  2D      & S-bul + e-d + G-bar & $6.6$  &  $2.3$ \\
  \hline
  GD07 1D maj.    & S-bul + e-d 	& $17.7$  &  $2.0$ \\
  S+11 2D    & S-bul + e-d + G-bar & $24.0$  &  $3.0$ \\
  L+14 2D    & S-bul + e-d + S-bar & $9.6$   &  $3.1$ \\
  \hline
  \end{tabular}
  \end{center}
  \label{tab:n1023}
  \tablecomments{
  S+11 obtained the largest value of the bulge effective radius, although they accounted for the bar in their model.
  }
  \end{table}

  \clearpage\newpage\noindent
  {\bf NGC 1300 \\}
  
  \begin{figure}[h]
  \begin{center}
  \includegraphics[width=\fitfigurewidth]{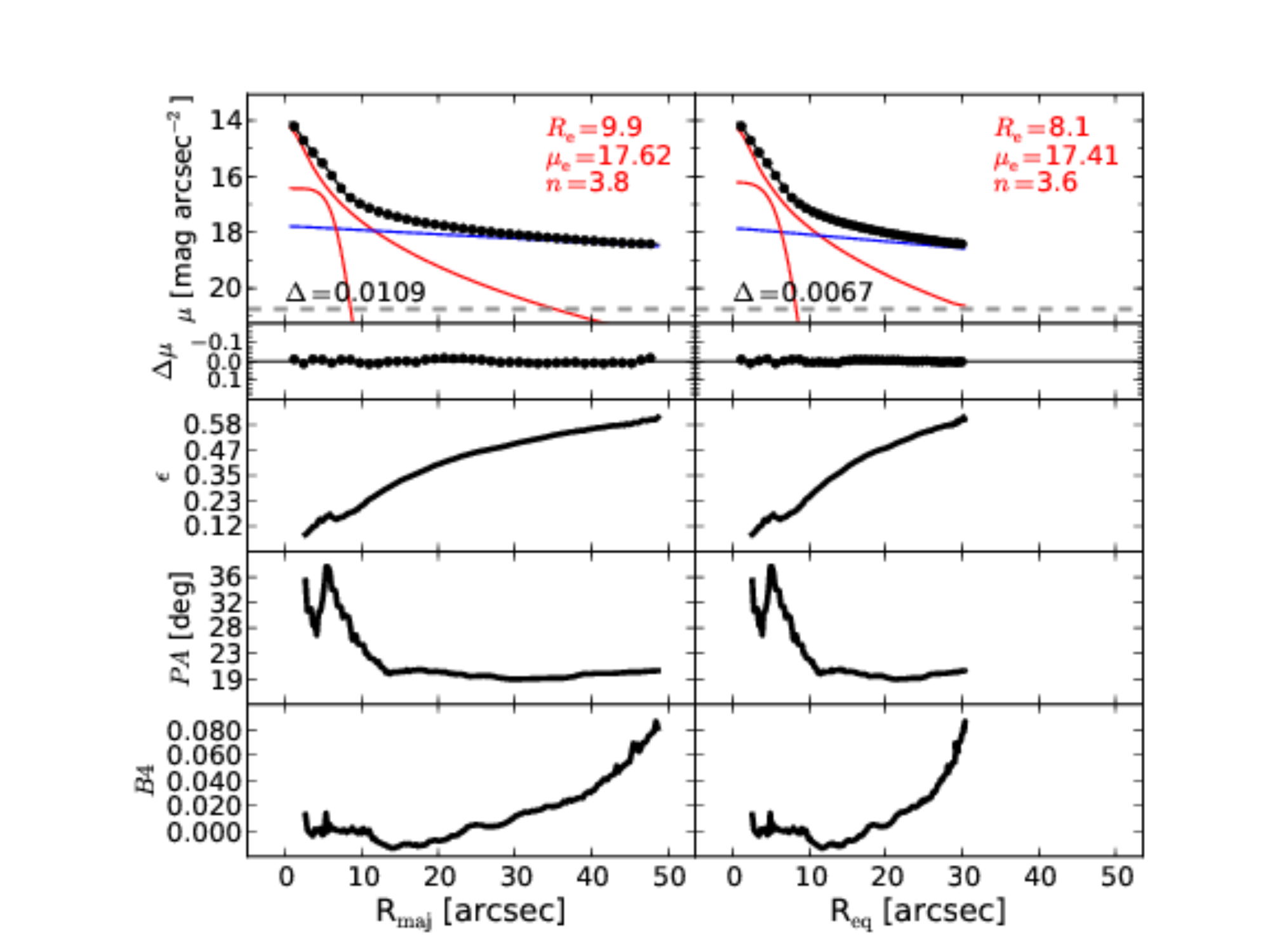}
  \caption{NGC 1300: 
  A face-on, barred spiral galaxy. 
  The morphology of NGC 1300 is quite complex. 
  In addition to a large-scale disk and a bulge, the galaxy is composed of 
  a large-scale bar that extends up to $R_{\rm maj} \lesssim 90''$, two prominent spiral arms 
  and an inner disk-like component ($R_{\rm maj} \lesssim 5''$), disclosed by the peaks in the ellipticity and $PA$ profiles.
  We truncate the light profile at $R_{\rm maj} \sim 50''$ and we fit the inner combination of the large-scale bar and the disk 
  as a single component, using an exponential function. 
  The embedded component is described with a low-$n$ S\'ersic profile. 
  }
  \end{center}
  \end{figure}

  \begin{table}[h]
  \small
  \caption{Best-fit parameters for the spheroidal component of NGC 1300.}
  \begin{center}
  \begin{tabular}{llcc}
  \hline
  {\bf Work} & {\bf Model}   & $\bm R_{\rm e,sph}$    & $\bm n_{\rm sph}$ \\
    &  &  $[\rm arcsec]$ & \\
  \hline
  1D maj. & S-bul + e-(d+bar) + S-id & $9.9$  &  $3.8$ \\
  1D eq.  & S-bul + e-(d+bar) + S-id & $8.1$  &  $3.6$ \\
  \hline
  S+11 2D    & S-bul + e-d & $85.4$  &  $3.0$ \\
  L+14 2D    & S-bul + e-d + PSF-n + e-id + S-bar + spiral arms & $10.4$  &  $4.3$ \\
  \hline
  \end{tabular}
  \end{center}
  \label{tab:n1300}
  \tablecomments{
  S+11 dramatically overestimated the bulge effective radius mainly because their model does not account for the large-scale bar.
  }
  \end{table}

  \clearpage\newpage\noindent
  {\bf NGC 1316 \\}

  \begin{figure}[h]
  \begin{center}
  \includegraphics[width=\fitfigurewidth]{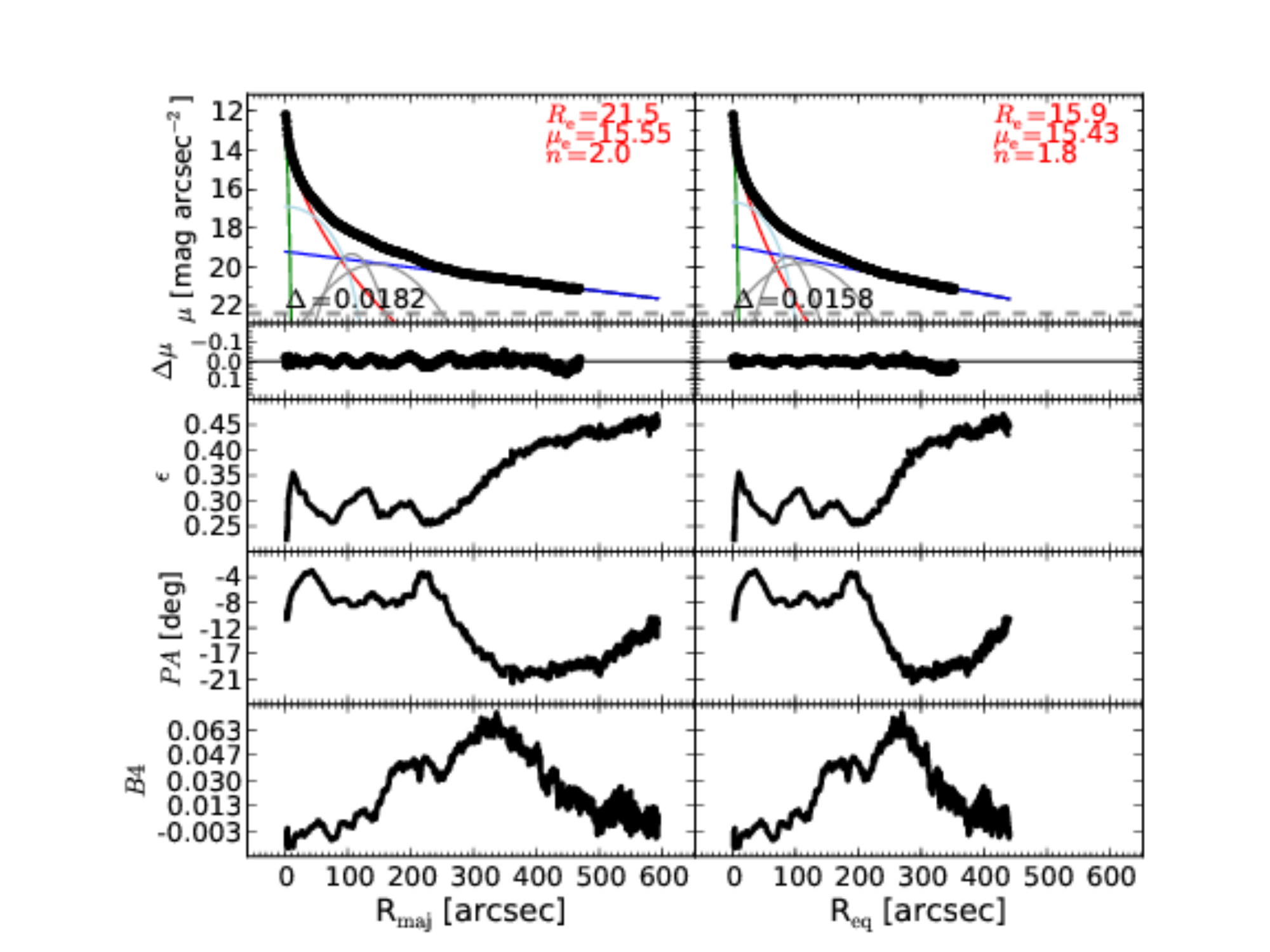}
  \caption{NGC 1316:  
  A merger. 
  The galaxy is composed of a bulge, an elongated structure that can be identified with a bar, 
  two obvious rings, an outer exponential disk or halo, and a bright nuclear component. }
  \end{center}
  \end{figure}

  \begin{table}[h]
  \small
  \caption{Best-fit parameters for the spheroidal component of NGC 1316.}
  \begin{center}
  \begin{tabular}{llcc}
  \hline
  {\bf Work} & {\bf Model}   & $\bm R_{\rm e,sph}$    & $\bm n_{\rm sph}$ \\
    &  &  $[\rm arcsec]$ & \\
  \hline
  1D maj. & S-bul + e-d + F-bar + 2 G-r + G-n & $21.5$  &  $2.0$ \\
  1D eq.  & S-bul + e-d + F-bar + 2 G-r + G-n & $15.9$  &  $1.8$ \\
  \hline
  S+11 2D      & S-bul + e-d + G-n & $93.0$  &  $5.0$ \\
  \hline
  \end{tabular}
  \end{center}
  \label{tab:n1316}
  \tablecomments{
  S+11 overestimated the bulge effective radius and S\'ersic index because their model does not take into account the bar.
  }
  \end{table}

  \clearpage\newpage\noindent
  {\bf NGC 1332 \\}

  \begin{figure}[h]
  \begin{center}
  \includegraphics[width=\fitfigurewidth]{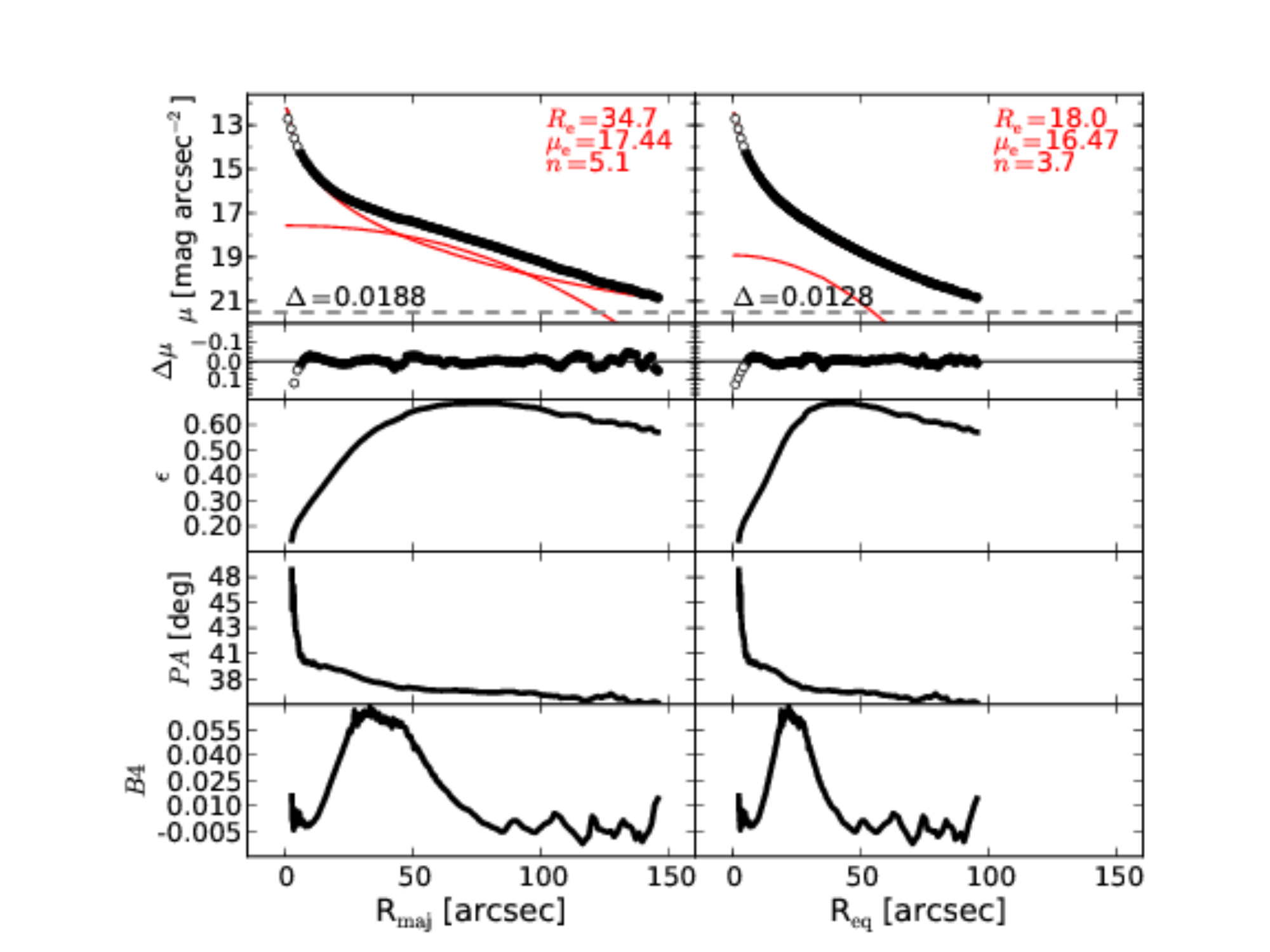}
  \caption{NGC 1332: 
  An edge-on elliptical/lenticular galaxy. 
  The identification of a disk is trivial due to its edge-on orientation, 
  although a visual inspection of the galaxy image is not enough to estabilish the radial extent of this disk. 
  The ellipticity profile indicates that the disk is indeed intermediate-sized, i.e.~it does not dominate at large radii. 
  Given the edge-on inclination of this disk, we model it with a low-$n$ S\'ersic profile. 
  The data within the innermost $6''.1$ are excluded from the fit.
  Our equivalent-axis decomposition returns a spheroidal component accounting for 95\% of the total light,
  and an embedded disk accounting for the remaining 5\%.
  In passing, we note that the bulge-disk decomposition performed by \citet{rusli2011} on NGC 1332 
  is significantly different from our best-fit model.
  \citet{rusli2011} did not identify the intermediate-scale embedded disk, 
  but instead proposed a model featuring a S\'ersic bulge ($n_{\rm sph} \sim 2.3$, $R_{\rm e,sph} \sim 8''$), accounting for 43\% of the total light, 
  and a large-scale exponential disk. 
  This result led them to the conclusion that NGC 1332 is a disk dominated lenticular galaxy 
  and is displaced from the $M_{\rm BH} - L_{\rm K,sph}$ (black hole mass vs. $K$-band spheroid luminosity) relation of \citet{marconihunt2003} 
  by an order of magnitude along the $M_{\rm BH}$ direction.
  Instead, the galaxy spheroid is a factor of two more luminous than claimed by \citet{rusli2011} and  
  NGC 1332 is not an outlier in our $M_{\rm BH} - L_{\rm sph}$ diagram. \\
  }
  \end{center}
  \end{figure}

  \begin{table}[h]
  \small
  \caption{Best-fit parameters for the spheroidal component of NGC 1332.}
  \begin{center}
  \begin{tabular}{llcc}
  \hline
  {\bf Work} & {\bf Model}   & $\bm R_{\rm e,sph}$    & $\bm n_{\rm sph}$ \\
    &  &  $[\rm arcsec]$ & \\
  \hline
  1D maj. & S-bul + S-id + m-c & $34.7$  &  $5.1$ \\
  1D eq.  & S-bul + S-id + m-c & $18.0$  &  $3.7$ \\
  \hline
  \end{tabular}
  \end{center}
  \label{tab:n1332}
  \end{table}

  \clearpage\newpage\noindent
  {\bf NGC 1374 \\}

  \begin{figure}[h]
  \begin{center}
  \includegraphics[width=\fitfigurewidth]{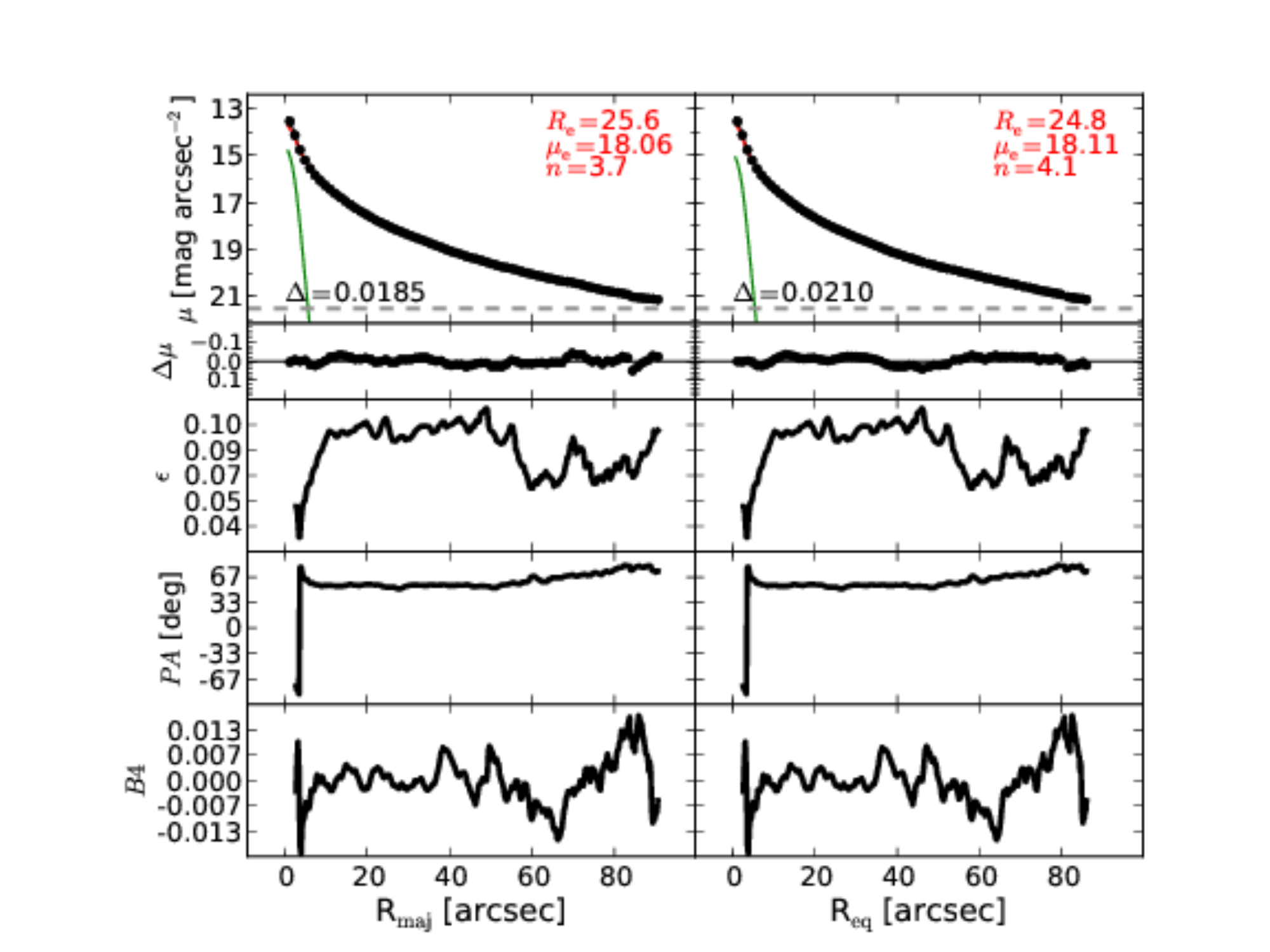}
  \caption{NGC 1374: 
  An elliptical galaxy. 
  A Gaussian function in our model accounts for the relatively faint additional nuclear component. 
  }
  \end{center}
  \end{figure}

  \begin{table}[h]
  \small
  \caption{Best-fit parameters for the spheroidal component of NGC 1374.}
  \begin{center}
  \begin{tabular}{llcc}
  \hline
  {\bf Work} & {\bf Model}   & $\bm R_{\rm e,sph}$    & $\bm n_{\rm sph}$ \\
    &  &  $[\rm arcsec]$ & \\
  \hline
  1D maj. & S-bul + G-n & $25.6$  &  $3.7$ \\
  1D eq.  & S-bul + G-n & $24.8$  &  $4.1$ \\
  2D      & S-bul + m-n & $25.2$  &  $3.7$ \\
  \hline
  \end{tabular}
  \end{center}
  \label{tab:n1374}
  \end{table}

  \clearpage\newpage\noindent
  {\bf NGC 1399 \\}
  
  \begin{figure}[h]
  \begin{center}
  \includegraphics[width=\fitfigurewidth]{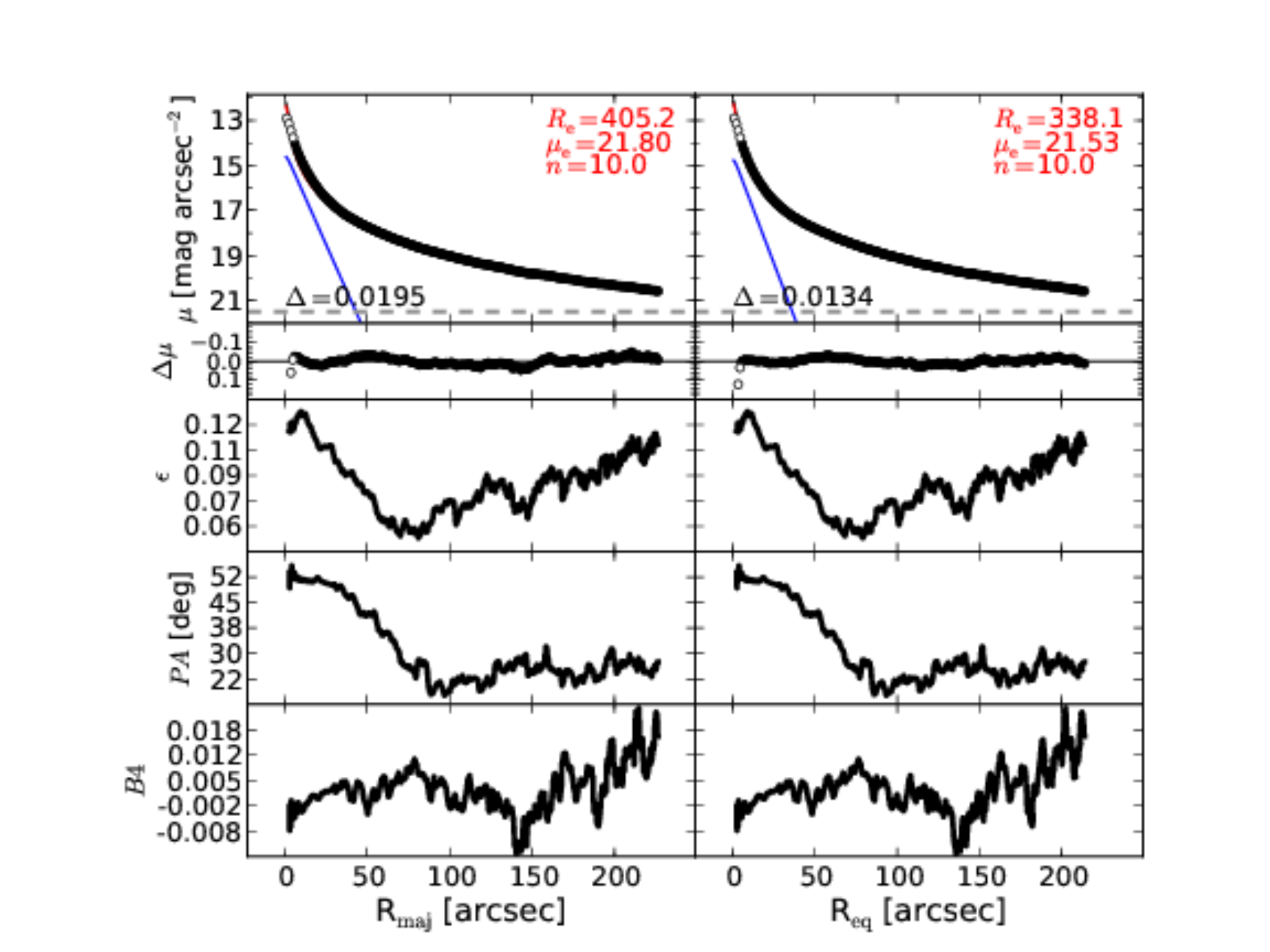}
  \caption{NGC 1399: 
  The central galaxy of the Fornax cluster, an elliptical galaxy with a slightly resolved partially depleted core \citep{rusli2013,dullograham2014cores}. 
  The nuclear activity of NGC 1399 is classified as Seyfert \citep{veroncettyveron2006},
  but the galaxy lacks dust emission \citep{tran2001}.
  The ellipticity and $PA$ profiles display a steep decline with increasing radius within $R_{\rm maj} \lesssim 60''$,
  suggesting the presence of an embedded disk.
  This inner component is also visible, although faint, in the unsharp mask.
  We note that, after excluding the innermost $6''.1$,
  a single S\'ersic profile is not sufficient to describe the whole galaxy light profile.
  The addition of an inner exponential function to model the disk notably improves the fit.
  }
  \end{center}
  \end{figure}

  \begin{table}[h]
  \small
  \caption{Best-fit parameters for the spheroidal component of NGC 1399.}
  \begin{center}
  \begin{tabular}{llcc}
  \hline
  {\bf Work} & {\bf Model}   & $\bm R_{\rm e,sph}$    & $\bm n_{\rm sph}$ \\
    &  &  $[\rm arcsec]$ & \\
  \hline
  1D maj. & S-bul + m-c + e-id & $405.2$  &  $10.0$ \\
  1D eq.  & S-bul + m-c + e-id & $338.1$  &  $10.0$ \\
  \hline
  GD07 1D maj.        & S-bul  		    & $-$      &  $16.8$ \\
  R+13 1D eq.         & core-S\'ersic + (S+e)-halo & $36.2$   &  $7.4$ \\
  L+14 2D             & S-bul + m-c		    & $154.0$  &  $11.1$ \\
  \hline
  \end{tabular}
  \end{center}
  \label{tab:n1399}
  \tablecomments{
  R+13 used a combination of a S\'ersic + exponential profile to model the galaxy's halo, 
  and obtained the smallest estimate of the effective radius, 
  and one which is at odds with the fact that central cluster galaxies typically have large sizes. 
  }
  \end{table}

  \clearpage\newpage\noindent
  {\bf NGC 2273 \\}

  \begin{figure}[h]
  \begin{center}
  \includegraphics[width=\fitfigurewidth]{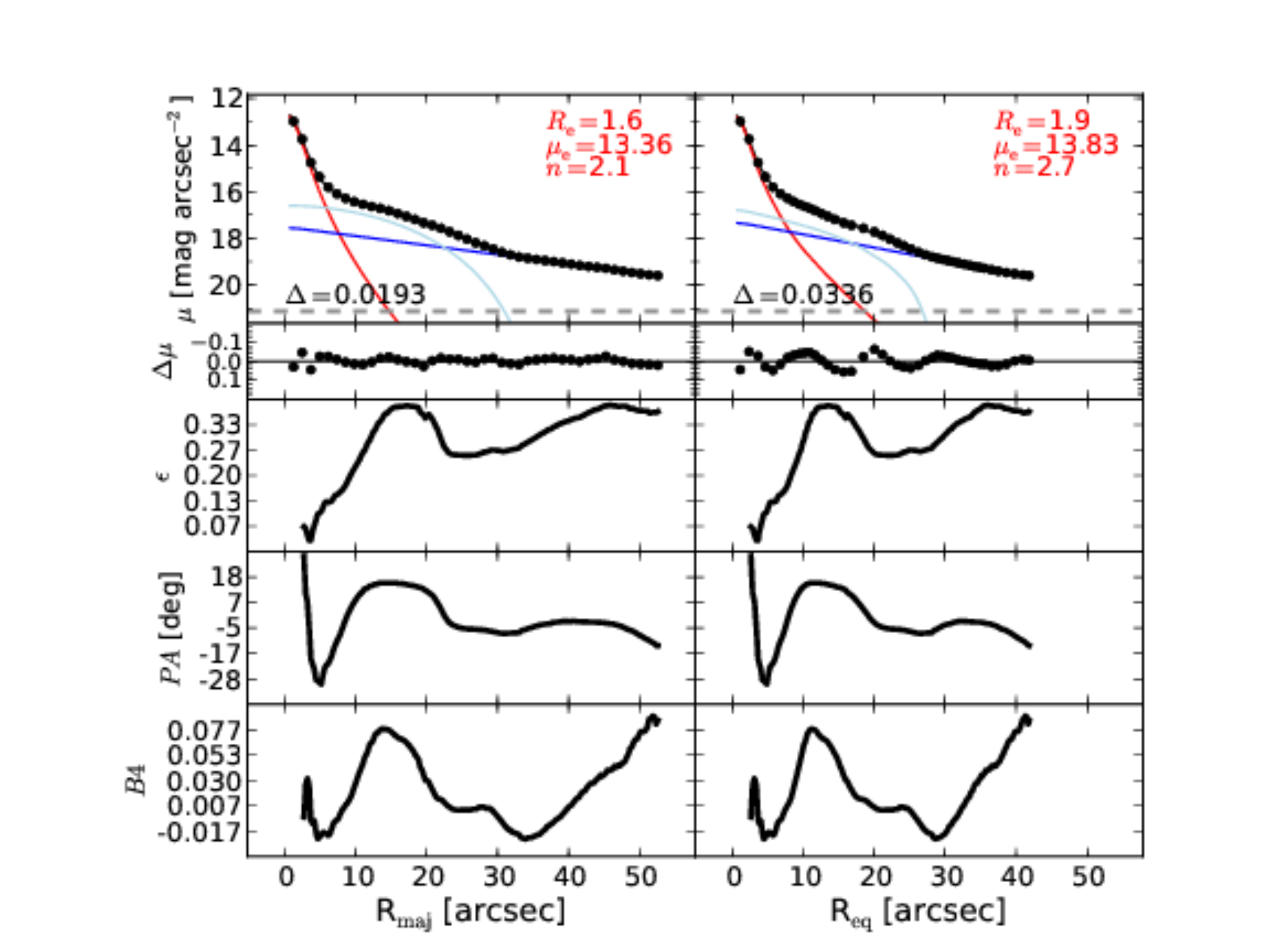}
  \caption{NGC 2273: 
  A barred spiral galaxy with a Seyfert AGN \citep{contini1998} 
  and circumnuclear dust \citep{simoeslopes2007}. 
  Its bar is surrounded by two tightly wound star-forming spiral arms that resemble a ring \citep{comeron2010}. 
  The bar of NGC 2273 extends out to $R_{\rm maj} \lesssim 25''$. 
  The pseudo-ring does not produce any evident swelling in the light profile, 
  therefore we do not account for it in the galaxy model.
  The isophotal parameters confirm the presence of a nuclear disky component within $R_{\rm maj} \lesssim 5''$.
  However, as noted by \cite{laurikainen2005}, any attempt to account for the embedded disk resulted in a degenerate model. 
  This is not surprising if one considers the poor spatial resolution of the galaxy image, 
  with the effective radius of the bulge comparable to the FWHM of the instrumental PSF.
  Although NGC 2273 hosts an optical AGN and nuclear dust, no central excess of light is observed in the 1D residuals.
  The addition of a nuclear component to the model does not significantly improve the fit nor change the bulge parameters. 
  }
  \end{center}
  \end{figure}

  \begin{table}[h]
  \small
  \caption{Best-fit parameters for the spheroidal component of NGC 2273.}
  \begin{center}
  \begin{tabular}{llccc}
  \hline
  {\bf Work} & {\bf Model}   & $\bm R_{\rm e,sph}$    & $\bm n_{\rm sph}$ \\
    &  &  $[\rm arcsec]$ & \\
  \hline
  1D maj. & S-bul + e-d + F-bar & $1.6$  &  $2.1$ \\
  1D eq.  & S-bul + e-d + F-bar & $1.9$  &  $2.7$ \\
  \hline
  L+10 2D         & S-bul + e-d + F-bar & $2.6$  &  $1.8$ \\
  \hline
  \end{tabular}
  \end{center}
  \label{tab:n2273}
  \end{table}

  \clearpage\newpage\noindent
  {\bf NGC 2549 \\}
  
  \begin{figure}[h]
  \begin{center}
  \includegraphics[width=\fitfigurewidth]{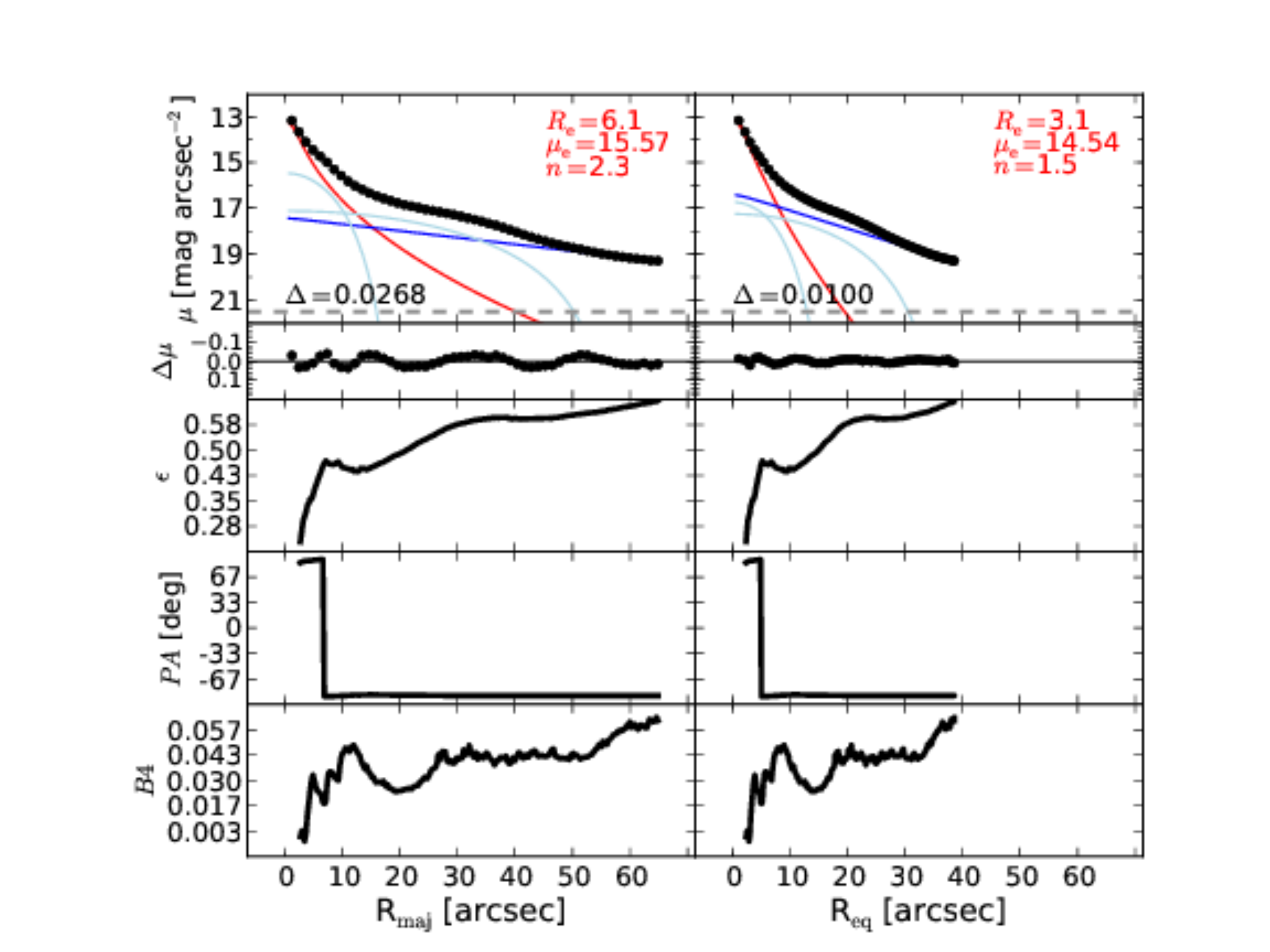}
  \caption{NGC 2549: 
  An edge-on barred lenticular galaxy. 
  Although the edge-on inclination of NGC 2549 complicates the identification of additional embedded components, 
  a large-scale bar ($R_{\rm maj} \lesssim 45''$) can be recognized in the galaxy image and -- more easily -- in the light profile.
  We model the large-scale bar with a Ferrer function.
  A peak in the ellipticity profile discloses the presence of a disky component
  embedded in the bulge ($R_{\rm maj} \lesssim 10''$).
  This inner component can be spotted also by looking at the velocity map (ATLAS$^{\rm 3D}$), and 
  we fit it with a Ferrer function.
  }
  \end{center}
  \end{figure}

  \begin{table}[h]
  \small
  \caption{Best-fit parameters for the spheroidal component of NGC 2549.}
  \begin{center}
  \begin{tabular}{llcc}
  \hline
  {\bf Work} & {\bf Model}   & $\bm R_{\rm e,sph}$    & $\bm n_{\rm sph}$ \\
    &  &  $[\rm arcsec]$ & \\
  \hline
  1D maj. & S-bul + e-d + F-bar + F-id & $6.1$  &  $2.3$ \\
  1D eq.  & S-bul + e-d + F-bar + F-id & $3.1$  &  $1.5$ \\
  2D      & S-bul + e-d + G-bar + G-id & $5.6$  &  $2.1$ \\
  \hline
  S+11 2D         & S-bul + e-d & $11.6$  &  $7.0$ \\
  \hline
  \end{tabular}
  \end{center}
  \label{tab:n2549}
  \tablecomments{
  The model of S+11 does not account for the large-scale bar and therefore largely overestimates the bulge S\'ersic index.
  }
  \end{table}

  \clearpage\newpage\noindent
  {\bf NGC 2778 \\}

  \begin{figure}[h]
  \begin{center}
  \includegraphics[width=\fitfigurewidth]{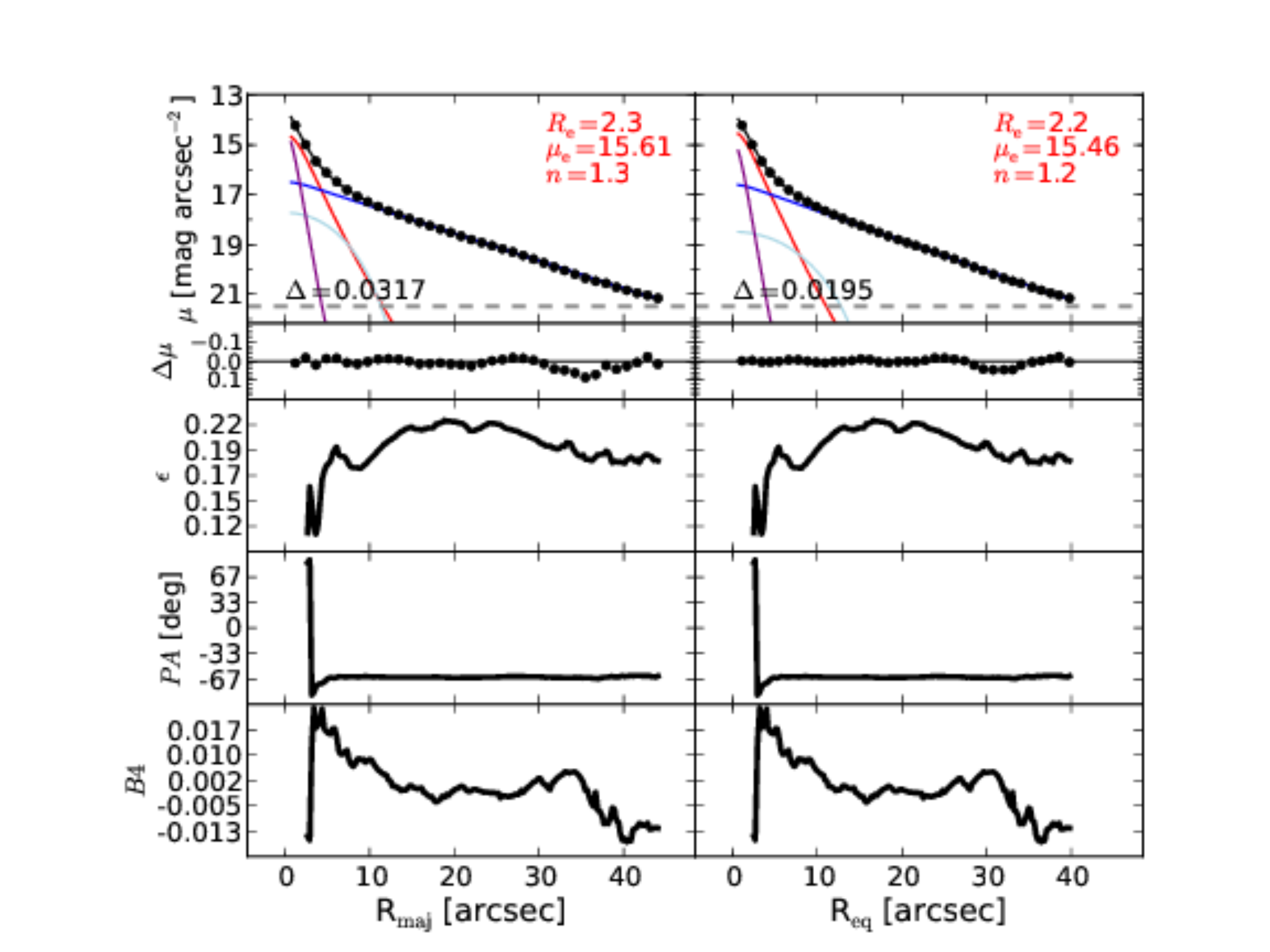}
  \caption{NGC 2778: 
  A face-on lenticular galaxy. 
  The peak in the ellipticity profile at $R_{\rm maj} \sim 5''$ reveals the existence of a nuclear component 
  embedded in the galaxy bulge.
  After an inspection of the unsharp mask, we identified this component with a small bar,
  that we model with a Ferrer function.
  We also account for some nuclear light excess by adding a PSF component to the model.
  We note that excluding the PSF component from our model does not significantly chiange the 
  bulge effective radius, but it does increase the bulge S\'ersic index to $n_{\rm sph} \sim 2$.
  }
  \end{center}
  \end{figure}

  \begin{table}[h]
  \small
  \caption{Best-fit parameters for the spheroidal component of NGC 2778.}
  \begin{center}
  \begin{tabular}{llcc}
  \hline
  {\bf Work} & {\bf Model}   & $\bm R_{\rm e,sph}$    & $\bm n_{\rm sph}$ \\
    &  &  $[\rm arcsec]$ & \\
  \hline
  1D maj. & S-bul + e-d + F-bar + PSF-n & $2.3$  &  $1.3$ \\
  1D eq.  & S-bul + e-d + F-bar + PSF-n & $2.2$  &  $1.2$ \\
  \hline
  GD07 1D maj.         & S-bul + e-d	     & $2.3$  &  $1.6$ \\
  S+11 2D         & S-bul + e-d	     & $2.5$  &  $2.5$ \\
  V+12 2D         & S-bul + e-d	     & $1.5$  &  $2.7$ \\
  L+14 2D         & S-bul + e-d + S-bar & $2.8$  &  $4.0$ \\
  \hline
  \end{tabular}
  \end{center}
  \label{tab:n2778}
  \end{table}

  \clearpage\newpage\noindent
  {\bf NGC 2787 \\}

  \begin{figure}[h]
  \begin{center}
  \includegraphics[width=\fitfigurewidth]{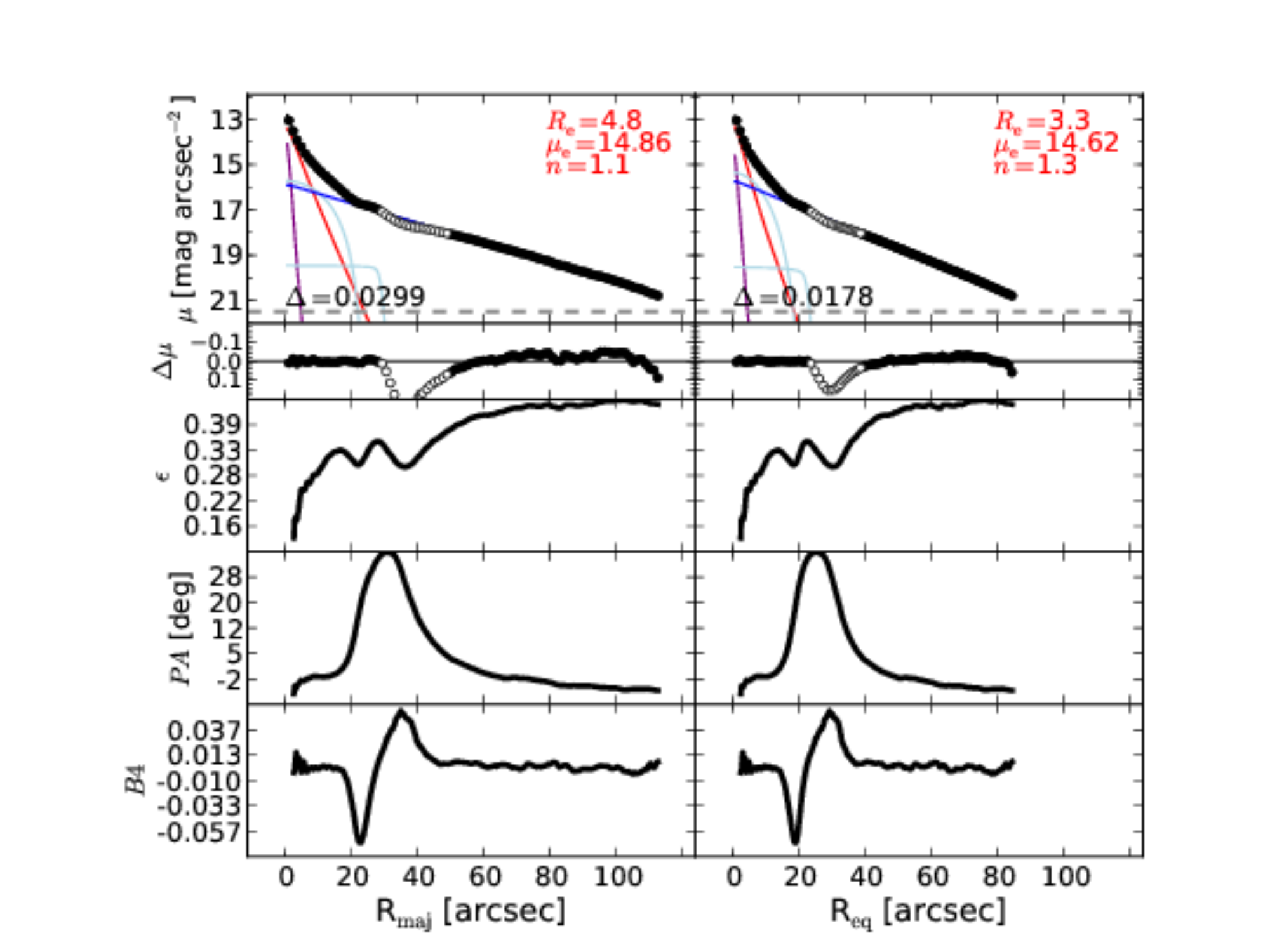}
  \caption{NGC 2787: 
  A barred lenticular galaxy with LINER nuclear activity \citep{veroncettyveron2006}.
  An inner disk is embedded in the bulge of this galaxy \citep{erwin2003n2787}.
  NGC 2787 features a spectacular structure of dust rings at its center \citep{erwinsparke2003}
  and a nuclear stellar disk (with size $\lesssim 1''.8$, \citealt{ledo2010}).
  NGC 2787 has an undoubtedly complex morphology. 
  This galaxy is composed of a bulge, a large-scale disk, a bar (see the peaks in the ellipticity, $PA$ and $B4$ profiles
  at $R_{\rm maj} \sim 30''$) with two ansae, an inner disk (or bar?) embedded in the bulge (see the peak in the ellipticity profile
  at $R_{\rm maj} \sim 17''$), and a bright nucleus.
  After masking the data affected by the ansae within $30'' \lesssim R_{\rm maj} \lesssim 50''$,
  we use a Ferrer function for the bar, a second Ferrer function for the inner disk (or bar) 
  and a PSF component for the nucleus. 
  }
  \end{center}
  \end{figure}

  \begin{table}[h]
  \small
  \caption{Best-fit parameters for the spheroidal component of NGC 2787.}
  \begin{center}
  \begin{tabular}{llcc}
  \hline
  {\bf Work} & {\bf Model}   & $\bm R_{\rm e,sph}$    & $\bm n_{\rm sph}$ \\
    &  &  $[\rm arcsec]$ & \\
  \hline
  1D maj. & S-bul + e-d + F-bar + F-id + PSF-nucleus & $4.8$  &  $1.1$ \\
  1D eq.  & S-bul + e-d + F-bar + F-id + PSF-nucleus & $3.3$  &  $1.3$ \\
  \hline
  GD07 1D maj.         & S-bul + e-d					  & $4.6$   &  $2.0$ \\
  L+10 2D         & S-bul + e-d + F-bar + F-l			  & $4.0$   &  $1.3$ \\
  S+11 2D         & S-bul + e-d + G-bar + G-n			  & $15.7$  &  $3.0$ \\
  L+14 2D         & S-bul + trunc. e-d + trunc. S-bar + S-id + PSF-n & $14.3$  &  $2.8$ \\
  \hline
  \end{tabular}
  \end{center}
  \label{tab:n2787}
  \tablecomments{
  S+11 found larger estimates of the effective radius and S\'ersic index because they did not account for the inner disk. 
  L+14 also reported a larger effective radius and S\'ersic index because they employed a truncated exponential disk and 
  truncated S\'ersic bar in their galaxy model. 
  }
  \end{table}

  \clearpage\newpage\noindent
  {\bf NGC 2974 \\}

  \begin{figure}[h]
  \begin{center}
  \includegraphics[width=\fitfigurewidth]{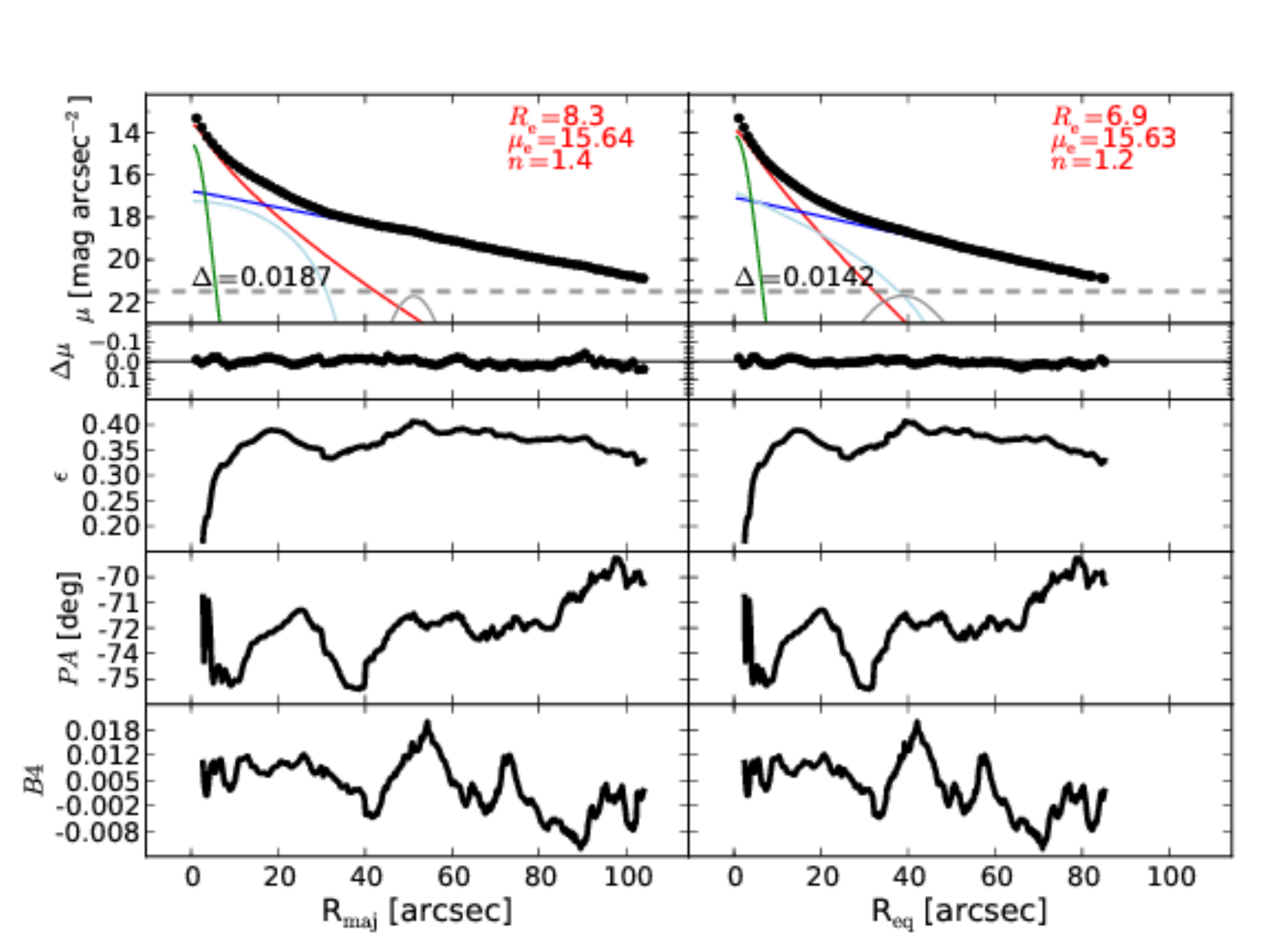}
  \caption{NGC 2974: 
  A spiral galaxy which has been misclassified as an elliptical galaxy in the RC3 catalog.
  The galaxy hosts a Seyfert AGN \citep{veroncettyveron2006} and filamentary dust in its center \citep{tran2001}.
  From an inspection of the unsharp mask, we identified a ring structure at $R_{\rm maj} \sim 50''$, 
  probably a residual of two tightly wound spiral arms,
  and an elongated bar-like component within $R_{\rm maj} \lesssim 30''$, that we fit with a Ferrer function.
  The pseudo-ring produces a peak in the $B4$ profile at $R_{\rm maj} \sim 50''$ and 
  the bar produces a peak in the ellipticity profile at $R_{\rm maj} \sim 20''$.
  Although the ring is extremely faint, it is important to account for it in the galaxy decomposition. 
  A model without the ring component results in a ``steeper'' exponential-disk (i.e. the disk would have a smaller scale length 
  and a brighter central surface brightness) and produces poor residuals within $R_{\rm maj} \lesssim 40''$.
  The nuclear component is fit with a Gaussian profile.
  }
  \end{center}
  \end{figure}

  \begin{table}[h]
  \small
  \caption{Best-fit parameters for the spheroidal component of NGC 2974.}
  \begin{center}
  \begin{tabular}{llcc}
  \hline
  {\bf Work} & {\bf Model}   & $\bm R_{\rm e,sph}$    & $\bm n_{\rm sph}$ \\
    &  &  $[\rm arcsec]$ & \\
  \hline
  1D maj. & S-bul + e-d + F-bar + G-n + G-r & $8.3$   &  $1.4$ \\
  1D eq.  & S-bul + e-d + F-bar + G-n + G-r & $6.9$   &  $1.2$ \\
  2D      & S-bul + e-d + G-bar + m-n	 & $10.6$  &  $1.3$ \\
  \hline
  S+11 2D         & S-bul + G-n & $27.2$  &  $3.0$ \\
  \hline
  \end{tabular}
  \end{center}
  \label{tab:n2974}
  \tablecomments{
  The model of S+11 does not account for the large-scale disk and thus overestimates the bulge effective radius and S\'ersic index.
  }
  \end{table}

  \clearpage\newpage\noindent
    
  {\bf NGC 3079 \\}

  \begin{figure}[h]
  \begin{center}
  \includegraphics[width=\fitfigurewidth]{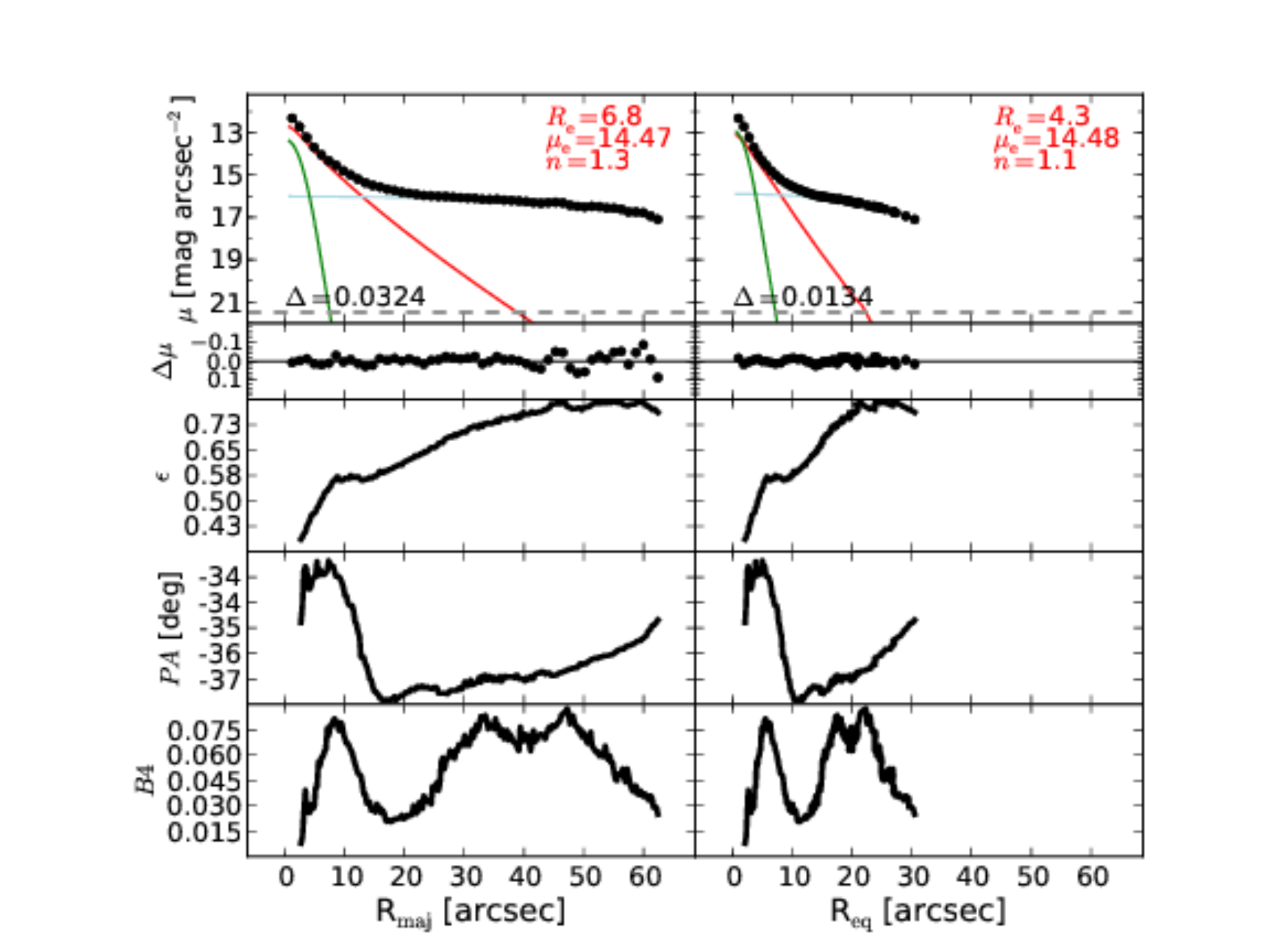}
  \caption{NGC 3079:
  An edge-on, late-type barred spiral galaxy 
  with a Seyfert AGN \citep{veroncettyveron2006} and circumnuclear dust \citep{martini2003}. 
  The bar extends out to $R_{\rm maj} \lesssim 70''$. 
  We truncate the light profile at $R_{\rm maj} \sim 65''$, before the ``edge'' of the bar, 
  and we successfully model the combination of the disk and the bar with a Ferrer function. 
  A Gaussian function accounts for the AGN emission. 
  }
  \end{center}
  \end{figure}
  
  \begin{table}[h]
  \small
  \caption{Best-fit parameters for the spheroidal component of NGC 2974.}
  \begin{center}
  \begin{tabular}{llcc}
  \hline
  {\bf Work} & {\bf Model}   & $\bm R_{\rm e,sph}$    & $\bm n_{\rm sph}$ \\
    &  &  $[\rm arcsec]$ & \\
  \hline
  1D maj. & S-bul + F-(bar+d) + G-n  & $6.8$  &  $1.3$ \\
  1D eq.  & S-bul + F-(bar+d) + G-n  & $4.3$  &  $1.1$ \\
  \hline
  S+11 2D         & S-bul + e-d + G-bar + G-n & $74.1$  &  $2.0$ \\
  \hline
  \end{tabular}
  \end{center}
  \label{tab:n2974}
  \tablecomments{
  It is not clear why S+11 obtained a dramatically larger bulge effective radius. 
  }
  \end{table}

  \clearpage\newpage\noindent
  {\bf NGC 3091 \\}

  \begin{figure}[h]
  \begin{center}
  \includegraphics[width=\fitfigurewidth]{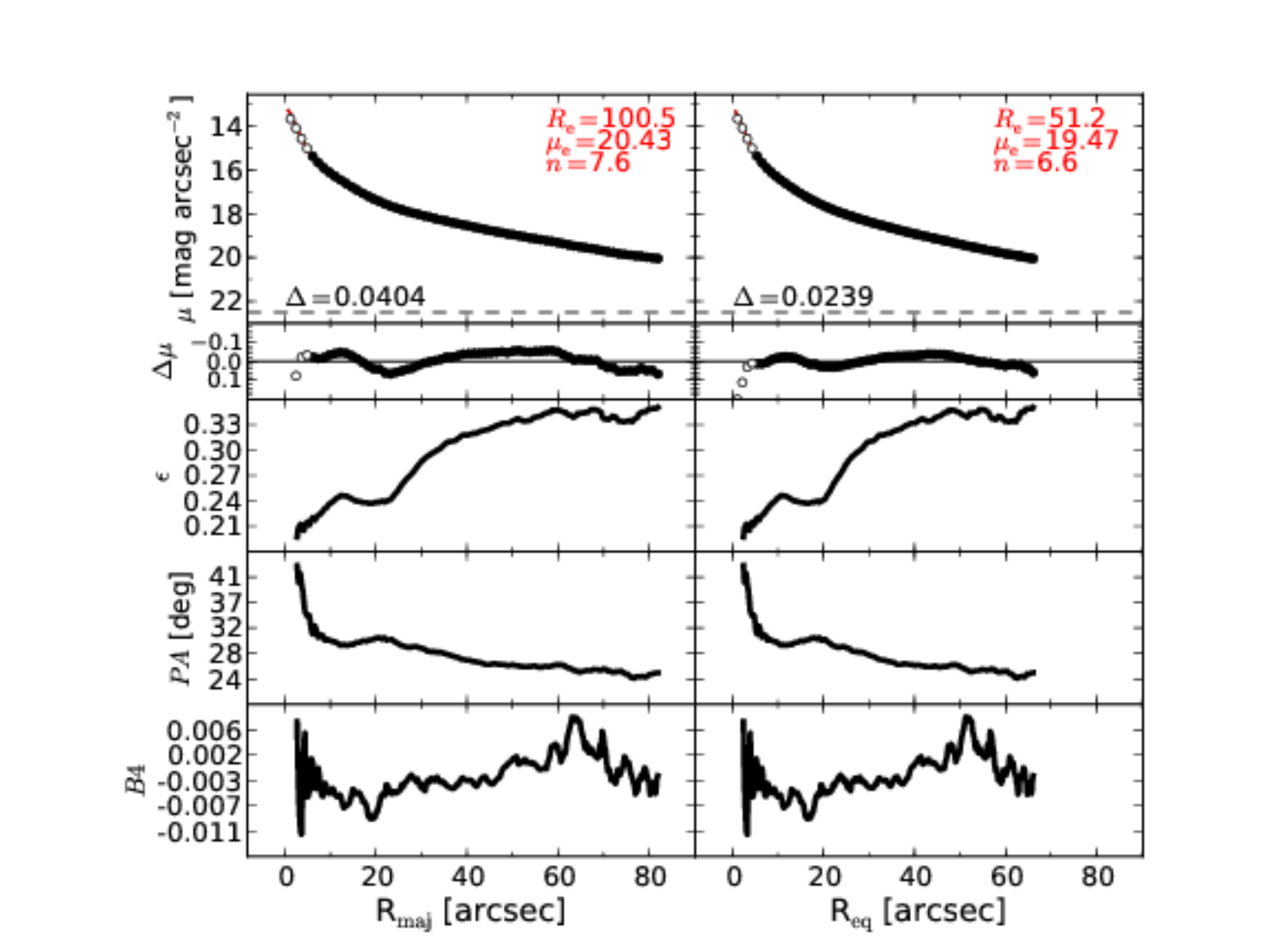}
  \caption{NGC 3091: 
  An elliptical galaxy with an unresolved partially depleted core \citep{rusli2013}. 
  The data within the innermost $6''.1$ are excluded from the fit.
  A faint embedded component ($R_{\rm maj} \lesssim 10''$) can be recognized in the unsharp mask and has a corresponding peak in the
  ellipticity profile.
  This extra component is clearly visible in the residual image obtained by subtracting a 2D S\'ersic model from the  
  galaxy image.
  However, any attempt to account for the embedded component resulted in an unsatisfactory fit and did not 
  significantly change the bulge parameters.
  We thus elect not to model the inner component. 
  }
  \end{center}
  \end{figure}

  \begin{table}[h]
  \small
  \caption{Best-fit parameters for the spheroidal component of NGC 3091.}
  \begin{center}
  \begin{tabular}{llcc}
  \hline
  {\bf Work} & {\bf Model}   & $\bm R_{\rm e,sph}$    & $\bm n_{\rm sph}$ \\
    &  &  $[\rm arcsec]$ & \\
  \hline
  1D maj. & S-bul + m-c & $100.5$  &  $7.6$ \\
  1D eq.  & S-bul + m-c & $51.2$	&  $6.6$ \\
  2D      & S-bul + m-c & $67.1$	&  $6.7$ \\
  \hline
  R+13 1D eq.         & core-S\'ersic & $91.0$  &  $9.3$ \\
  \hline
  \end{tabular}
  \end{center}
  \label{tab:n3091}
  \end{table}

  \clearpage\newpage\noindent
  {\bf NGC 3115 \\}

  \begin{figure}[h]
  \begin{center}
  \includegraphics[width=\fitfigurewidth]{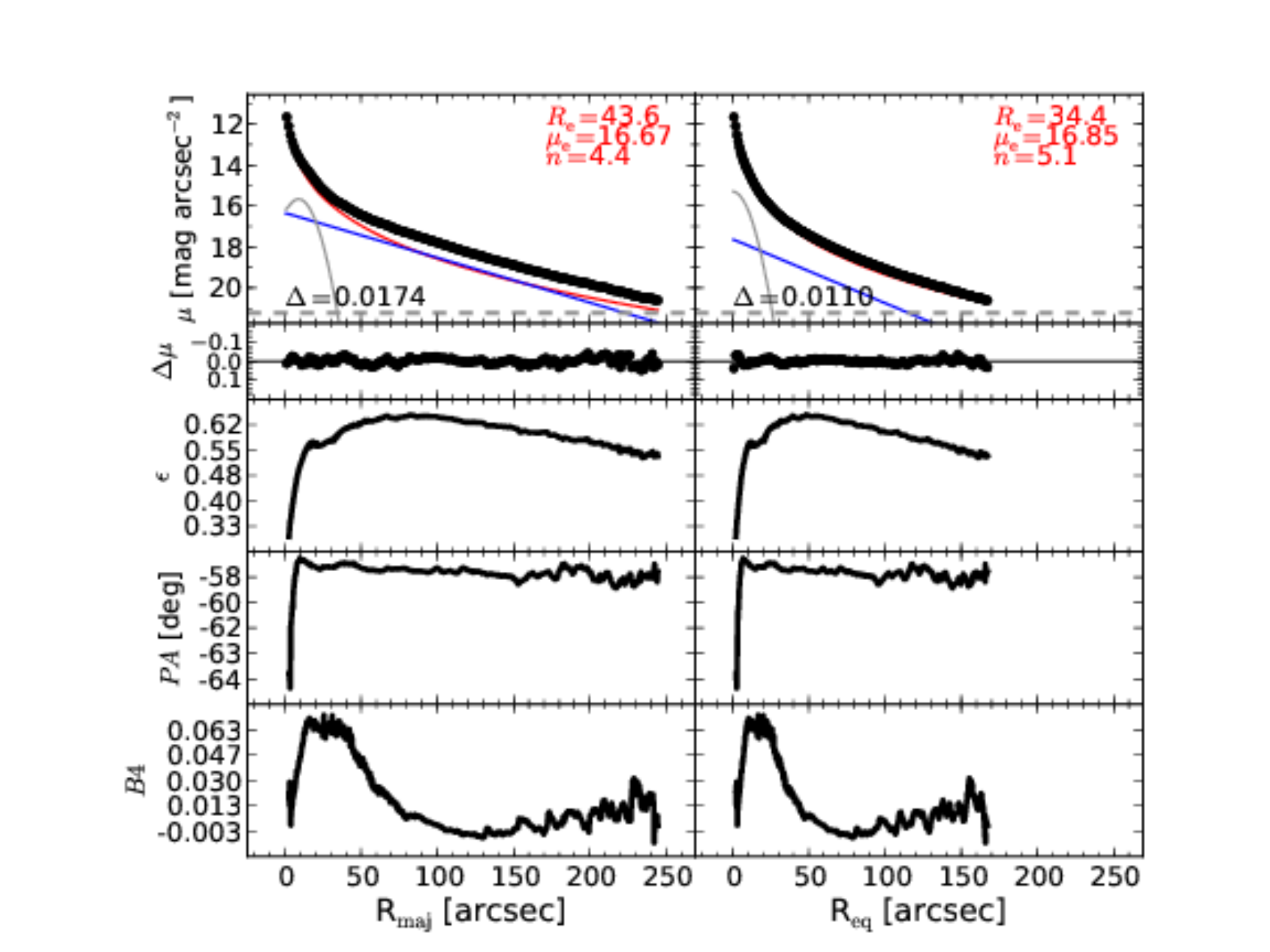}
  \caption{NGC 3115:
  An edge-on elliptical/lenticular galaxy. 
  This galaxy features an intermediate-scale disk and a nuclear stellar disk (with size $\sim 1''$, \citealt{scorzabender1995,ledo2010}).
  The presence of the intermediate-scale disk 
  is immediately evident by looking at the ellipticity and $B4$ profiles.
  From the galaxy center to the outskirts, the ellipticity increases until it reaches a maximum 
  ($\epsilon \sim 0.65$) at $R_{\rm maj} \sim 90''$. 
  After this point, it starts declining, being as low as $\epsilon \sim 0.55$ at $R_{\rm maj} \sim 240''$. 
  The galaxy isophotes display a positive diskyness within $R_{\rm maj} \sim 90''$, but become perfect ellipses 
  ($B4 \sim 0$) beyond that point.
  Because galaxy disks typically have fixed ellipticity, reflecting their inclination to our line of sight,
  the previous observations tell us that, going from the galaxy center to the outer regions,
  the disk of NGC 3115 becomes increasingly important relative to the spheroid light, reaching its maximum at $R \sim 90''$.
  Beyond $R_{\rm maj} \gtrsim 90''$, the contribution from the disk light starts declining more rapidly than the spheroid light.
  This interpretation is supported by the results of \cite{arnold2011n3115},
  who found that, within $R_{\rm maj} \sim 110''$, the bulge of NGC 3115 is flattened and rotates rapidly ($v/\sigma \gtrsim 1.5$),
  whereas at larger radii the rotation declines dramatically (to $v/\sigma \sim 0.7$), 
  but remains well aligned with the inner regions. 
  From an inspection of the unsharp mask, we identify a faint nuclear edge-on ring ($R_{\rm maj} \sim 15''$), 
  which produces a corresponding peak in the ellipticity profile. 
  We do not observe any nuclear excess of light in the 1D residuals. 
  The addition of a PSF component to the final model does not significantly change the outcome of the fit. 
  }
  \end{center}
  \end{figure}

  \begin{table}[h]
  \small
  \caption{Best-fit parameters for the spheroidal component of NGC 3115.}
  \begin{center}
  \begin{tabular}{llcc}
  \hline
  {\bf Work} & {\bf Model}   & $\bm R_{\rm e,sph}$    & $\bm n_{\rm sph}$ \\
    &  &  $[\rm arcsec]$ & \\
  \hline
  1D maj. & S-bul + e-d + G-r & $43.6$  &  $4.4$ \\
  1D eq.  & S-bul + e-d + G-r & $34.4$  &  $5.1$ \\
  \hline
  S+11 2D         & S-bul + e-d	       & $27.1$ &  $3.0$ \\
  L+14 2D         & S-bul + e-d + S-halo  & $3.9$  &  $3.0$ \\
  \hline
  \end{tabular}
  \end{center}
  \label{tab:n3115}
  \tablecomments{
  L+14 used a model with a bulge encased in a larger disk, and attributed the excess of light at large radii to a halo. 
  In doing so, they obviously obtained a smaller bulge effective radius.
  }
  \end{table}

  \clearpage\newpage\noindent
  {\bf NGC 3227 \\}

  \begin{figure}[h]
  \begin{center}
  \includegraphics[width=\fitfigurewidth]{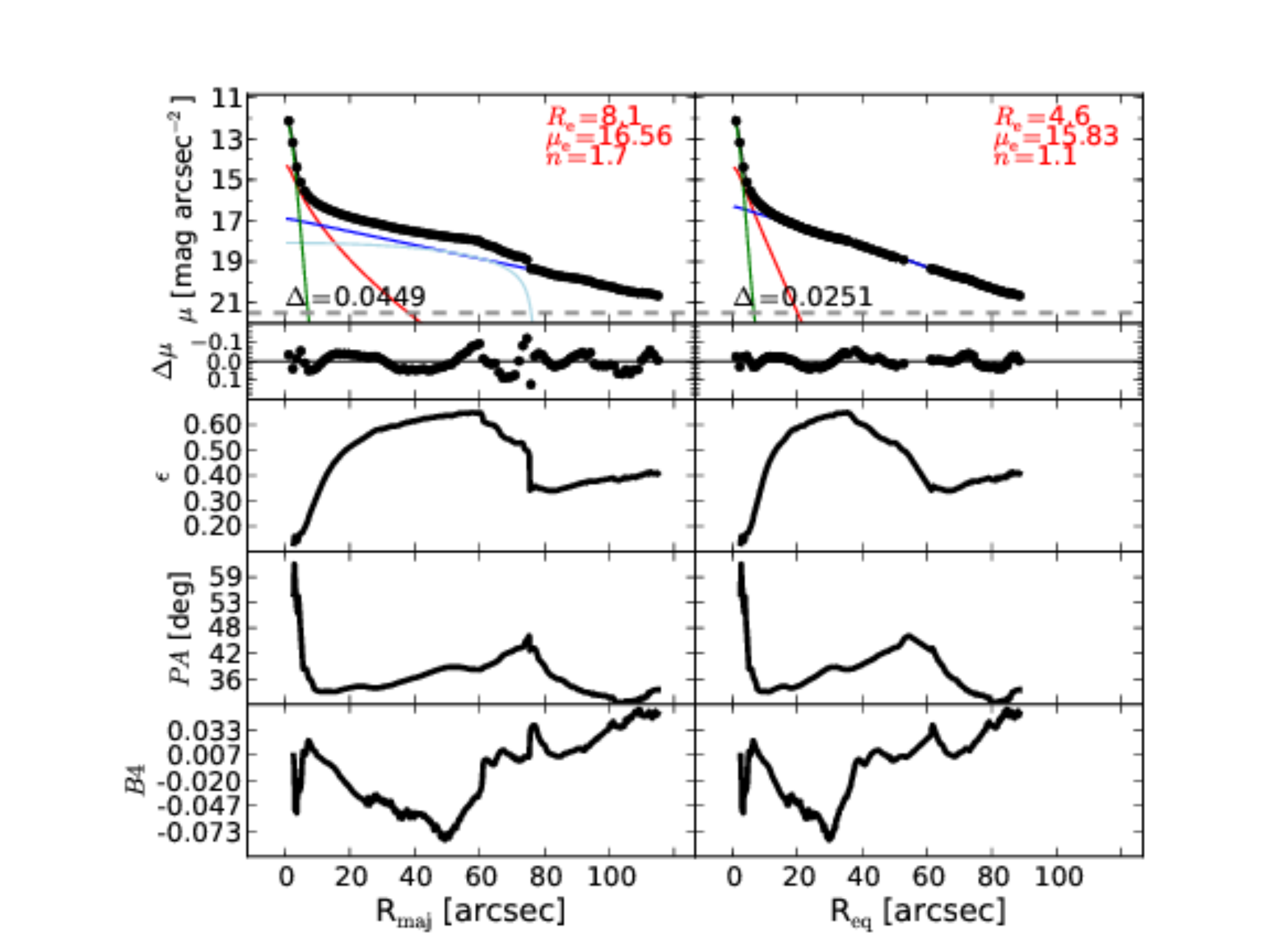}
  \caption{NGC 3227: 
  A spiral galaxy with a large-scale bar. 
  This galaxy hosts a Seyfert AGN \citep{khachikian1974} and circumnuclear dust \citep{martini2003}. 
  The large-scale bar produces an evident bump in the major-axis surface brightness profile,
  which is absent in the equivalent-axis profile. 
  We model the bar with a Ferrer function and the nucleus with a Gaussian function. 
  The Ferrer component is rejected by the equivalent-axis fit. 
  }
  \end{center}
  \end{figure}

  \begin{table}[h]
  \small
  \caption{Best-fit parameters for the spheroidal component of NGC 3227.}
  \begin{center}
  \begin{tabular}{llcc}
  \hline
  {\bf Work} & {\bf Model}   & $\bm R_{\rm e,sph}$    & $\bm n_{\rm sph}$ \\
    &  &  $[\rm arcsec]$ & \\
  \hline
  1D maj. & S-bul + e-d + F-bar + G-n   & $8.1$  &  $1.7$ \\
  1D eq.  & S-bul + e-d + [F-bar] + G-n & $4.6$  &  $1.1$ \\
  \hline
  L+10 2D         & S-bul + e-d + F-bar & $1.8$  &  $2.2$ \\
  S+11 2D         & S-bul + e-d	     & $82.9$ &  $4.0$ \\
  L+14 2D         & S-bul + e-d + S-bar & $0.7$  &  $4.1$ \\
  \hline
  \end{tabular}
  \end{center}
  \label{tab:n3227}
  \tablecomments{
  The models of L+10 and L+14 do not account for the bright nulcear component, 
  and thus underestimate the bulge effective radius and overestimate the bulge S\'ersic index.
  The bulge effective radius obtained by S+11 is larger because they did not model the bar. 
  }
  \end{table}

  \clearpage\newpage\noindent
  {\bf NGC 3245 \\}

  \begin{figure}[h]
  \begin{center}
  \includegraphics[width=\fitfigurewidth]{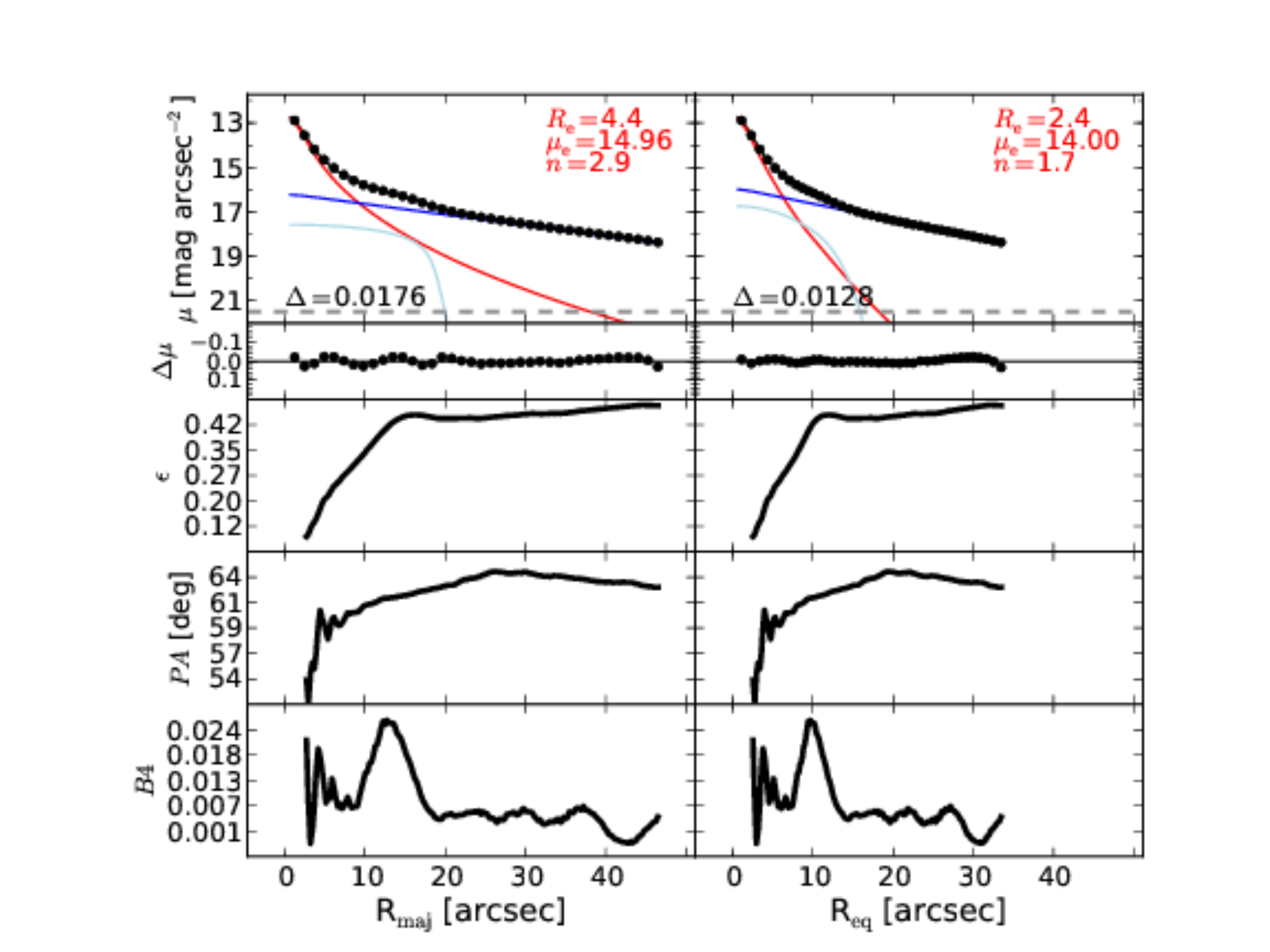}
  \caption{NGC 3245:
  A lenticular galaxy.
  The unsharp mask reveals the presence of an embedded disk ($R_{\rm maj} \lesssim 15''$), 
  confirmed by the corresponding peaks in the ellipticity and $B4$ profiles.
  We model this inner component with a Ferrer function. 
  }
  \end{center}
  \end{figure}

  \begin{table}[h]
  \small
  \caption{Best-fit parameters for the spheroidal component of NGC 3245.}
  \begin{center}
  \begin{tabular}{llcc}
  \hline
  {\bf Work} & {\bf Model}   & $\bm R_{\rm e,sph}$    & $\bm n_{\rm sph}$ \\
    &  &  $[\rm arcsec]$ & \\
  \hline
  1D maj. & S-bul + e-d + F-id & $4.4$  &  $2.9$ \\
  1D eq.  & S-bul + e-d + F-id & $2.4$  &  $1.7$ \\
  2D      & S-bul + e-d + G-id & $1.9$  &  $1.8$ \\
  \hline
  GD07 1D maj.         & S-bul + e-d		& $11.3$ &  $4.3$ \\
  L+10 2D         & S-bul + e-(d+l) + F-id & $4.0$  &  $2.4$ \\
  S+11 2D         & S-bul + e-d		& $4.6$  &  $2.5$ \\
  V+12 2D         & S-bul + e-d		& $3.5$  &  $2.6$ \\
  B+12 2D         & S-bul + e-d		& $4.0$  &  $1.6$ \\
  L+14 2D         & S-bul + e-d + S-bar	& $2.0$  &  $1.6$ \\
  \hline
  \end{tabular}
  \end{center}
  \label{tab:n3245}
  \end{table}

  \clearpage\newpage\noindent

  {\bf NGC 3377 \\}

  \begin{figure}[h]
  \begin{center}
  \includegraphics[width=\fitfigurewidth]{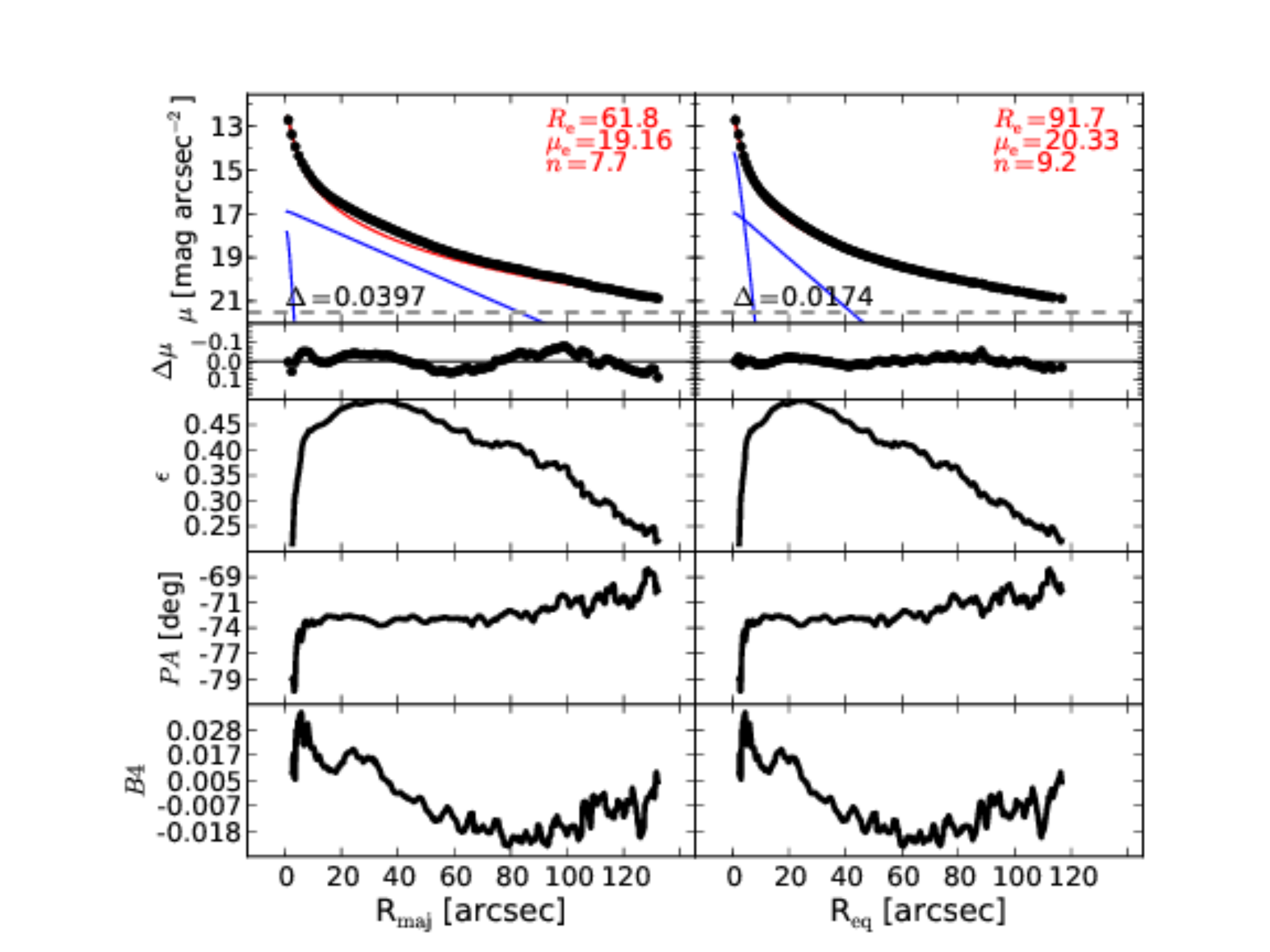}
  \caption{NGC 3377: 
  An elliptical galaxy with an intermediate-scale disk and a nuclear stellar disk \citep{ledo2010}. 
  For this galaxy, \citet{arnold2014} reported a strong decline in local specific angular momentum beyond $\sim 35''$, 
  with a disk-to-bulge transition occurring at about $R_{\rm maj} \sim 50''$.
  \citet{arnold2014} attempted a three-component decomposition (accounting for a bulge, an embedded disk and a central cusp), 
  but unfortunately they fit the spheroidal component with a limited two-parameter $R^{1/4}$ function 
  rather than a three-parameter $R^{1/n}$ S\'ersic function.
  Here we use a surface brightness profile twice as extended as that used by \citet{arnold2014} 
  and we identify a nuclear disk and an intermediate-scale disk, both embedded in a larger spheroidal component.
  The presence of the intermediate-scale disk is also evident in the unsharp mask of the image mosaic 
  and also from the ellipticity, which increases with increasing radius until $R_{\rm maj} \sim 45''$ 
  and then drops at larger radii.
  The nuclear stellar disk can be spotted from the peak at $R_{\rm maj} \sim 5''$ in the $B4$ profile.
  Each of the two disks is modelled with an exponential profile. 
  }
  \end{center}
  \end{figure}
  
  \begin{table}[h]
  \small
  \caption{Best-fit parameters for the spheroidal component of NGC 3377.}
  \begin{center}
  \begin{tabular}{llcc}
  \hline
  {\bf Work} & {\bf Model}   & $\bm R_{\rm e,sph}$    & $\bm n_{\rm sph}$ \\
    &  &  $[\rm arcsec]$ & \\
  \hline
  1D maj. & S-bul + 2 e-id     & $61.8$  &  $7.7$ \\
  1D eq.  & S-bul + 2 e-id     & $91.7$  &  $9.2$ \\
  2D      & S-bul + e-id + m-n & $71.8$  &  $3.7$ \\
  \hline
  GD07 1D maj.    & S-bul     		     & $44.1$  &  $3.0$ \\
  S+11 2D         & S-bul     		     & $55.2$  &  $6.0$ \\
  B+12 2D         & S-bul     		     & $43.5$  &  $3.5$ \\
  L+14 2D         & S-bul + e-id + e-d + S-halo & $10.1$  &  $6.0$ \\
  \hline
  \end{tabular}
  \end{center}
  \label{tab:n3377}
  \tablecomments{L+14 obtained the smallest estimate of the effective radius because they oversubtracted a halo. 
  In our 2D fit, we were not successful in modelling the nuclear disk and opted for masking the nuclear region of the galaxy. 
  Such 2D model resulted in a significantly lower S\'ersic index, that we trust being underestimated.  
  }
  \end{table}

  \clearpage\newpage\noindent
  {\bf NGC 3384 \\}

  \begin{figure}[h]
  \begin{center}
  \includegraphics[width=\fitfigurewidth]{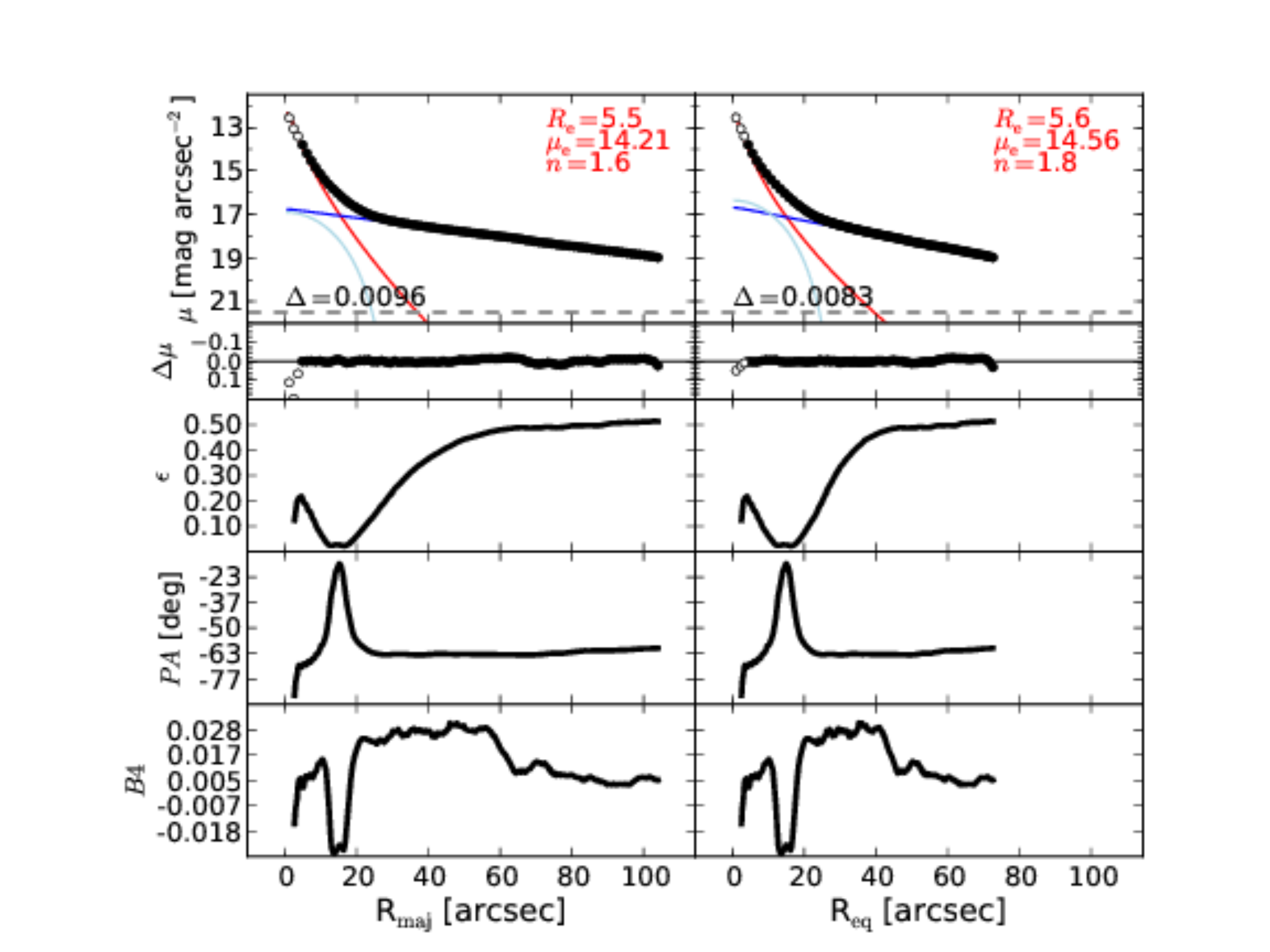}
  \caption{NGC 3384: 
  A barred lenticular galaxy that hosts 
  an embedded disk (with size $\lesssim 11''$, \citealt{erwin2004}) and 
  a nuclear stellar disk (with size $\lesssim 0''.8$, \citealt{ledo2010}).
  The isophotal parameters show clearly the presence of a boxy bar that extends out to $R_{\rm maj} \lesssim 18''$.
  We model the bar with a Ferrer function.
  An embedded disk can be seen in the unsharp mask and produces the peak in the ellipticity profile at $R_{\rm maj} \sim 4''$.
  However, any attempt to account for this component resulted in a degenerate model.
  We elect to fit neither the embedded disk nor the nuclear disk, by excluding from the fit the data within $R_{\rm maj} < 3''.7$. 
  }
  \end{center}
  \end{figure}

  \begin{table}[h]
  \small
  \caption{Best-fit parameters for the spheroidal component of NGC 3384.}
  \begin{center}
  \begin{tabular}{llcc}
  \hline
  {\bf Work} & {\bf Model}   & $\bm R_{\rm e,sph}$    & $\bm n_{\rm sph}$ \\
    &  &  $[\rm arcsec]$ & \\
  \hline
  1D maj. & S-bul + e-d + F-bar + m-n & $5.5$  &  $1.6$ \\
  1D eq.  & S-bul + e-d + F-bar + m-n & $5.6$  &  $1.8$ \\
  \hline
  GD07 1D maj.         & S-bul + e-d		      & $2.5$  &  $1.7$ \\
  L+10 2D         & S-bul + e-d + 2 F-bar        & $4.0$  &  $1.5$ \\
  S+11 2D         & S-bul + e-d + G-bar	      & $4.4$  &  $2.5$ \\
  B+12 2D         & S-bul + e-d		      & $8.3$  &  $2.3$ \\
  L+14 2D         & S-bul + e-d + 2 S-id + S-bar & $5.9$  &  $2.5$ \\
  \hline
  \end{tabular}
  \end{center}
  \label{tab:n3384}
  \end{table}

  \clearpage\newpage\noindent
  {\bf NGC 3393 \\}

  \begin{figure}[h]
  \begin{center}
  \includegraphics[width=\fitfigurewidth]{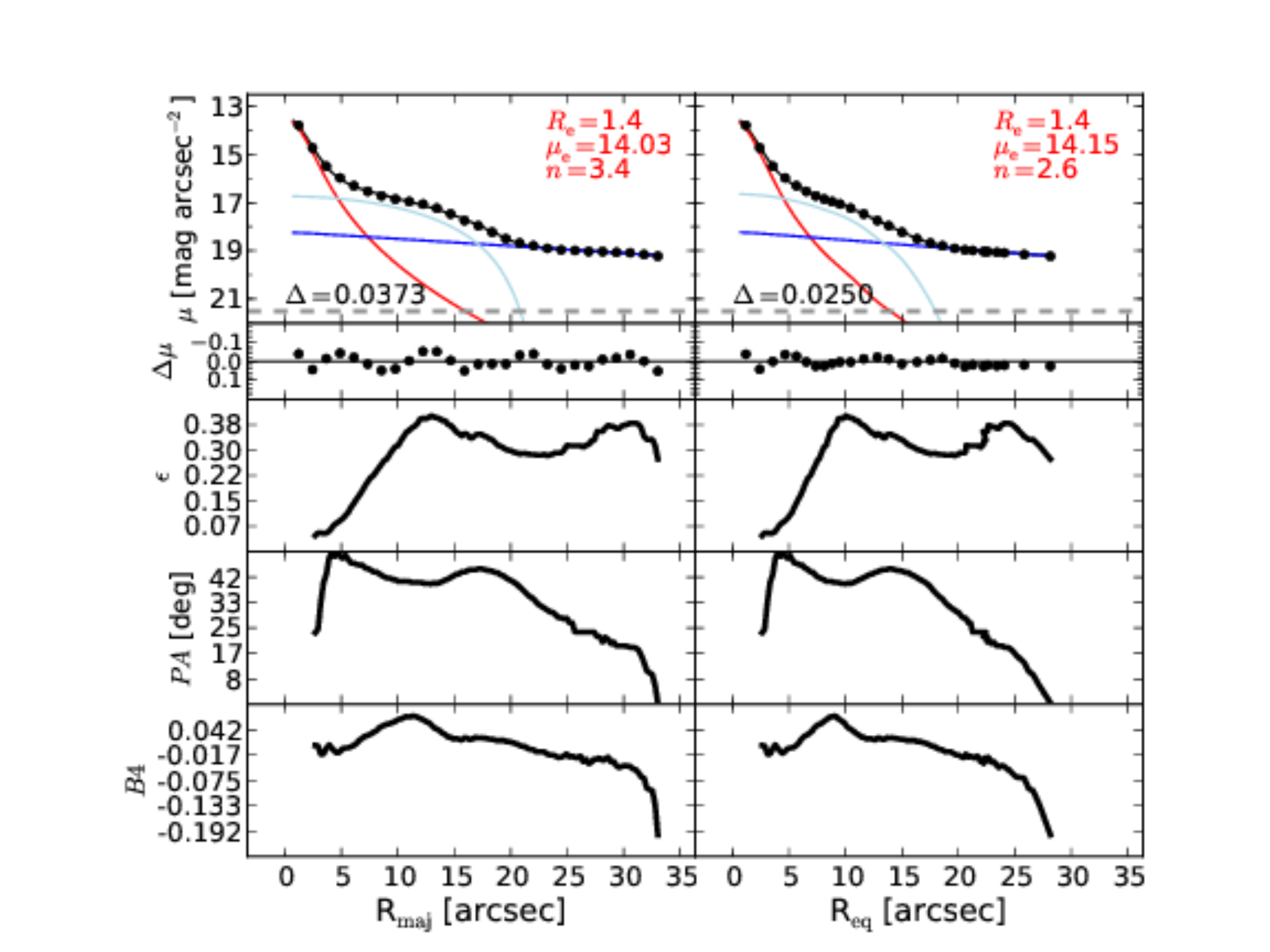}
  \caption{NGC 3393: 
  An edge-on spiral galaxy hosting a Seyfert AGN \citep{diaz1988n3393} and 
  circumnuclear dust \citep{martini2003}.
  The galaxy has a large-scale bar and a nuclear bar (with size $\lesssim 3''$, \citealt{erwin2004}).  
  The large-scale bar ($R_{\rm maj} \lesssim 20''$) is modeled with a Ferrer function.
  We do not model the nuclear bar and the AGN component because the poor spatial resolution of the data does not allow to fit them.
  }
  \end{center}
  \end{figure}

  \begin{table}[h]
  \small
  \caption{Best-fit parameters for the spheroidal component of NGC 3393.}
  \begin{center}
  \begin{tabular}{llcc}
  \hline
  {\bf Work} & {\bf Model}   & $\bm R_{\rm e,sph}$    & $\bm n_{\rm sph}$ \\
    &  &  $[\rm arcsec]$ & \\
  \hline
  1D maj. & S-bul + e-d + F-bar & $1.4$  &  $3.4$ \\
  1D eq.  & S-bul + e-d + F-bar & $1.4$  &  $2.6$ \\
  2D      & S-bul + e-d + G-bar & $1.2$  &  $1.9$ \\
  \hline
  \end{tabular}
  \end{center}
  \label{tab:n3393}
  \end{table}

  \clearpage\newpage\noindent
  {\bf NGC 3414 \\}

  \begin{figure}[h]
  \begin{center}
  \includegraphics[width=\fitfigurewidth]{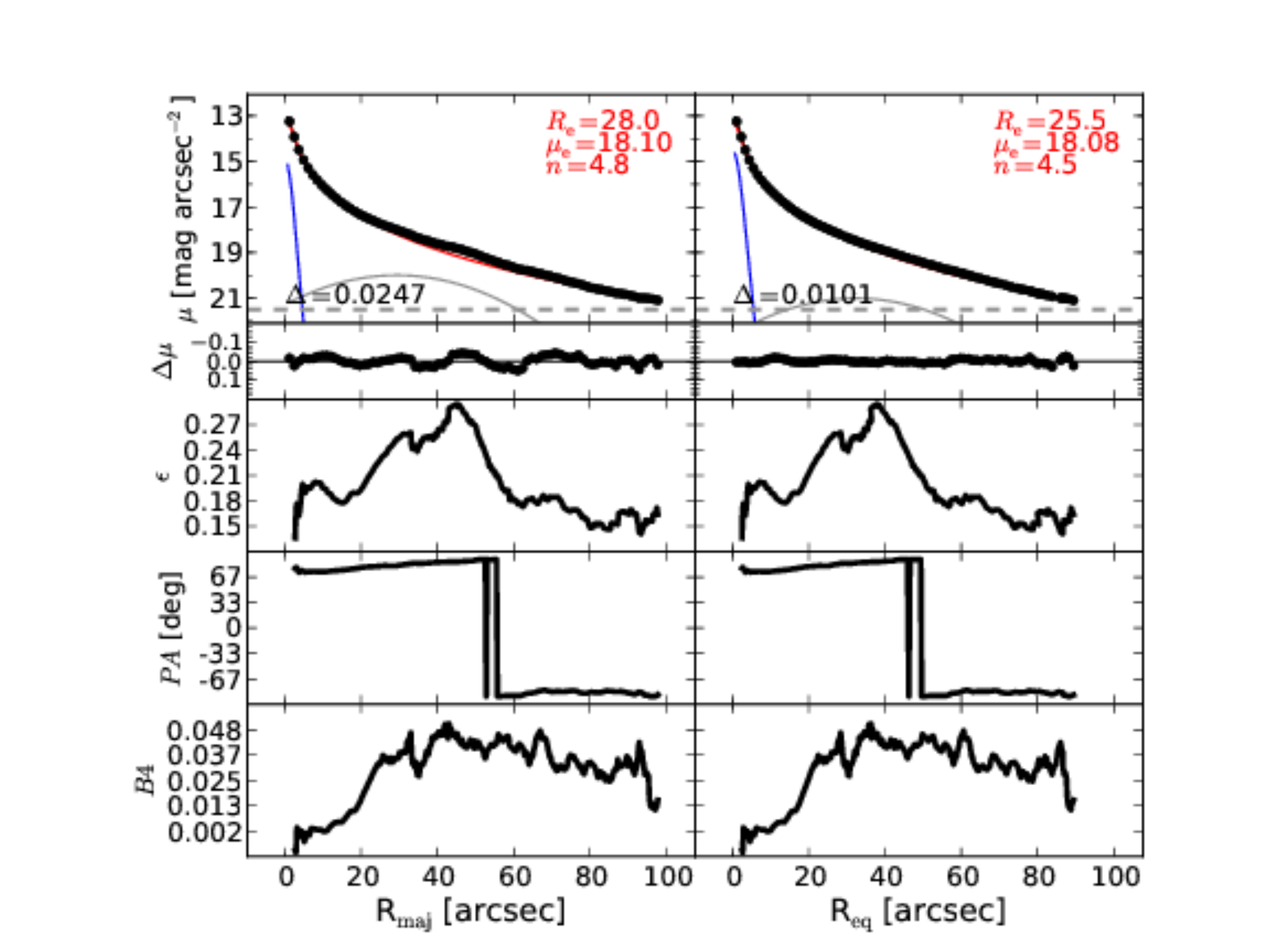}
  \caption{NGC 3414: 
  A peculiar lenticular/elliptical galaxy. 
  Different interpretations have been proposed about the morphology of this galaxy.
  \cite{whitmore1990} suggested that NGC 3414 is a spheroidal galaxy 
  with a large-scale edge-on polar ring, rather than a face-on barred lenticular galaxy.
  \cite{laurikainen2010} decomposed NGC 3414 as a barred lenticular galaxy, but 
  they cautioned against the uncertainty of their solution due to the possible misinterpretation of the galaxy morphology.
  The unsharp mask clearly shows that an embedded component does not extend all the way through the 
  center of the galaxy, but instead it is truncated at $R_{\rm maj} \sim 20''$.
  However, an elongated structure in the galaxy center resembles a nuclear edge-on disk ($R_{\rm maj} \lesssim 5''$).
  The velocity map (ATLAS$^{\rm 3D}$) shows rotation within $R_{\rm maj} \lesssim 5''$, no rotation within $5'' \lesssim R_{\rm maj} \lesssim 15''$,
  and counterrotation beyond $R_{\rm maj} \gtrsim 15''$.
  The velocity pattern is consistent with the galaxy being a spheroidal system containing an edge-on polar ring and an edge-on nuclear disk.
  }
  \end{center}
  \end{figure}

  \begin{table}[h]
  \small
  \caption{Best-fit parameters for the spheroidal component of NGC 3414.}
  \begin{center}
  \begin{tabular}{llcc}
  \hline
  {\bf Work} & {\bf Model}   & $\bm R_{\rm e,sph}$    & $\bm n_{\rm sph}$ \\
    &  &  $[\rm arcsec]$ & \\
  \hline
  1D maj. & S-bul + e-id + G-r & $28.0$  &  $4.8$ \\
  1D eq.  & S-bul + e-id + G-r & $25.5$  &  $4.5$ \\
  \hline
  L+10 2D         & S-bul + e-d + F-bar	  & $5.0$  &  $2.6$ \\
  \hline
  \end{tabular}
  \end{center}
  \label{tab:n3414}
  \tablecomments{
  L+10 used a model with a large-scale exponential disk, thus they obtained a smaller bulge effective radius and S\'ersic index.
  }
  \end{table}

  \clearpage\newpage\noindent
  {\bf NGC 3489 \\}

  \begin{figure}[h]
  \begin{center}
  \includegraphics[width=\fitfigurewidth]{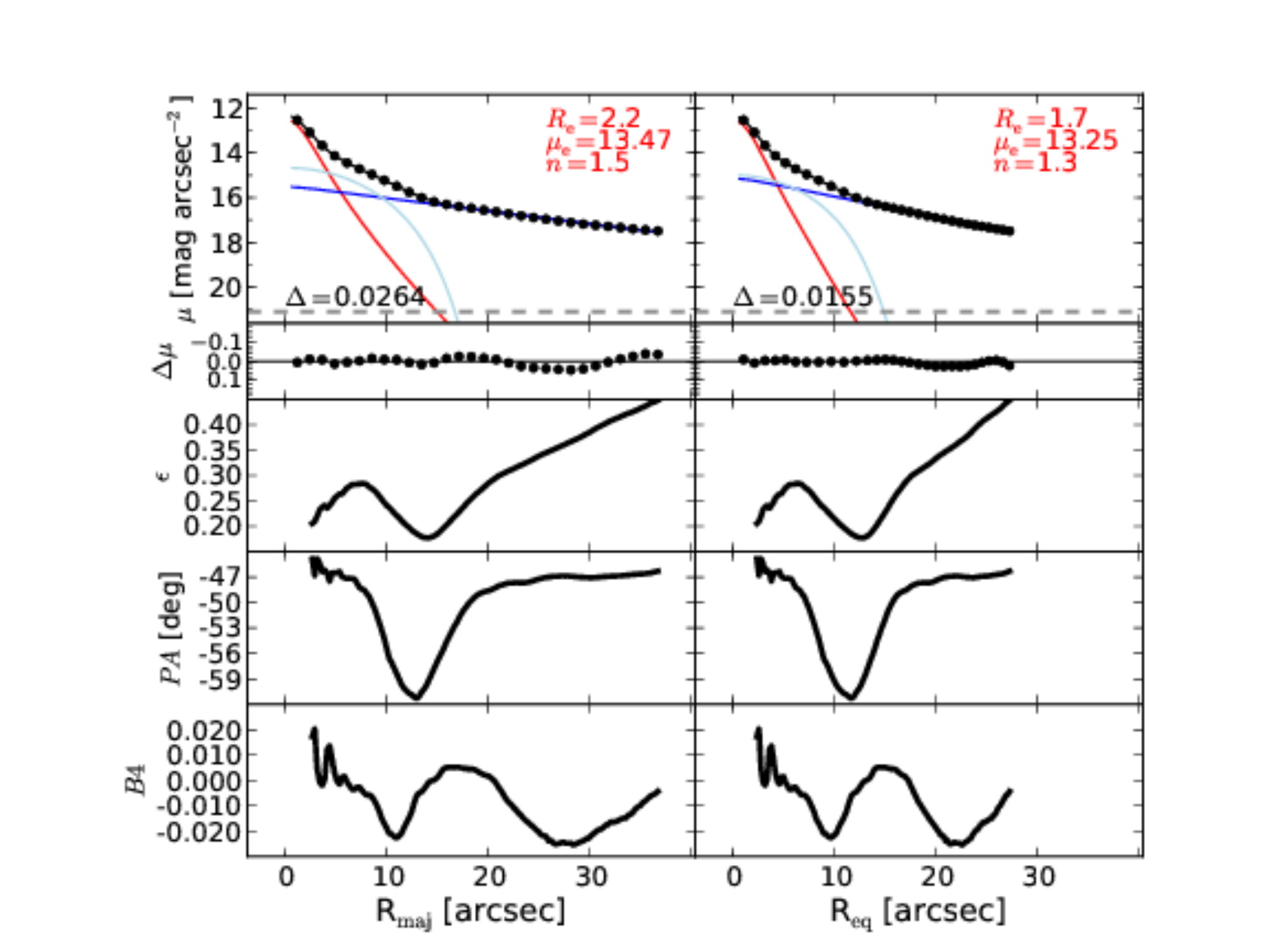}
  \caption{NGC 3489: 
  A barred lenticular galaxy. 
  Its disk has a smooth anti-truncation at $R_{\rm maj} \sim 85''$ \citep{erwin2008}.
  The bar is fit with a Ferrer function.
  }
  \end{center}
  \end{figure}

  \begin{table}[h]
  \small
  \caption{Best-fit parameters for the spheroidal component of NGC 3489.}
  \begin{center}
  \begin{tabular}{llcc}
  \hline
  {\bf Work} & {\bf Model}   & $\bm R_{\rm e,sph}$    & $\bm n_{\rm sph}$ \\
    &  &  $[\rm arcsec]$ & \\
  \hline
  1D maj. & S-bul + e-d + F-bar & $2.2$  &  $1.5$ \\
  1D eq.  & S-bul + e-d + F-bar & $1.7$  &  $1.3$ \\
  2D      & S-bul + e-d + G-bar & $1.7$  &  $2.1$ \\
  \hline
  L+10 2D         & S-bul + e-d + F-bar & $2.0$  &  $2.1$ \\
  S+11 2D         & S-bul + e-d & $4.6$  &  $1.5$ \\
  \hline
  \end{tabular}
  \end{center}
  \label{tab:n3489}
  \end{table}

  \clearpage\newpage\noindent

  {\bf NGC 3585 \\}

  \begin{figure}[h]
  \begin{center}
  \includegraphics[width=\fitfigurewidth]{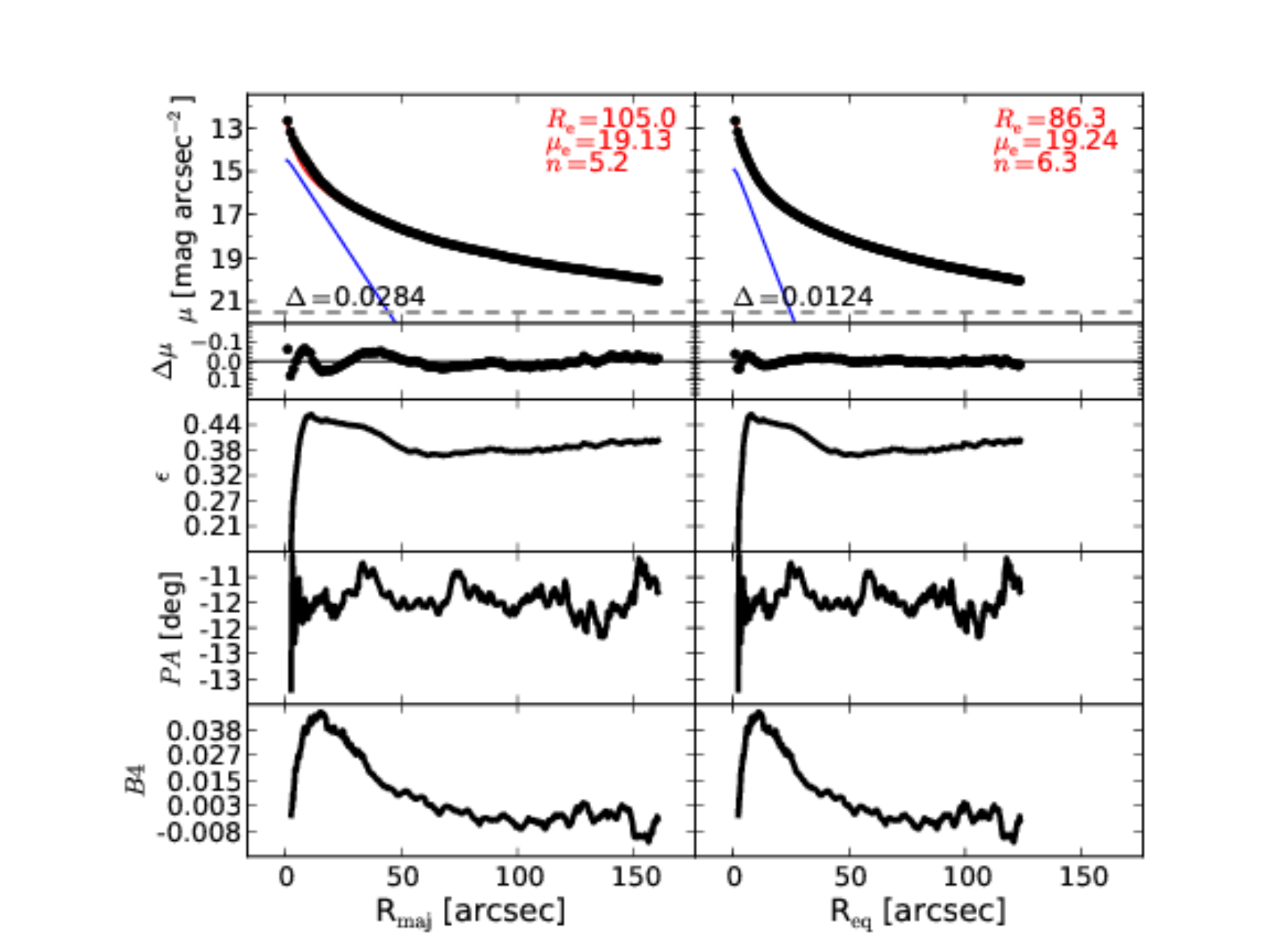}
  \caption{NGC 3585: 
  An elliptical galaxy with an embedded disk \citep{scorzabender1995}. 
  The embedded disk is visible in the unsharp mask and modeled with an exponential function. }
  \end{center}
  \end{figure}

  \begin{table}[h]
  \small
  \caption{Best-fit parameters for the spheroidal component of NGC 3585.}
  \begin{center}
  \begin{tabular}{llcc}
  \hline
  {\bf Work} & {\bf Model}   & $\bm R_{\rm e,sph}$    & $\bm n_{\rm sph}$ \\
    &  &  $[\rm arcsec]$ & \\
  \hline
  1D maj. & S-bul + e-id & $105.0$  &  $5.2$ \\
  1D eq.  & S-bul + e-id & $86.3$  &  $6.3$ \\
  \hline
  S+11 2D         & S-bul + e-d      & $15.5$   &  $2.5$ \\
  \hline
  \end{tabular}
  \end{center}
  \label{tab:n3585}
  \tablecomments{
  S+11 2D obtained smaller estimates of the effective radius and S\'ersic index because
  they included a large-scale disk in their model.
  }
  \end{table}

  \clearpage\newpage\noindent

  {\bf NGC 3607 \\}

  \begin{figure}[h]
  \begin{center}
  \includegraphics[width=\fitfigurewidth]{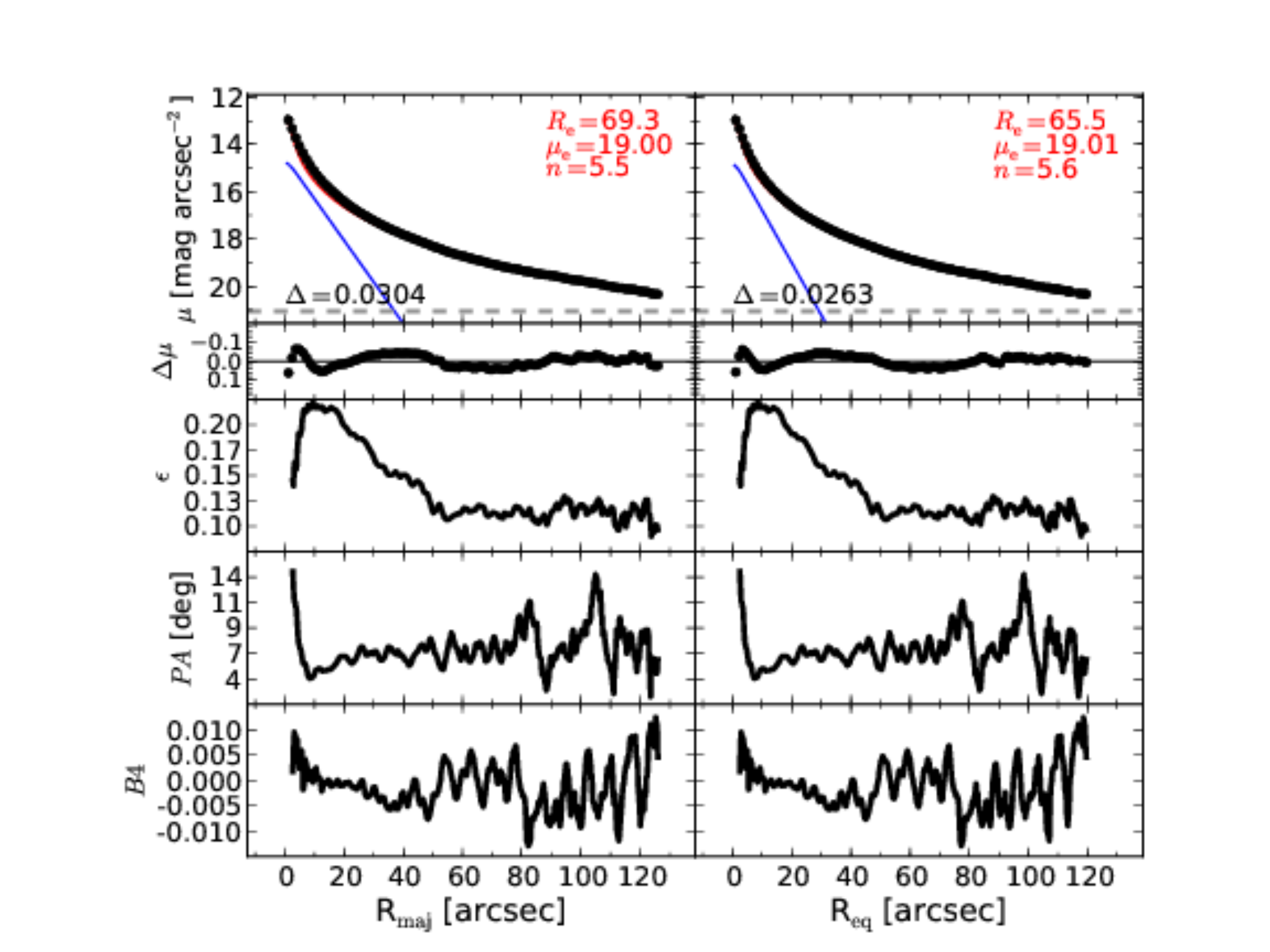}
  \caption{NGC 3607: 
  An elliptical galaxy with an embedded disk. 
  The velocity map (ATLAS$^{\rm 3D}$) shows rotation only within $R_{\rm maj} \lesssim 20''$. 
   }
  \end{center}
  \end{figure}
  
  \begin{table}[h]
  \small
  \caption{Best-fit parameters for the spheroidal component of NGC 3607.}
  \begin{center}
  \begin{tabular}{llcc}
  \hline
  {\bf Work} & {\bf Model}   & $\bm R_{\rm e,sph}$    & $\bm n_{\rm sph}$ \\
    &  &  $[\rm arcsec]$ & \\
  \hline
  1D maj. & S-bul + e-id & $69.3$  &  $5.5$ \\
  1D eq.  & S-bul + e-id & $65.5$  &  $5.6$ \\
  2D      & S-bul + e-id & $60.0$  &  $5.3$ \\
  \hline
  L+10 2D	   & S-bul + e-d      & $6.5$   &  $1.5$ \\
  S+11 2D	   & S-bul + G-n      & $44.6$  &  $5.0$ \\
  B+12 2D	   & S-bul	      & $56.3$  &  $4.7$ \\
  \hline
  \end{tabular}
  \end{center}
  \label{tab:n3607}
  \tablecomments{L+10 obtained the smallest estimates of the effective radius and S\'ersic index because
  they included a large-scale disk in their model.
  }
  \end{table}

  \clearpage\newpage\noindent

  {\bf NGC 3608 \\}

  \begin{figure}[h]
  \begin{center}
  \includegraphics[width=\fitfigurewidth]{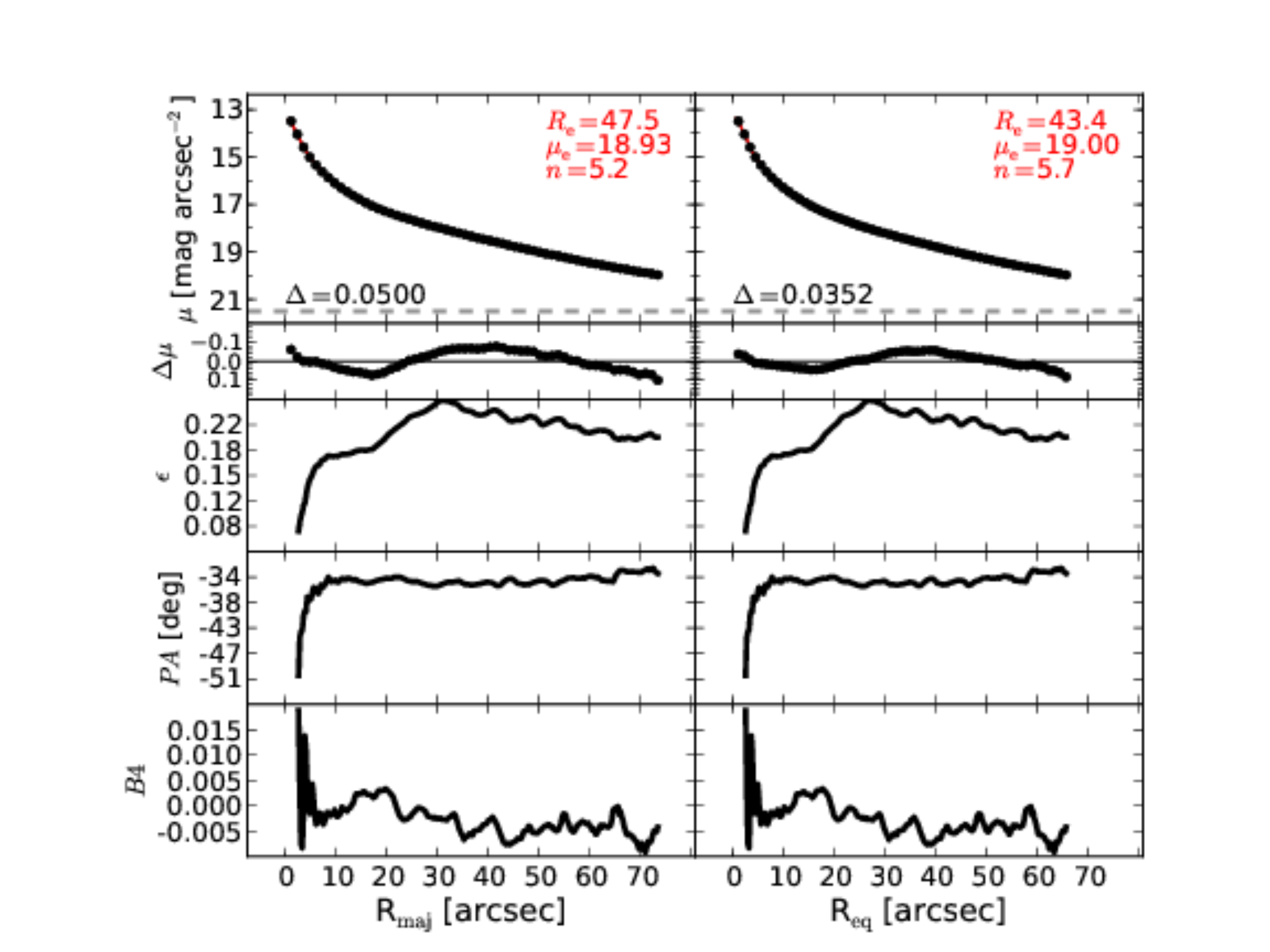}
  \caption{NGC 3608: 
  An elliptical galaxy with an unresolved partially depleted core \citep{rusli2013} 
  and a kinematically decoupled core (ATLAS$^{\rm 3D}$, SLUGGS). 
  We find that masking the nuclear region of this galaxy is unnnecessary and we fit the light profile with a single S\'ersic model.
  }
  \end{center}
  \end{figure}
  
  \begin{table}[h]
  \small
  \caption{Best-fit parameters for the spheroidal component of NGC 3608.}
  \begin{center}
  \begin{tabular}{llcc}
  \hline
  {\bf Work} & {\bf Model}   & $\bm R_{\rm e,sph}$    & $\bm n_{\rm sph}$ \\
    &  &  $[\rm arcsec]$ & \\
  \hline
  1D maj. & S-bul & $47.5$  &  $5.2$ \\
  1D eq.  & S-bul & $43.4$  &  $5.7$ \\
  2D      & S-bul & $62.0$  &  $7.0$ \\
  \hline
  S+11 2D         & S-bul            & $56.4$  &  $6.0$ \\
  B+12 2D         & S-bul            & $182.2$  &  $9.0$ \\
  R+13 1D eq.         & core-S\'ersic    & $56.9$  &  $6.3$ \\
  L+14 2D         & S-bul            & $48.9$  &  $6.6$ \\
  \hline
  \end{tabular}
  \end{center}
  \label{tab:n3608}
  \tablecomments{B+12 obtained the largest estimates of the effective radius and S\'ersic index, 
  possibly due to incorrect sky subtraction. }
  \end{table}

  \clearpage\newpage\noindent
  {\bf NGC 3842 \\}
  
  \begin{figure}[h]
  \begin{center}
  \includegraphics[width=\fitfigurewidth]{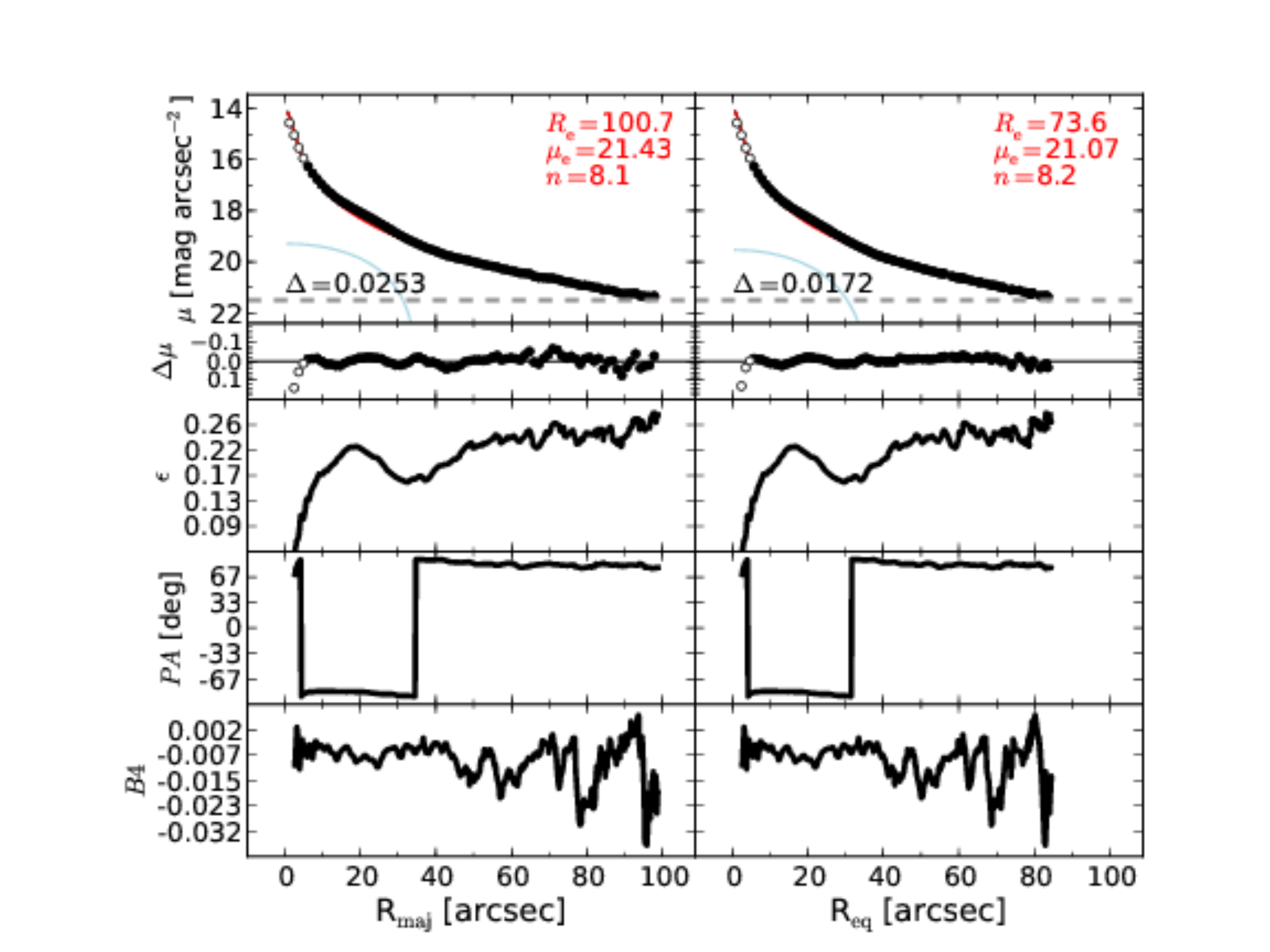}
  \caption{NGC 3842: 
  The brightest member of the Leo Cluster (Abell 1367), 
  an elliptical galaxy with an unresolved partially depleted core \citep{rusli2013,dullograham2014cores}. 
  The light profile presents a swelling at $R_{\rm maj} \sim 20''$ and the ellipticity profile has a peak in the same position.
  One can also glimpse a faint elongated oval in the unsharp mask.
  The innermost $6''.1$ are excluded from the fit.
  A single S\'ersic model is not sufficient to provide a good description of the galaxy light profile, 
  therefore we add a second component (Ferrer function) to account for the embedded lens.
  }
  \end{center}
  \end{figure}

  \begin{table}[h]
  \small
  \caption{Best-fit parameters for the spheroidal component of NGC 3842.}
  \begin{center}
  \begin{tabular}{llcc}
  \hline
  {\bf Work} & {\bf Model}   & $\bm R_{\rm e,sph}$    & $\bm n_{\rm sph}$ \\
    &  &  $[\rm arcsec]$ & \\
  \hline
  1D maj. & S-bul + F-l + m-c  & $100.7$  &  $8.1$ \\
  1D eq.  & S-bul + F-l + m-c  & $73.6$   &  $8.2$ \\
  \hline
  R+13 1D eq.         & core-S\'ersic & $58.8$  &  $6.3$ \\
  \hline
  \end{tabular}
  \end{center}
  \label{tab:n3842}
  \end{table}

  \clearpage\newpage\noindent
  {\bf NGC 3998 \\}

  \begin{figure}[h]
  \begin{center}
  \includegraphics[width=\fitfigurewidth]{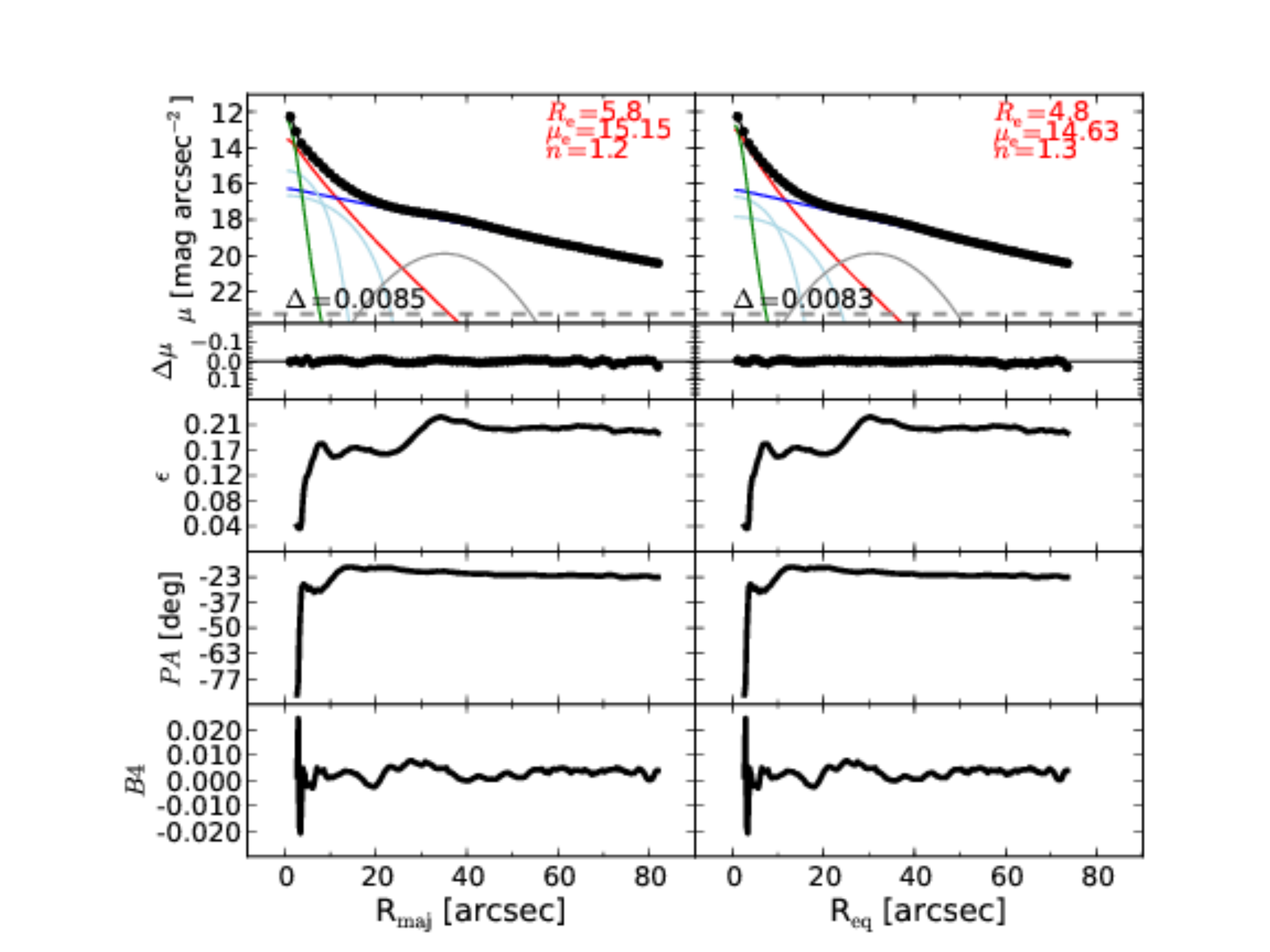}
  \caption{NGC 3998: 
  A barred lenticular galaxy with a Seyfert AGN and nuclear dust \citep{knapp1996n3998}.
  Despite its large stellar velocity dispersion, this galaxy does not have a partially depleted core. 
  \cite{gutierrez2011} identified a bar at $R_{\rm maj} \lesssim 8''$, a ring between $30'' \lesssim R_{\rm maj} \lesssim 50''$, and  
  an anti-truncation in the light profile of the disk at $R_{\rm maj} \sim 122''$.
  \cite{laurikainen2010} found that NGC 3998 features a weak bar at $R_{\rm maj} < 8''$,
  a bright lens at $R_{\rm maj} < 15''$ and a weak bump in the surface brightness profile at $R_{\rm maj} \sim 40''$.
  We see three distinct peaks in the ellipticity profile.
  The first two peaks occur at $R_{\rm maj} \sim 7''$ and $R_{\rm maj} \sim 15''$, and they correspond to a weak bar and to a faint 
  oval component, respectively.
  These components are fit with two Ferrer functions.
  The third peak at $R_{\rm maj} \sim 35''$ coincides with a bump in the surface brightness profile and is produced by a ring.
  The AGN component is modeled with a Gaussian profile.
  }
  \end{center}
  \end{figure}

  \begin{table}[h]
  \small
  \caption{Best-fit parameters for the spheroidal component of NGC 3998.}
  \begin{center}
  \begin{tabular}{llcc}
  \hline
  {\bf Work} & {\bf Model}   & $\bm R_{\rm e,sph}$    & $\bm n_{\rm sph}$ \\
    &  &  $[\rm arcsec]$ & \\
  \hline
  1D maj. & S-bul + e-d + F-bar + F-l + G-n + G-r & $5.8$  &  $1.2$ \\
  1D eq.  & S-bul + e-d + F-bar + F-l + G-n + G-r & $4.8$  &  $1.3$ \\
  \hline
  L+10 2D         & S-bul + e-d + F-bar + F-l & $5.0$  &  $2.0$ \\
  S+11 2D         & S-bul + e-d + G-bar + G-n & $4.7$  &  $1.5$ \\
  B+12 2D         & S-bul + e-d & $5.7$  &  $2.3$ \\
  L+14 2D         & S-bul + trunc. e-d + PSF-n + S-bar + S-id & $2.0$  &  $1.1$ \\
  \hline
  \end{tabular}
  \end{center}
  \label{tab:n3998}
  \end{table}

  \clearpage\newpage\noindent
  {\bf NGC 4026 \\}

  \begin{figure}[h]
  \begin{center}
  \includegraphics[width=\fitfigurewidth]{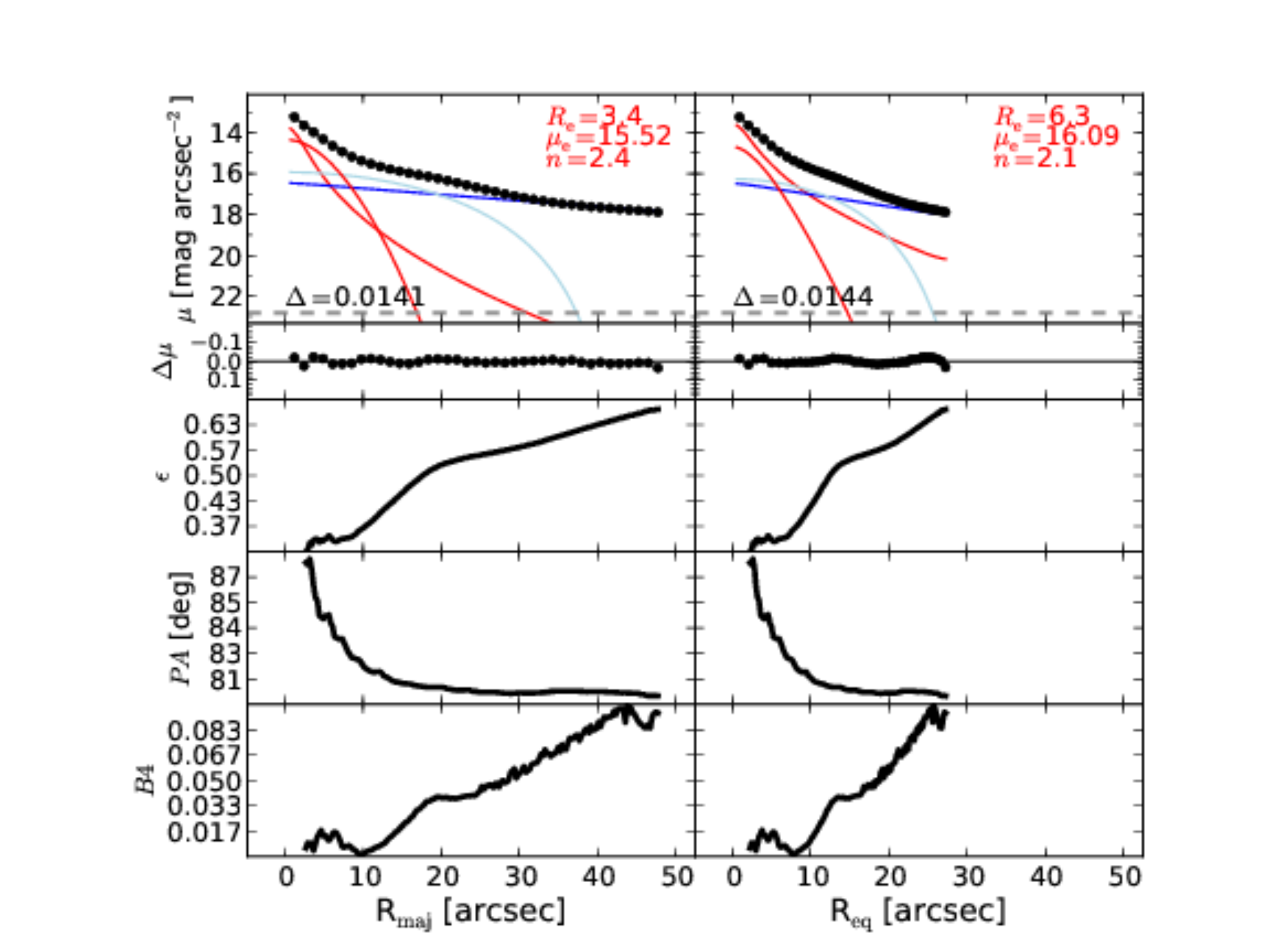}
  \caption{NGC 4026: 
  An edge-on lenticular galaxy with a nuclear stellar disk (with size $\lesssim 0''.5$, \citealt{ledo2010}).
  The unsharp mask reveals the presence of a bar ($R_{\rm maj} \lesssim 30''$) and a disky component 
  embedded in the bulge that is responsible for the peak at $R_{\rm maj} \sim 5''$ in the $B4$ profile.
  The disky component can be also recognized in the velocity map (ATLAS$^{\rm 3D}$).
  The bar is fit with a Ferrer function and the inner disk with a low-$n$ S\'ersic profile.
  We do not model the nuclear component to avoid degeneracies.
  }
  \end{center}
  \end{figure}

  \begin{table}[h]
  \small
  \caption{Best-fit parameters for the spheroidal component of NGC 4026.}
  \begin{center}
  \begin{tabular}{llcc}
  \hline
  {\bf Work} & {\bf Model}   & $\bm R_{\rm e,sph}$    & $\bm n_{\rm sph}$ \\
    &  &  $[\rm arcsec]$ & \\
  \hline
  1D maj. & S-bul + e-d + F-bar + S-id & $3.4$  &  $2.4$ \\
  1D eq.  & S-bul + e-d + F-bar + S-id & $6.3$  &  $2.1$ \\
  \hline
  S+11 2D         & S-bul + e-d & $11.4$  &  $3.5$ \\
  \hline
  \end{tabular}
  \end{center}
  \label{tab:n4026}
  \tablecomments{S+11 obtained larger estimates of the bulge effective radius and S\'ersic index because they did not model the bar.}
  \end{table}

  \clearpage\newpage\noindent
  {\bf NGC 4151 \\}

  \begin{figure}[h]
  \begin{center}
  \includegraphics[width=\fitfigurewidth]{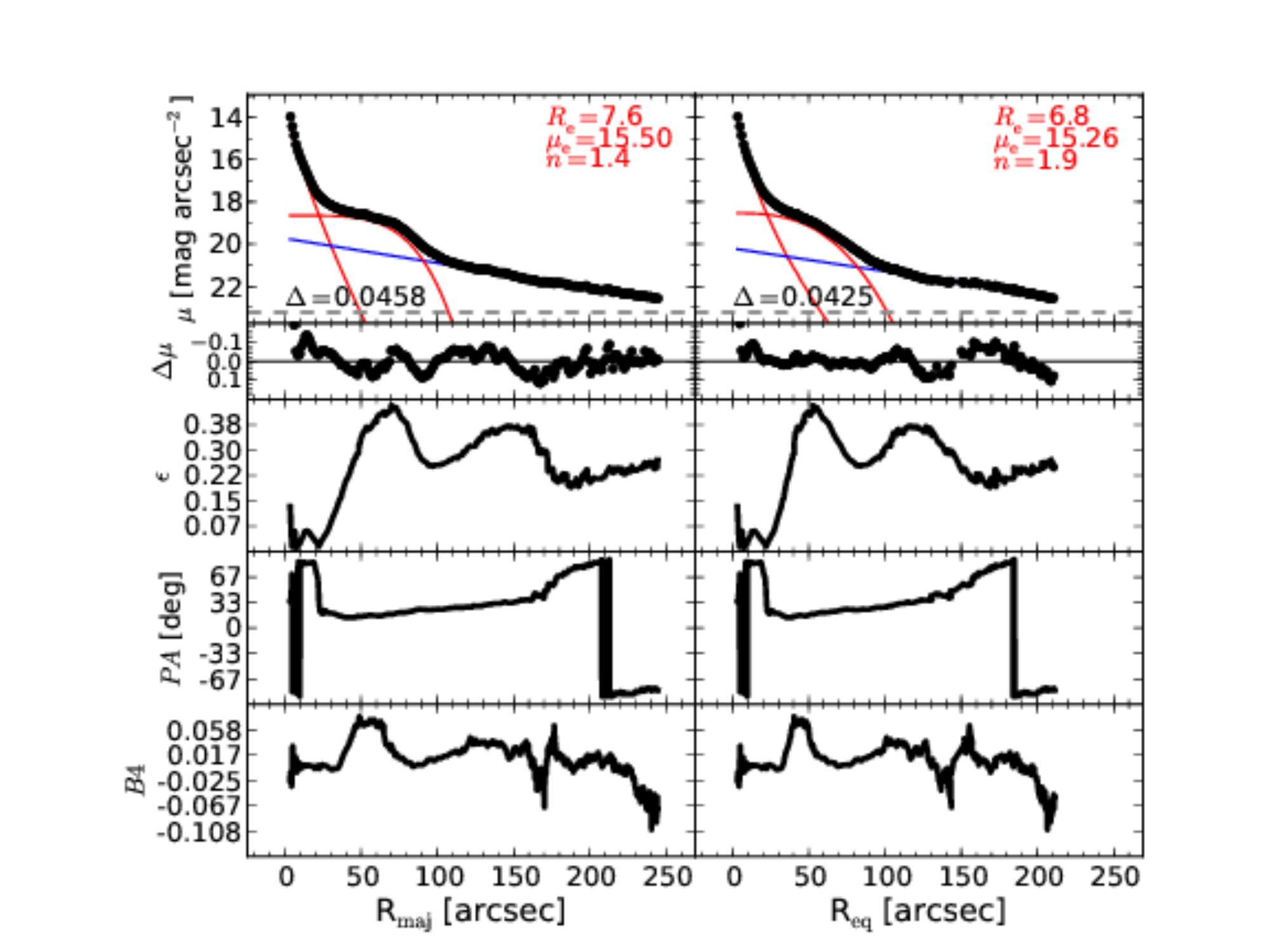}
  \caption{NGC 4151:
  A face-on barred spiral galaxy that hosts a Seyfert AGN \citep{veroncettyveron2006} 
  and circumnuclear dust \citep{pott2010}.
  The nucleus of this galaxy is very bright and the IRAf task {\tt ellipse} fails at fitting the isophotes within $R_{\rm maj} < 4''$, 
  thus our light profile starts at $R_{\rm maj} \sim 4''$. 
  The bar is fit with a low-$n$ S\'ersic profile.  
  }
  \end{center}
  \end{figure}

  \begin{table}[h]
  \small
  \caption{Best-fit parameters for the spheroidal component of NGC 4151.}
  \begin{center}
  \begin{tabular}{llcc}
  \hline
  {\bf Work} & {\bf Model}   & $\bm R_{\rm e,sph}$    & $\bm n_{\rm sph}$ \\
    &  &  $[\rm arcsec]$ & \\
  \hline
  1D maj. & S-bul + e-d + S-bar [+ m-n] & $7.6$  &  $1.4$ \\
  1D eq.  & S-bul + e-d + S-bar [+ m-n] & $6.8$  &  $1.9$ \\
  \hline
  S+11 2D         & S-bul + e-d + G-n         & $5.4$  &  $3.5$ \\
  \hline
  \end{tabular}
  \end{center}
  \label{tab:n4151}
  \end{table}

  \clearpage\newpage\noindent
  {\bf NGC 4261 \\}

  \begin{figure}[h]
  \begin{center}
  \includegraphics[width=\fitfigurewidth]{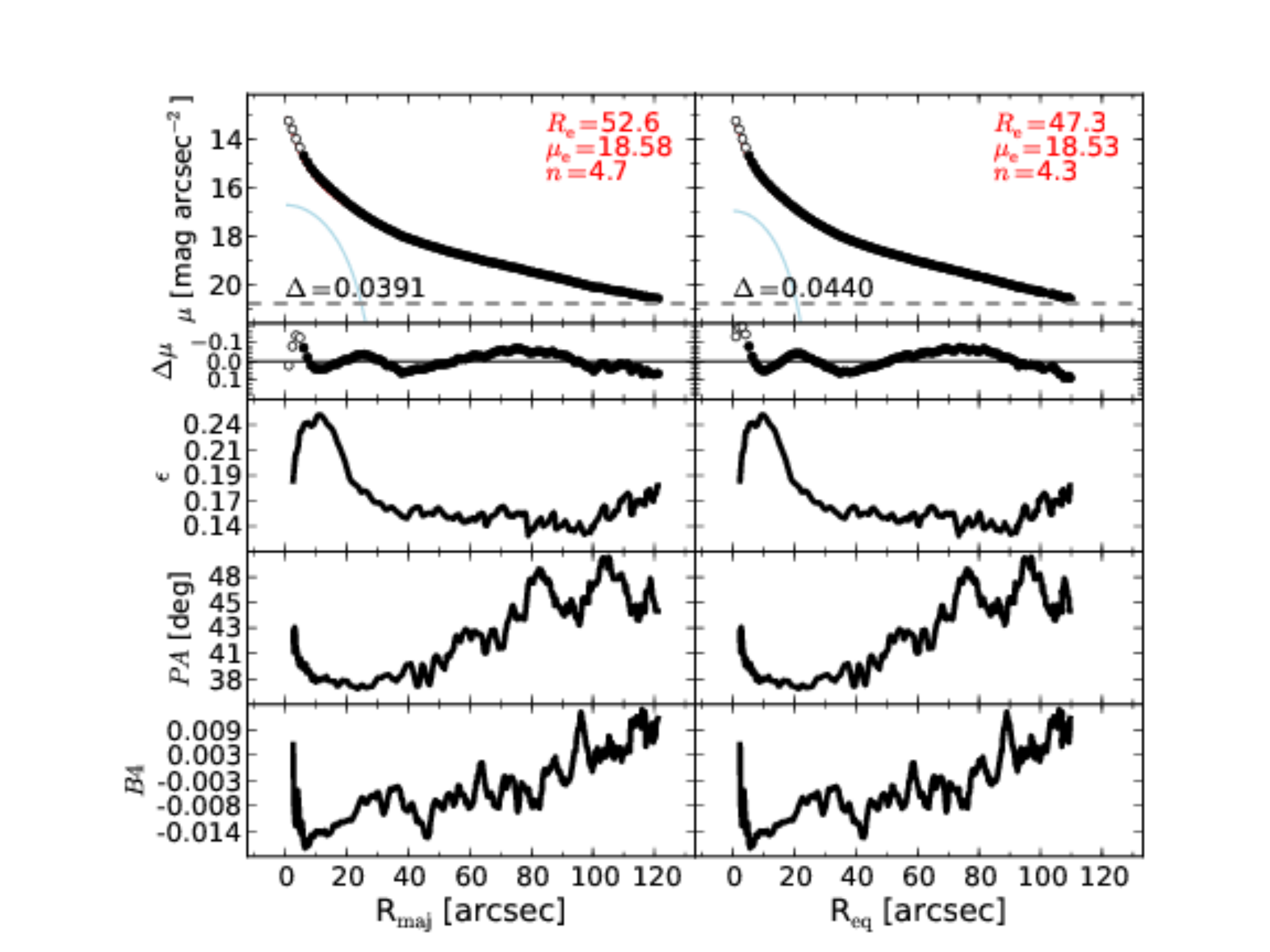}
  \caption{NGC 4261: 
  An elliptical galaxy with a LINER nucleus \citep{veroncettyveron2006}
  and a dusty nuclear disk \citep{tran2001}.
  The galaxy features an unresolved partially depleted core \citep{rusli2013}. 
  The ellipticity profile has a peak at $R_{\rm maj} \sim 10''$, revealing the presence of an embedded component, 
  that we model with a Ferrer function.
  The data within the innermost $6''.1$ are excluded from the fit.
  }
  \end{center}
  \end{figure}

  \begin{table}[h]
  \small
  \caption{Best-fit parameters for the spheroidal component of NGC 4261.}
  \begin{center}
  \begin{tabular}{llcc}
  \hline
  {\bf Work} & {\bf Model}   & $\bm R_{\rm e,sph}$    & $\bm n_{\rm sph}$ \\
    &  &  $[\rm arcsec]$ & \\
  \hline
  1D maj. & S-bul + m-c + F-id & $52.6$  &  $4.7$ \\
  1D eq.  & S-bul + m-c + F-id & $47.3$  &  $4.3$ \\
  2D      & S-bul + m-c + e-id & $50.4$  &  $4.4$ \\
  \hline
  GD07 1D maj.      & S-bul		& $88.6$  &  $7.3$ \\
  S+11 2D      & S-bul + e-d + G-n & $22.6$  &  $4.0$ \\
  V+12 2D      & S-bul + m-c	& $24.2$  &  $3.5$ \\
  B+12 2D      & S-bul		& $48.8$  &  $4.3$ \\
  R+13 1D eq.      & core-S\'ersic	& $77.1$  &  $6.3$ \\
  L+14 2D      & S-bul + m-c	& $68.4$  &  $6.5$ \\
  \hline
  \end{tabular}
  \end{center}
  \label{tab:n4261}
  \tablecomments{
  S+11 found the smallest estimate of the effective radius because they added a large-scale disk to their model.
  V+12 obtained a small estimate of the effective radius because of the limited radial extent of their data.
  }
  \end{table}

  \clearpage\newpage\noindent
  {\bf NGC 4291 \\}

  \begin{figure}[h]
  \begin{center}
  \includegraphics[width=\fitfigurewidth]{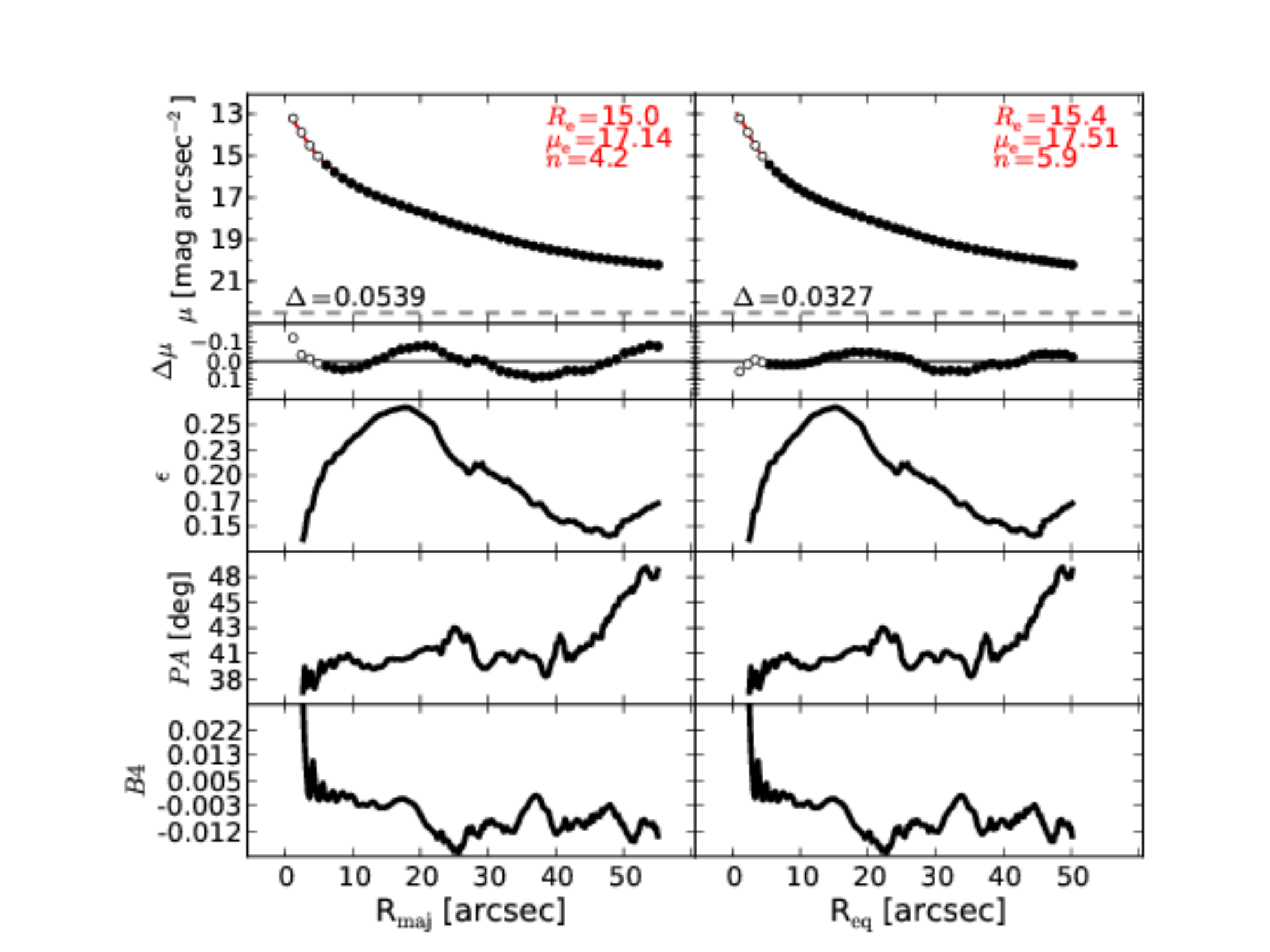}
  \caption{NGC 4291: 
  An elliptical galaxy with an unresolved partially depleted core \citep{rusli2013}. 
  After excluding the data within the innermost $6''.1$ from the fit, 
  we observe that NGC 4291 can be reasonably well modeled with a S\'ersic profile.
  }
  \end{center}
  \end{figure}

  \begin{table}[h]
  \small
  \caption{Best-fit parameters for the spheroidal component of NGC 4291.}
  \begin{center}
  \begin{tabular}{llcc}
  \hline
  {\bf Work} & {\bf Model}   & $\bm R_{\rm e,sph}$    & $\bm n_{\rm sph}$ \\
    &  &  $[\rm arcsec]$ & \\
  \hline
  1D maj. & S-bul + m-c & $15.0$  &  $4.2$ \\
  1D eq.  & S-bul + m-c & $15.4$  &  $5.9$ \\
  2D      & S-bul + m-c & $20.8$  &  $7.7$ \\
  \hline
  GD07 1D maj.         & S-bul         & $14.8$  &  $4.0$ \\
  R+13 1D eq.         & core-S\'ersic & $15.3$  &  $5.6$ \\
  L+14 2D         & S-bul + m-c   & $21.3$  &  $8.6$ \\
  \hline
  \end{tabular}
  \end{center}
  \label{tab:n4291}
  \end{table}

  \clearpage\newpage\noindent
  {\bf NGC 4388 \\}

  \begin{figure}[h]
  \begin{center}
  \includegraphics[width=\fitfigurewidth]{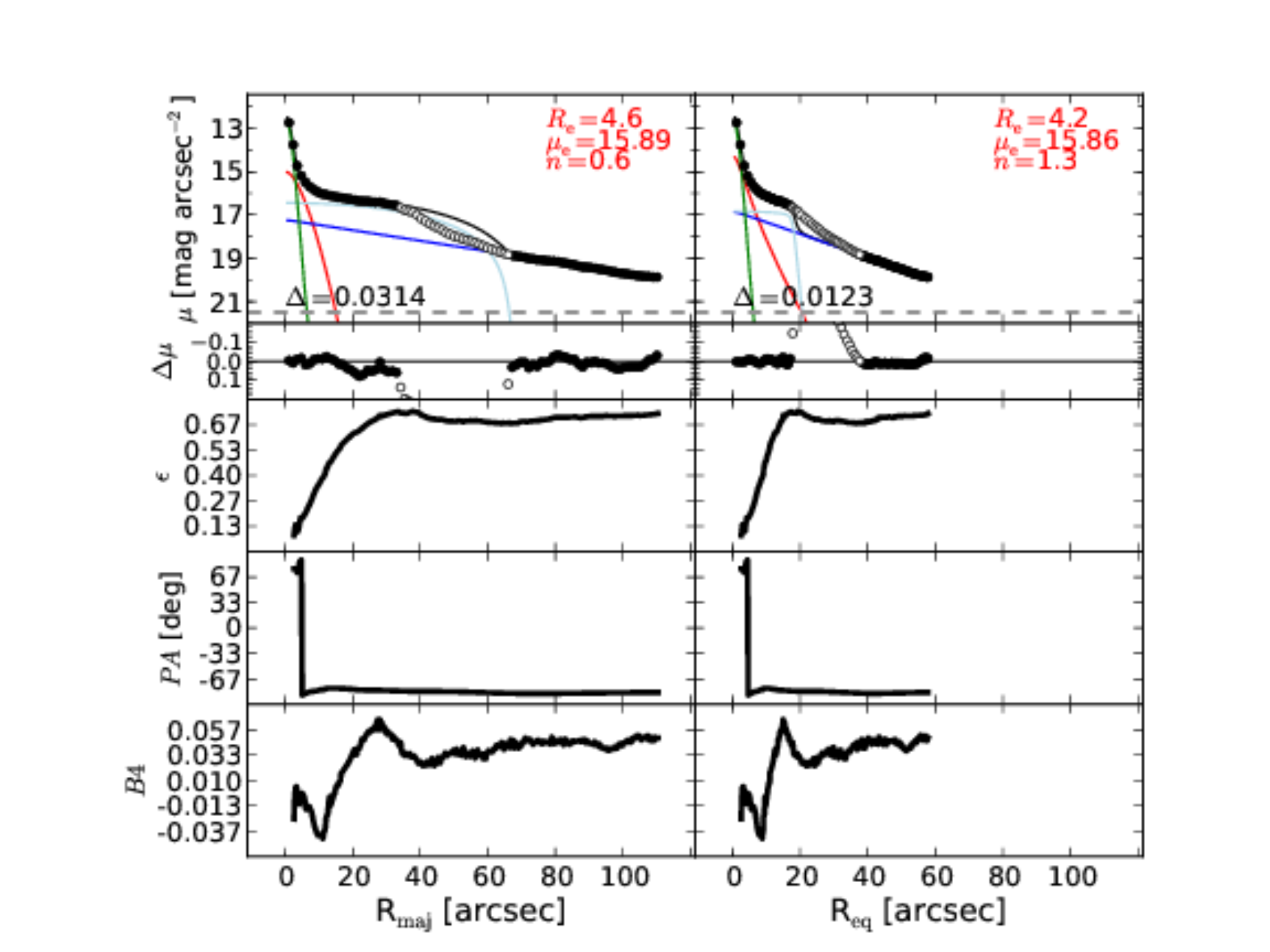}
  \caption{NGC 4388: 
  An edge-on spiral galaxy with a Seyfert AGN \citep{veroncettyveron2006} 
  and copious nuclear dust \citep{martini2003}.
  The peaks at $R_{\rm maj} \sim 30''$ in the ellipticity and $B4$ profiles signal the presence of a bar.
  The data between $35'' \lesssim R_{\rm maj} \lesssim 65''$ are excluded from the fit.
  The AGN component is fit with a Gaussian profile.
  }
  \end{center}
  \end{figure}

  \begin{table}[h]
  \small
  \caption{Best-fit parameters for the spheroidal component of NGC 4388.}
  \begin{center}
  \begin{tabular}{llccc}
  \hline
  {\bf Work} & {\bf Model}   & $\bm R_{\rm e,sph}$    & $\bm n_{\rm sph}$ \\
    &  &  $[\rm arcsec]$ & \\
  \hline
  1D maj. & S-bul + e-d + F-bar + G-n & $4.6$  &  $0.6$ \\
  1D eq.  & S-bul + e-d + F-bar + G-n & $4.2$  &  $1.3$ \\
  \hline
  \end{tabular}
  \end{center}
  \label{tab:n4388}
  \end{table}

  \clearpage\newpage\noindent
  {\bf NGC 4459 \\}

  \begin{figure}[h]
  \begin{center}
  \includegraphics[width=\fitfigurewidth]{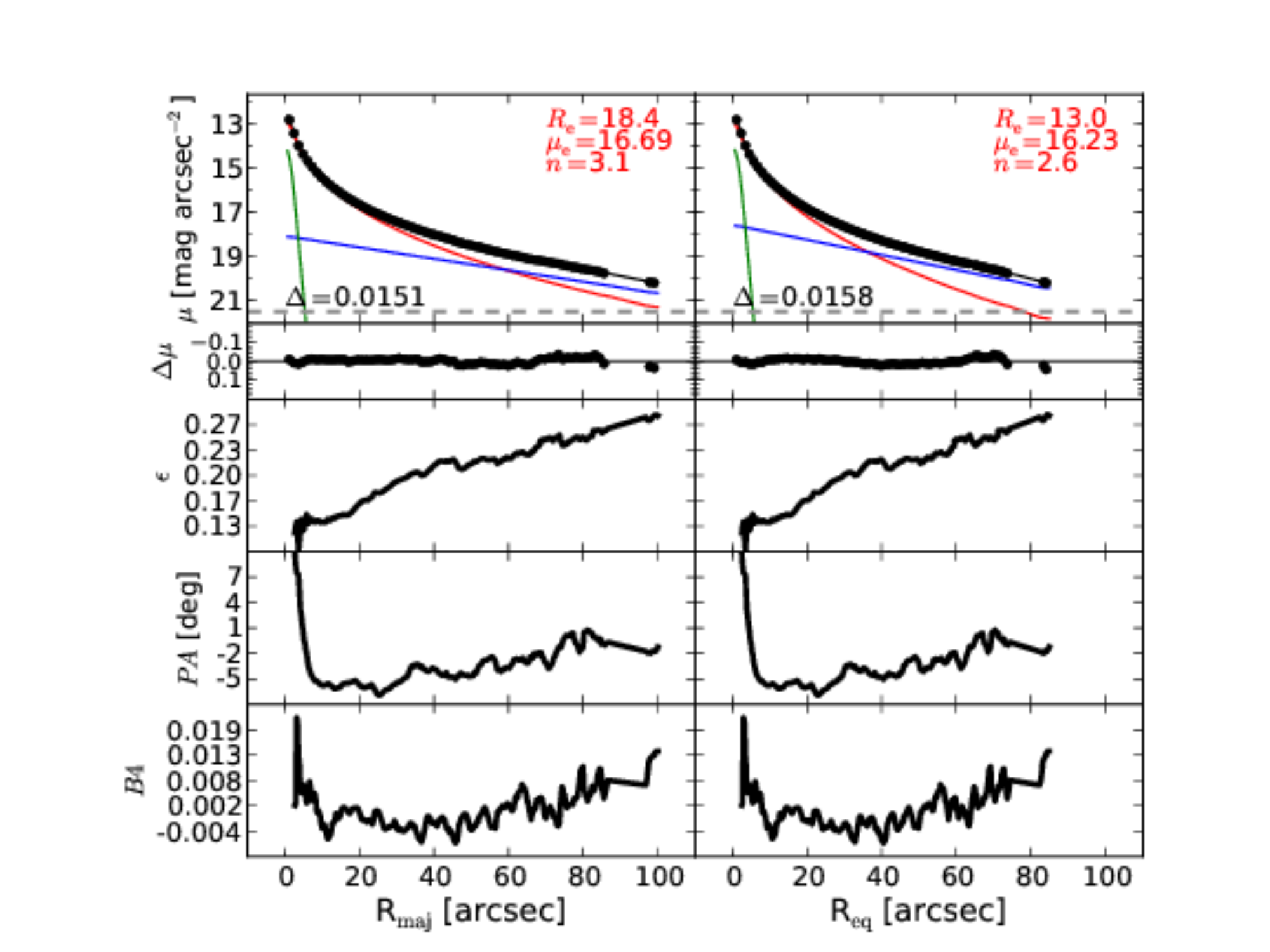}
  \caption{NGC 4459: 
  A lenticular galaxy, 
  whose disk profile has been reported to have an anti-truncation at $R_{\rm maj} \sim 119''$ \citep{gutierrez2011}.
  The ellipticity constantly increases across all the observed radial range ($R_{\rm maj} \lesssim 100''$).
  This is an indication that, going from the galaxy center to the outskirts,
  the disk component becomes increasingly more important over the spheroidal component.
  However, the lack of a plateau in the ellipticity profile at large radii implies that at $R_{\rm maj} \sim 100''$ 
  the contribution from the bulge is still significant compared to that of the disk.
  We note that the antitruncation reported by \cite{gutierrez2011} could be an artificial feature 
  produced by the transition from the S\'ersic bulge to the exponential disk. 
  According to their analysis of the surface brightness profile of NGC 4459 (their Figure 14), 
  the contribution from the disk completely overcomes that of the bulge beyond $R_{\rm maj} \gtrsim 60''$. 
  In the surface brightness profile, they identified two exponential declines with different scale lengths 
  (the first between $60'' \lesssim R_{\rm maj} \lesssim 110''$, and the second beyond $R_{\rm maj} \gtrsim 120''$). 
  However, we checked that in the radial range $60'' \lesssim R_{\rm maj} \lesssim 110''$ 
  the surface brightness profile is not a perfect exponential, but presents a curvature. 
  This can be securely assessed only by fitting the data within the mentioned radial range with a single exponential function 
  and plotting the residuals; 
  if the residuals betray a curvature, the data cannot be accommodated by a single exponential function. 
  According to our decomposition, the disk of NGC 4459 starts dominating beyond $R_{\rm maj} \gtrsim 100''$. 
  The exponential function of our model seems to nicely match the ``second'' exponential decline identified by \cite{gutierrez2011}.  
  We do not find evidence for any embedded components in our data.
  A nuclear light excess is modeled with a Gaussian profile.
  }
  \end{center}
  \end{figure}

  \begin{table}[h]
  \small
  \caption{Best-fit parameters for the spheroidal component of NGC 4459.}
  \begin{center}
  \begin{tabular}{llccc}
  \hline
  {\bf Work} & {\bf Model}   & $\bm R_{\rm e,sph}$    & $\bm n_{\rm sph}$ \\
    &  &  $[\rm arcsec]$ & \\
  \hline
  1D maj. & S-bul + e-d + G-n & $18.4$  &  $3.1$ \\
  1D eq.  & S-bul + e-d + G-n & $13.0$  &  $2.6$ \\
  \hline
  L+10 2D      & S-bul + e-d + F-l & $7.0$   &  $3.0$ \\
  S+11 2D      & S-bul + e-d	& $10.3$  &  $2.5$ \\
  V+12 2D      & S-bul + M-n	& $25.0$  &  $3.9$ \\
  B+12 2D      & S-bul		& $155.2$ &  $7.4$ \\
  \hline
  \end{tabular}
  \end{center}
  \label{tab:n4459}
  \tablecomments{
  B+12 did not model the large-scale disk and thus overestimated the bulge effective radius and S\'ersic index.
  }
  \end{table}

  \clearpage\newpage\noindent
  {\bf NGC 4473 \\}

  \begin{figure}[h]
  \begin{center}
  \includegraphics[width=\fitfigurewidth]{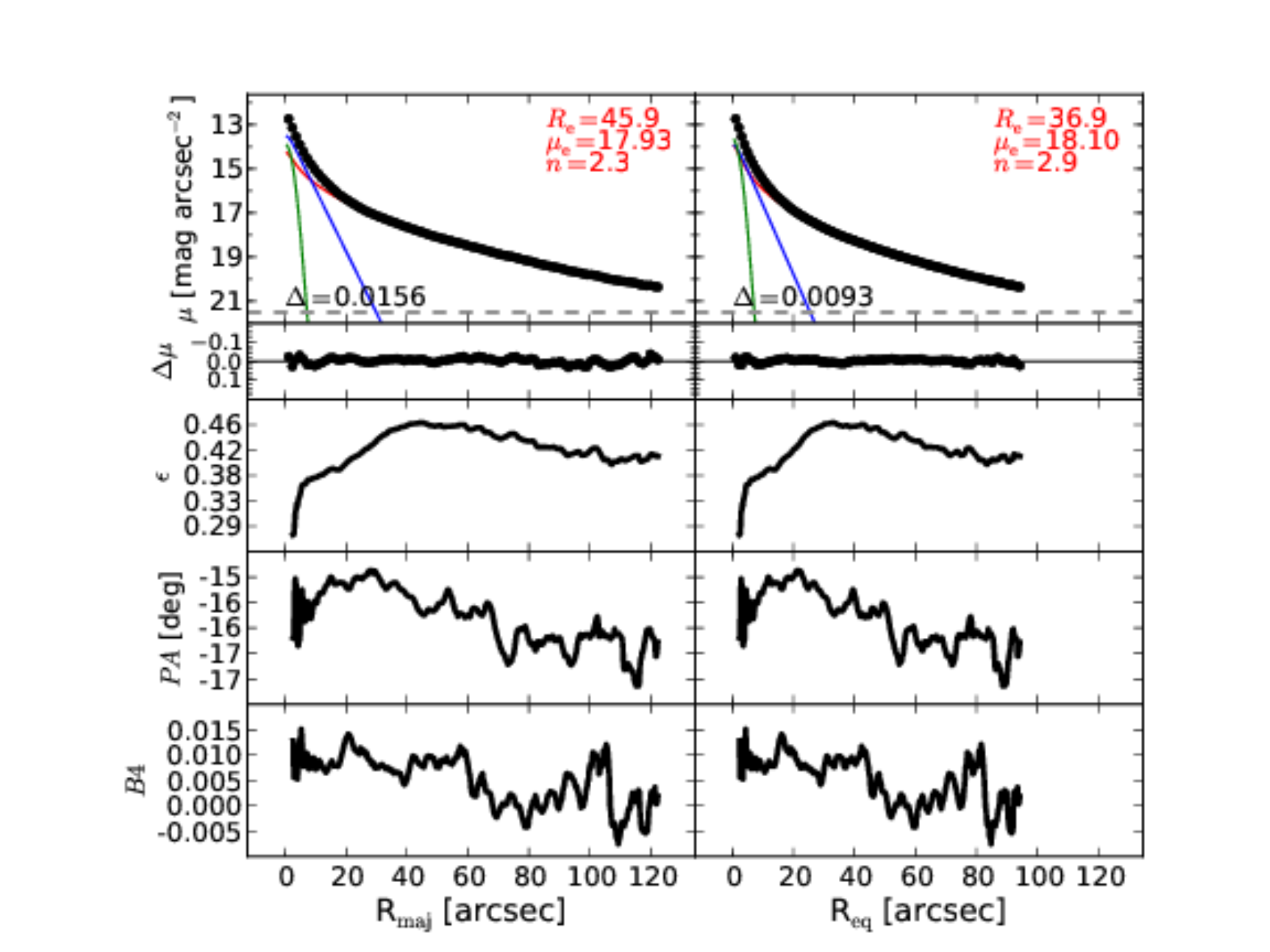}
  \caption{NGC 4473: 
  An elliptical galaxy with an embedded disk \citep{ledo2010}.
  The disk is clearly visible in the velocity map (ATLAS$^{\rm 3D}$, SLUGGS). 
  We account for a nuclear excess of light by adding a narrow Gaussian function to the model.
  }
  \end{center}
  \end{figure}
  
  \begin{table}[h]
  \small
  \caption{Best-fit parameters for the spheroidal component of NGC 4473.}
  \begin{center}
  \begin{tabular}{llccc}
  \hline
  {\bf Work} & {\bf Model}   & $\bm R_{\rm e,sph}$    & $\bm n_{\rm sph}$ \\
    &  &  $[\rm arcsec]$ & \\
  \hline
  1D maj. & S-bul + e-id + G-n & $45.9$  &  $2.3$ \\
  1D eq.  & S-bul + e-id + G-n & $36.9$  &  $2.9$ \\
  \hline
  GD07 1D maj.      & S-bul                 & $39.6$   &  $2.7$ \\
  S+11 2D      & S-bul                 & $49.3$   &  $7.0$ \\
  V+12 2D      & S-bul + m-c           & $21.3$   &  $4.3$ \\
  B+12 2D      & S-bul + e-d           & $10.6$   &  $2.2$ \\
  L+14 2D      & S-bul                 & $27.9$   &  $5.1$ \\
  \hline
  \end{tabular}
  \end{center}
  \label{tab:n4459}
  \tablecomments{
  B+12 obtained the smallest estimates of the effective radius and S\'ersic index because they included a large-scale disk in their model. 
  }
  \end{table}

  \clearpage\newpage\noindent
  {\bf NGC 4564 \\}

  \begin{figure}[h]
  \begin{center}
  \includegraphics[width=\fitfigurewidth]{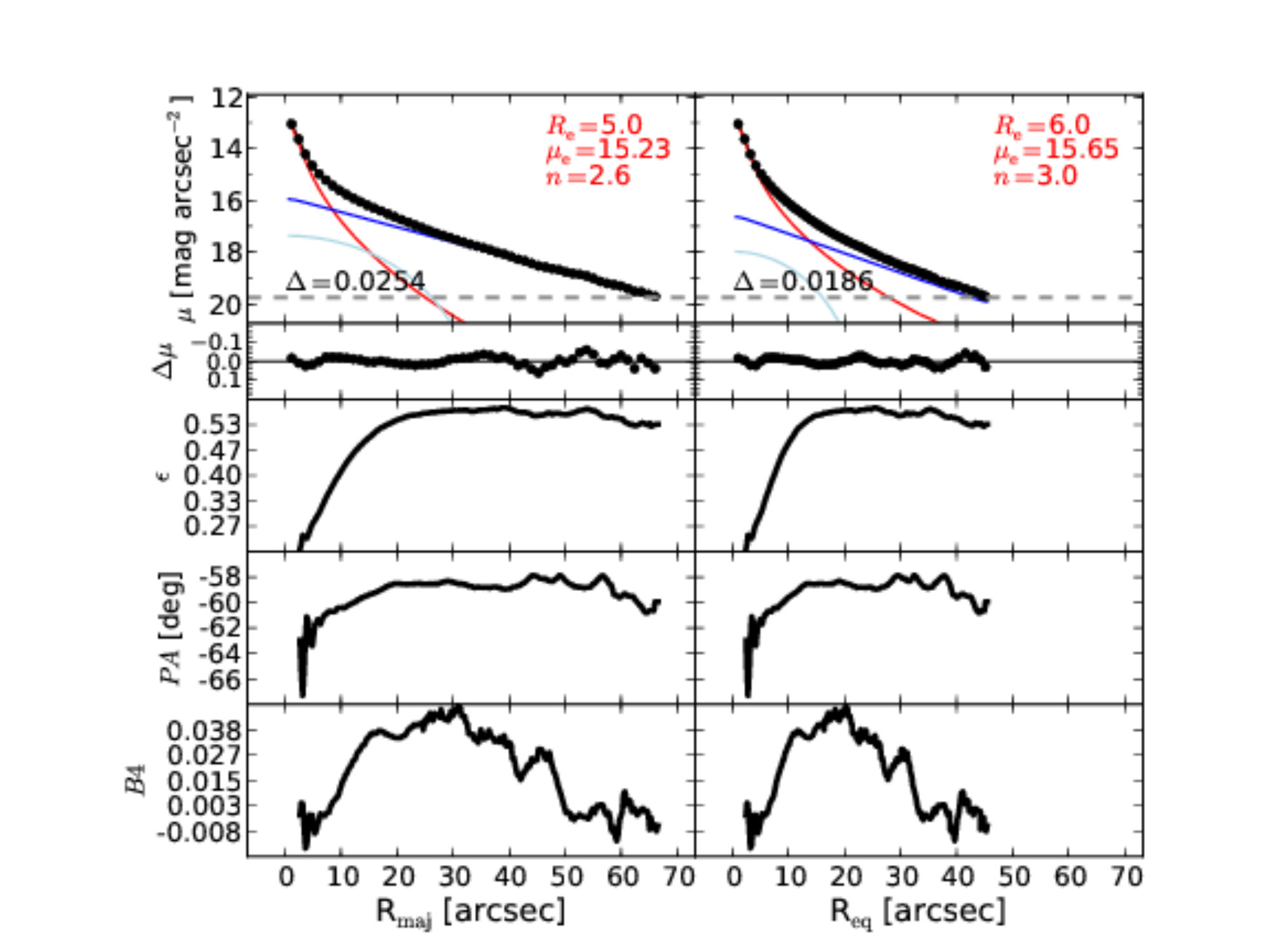}
  \caption{NGC 4564: 
  An edge-on lenticular galaxy.
  In the unsharp mask, one can glimpse an oval structure extending out to $R_{\rm maj} \lesssim 15''$ (see the peak in the $B4$ profile).
  Fitting NGC 4564 with a S\'ersic + exponential model produces poor residuals.
  However, the addition of a Ferrer function to the model dramatically improves the fit and smoothes the residuals.
  }
  \end{center}
  \end{figure}

  \begin{table}[h]
  \small
  \caption{Best-fit parameters for the spheroidal component of NGC 4564.}
  \begin{center}
  \begin{tabular}{llcc}
  \hline
  {\bf Work} & {\bf Model}   & $\bm R_{\rm e,sph}$    & $\bm n_{\rm sph}$ \\
    &  &  $[\rm arcsec]$ & \\
  \hline
  1D maj. & S-bul + e-d + F-l & $5.0$  &  $2.6$ \\
  1D eq.  & S-bul + e-d + F-l & $6.0$  &  $3.0$ \\
  \hline
  GD07 1D maj.      & S-bul + e-d & $4.3$  &  $3.2$ \\
  S+11 2D      & S-bul + e-d & $25.0$ &  $7.0$ \\
  V+12 2D      & S-bul + e-d & $3.0$  &  $3.7$ \\
  \hline
  \end{tabular}
  \end{center}
  \label{tab:n4564}
  \tablecomments{The most dicrepant results are obtained by S+11, although they used the same model as GD07 and V+12. }
  \end{table}

  \clearpage\newpage\noindent
  {\bf NGC 4596 \\}

  \begin{figure}[h]
  \begin{center}
  \includegraphics[width=\fitfigurewidth]{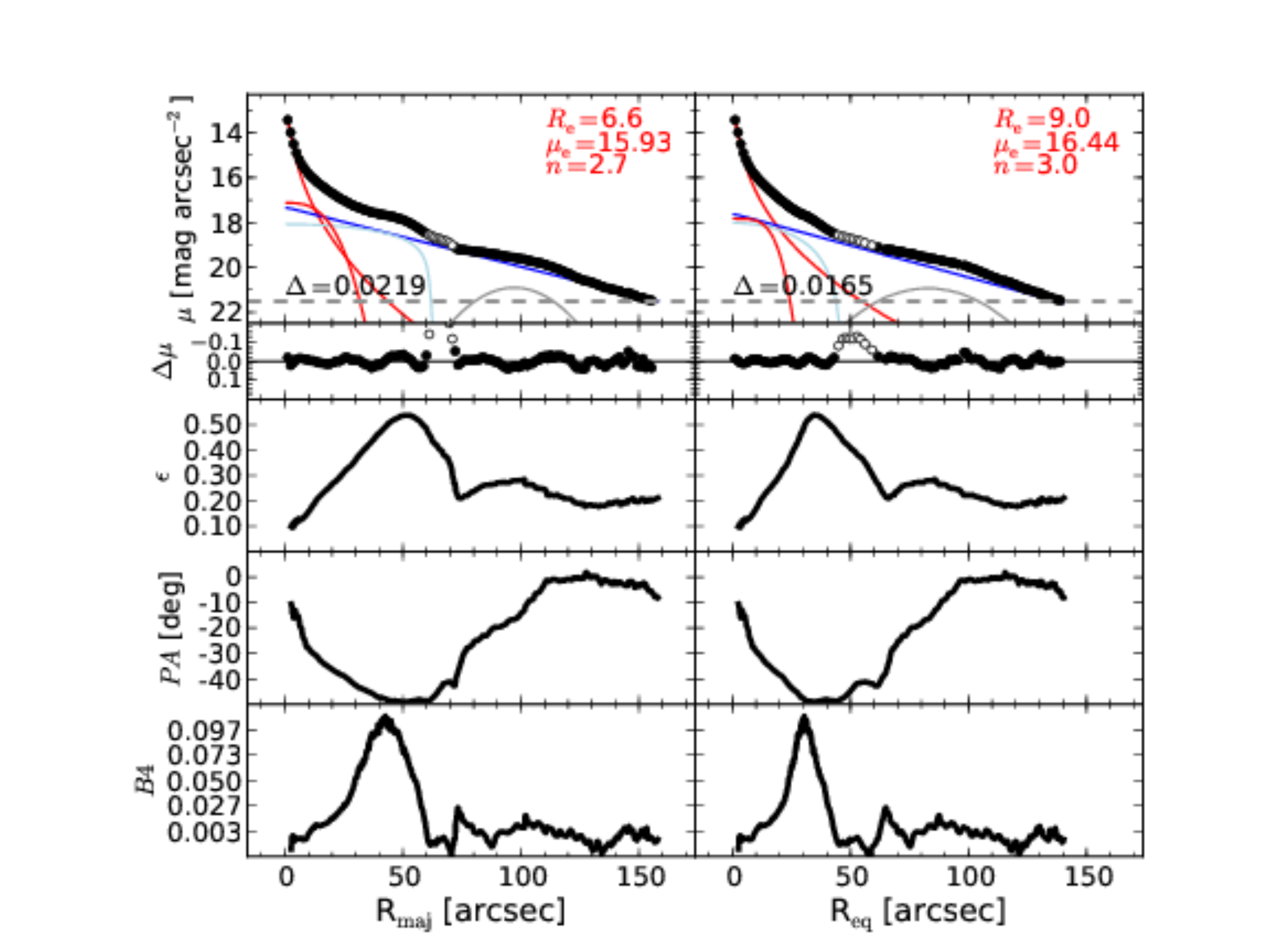}
  \caption{NGC 4596: 
  An edge-on barred lenticular galaxy.
  The morphology of this galaxy is quite complex. 
  The bar extends out to $R_{\rm maj} \lesssim 50''$ and concludes in two evident ansae.
  The large-scale disk features a wide ring that is responsible for the curvature in the light profile observed within 
  $60'' \lesssim R_{\rm maj} \lesssim 130''$ (see also \citealt{comeron2014}).
  An additional embedded disky component ($R_{\rm maj} \lesssim 15''$) can be recognized in the $B4$ profile and in the velocity map.
  The bump in the light profile within $60'' \lesssim R_{\rm maj} \lesssim 70''$ corresponds to the ansae of the bar and 
  this data range is therefore excluded from the fit.
  The large scale bar is fit with a Ferrer function and the inner disk with a low-$n$ S\'ersic profile.
  }
  \end{center}
  \end{figure}

  \begin{table}[h]
  \small
  \caption{Best-fit parameters for the spheroidal component of NGC 4596.}
  \begin{center}
  \begin{tabular}{llcc}
  \hline
  {\bf Work} & {\bf Model}   & $\bm R_{\rm e,sph}$    & $\bm n_{\rm sph}$ \\
    &  &  $[\rm arcsec]$ & \\
  \hline
  1D maj. & S-bul + e-d + F-bar + G-r + S-id & $6.6$  &  $2.7$ \\
  1D eq.  & S-bul + e-d + F-bar + G-r + S-id & $9.0$  &  $3.0$ \\
  \hline
  L+10 2D      & S-bul + e-d + F-bar + F-l & $2.8$  &  $1.4$ \\
  S+11 2D      & S-bul + e-d + G-bar & $28.0$  &  $3.0$ \\
  V+12 2D      & S-bul + e-d + S-bar & $13.2$  &  $3.6$ \\
  B+12 2D      & S-bul + e-d & $44.9$  &  $4.4$ \\
  \hline
  \end{tabular}
  \end{center}
  \label{tab:n4596}
  \tablecomments{
  B+12 fit neither the bar nor the inner disk and obtained the largest estimates of the bulge effective radius and S\'ersic index.
  The models of S+11 and V+12 do not account for the inner disk and thus result in larger estimates of the bulge effective radius.
  }
  \end{table}

  \clearpage\newpage\noindent
  {\bf NGC 4697 \\}

  \begin{figure}[h]
  \begin{center}
  \includegraphics[width=\fitfigurewidth]{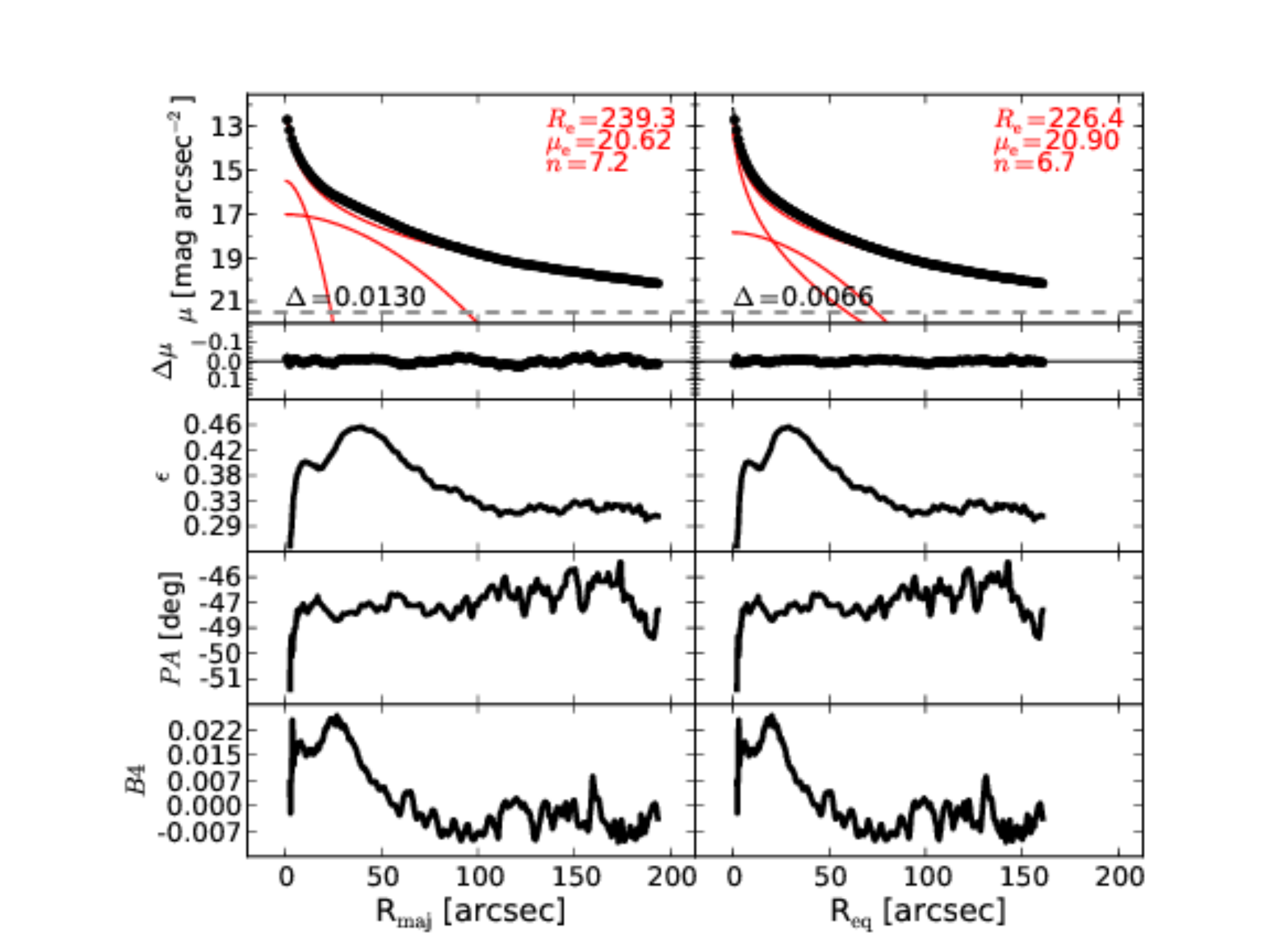}
  \caption{NGC 4697: 
  An elliptical galaxy with an embedded disk \citep{scorzabender1995}.
  The velocity map (ATLAS$^{\rm 3D}$, SLUGGS) and the unsharp mask of NGC 4697 clearly show 
  the presence of an intermediate-size disk embedded in the galaxy's spheroidal component.
  However, the ellipticity profile presents two peaks. 
  The peak at $R_{\rm maj} \sim 40''$ corresponds to the intermediate-size embedded disk just mentioned,
  while the peak at $R_{\rm maj} \sim 10''$ pertains to a smaller inner disk.
  After testing different decomposition models, 
  in which we fit the two embedded disks with different functions (exponential, S\'ersic, Ferrer),
  while always describing the main spheroidal component with a S\'ersic profile,
  we noticed that the spheroid parameters do not significantly vary among the various decompositions.
  Our preferred model for NGC 4697 consists of a S\'ersic-bulge + 2 S\'ersic-inner disks.
  }
  \end{center}
  \end{figure}

  \begin{table}[h]
  \small
  \caption{Best-fit parameters for the spheroidal component of NGC 4697.}
  \begin{center}
  \begin{tabular}{llcc}
  \hline
  {\bf Work} & {\bf Model}   & $\bm R_{\rm e,sph}$    & $\bm n_{\rm sph}$ \\
    &  &  $[\rm arcsec]$ & \\
  \hline
  1D maj. & S-bul + 2 S-id & $239.3$  &  $7.2$ \\
  1D eq.  & S-bul + 2 S-id & $226.4$  &  $6.7$ \\
  2D      & S-bul + e-id	& $121.4$  &  $5.0$ \\
  \hline
  GD07 1D maj.  & S-bul & $-$  &  $4.0$ \\
  S+11 2D  & S-bul & $100.5$  &  $5.0$ \\
  V+12 (1) 2D  & S-bul & $39.1$  &  $3.8$ \\
  V+12 (2) 2D  & S-bul + e-d & $10.0$  &  $2.9$ \\
  L+14 2D  & S-bul + e-d + PSF-n + S-halo & $6.3$  &  $2.1$ \\
  \hline
  \end{tabular}
  \end{center}
  \label{tab:n4697}
  \tablecomments{
  In both their models, V+12 underestimated the effective radius and the S\'ersic index because of the small radial range of their data. 
  In an effort to model the curved light profile, 
  L+14 included a large-scale disk plus a halo, and thus underestimated the spheroid effective radius and the S\'ersic index. 
  }
  \end{table}

  \clearpage\newpage\noindent
  {\bf NGC 4889 \\}

  \begin{figure}[h]
  \begin{center}
  \includegraphics[width=\fitfigurewidth]{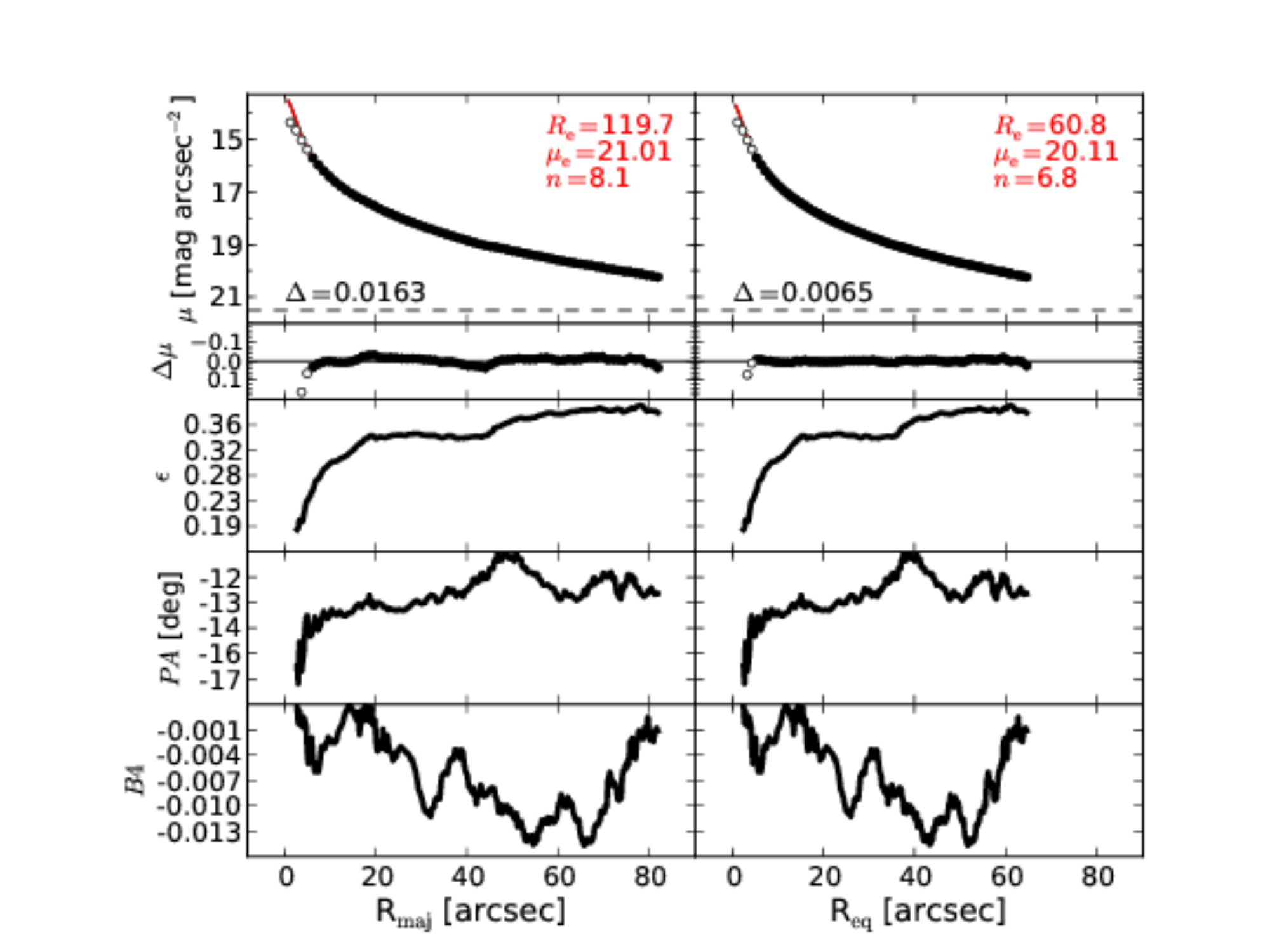}
  \caption{NGC 4889: 
  The brightest member of the Coma cluster, 
  an elliptical galaxy with an unresolved partially depleted core \citep{rusli2013}. 
  We exclude the innermost $6''.1$ from the fit and
  successfully model the galaxy with a single S\'ersic profile.
  }
  \end{center}
  \end{figure}

  \begin{table}[h]
  \small
  \caption{Best-fit parameters for the spheroidal component of NGC 4889.}
  \begin{center}
  \begin{tabular}{llcc}
  \hline
  {\bf Work} & {\bf Model}   & $\bm R_{\rm e,sph}$    & $\bm n_{\rm sph}$ \\
    &  &  $[\rm arcsec]$ & \\
  \hline
  1D maj. & S-bul + m-c & $119.7$  &  $8.1$ \\
  1D eq.  & S-bul + m-c & $60.8$	&  $6.8$ \\
  2D      & S-bul + m-c & $104.3$  &  $7.8$ \\
  \hline
  R+13 1D eq.         & core-S\'ersic & $169.2$  &  $9.8$ \\
  \hline
  \end{tabular}
  \end{center}
  \label{tab:n4889}
  \end{table}
    
  \clearpage\newpage\noindent
  {\bf NGC 4945 \\}

  \begin{figure}[h]
  \begin{center}
  \includegraphics[width=\fitfigurewidth]{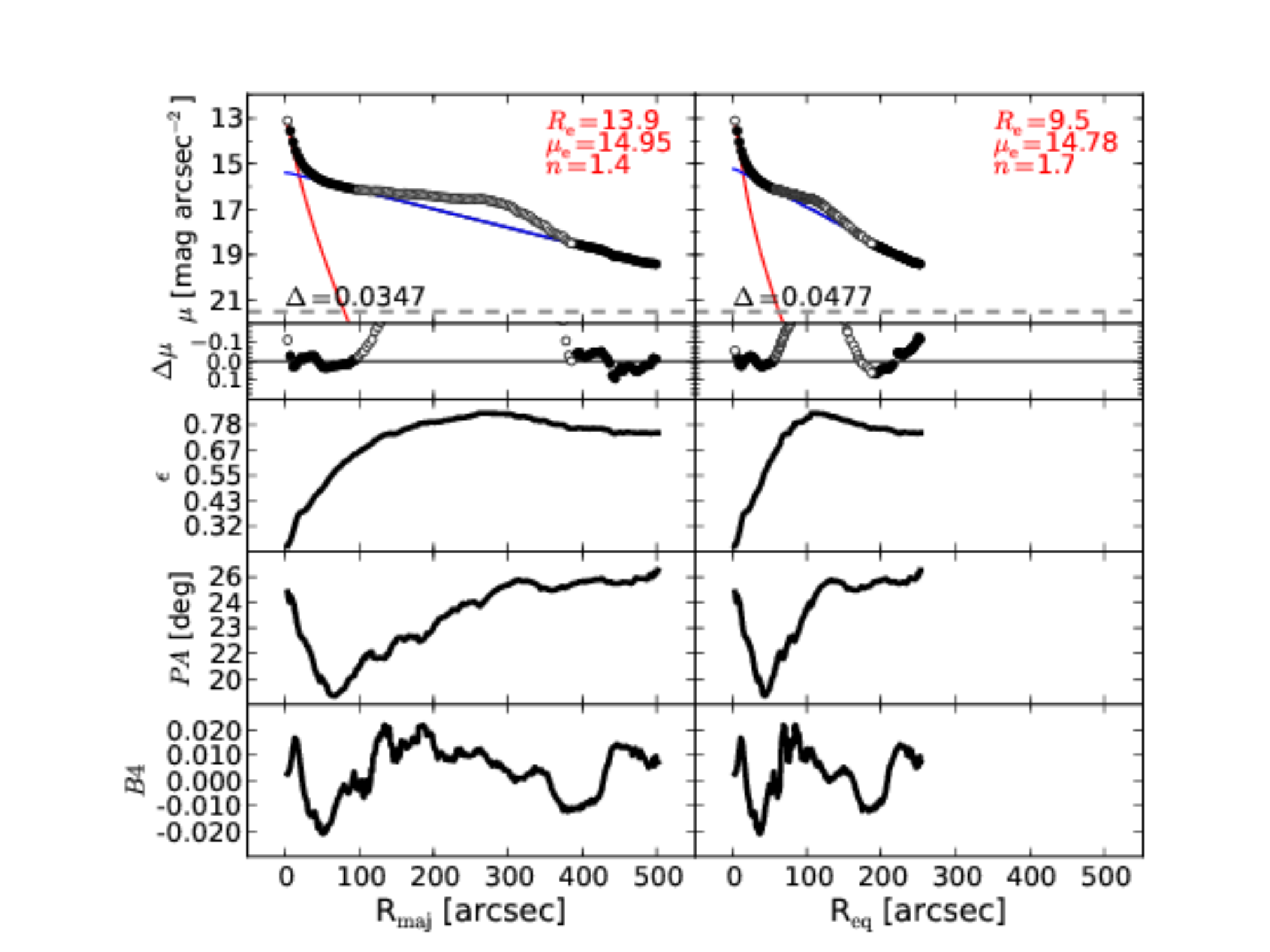}
  \caption{NGC 4945: 
  An edge-on, dusty spiral galaxy that hosts a Seyfert AGN \citep{lin2011}.
  The light profile has an obvious bump that can be ascribed to the bar.
  This bumb cannot be easily modeled with a Ferrer function {\bf nor with a S\'ersic profile}, 
  thus we exclude the data in the range $100'' \lesssim R_{\rm maj} \sim 400''$ from the fit.
  We also exclude the data within the innermost $6''.4$ due to the contribution from the AGN.
  }
  \end{center}
  \end{figure}

  \begin{table}[h]
  \small
  \caption{Best-fit parameters for the spheroidal component of NGC 4945.}
  \begin{center}
  \begin{tabular}{llcc}
  \hline
  {\bf Work} & {\bf Model}   & $\bm R_{\rm e,sph}$    & $\bm n_{\rm sph}$ \\
    &  &  $[\rm arcsec]$ & \\
  \hline
  1D maj. & S-bul + e-d & $13.9$  &  $1.4$ \\
  1D eq.  & S-bul + e-d & $9.5$   &  $1.7$ \\
  2D      & S-bul + e-d & $16.2$  &  $0.8$ \\
  \hline
  \end{tabular}
  \end{center}
  \label{tab:n4945}
  \end{table}

  \clearpage\newpage\noindent
  {\bf NGC 5077 \\}

  \begin{figure}[h]
  \begin{center}
  \includegraphics[width=\fitfigurewidth]{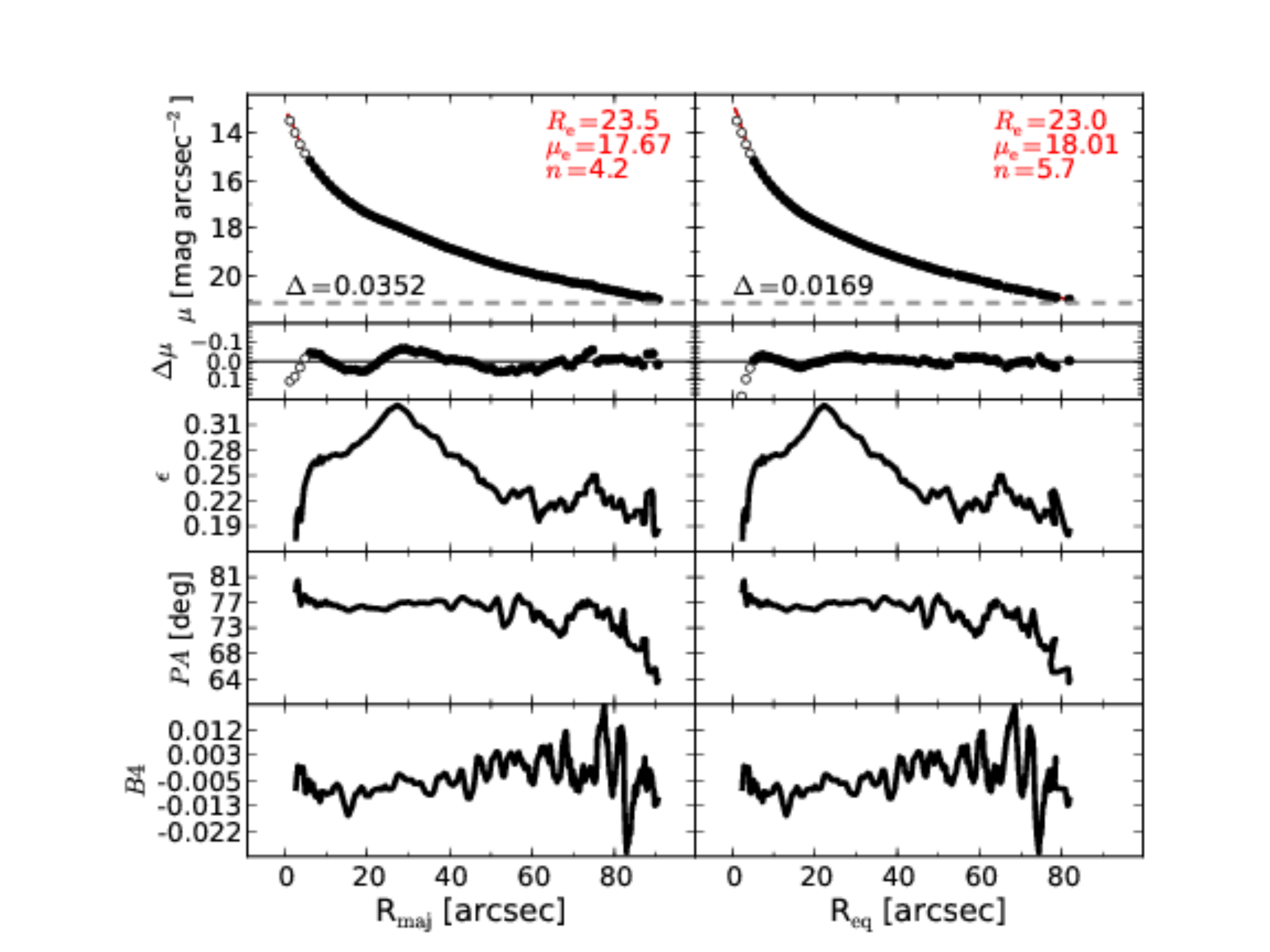}
  \caption{NGC 5077:  
  An elliptical galaxy with an unresolved partially depleted core \citep{trujillo2004coresersicmodel}. 
  We mask the data within the innermost $6''.1$ and fit the galaxy with a S\'ersic profile. }
  \end{center}
  \end{figure}

  \begin{table}[h]
  \small
  \caption{Best-fit parameters for the spheroidal component of NGC 5077.}
  \begin{center}
  \begin{tabular}{llcc}
  \hline
  {\bf Work} & {\bf Model}   & $\bm R_{\rm e,sph}$    & $\bm n_{\rm sph}$ \\
    &  &  $[\rm arcsec]$ & \\
  \hline
  1D maj. & S-bul + m-c & $23.5$  &  $4.2$ \\
  1D eq.  & S-bul + m-c & $23.0$  &  $5.7$ \\
  2D      & S-bul + m-c & $30.5$  &  $6.8$ \\
  \hline
  S+11 2D         & S-bul + G-n & $29.2$  &  $6.0$ \\
  \hline
  \end{tabular}
  \end{center}
  \label{tab:n5077}
  \end{table}

  \clearpage\newpage\noindent

  {\bf NGC 5128 - Centaurus A \\}

  \begin{figure}[h]
  \begin{center}
  \includegraphics[width=\fitfigurewidth]{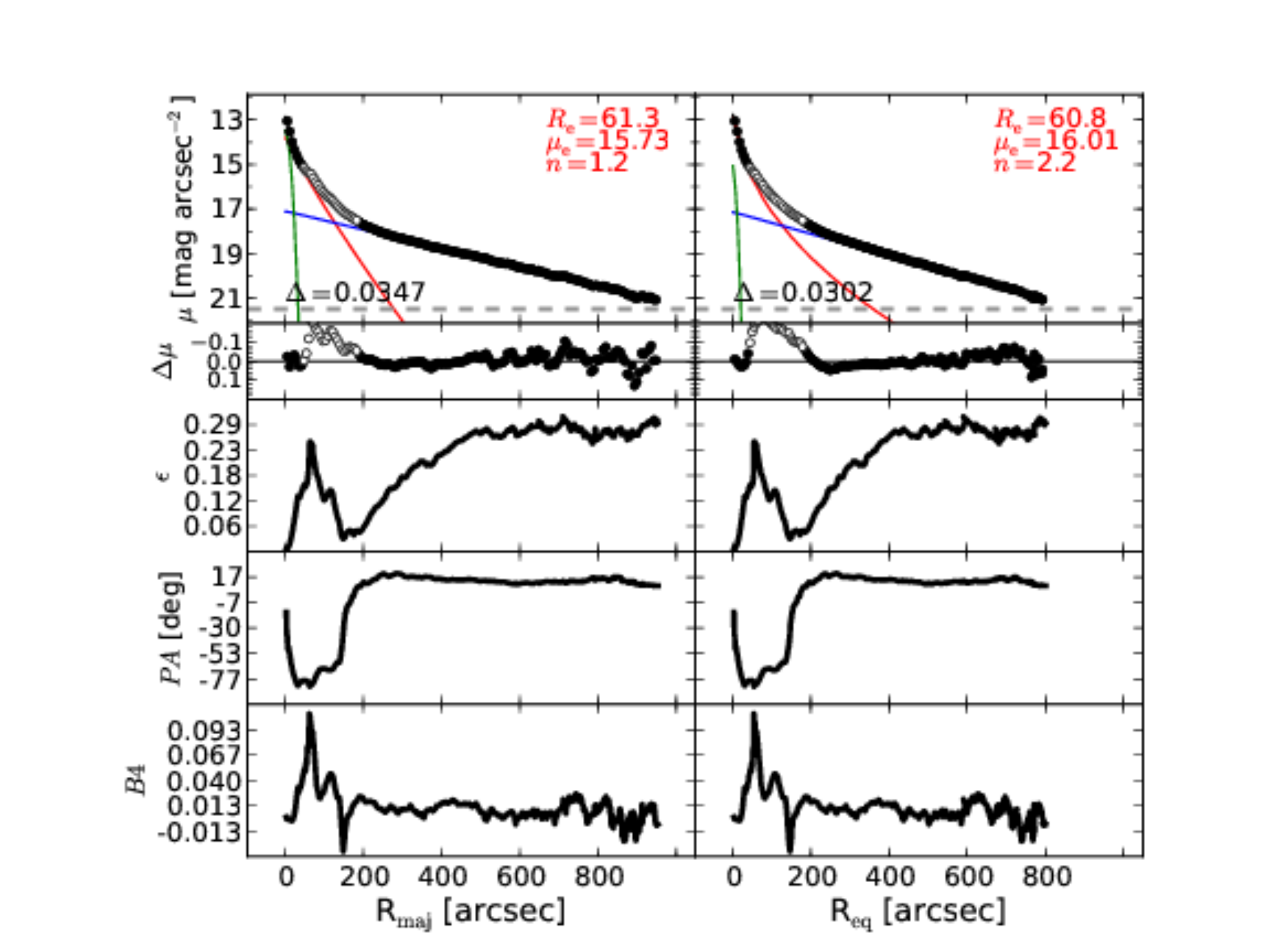}
  \caption{NGC 5128: 
  A merging system. 
  The data within $50'' \lesssim R_{\rm maj} \lesssim 200''$ are excluded from the fit.}
  \end{center}
  \end{figure}

  \begin{table}[h]
  \small
  \caption{Best-fit parameters for the spheroidal component of NGC 5128.}
  \begin{center}
  \begin{tabular}{llccc}
  \hline
  {\bf Work} & {\bf Model}   & $\bm R_{\rm e,sph}$    & $\bm n_{\rm sph}$ \\
    &  &  $[\rm arcsec]$ & \\
  \hline
  1D maj. & S-bul + e-halo + G-n & $61.3$  &  $1.2$ \\
  1D eq.  & S-bul + e-halo + G-n & $60.8$  &  $2.2$ \\
  \hline
  S+11 2D         & S-bul + e-d + G-n    & $103.6$  &  $3.5$ \\
  \hline
  \end{tabular}
  \end{center}
  \label{tab:n5128}
  \end{table}

  \clearpage\newpage\noindent

  {\bf NGC 5576 \\}

  \begin{figure}[h]
  \begin{center}
  \includegraphics[width=\fitfigurewidth]{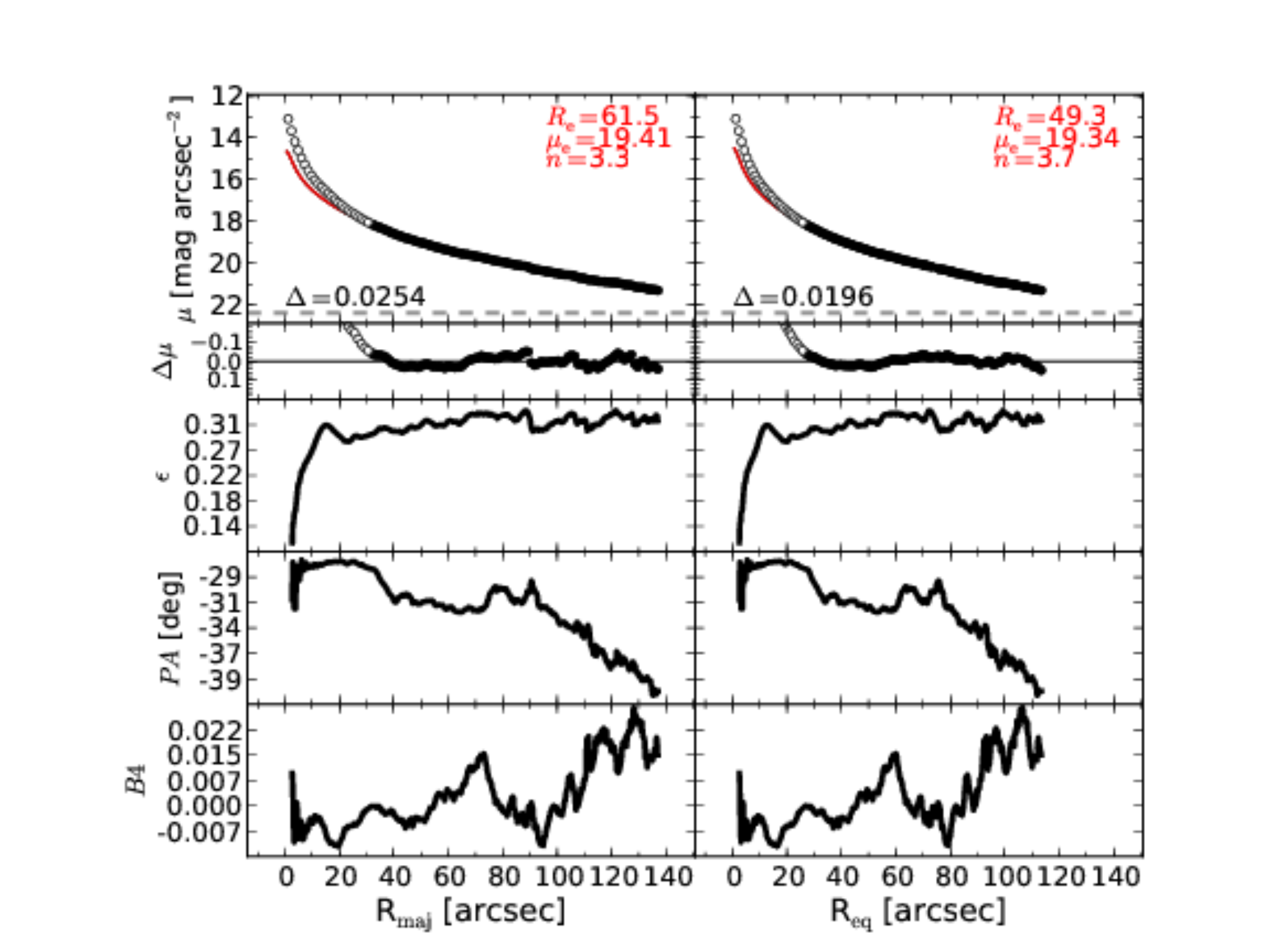}
  \caption{NGC 5576: 
  An elliptical galaxy with a disturbed morphology. 
  The isophotal parameters suggest the presence of an embedded disk ($R_{\rm maj} \lesssim 20''$), 
  but attempts to account for such a component were unsuccessful. 
  The data within $R_{\rm maj} \lesssim 36''$ are excluded from the fit 
  and the galaxy is modeled with a S\'ersic profile.
  } 
  \end{center}
  \end{figure}
  
  \begin{table}[h]
  \small
  \caption{Best-fit parameters for the spheroidal component of NGC 5576.}
  \begin{center}
  \begin{tabular}{llcc}
  \hline
  {\bf Work} & {\bf Model}   & $\bm R_{\rm e,sph}$    & $\bm n_{\rm sph}$ \\
    &  &  $[\rm arcsec]$ & \\
  \hline
  1D maj. & S-bul  & $61.5$  &  $3.3$ \\
  1D eq.  & S-bul  & $49.3$  &  $3.7$ \\
  2D      & S-bul  & $45.9$  &  $8.3$ \\
  \hline
  S+11 2D      & S-bul & $34.3$  &  $7.0$ \\
  V+12 2D      & S-bul & $16.9$  &  $5.1$ \\
  B+12 2D      & S-bul & $77.6$  &  $8.7$ \\
  \hline
  \end{tabular}
  \end{center}
  \label{tab:n5845}
  \tablecomments{The results obtained by S+11 and B+12 agree best with the results from our 2D model, 
  in which we did not mask the inner region of the galaxy. 
  The small radial extent of the data used by V+12 led them to underestimate the effective radius.}
  \end{table}

  \clearpage\newpage\noindent
  {\bf NGC 5845 \\}

  \begin{figure}[h]
  \begin{center}
  \includegraphics[width=\fitfigurewidth]{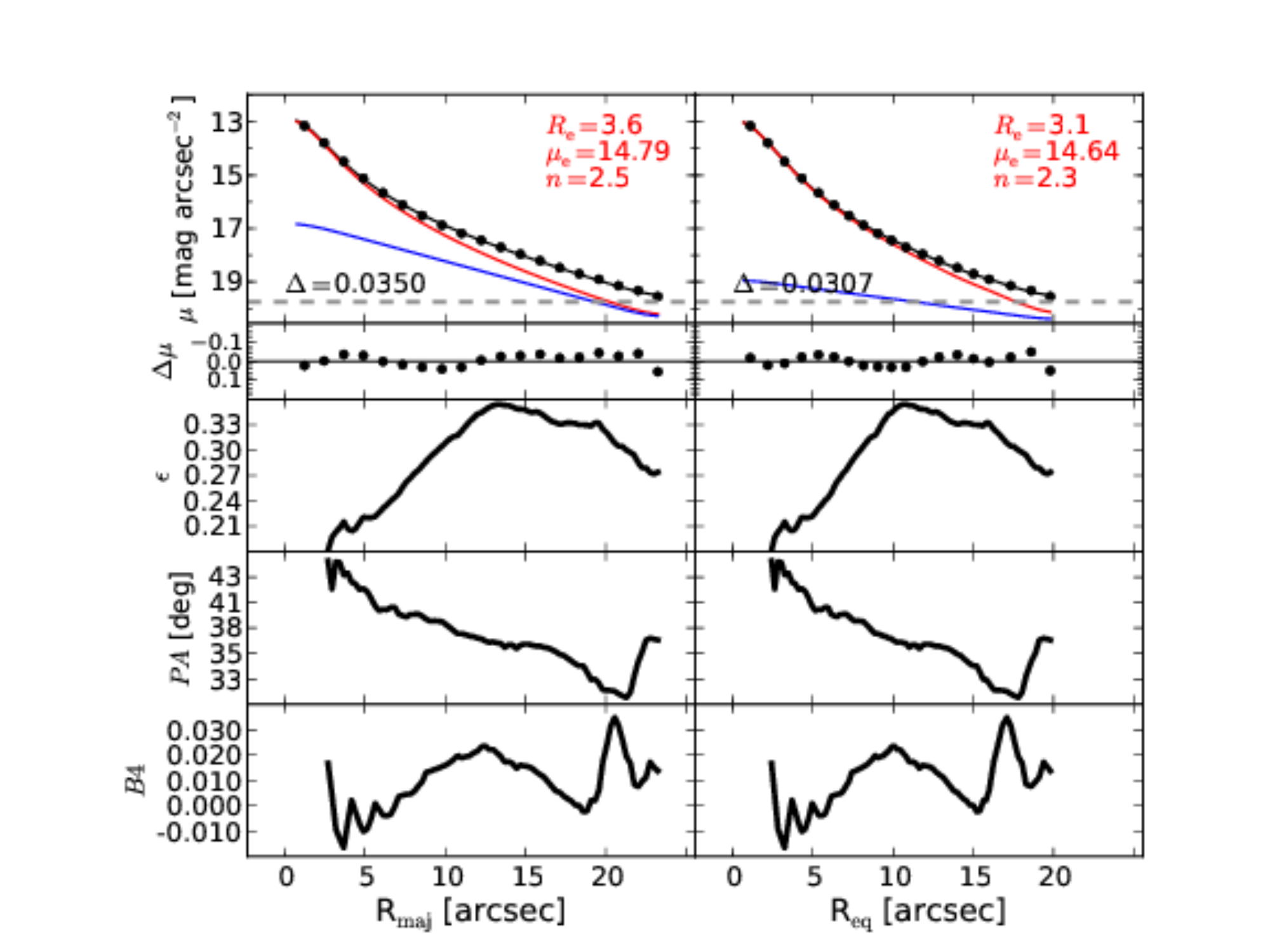}
  \caption{NGC 5845: 
  A lenticular galaxy. 
  From the unsharp mask and the velocity map (ATLAS$^{\rm 3D}$), we identify a disk, but it is not clear whether the disk is large-scale or intermediate-scale.
  The ellipticity profile has a peak at $R_{\rm maj} \sim 13''$ which suggests that the disk is indeed intermediate-scale, 
  i.e.~it does not dominate at large radii.
  }
  \end{center}
  \end{figure}

  \begin{table}[h]
  \small
  \caption{Best-fit parameters for the spheroidal component of NGC 5845.}
  \begin{center}
  \begin{tabular}{llcc}
  \hline
  {\bf Work} & {\bf Model}   & $\bm R_{\rm e,sph}$    & $\bm n_{\rm sph}$ \\
    &  &  $[\rm arcsec]$ & \\
  \hline
  1D maj. & S-bul + e-d & $3.6$  &  $2.5$ \\
  1D eq.  & S-bul + e-d & $3.1$  &  $2.3$ \\
  2D      & S-bul + e-d & $2.8$  &  $2.4$ \\
  \hline
  GD07 1D maj.      & S-bul & $4.1$  &  $3.2$ \\
  S+11 2D      & S-bul & $3.7$  &  $3.0$ \\
  V+12 2D      & S-bul & $3.5$  &  $2.6$ \\
  B+12 2D      & S-bul & $4.1$  &  $3.5$ \\
  L+14 2D      & S-bul & $3.5$  &  $2.8$ \\
  \hline
  \end{tabular}
  \end{center}
  \label{tab:n5845}
  \end{table}

  \clearpage\newpage\noindent
  {\bf NGC 5846 \\}

  \begin{figure}[h]
  \begin{center}
  \includegraphics[width=\fitfigurewidth]{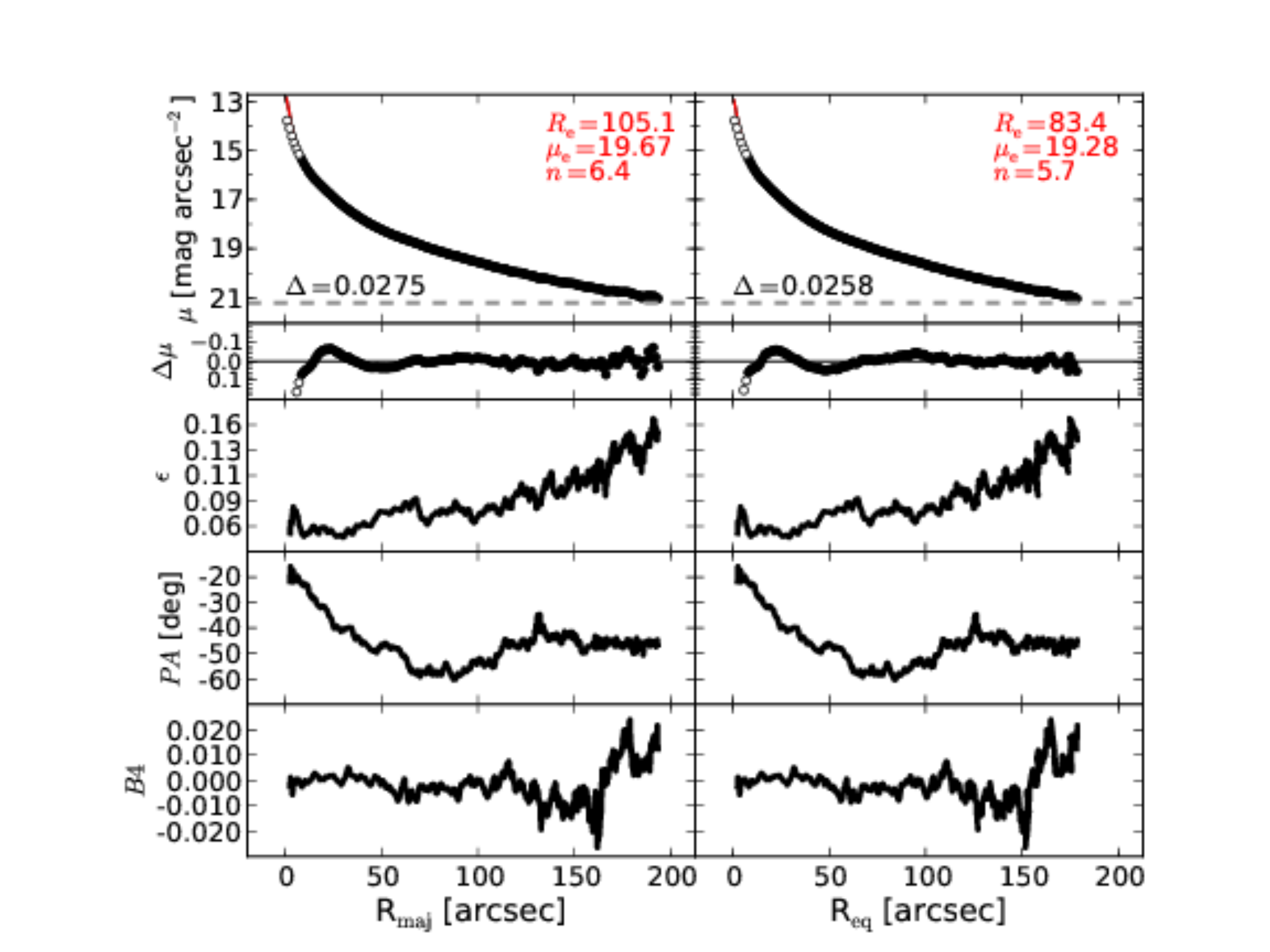}
  \caption{NGC 5846: 
  An elliptical galaxy with an unresolved partially depleted core \citep{rusli2013}. 
  This galaxy has a LINER nucleus \citep{carrillo1999} and filamentary nuclear dust \citep{tran2001}.
  It also displays a strong isophotal twist between its center and $R_{\rm maj} \sim 70''$.
  The light profile presents a slight bump at $R_{\rm maj} \sim 20''$.
  However, the isophotal parameters, the unsharp mask and the velocity map (ATLAS$^{\rm 3D}$) lack of clear evidence for an embedded component.
  We thus model NGC 5846 with a single S\'ersic profile, after masking the data within innermost $6''.1$.
  }
  \end{center}
  \end{figure}

  \begin{table}[h]
  \small
  \caption{Best-fit parameters for the spheroidal component of NGC 5846.}
  \begin{center}
  \begin{tabular}{llcc}
  \hline
  {\bf Work} & {\bf Model}   & $\bm R_{\rm e,sph}$    & $\bm n_{\rm sph}$ \\
    &  &  $[\rm arcsec]$ & \\
  \hline
  1D maj. & S-bul + m-c & $105.1$  &  $6.4$ \\
  1D eq.  & S-bul + m-c & $83.4$	&  $5.7$ \\
  2D      & S-bul + m-c & $85.1$	&  $5.2$ \\
  \hline
  S+11 2D      & S-bul	    & $36.4$   &  $3.0$ \\
  V+12 2D      & S-bul + m-c   & $46.3$   &  $3.7$ \\
  R+13 1D eq.      & core-S\'ersic & $113.2$  &  $5.3$ \\
  \hline
  \end{tabular}
  \end{center}
  \label{tab:n5846}
  \tablecomments{S+11 did not mask the core and thus underestimated the effective radius and S\'ersic index. }
  \end{table}

  \clearpage\newpage\noindent
  {\bf NGC 6251 \\}

  \begin{figure}[h]
  \begin{center}
  \includegraphics[width=\fitfigurewidth]{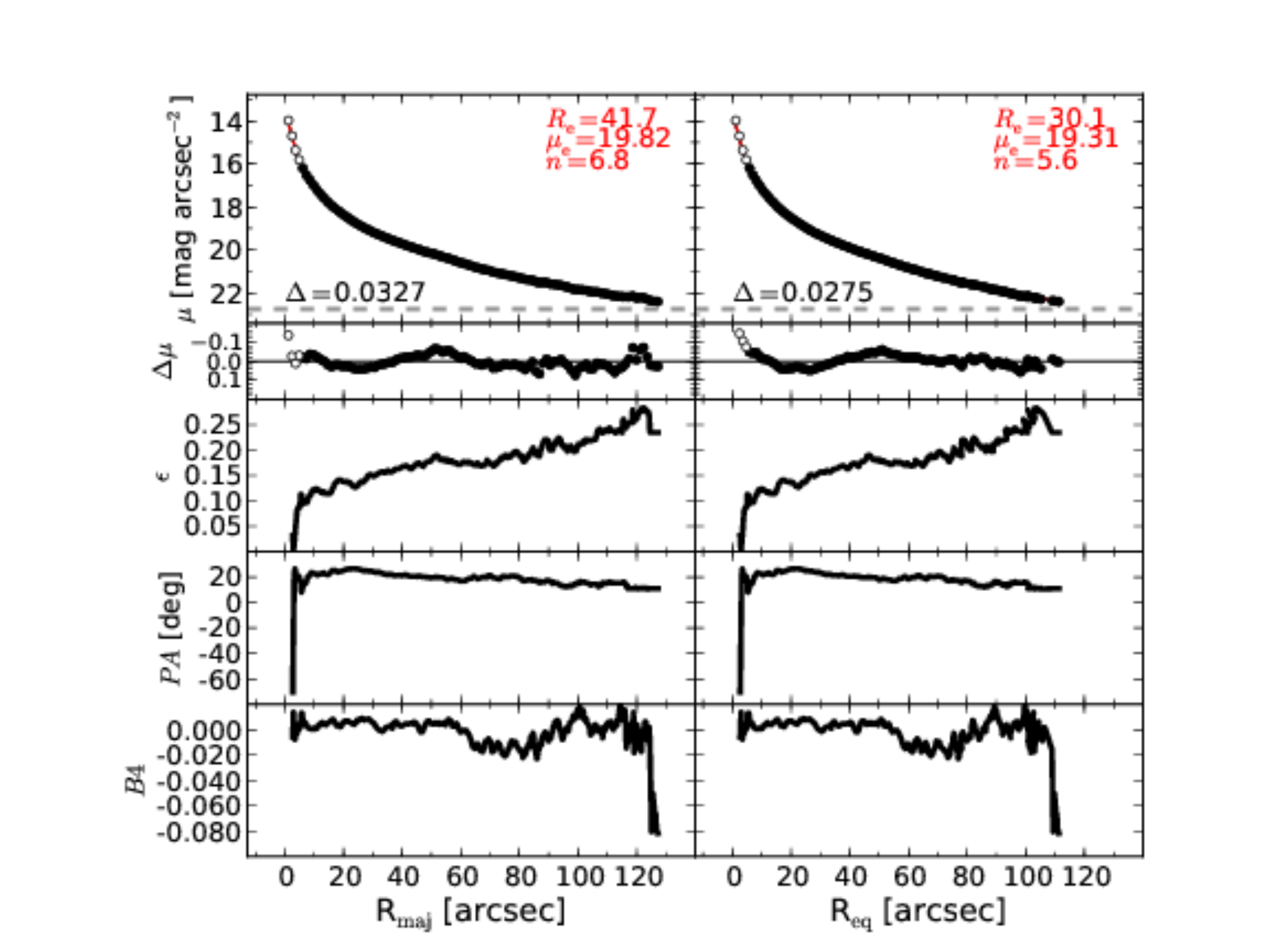}
  \caption{NGC 6251: 
  An elliptical galaxy. 
  Its large stellar velocity dispersion suggests the presence of a partially depleted core.
  The galaxy features a nuclear disk of dust \citep{ferrareseford1999n6251} and a Seyfert AGN \citep{panessabassani2002}.
  We mask the data within the innermost $6''.1$. 
  A single S\'ersic profile provides a good description of this galaxy.
  }
  \end{center}
  \end{figure}

  \begin{table}[h]
  \small
  \caption{Best-fit parameters for the spheroidal component of NGC 6251.}
  \begin{center}
  \begin{tabular}{llcc}
  \hline
  {\bf Work} & {\bf Model}   & $\bm R_{\rm e,sph}$    & $\bm n_{\rm sph}$ \\
    &  &  $[\rm arcsec]$ & \\
  \hline
  1D maj. & S-bul + m-c & $41.7$  &  $6.8$ \\
  1D eq.  & S-bul + m-c & $30.1$  &  $5.6$ \\
  2D      & S-bul + m-c & $39.3$  &  $7.1$ \\
  \hline
  GD07 1D maj.      & S-bul	     & $173.9$  &  $11.8$ \\
  S+11 2D      & S-bul + G-n    & $42.4$	&  $7.0$ \\
  L+14 2D      & S-bul	     & $20.6$	&  $5.0$ \\
  \hline
  \end{tabular}
  \end{center}
  \label{tab:n6251}
  \tablecomments{It is not clear why GD07 obtained the largest estimates of the effective radius and S\'ersic index 
  (possibly the AGN was bright in their $R$-band image and added to the central cusp). }
  \end{table}

  \clearpage\newpage\noindent
  {\bf NGC 7052 \\}

  \begin{figure}[h]
  \begin{center}
  \includegraphics[width=\fitfigurewidth]{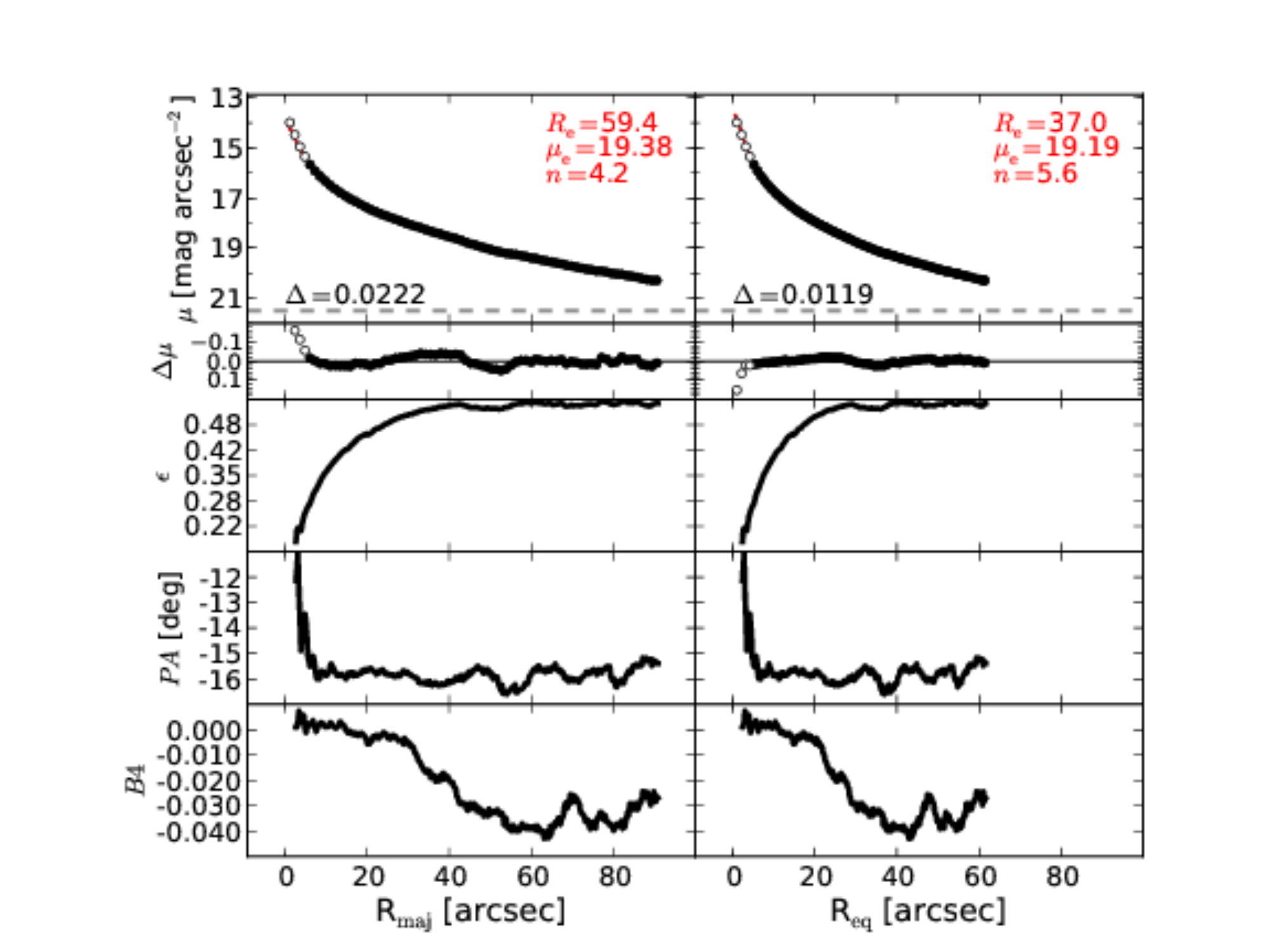}
  \caption{NGC 7052: 
  An elliptical galaxy with an unresolved partially depleted core \citep{quillen2000}. 
  We mask the data within the innermost $6''.1$ and model the galaxy with a single S\'ersic profile.
  }
  \end{center}
  \end{figure}

  \begin{table}[h]
  \small
  \caption{Best-fit parameters for the spheroidal component of NGC 7052.}
  \begin{center}
  \begin{tabular}{llcc}
  \hline
  {\bf Work} & {\bf Model}   & $\bm R_{\rm e,sph}$    & $\bm n_{\rm sph}$ \\
    &  &  $[\rm arcsec]$ & \\
  \hline
  1D maj. & S-bul + m-c & $59.4$  &  $4.2$ \\
  1D eq.  & S-bul + m-c & $37.0$  &  $5.6$ \\
  2D      & S-bul + m-c & $36.2$  &  $4.0$ \\
  \hline
  GD07 1D maj.      & S-bul & $70.4$  &  $4.6$ \\
  S+11 2D      & S-bul & $39.3$  &  $5.0$ \\
  V+12 2D      & S-bul + e-d & $4.3$  &  $1.8$ \\
  L+14 2D      & S-bul & $26.6$  &  $4.2$ \\
  \hline
  \end{tabular}
  \end{center}
  \label{tab:n7052}
  \tablecomments{The model of V+12 2D includes an artificial large-scale disk 
  and thus results in the lowest estimates of the effective radius and S\'ersic index.}
  \end{table}

  \clearpage\newpage\noindent
  {\bf NGC 7619 \\}

  \begin{figure}[h]
  \begin{center}
  \includegraphics[width=\fitfigurewidth]{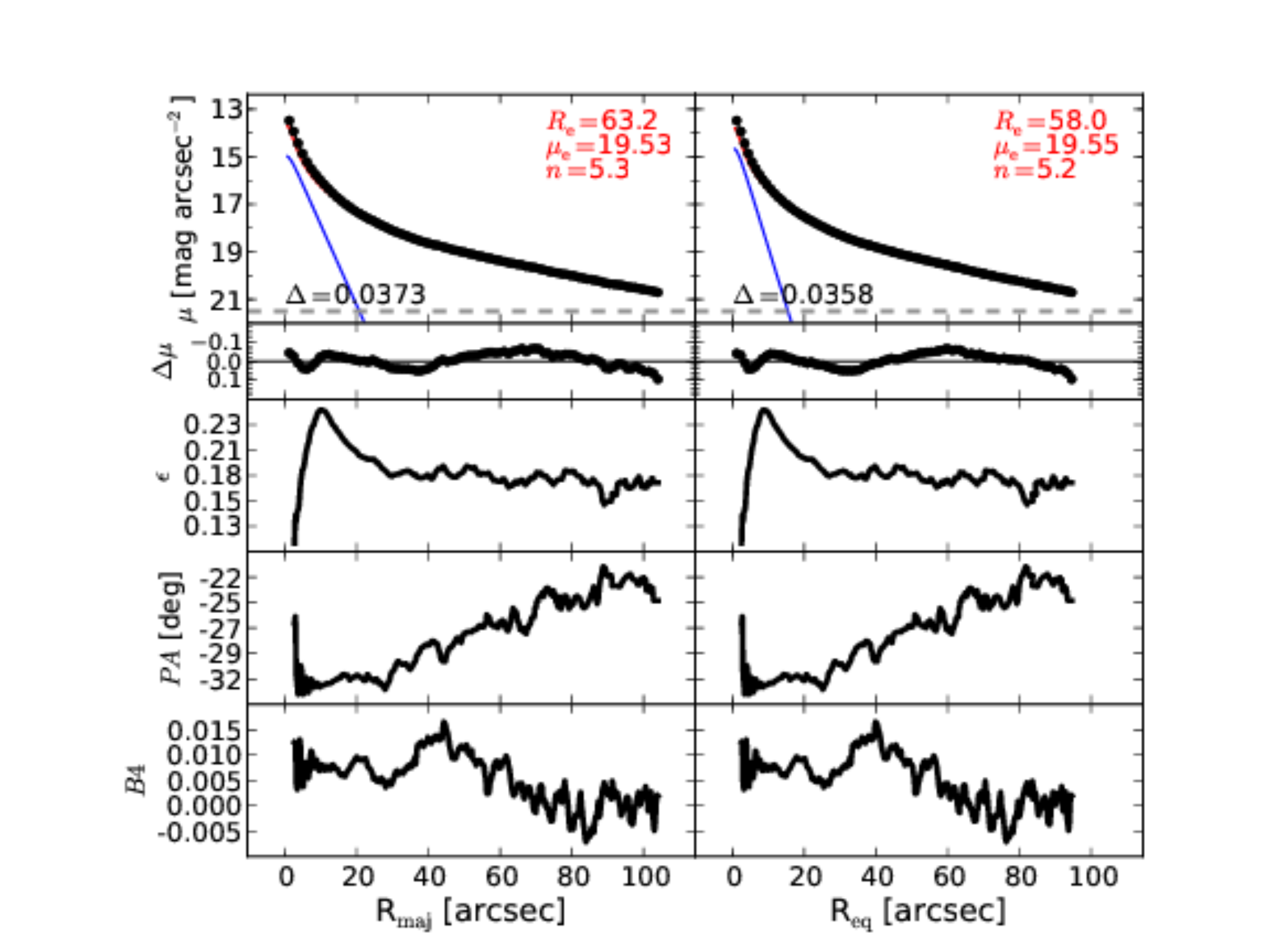}
  \caption{NGC 7619: 
  An elliptical galaxy with an unresolved partially depleted core \citep{rusli2013}. 
  We identified an embedded disk signaled by the peak at $R_{\rm maj} \sim 10''$ in the ellipticity profile.
  The velocity map of this galaxy confirms the presence of a fast rotating component (Falcon-Barroso, private comm.).
  We note that the residuals obtained from our bulge + inner-disk model do not suggest the presence of a partially depleted core.
  }
  \label{fig:n7619}
  \end{center}
  \end{figure}

  \begin{table}[h]
  \small
  \caption{Best-fit parameters for the spheroidal component of NGC 7619.}
  \begin{center}
  \begin{tabular}{llcc}
  \hline
  {\bf Work} & {\bf Model}   & $\bm R_{\rm e,sph}$    & $\bm n_{\rm sph}$ \\
    &  &  $[\rm arcsec]$ & \\
  \hline
  1D maj. & S-bul + e-id & $63.2$  &  $5.3$ \\
  1D eq.  & S-bul + e-id & $58.0$  &  $5.2$ \\
  \hline
  R+13 1D eq.      & core-S\'ersic & $100.1$  &  $9.3$ \\
  \hline
  \end{tabular}
  \end{center}
  \label{tab:n7619}
  \end{table}

  \clearpage\newpage\noindent
  {\bf NGC 7768 \\}

  \begin{figure}[h]
  \begin{center}
  \includegraphics[width=\fitfigurewidth]{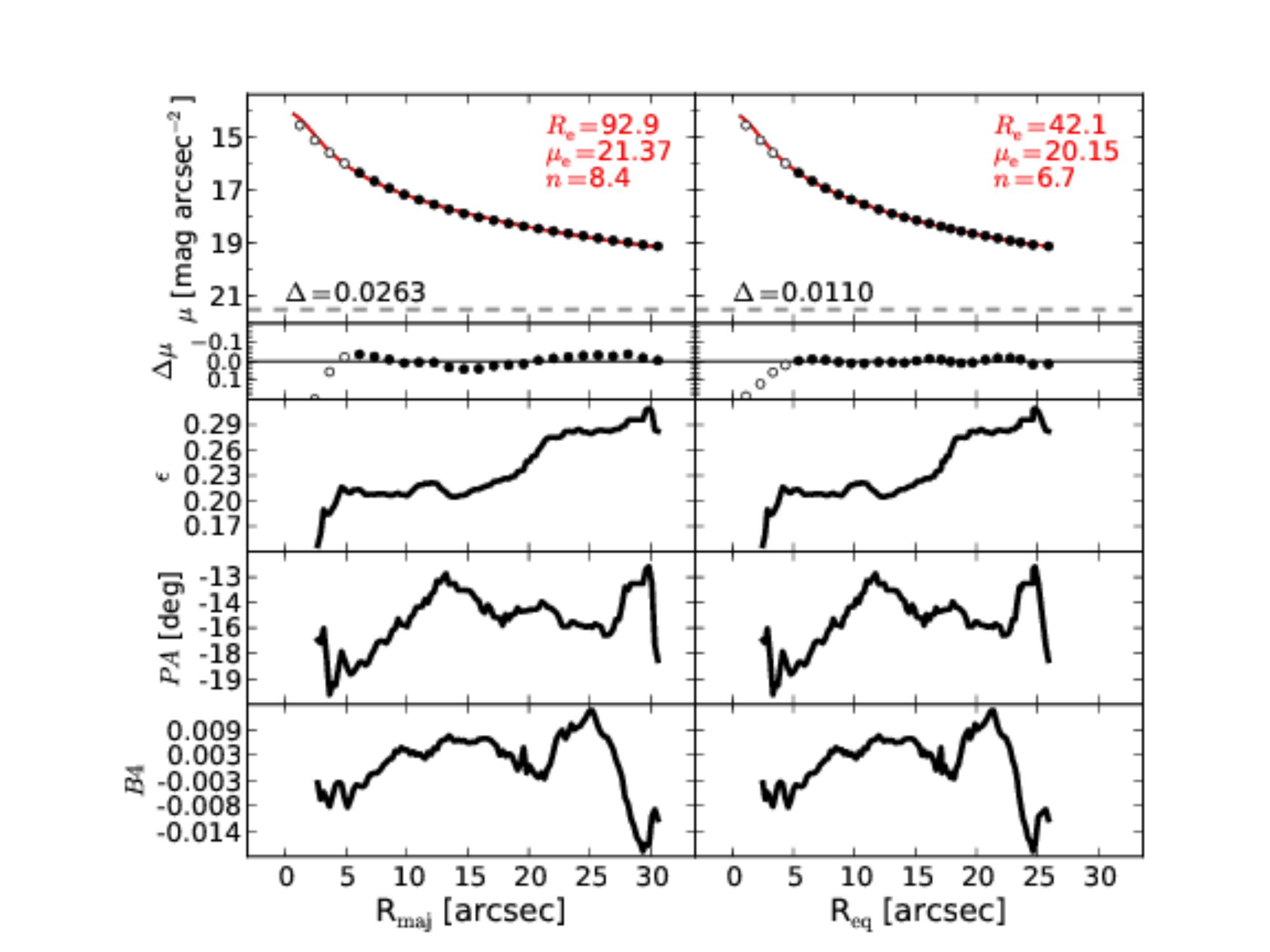}
  \caption{NGC 7768: 
  An elliptical galaxy with an unresolved partially depleted core \citep{rusli2013}. 
  The image of NGC 7768 is corrupted by a saturated star, which lies close to the galaxy.
  The sky background is not constant across the image.
  To be safe, we fit only the data within the innermost $R_{\rm maj} \lesssim 30''$, where the contribution from the background is negligible.
  The data within $R_{\rm maj} < 6''.1$ are excluded from the fit.
  }
  \end{center}
  \end{figure}

  \begin{table}[h]
  \small
  \caption{Best-fit parameters for the spheroidal component of NGC 7768.}
  \begin{center}
  \begin{tabular}{llcc}
  \hline
  {\bf Work} & {\bf Model}   & $\bm R_{\rm e,sph}$    & $\bm n_{\rm sph}$ \\
    &  &  $[\rm arcsec]$ & \\
  \hline
  1D maj. & S-bul + m-c & $92.9$  &  $8.4$ \\
  1D eq.  & S-bul + m-c & $42.1$  &  $6.7$ \\
  \hline
  R+13 1D eq.      & core-S\'ersic & $46.1$  &  $6.2$ \\
  \hline
  \end{tabular}
  \end{center}
  \label{tab:n7768}
  \end{table}
   
  \clearpage\newpage\noindent
  {\bf UGC 03789 \\}

  \begin{figure}[h]
  \begin{center}
  \includegraphics[width=\fitfigurewidth]{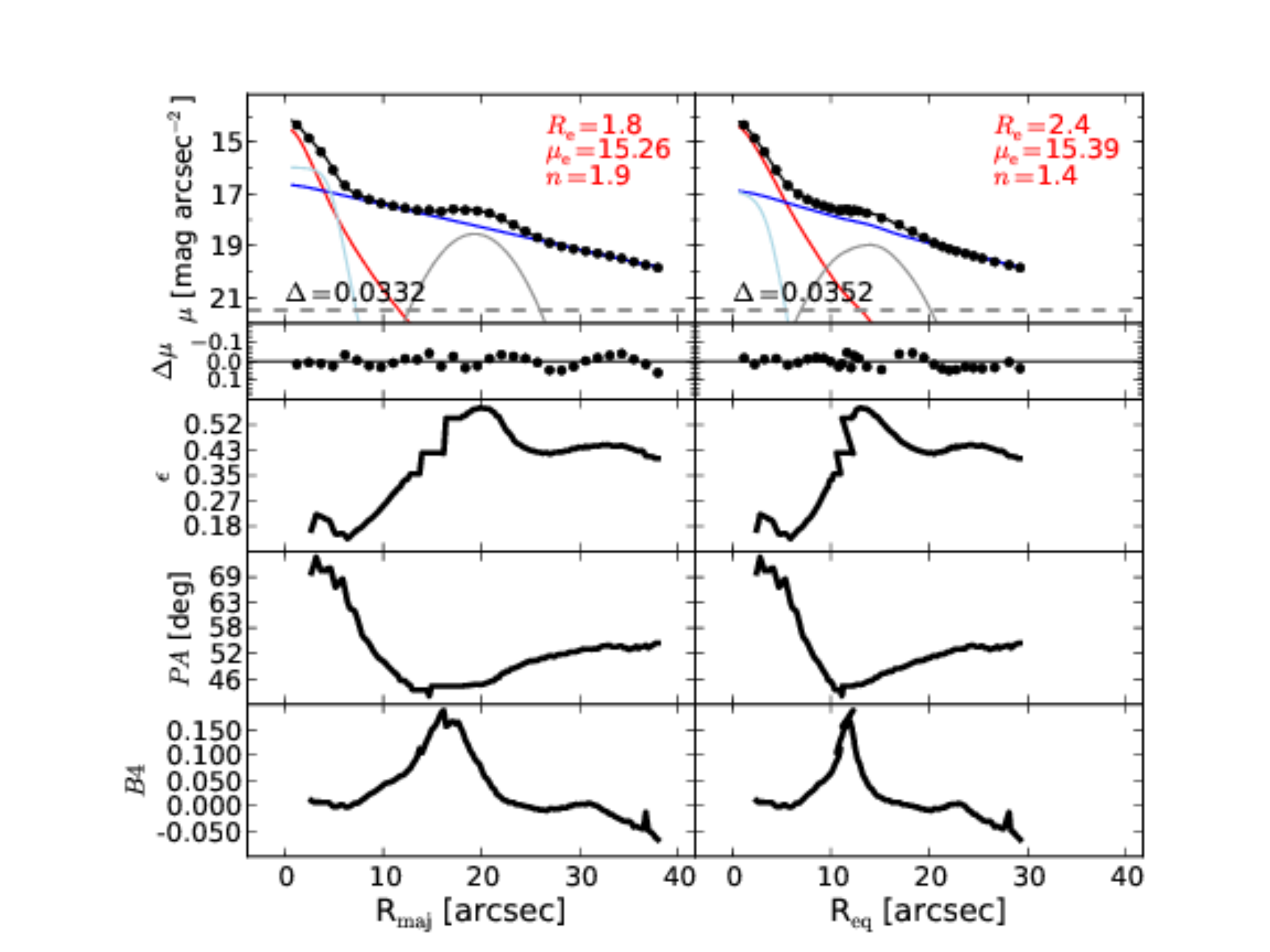}
  \caption{UGC 03789:
  A face-on spiral galaxy, 
  featuring a ring ($R_{\rm maj} \sim 20''$) and a nuclear bar ($R_{\rm maj} \lesssim 4''$),
  which can be seen in the unsharp mask and produces corresponding peaks in the ellipticity and $PA$ profiles.
  The bar is fit with a Ferrer function.
  }
  \end{center}
  \end{figure}

  \begin{table}[h]
  \small
  \caption{Best-fit parameters for the spheroidal component of UGC 03789.}
  \begin{center}
  \begin{tabular}{llcc}
  \hline
  {\bf Work} & {\bf Model}   & $\bm R_{\rm e,sph}$    & $\bm n_{\rm sph}$ \\
    &  &  $[\rm arcsec]$ & \\
  \hline
  1D maj. & S-bul + e-d + F-bar + G-r & $1.8$  &  $1.9$ \\
  1D eq.  & S-bul + e-d + F-bar + G-r & $2.4$  &  $1.4$ \\
  \hline
  \end{tabular}
  \end{center}
  \label{tab:ugc3789}
  \end{table}


\clearpage

\appendix

\section{1D analytical functions}
\label{sec:app1}
Here we provide the mathematical expressions of the analytical functions used to model 
the observed surface brightness profiles, $\mu(R)$, of galaxies.
The projected galactic radius, $R$,
corresponds to the distance of the isophotes from the galaxy center (along either the major- or equivalent-axis). \\

The \citeauthor{sersic1963} (\citeyear{sersic1963,sersic1968}) model is a three-parameter function of the following form:
\begin{equation}
\mu_{\rm S\acute{e}rsic}(\mu_{\rm e},R_{\rm e},n;R) = \mu_{\rm e} + \frac{2.5~b_{\rm n}}{\ln(10)} 
\Biggl[\biggl(\frac{R}{R_{\rm e}}\biggr)^{1/n}-1\Biggr] ,
\end{equation}
\citep{caon1993,andredakis1995,grahamdriver2005}
where $\mu_{\rm e}$ is the surface brightness at the effective radius $R_{\rm e}$ that encloses 
half of the total light from the model. 
The S\'ersic index $n$ is the parameter that measures the curvature of the radial light profile,
and $b_{\rm n}$ is a scalar value defined in terms of the S\'ersic index $n$ such that:
\begin{equation}
\Gamma(2n) = 2\gamma(2n,b_{\rm n}),
\end{equation}
where $\Gamma$ is the complete gamma function \citep{ciotti1991} 
and $\gamma$ is the incomplete gamma function defined by 
\begin{equation}
\gamma(2n,x) = \int_{0}^{x} {\rm e}^{-t} t^{2n-1} {\rm d}t .
\end{equation} \\

The exponential model is a special case ($n=1$) of the S\'ersic model.
It can therefore be written as a two-parameter function such that:
\begin{equation}
\mu_{\rm exponential}(\mu_{\rm 0},h;R) = \mu_{\rm 0} + \frac{2.5}{\ln(10)} \biggl(\frac{R}{h} \biggr) ,
\end{equation}
where $\mu_{\rm 0}$ is the central surface brightness and $h$ is the scale length equal to $R_{\rm e}/1.678$. \\

The Gaussian model is another special case ($n=0.5$) of the S\'ersic model, and thus also a two-parameter function of the following form:
\begin{equation}
\mu_{\rm Gaussian}(\mu_{\rm 0},FWHM;R) = \mu_{\rm 0} + \frac{2.5}{\ln(10)} 
\biggl[\frac{R^2}{2(FWHM/2.355)^2} \biggr] ,
\label{eq:gauss}
\end{equation}
where $\mu_{\rm 0}$ is the central surface brightness 
and $FWHM$ is the full width at half maximum of the Gaussian profile. \\

The \cite{moffat1969} model is a three-parameter function that can be expressed as:
\begin{equation}
\mu_{\rm Moffat}(\mu_{\rm 0},\alpha,\beta;R) = \mu_{\rm 0} - 2.5 
\log \Biggl[ 1 + \biggl( \frac{R}{\alpha} \biggr)^2 \Biggr]^{-\beta} ,
\end{equation}
where $\mu_{\rm 0}$ is the central surface brightness,
$\alpha$ is related to the $FWHM$ through 
\begin{equation}
FWHM = 2\alpha \sqrt{2^{1/\beta}-1} ,
\end{equation}
and $\beta$ regulates the shape of the profile at large radii. \\

The Ferrer model is a four-parameter function defined as:
\begin{equation}
\mu_{\rm Ferrer}(\mu_{\rm 0},R_{\rm out},\alpha,\beta;R) = \left\{
  \begin{array}{l l}
    \mu_{\rm 0} - 2.5 \log \Biggl[ 1 - \biggl(\frac{R}{R_{\rm out}} \biggr)^{2-\beta} \Biggr]^{\alpha} 
    & \quad \text{for $R<R_{\rm out}$}\\
    +\infty & \quad \text{for $R\geq R_{\rm out}$}
  \end{array} \right. ,
\end{equation}
where $\mu_{\rm 0}$ is the central surface brightness,
$\alpha$ controls the sharpness of the truncation,
$\beta$ is related to the central slope,
and $R_{\rm out}$ is the outer radial limit within which the function is defined. \\

The symmetric Gaussian ring is a three-parameter function of the following form:
\begin{equation}
\mu_{\rm Gaussian}(\mu_{\rm 0},R_{\rm 0},FWHM;R) = \mu_{\rm 0} + \frac{2.5}{\ln(10)} 
\biggl[\frac{(R-R_{\rm 0})^2}{2(FWHM/2.355)^2} \biggr] ,
\end{equation}
where $\mu_{\rm 0}$ and $FWHM$ have the same meaning as in equation \ref{eq:gauss}, 
and $R_{\rm 0}$ is the radius at which the Gaussian profile is centered. \\

\bibliography{SMBHbibliography}

\clearpage

\end{document}